\documentclass[12pt,american]{article}

\usepackage{cancel}
\usepackage{geometry}
\usepackage[hyperlink]{}

\usepackage{hyperref}
\hypersetup{
     colorlinks   = true,
     linkcolor    = blue,
     urlcolor     = blue,
     citecolor    = blue
}

\usepackage{graphicx}
\usepackage{bm}
\usepackage{dcolumn}
\usepackage{amssymb}
\usepackage{amsmath}
\usepackage{slashed}
\usepackage{amsfonts}
\usepackage{epsfig}
\usepackage{stackrel}
\usepackage{paralist}
\usepackage[rgb]{xcolor}
\usepackage{pdfcomment}
\usepackage{epstopdf}

\usepackage{units}
\usepackage{amsthm}

\usepackage{amsmath}
\numberwithin{equation}{section}

\usepackage{amssymb}

\usepackage{centernot}

\usepackage{multirow}

\usepackage[caption=false]{subfig}

\usepackage[numbers,compress]{natbib}
\usepackage{setspace}
\usepackage[toc,page,header]{appendix} 
\usepackage{minitoc} 

\begin{document}

\global\long\def\ket#1{\left|#1\right\rangle }%
\global\long\def\bra#1{\left\langle #1\right|}%
\global\long\def\braket#1#2{\left\langle #1\right|\left.#2\right\rangle }%
\global\long\def\slashed#1{\not\mathrlap{#1}}%

\setcounter{secnumdepth}{4} 

\begin{centering}
\includegraphics[width=0.91\columnwidth]{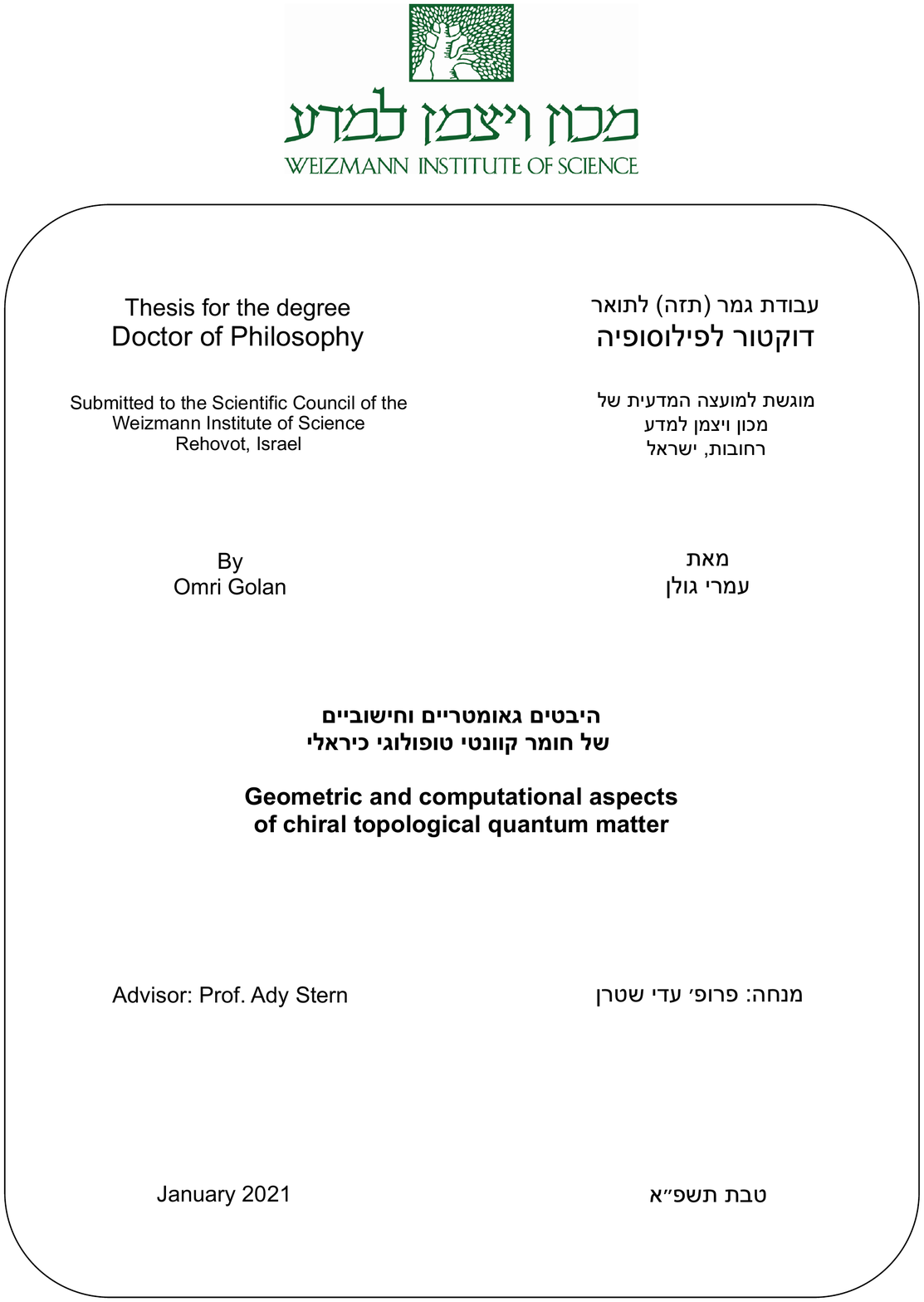}
\par\end{centering}

\thispagestyle{empty}

\pagebreak{}

\begin{abstract}
    
In this thesis, we study chiral topological phases of 2+1 dimensional quantum matter. Such phases are abstractly characterized
by their non-vanishing chiral central charge $c$, a topological invariant which appears as the  coefficient of a gravitational Chern-Simons (gCS) action in bulk, and of corresponding gravitational anomalies at boundaries. The chiral
central charge is of particular importance in chiral superfluids and
superconductors (CSF/Cs), where $U\left(1\right)$ particle-number symmetry
is broken, and $c$ is, in some cases,
the only topological invariant characterizing the system.
However, as opposed to invariants which can be probed by gauge fields in place of gravity, the concrete physical
implications of $c$ in the context
of condensed matter physics is quite subtle, and has been the subject
of ongoing research and controversy. The first two parts of this thesis are devoted to the  physical interpretation of
the gCS action and gravitational anomalies in the context of 
CSF/Cs, where they are of particular importance, but
have nevertheless remained poorly understood. 

 As a first approach, we demonstrate that, at low energy, fermionic excitations in $p$-wave CSF/Cs
     experience an \textit{emergent} relativistic Riemann-Cartan geometry, described by the superconducting order parameter and background $U(1)$ gauge field.  This description is then used to infer gCS energy-momentum responses to a space-time dependent order parameter, and relate these to an order parameter induced gravitational anomaly.    The presence of torsion in the emergent geometry, as well as the  multiplicity of low energy spinors in the case of lattice systems, lead to additional effects which are not of topological origin, but nevertheless mimic closely the above gCS physics. We show how these different phenomena can be disentangled.  
     
     We then take on a fully non-relativistic analysis of CSFs, obtaining a low energy effective field theory that consistently captures
both their chiral Goldstone mode and their non-relativistic gCS action. Using the theory  we find that
$c$ cannot be extracted from a measurement of the odd viscosity tensor alone, despite naive expectation based on previous work. Nevertheless, a related observable,
termed `improved odd viscosity', does allow for the bulk measurement
of $c$. Additional results of the same spirit are found in Galilean
invariant CSFs.  

Finally, we turn to a seemingly unrelated aspect of chiral topological phases  -  their computational complexity. The  infamous \textit{sign problem} leads  to  an  exponential  complexity  in  Monte  Carlo  simulations  of generic many-body quantum systems. Nevertheless, many phases of matter are known to admit a sign-problem-free representative, allowing an efficient classical simulation. Motivated by long standing open problems in many-body  physics, as well as fundamental questions in quantum complexity, the possibility of \textit{intrinsic} sign problems, where a phase of matter admits no sign-problem-free representative, was recently raised but remains largely unexplored. Here, we establish the existence of an intrinsic sign problem in
a broad class of  chiral topological phases, defined by the requirement that $e^{2\pi i c/24}$ is \textit{not} the topological spin of an anyon. Within
this class, we exclude the possibility of `stoquastic' Hamiltonians
for bosons, and of sign-problem-free determinantal Monte
Carlo algorithms for fermions. We obtain analogous results for phases
that are \textit{spontaneously} chiral, and present evidence for an
extension of  our results that applies to both chiral and non-chiral
topological matter. 

\end{abstract}
 
\thispagestyle{empty}

\pagebreak{}

\thispagestyle{empty}

\setcounter{tocdepth}{0} 

\doparttoc 
\faketableofcontents 

\part{} 

\parttoc 


\thispagestyle{empty}

\pagebreak{}

\clearpage

\addcontentsline{toc}{section}{Acknowledgments} 
\section*{Acknowledgments}

Adding a grain of sand to one of the very many and ever growing summits of the mountain range known as `human knowledge' is truly a great privilege. The climb, however, is usually no easy task, and the one that produced this thesis was no exception. Here, I would like to pause and thank those who  supported me en route.

First, I am grateful to my advisor Ady Stern, who suggested an unorthodox but tailor made research direction, and collaborated with me on the first and most challenging part of the work. Ady is a formidable physicist as well as a generous advisor, and provided a rare combination of support and freedom in my research. It was also wonderful to work under someone who views laughter as a way of life. 

Subsequent work was performed in mostly long distance but close collaborations with Sergej Moroz, Carlos Hoyos, Félix Rose, Zohar Ringel and Adam Smith. In particular, Sergej, Carlos and Zohar served as additional mentors, each with his own unique style of doing research. I also benefited greatly from extensive discussions with Paul Wiegmann, Andrey Gromov, Ryan Thorngren, Weihan Hsiao, Semyon Klevtsov, Barry Bradlyn and Thomas Kvorning. 

The members of my PhD advisory committee, Micha Berkooz and David Mross, as well as my senior group members, Yuval Baum, Ion Cosma Fulga, Jinhong Park, Raquel Queiroz and Tobias Holder, provided much needed physical and meta-physical advice along the way.
Our administrative staff, Hava Shapira, Merav Laniado, Einav Yaish, Inna Dombrovsky, Yuval Toledo and Yuri Magidov, sustained an incredibly efficient and warm work environment. I also thank my fellow graduate students Eyal Leviatan, Ori Katz, Dan Klein, Avraham Moriel, Shaked Rozen, Asaf Miron, Yotam Shpira, Adar Sharon, Dan Dviri and, of course,  Yuval Rosenberg, who dragged me to the Weizmann institute when we were kids, and got me hooked on physics. 

Zooming out, I am grateful to my parents Sharona and Gabi, for their continued support in whatever I choose to do, and to my wife and best friend Adi, for making my life happy and balanced. Since we became parents, my work would not have been possible without Adi's backing, in particular since the spreading of Coronavirus, which eliminated some of our support systems, as well as the distinction between work and home. Finally, I thank our boys Adam and Shlomi for their smiles, laughter, and curiosity - a reminder of why I was drawn to science in the first place.

\pagebreak{}

\clearpage

\addcontentsline{toc}{section}{Publications} 
\section*{Publications \label{sec:List-of-publications}}

This thesis is based on the following publications: 
\begin{itemize}
\item Reference \citep{PhysRevB.98.064503}: Omri Golan and Ady Stern. Probing
topological superconductors with emergent gravity. \href{https://journals.aps.org/prb/abstract/10.1103/PhysRevB.98.064503}{Phys. Rev. B, 98:064503},
2018.
\item Reference \citep{PhysRevB.100.104512}: Omri Golan, Carlos Hoyos,
and Sergej Moroz. Boundary central charge from bulk odd viscosity:
Chiral superfluids. \href{https://journals.aps.org/prb/abstract/10.1103/PhysRevB.100.104512}{Phys. Rev. B, 100:104512},
2019.
\item Reference \citep{PhysRevResearch.2.043032}: Omri Golan, Adam Smith,
and Zohar Ringel. Intrinsic sign problem in fermionic and bosonic
chiral topological matter. \href{https://journals.aps.org/prresearch/abstract/10.1103/PhysRevResearch.2.043032}{Phys. Rev. Research, 2:043032},
2020. 
\end{itemize}
Complementary results are obtained in:
\begin{itemize}
\item Reference \citep{10.21468/SciPostPhys.9.1.006}: Félix Rose, Omri
Golan, and Sergej Moroz. Hall viscosity and conductivity of two-dimensional
chiral superconductors. \href{https://scipost.org/10.21468/SciPostPhys.9.1.006}{SciPost Phys., 9:6},
2020.
\item Reference \citep{PhysRevResearch.2.033515}: Adam Smith, Omri Golan,
and Zohar Ringel. Intrinsic sign problems in topological quantum field
theories. \href{https://journals.aps.org/prresearch/abstract/10.1103/PhysRevResearch.2.033515}{Phys. Rev. Research, 2:033515},
2020. 
\end{itemize}

\thispagestyle{empty}

\pagebreak{}

\pagenumbering{arabic} 

\section{Introduction and summary\label{sec:Introduction}}

\subsection{Overview}

The study of \textit{topological phases of matter} began in 1980,
when the Hall conductivity in a two-dimensional electron gas was measured
to be an integer multiple of $e^{2}/h$, to within a relative error
of $10^{-7}$ \citep{RevModPhys.58.519}, subsequently reduced below $10^{-10}$ \cite{vonKlitzing2017}. Following this discovery,
it was theoretically understood that in many-body quantum systems,
certain physical observables must be \textit{precisely} quantized,
under the right circumstances \citep{avron1983homotopy,thouless1982quantized}.
Around the same time, quantum field theorists extensively studied
the phenomena of \textit{anomalies} \citep{alvarez1984gravitational,alvarez1985anomalies,bertlmann2000anomalies},
where classical symmetries and conservation laws are quantum mechanically
violated, and discovered the seemingly exotic \textit{anomaly inflow
mechanism} \citep{callan1985anomalies,naculich1988axionic}, which
physically interprets anomalies in terms of \textit{topological effective
actions} in higher space-time dimensions. It was only later understood
that topological effective actions and anomalies actually capture
the essential physics of topological phases of matter, and even classify
them \citep{read2000paired,wen2013classifying,ryu2012electromagnetic,RevModPhys.88.035001,freed2016reflection,PhysRevB.98.035151}.

In particular, 2+1D gapped chiral topological phases are characterized
by a gravitational Chern-Simons (gCS) action \citep{Chern-Simons,jackiw2003chern,kraus2006holographic,witten2007three,stone2012gravitational} and corresponding 1+1D
gravitational anomalies \citep{alvarez1984gravitational,bertlmann2000anomalies,bastianelli2006path}, having the chiral central charge $c$ as
a precisely quantized coefficient, or topological invariant. The chiral
central charge is of particular importance in chiral superfluids and
superconductors \citep{read2000paired,volovik2009universe}, where $U\left(1\right)$ particle-number symmetry
is broken spontaneously or explicitly, and $c$ is, in some cases,
the only topological invariant characterizing the system at low energy.
However, as opposed to topological invariants related to gauge fields
for internal symmetries in place of gravity, the concrete physical
implications of $c$ (and even its very definition) in the context
of condensed matter physics is quite subtle, and has been the subject
of ongoing research and controversy \citep{volovik1990gravitational,read2000paired,haldane2009hall,haldane2011geometrical,wang2011topological,qin2011energy,ryu2012electromagnetic,you2014theory,abanov2014electromagnetic,gromov2014density,shitade2014bulk,shitade2014heat,PhysRevB.90.045123,gromov2015framing,gromov2015thermal,bradlyn2015low,bradlyn2015topological,klevtsov2015geometric,gromov2016boundary,gromov2017bimetric,gromov2017investigating,klevtsov2017laughlin,klevtsov2017lowest,klevtsov2017quantum,nakai2017laughlin,wiegmann2018inner,Cappelli_2018,schine2018measuring,kapustin2019thermal,hu2020microscopic}.
The first goal of  this thesis is to physically
interpret the chiral central charge in the context of chiral superfluids
and superconductors, where it is of particular importance, but has
nevertheless remained poorly understood. This goal is pursued in Sec.\ref{sec:Main-section-1:}-\ref{sec:Main-section-2:}.

A seemingly unrelated aspect of chiral topological phases is the complexity
of simulating them on classical (as opposed to quantum) computers.
 It is generally believed that chiral topological matter is 'hard'
to simulate efficiently with classical resources. Concretely, it is
known that chiral topological phases do not admit local commuting
projector Hamiltonians \citep{PhysRevB.89.195130,potter2015protection,PhysRevB.98.165104,PhysRevB.97.245144,kapustin2019thermal},
nor do they admit local Hamiltonians with a PEPS state as an exact
ground state \citep{PhysRevLett.111.236805,PhysRevB.90.115133,PhysRevB.92.205307,PhysRevB.98.184409}.
We will be interested in quantum Monte Carlo (QMC) simulations, arguably
the most important tools in computational many-body quantum physics
\citep{RevModPhys.67.279,Assaad,PhysRevX.3.031010,PhysRevD.88.021701,kaul2013bridging,GazitE6987,berg2019monte,li2019sign},
and in the infamous \textit{sign problem}, which is the generic obstruction
to an efficient QMC simulation \citep{troyer2005computational,marvian2018computational,klassen2019hardness}.

The accumulated experience of QMC practitioners suggests that the
sign problem was never solved in chiral topological matter. Since
these phases are abstractly defined by their non-vanishing chiral
central charge, one may suspect that the chiral central charge and
related gravitational phenomena pose an obstruction to sign-problem-free
QMC.  Such an obstruction is termed \textit{intrinsic sign problem}
\citep{hastings2016quantum,ringel2017quantized}, and is of interest
beyond the context of chiral topological matter, as it is widely accepted
that long-standing open problems in many-body quantum physics, such
as the nature of high-temperature superconductivity \citep{Santos_2003,PhysRevB.80.075116,PhysRevX.5.041041,kantian2019understanding},
dense nuclear matter \citep{Hands_2000,PhysRevD.66.074507,10.1093/ptep/ptx018},
and the fractional quantum Hall state at filling $5/2$ \citep{banerjee2018observation,PhysRevB.98.045112,PhysRevLett.121.026801,PhysRevB.97.121406,PhysRevB.99.085309,hu2020microscopic},
remain open because no solution to the sign problem in a relevant
model has thus far been found. Since the aforementioned open problems
are all fermionic, we are particularly motivated to study the possibility
of intrinsic sign problems in fermionic matter. The second goal of
 this thesis, pursued in Sec.\ref{sec:Main-section-3:},
is to establish the existence of an intrinsic sign problem in chiral
topological phases of matter, based on their non-vanishing chiral
central charge, and with an emphasis on fermionic systems. 

The following Sec.\ref{subsec:Chiral-topological-matter}-\ref{subsec:Complexity-of-simulating}
introduce the central concepts described above in more detail, pose
the main questions we address in this thesis, and summarize the answers
we find.

\subsection{Chiral topological matter \label{subsec:Chiral-topological-matter}}

A gapped local many-body quantum Hamiltonian is said to be in a topological
phase of matter if it cannot be deformed to a trivial reference Hamiltonian,
without closing the energy gap or violating locality. If a symmetry
is enforced, only symmetric deformations are considered, and it is
additionally required that the symmetry is not spontaneously broken
\citep{wang2017symmetric,zeng2019quantum}. For Hamiltonians defined
on a lattice, a natural trivial Hamiltonian is given by the atomic
limit of decoupled lattice sites, where the symmetry acts independently
on each site. In this thesis we consider both lattice and continuum
models.

Topological phases with a unique ground state on the 2-dimensional
torus exist only with a prescribed symmetry group\footnote{A subtle point is that the minimal symmetry group for fermionic systems is fermion parity - the $\mathbb{Z}_2$ group generated by $(-1)^N$, where $N$ is the fermion number. This should be contrasted with bosonic systems, which may have no symmetries.}
and are termed symmetry protected topological phases (SPTs) \citep{chen2011symmetry,Kapustin_2015,Kapustin_2017}. 
When such phases are placed on the cylinder, they support anomalous
boundary degrees of freedom which cannot be realized on isolated 1-dimensional
spatial manifolds, as well as corresponding quantized bulk response
coefficients. Notable examples are the integer quantum Hall states,
topological insulators, and topological superconductors \citep{qi2011topological}.
Topological superconductivity and superfluidity will be discussed
in detail in Sec.\ref{subsec:Chiral-superfluids-and}.

Topological phases with a degenerate  ground state subspace  on
the torus are termed topologically ordered, or symmetry enriched if
a symmetry is enforced \citep{doi:10.1142/S0217979290000139,PhysRevB.82.155138}.
Beyond the phenomena exhibited by SPTs, these support localized quasiparticle
excitations with anyonic statistics and fractional charge under the
symmetry group. Notable examples are fractional quantum Hall states
\citep{nayak2008non,PhysRevLett.110.067208}, quantum spin liquids
\citep{Savary_2016}, and fractional topological insulators \citep{PhysRevLett.103.196803,doi:10.1146/annurev-conmatphys-031115-011559}.

\begin{figure}[!th]
\begin{centering}
\includegraphics[width=0.6\columnwidth]{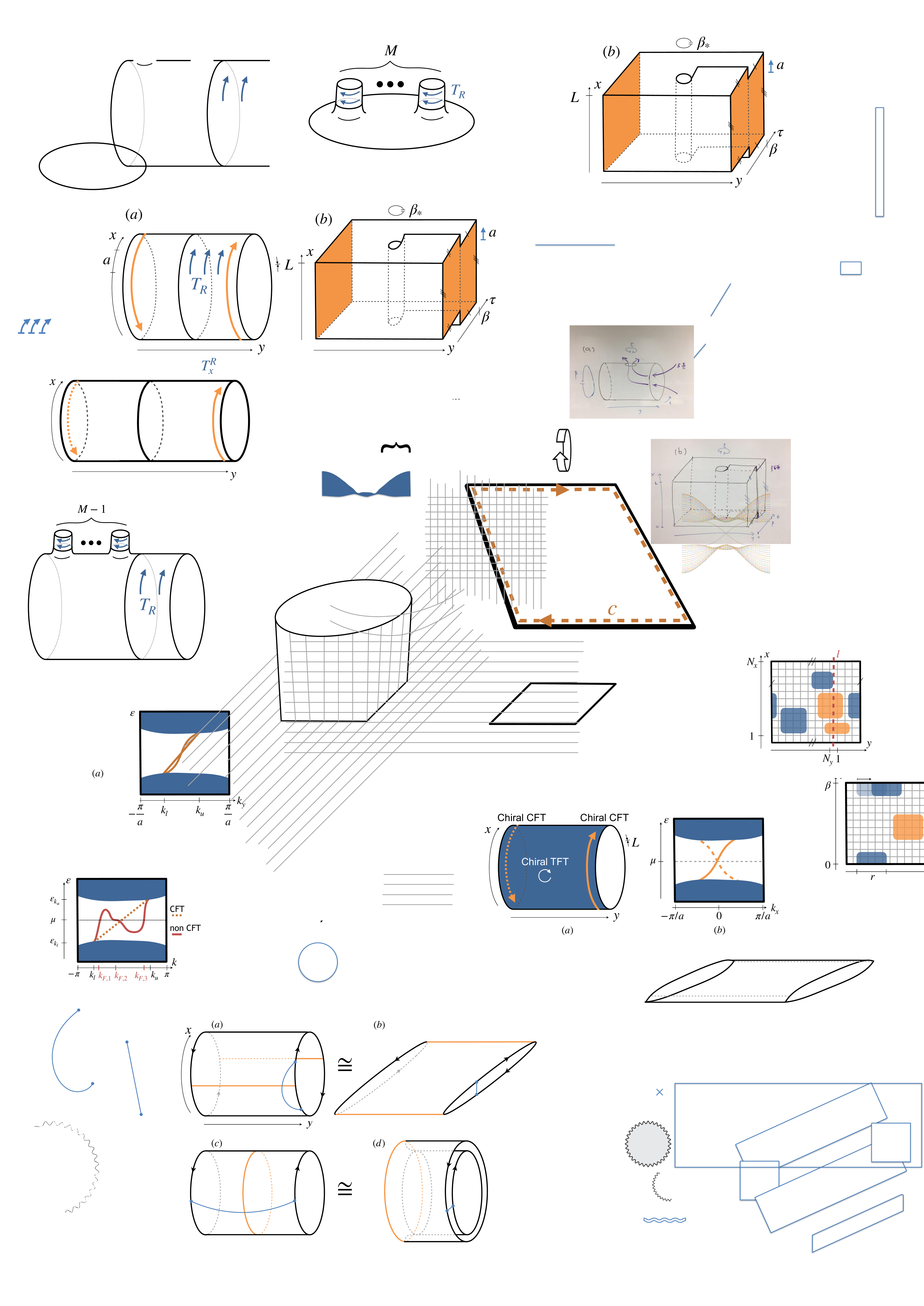}
\par\end{centering}
\caption{Chiral topological phases of matter on the cylinder. (a) The low energy
description of a chiral topological phase is comprised of two, counter
propagating, chiral conformal field theories (CFTs) on the boundary,
and a chiral topological field theory (TFT) in the bulk.   (b)
Examples: schematic single-particle spectrum of a Chern insulator
and of the Majorana fermions describing a Kitaev spin liquid. Assuming
discrete translational symmetry with spacing $a$ in the $x$ direction,
one can plot the single-particle eigen-energies $\varepsilon$ on
the cylinder as a function of (quasi) momentum $k_{x}$.  This reveals
an integer number of chiral dispersion branches whose eigen-states
are supported on one of the two boundary components. In the Chern
insulator (Kitaev spin liquid) these correspond to the Weyl (Majorana-Weyl)
fermion CFT, with $c=\pm1$ ($c=\pm1/2$) per branch. The velocity,
$v=\left|\partial\varepsilon/\partial k_{x}\right|$ at the chemical
potential $\mu$, is a non-universal parameter. \label{fig:Chiral-topological-phases}
}
\end{figure}

In this thesis, we consider \textit{chiral} topological phases, where
the boundary degrees of freedom that appear on the cylinder propagate
unidirectionally. At energies small compared with the bulk gap, the
boundary can be described by a chiral conformal field theory (CFT)
\citep{ginsparg1988applied,di1996conformal}, while the bulk reduces
to a chiral topological field theory (TFT) \citep{kitaev2006anyons,freed2016reflection},
see Fig.\ref{fig:Chiral-topological-phases}(a). Such phases may be
bosonic and fermionic, and may be protected or enriched by an on-site
symmetry, but we will not make use of this symmetry in our analysis
- only the chirality of the phase will be used.

A notable example for chiral topological phases is given by Chern
insulators \citep{PhysRevLett.61.2015,qi2008topological,ryu2010topological}:
SPTs protected by the $U\left(1\right)$ fermion number symmetry,
which admit free-fermion Hamiltonians. The single particle spectrum
of a Chern insulator on the cylinder is depicted in Fig.\ref{fig:Chiral-topological-phases}(b).
Another notable example are the topologically ordered Kitaev spin
liquids \citep{kitaev2006anyons,takagi2019concept}, which can be
described by Majorana fermions with a single particle spectrum similar
to Fig.\ref{fig:Chiral-topological-phases}(b), coupled to a $\mathbb{Z}_{2}$
(fermion-parity) gauge field.  Note that the velocity $v$ of the
boundary CFT is a non-universal parameter which generically changes
as the microscopic Hamiltonian is deformed.  More generally, different
chiral branches may have different velocities.

The chirality of the boundary CFT and bulk TFT is manifested by their
non-vanishing chiral central charge $c$, which is rational and \textit{universal}
- it is a topological invariant with respect to continuous deformations
of the Hamiltonian which preserve locality and the bulk energy gap,
and therefore constant throughout a topological phase \citep{Witten_1989,kitaev2006anyons,gromov2015framing,bradlyn2015topological}.
On the boundary $c$ is defined with respect to an orientation of
the cylinder, so the two boundary components have opposite chiral
central charges. Since, as described below, $c$ is much better understood
from the boundary perspective, we sometimes refer to it as the \textit{boundary}
chiral central charge. A main theme of this thesis is the study of
$c$ from the \textit{bulk} perspective, and the relation between
the two perspectives implied by the anomaly inflow mechanism.

\subsection{Geometric physics in chiral topological matter\label{subsec:Geometric-physics-in}}

The non-vanishing of $c$ implies a number of geometric, or 'gravitational',
physical phenomena \citep{ginsparg1988applied,di1996conformal,abanov2014electromagnetic,klevtsov2015geometric,bradlyn2015topological,gromov2016boundary}.
In particular, the boundary supports a non-vanishing energy current
$J_{E}$, which receives a correction 
\begin{align}
J_{E}\left(T\right) & =J_{E}\left(0\right)+2\pi T^{2}\frac{c}{24},\label{eq:1-2}
\end{align}
\textcolor{black}{at a temperature $T>0$, and in the thermodynamic
limit $L=\infty$, where $L$ is the circumference of the cylinder.
Note that we set $K_{\text{B}}=1$ and $\hbar=1$ throughout. }Within
CFT, this correction is universal since it is independent of $v$.
Taking the two counter propagating boundary components of the cylinder
into account, and placing these at slightly different temperatures,
leads to a thermal Hall conductance $K_{H}=c\pi T/6$ \citep{kane1997quantized,read2000paired,cappelli2002thermal},
a prediction that recently led to the first measurements of $c$ 
\citep{jezouin2013quantum,banerjee2017observed,banerjee2018observation,Kasahara:2018aa}. 

In analogy with Eq.\eqref{eq:1-2}, the boundary of a chiral topological
phase also supports a non-vanishing ground state (or $T=0$) momentum
density $p\left(L\right)$, which receives a universal correction
on a cylinder with finite circumference $L<\infty$. The details
of this finite-size correction will be described in Sec.\ref{sec:Main-section-3:},
where it is used to relate the chiral central charge (as well as the
topological spins of anyon excitations) to the complexity of simulating
chiral topological matter on classical computers. 

Abstractly, both $T>0$ and $L<\infty$ corrections described above
follow directly from the (chiral) Virasoro anomaly, or Virasoro central
extension, which defines $c$ in 2D CFT. Equivalently, these corrections
can be understood in terms of the 'global' gravitational anomaly -
the complex phase accumulated by a CFT partition function on the torus
under a Dehn twist \citep{ginsparg1988applied,di1996conformal}. This
anomaly is termed 'global' since the Dehn twist is a large coordinate
transformation, or more accurately, an element of the diffeomorphism
group of the torus, which lies outside of the connected component
of the identity. The Dehn twist is therefore the geometric analog
of the large $U\left(1\right)$ gauge transformation used in the celebrated
Laughlin argument, and an attempt has been made to follow this analogy
and produce a 'thermal Laughlin argument' \citep{nakai2017laughlin}. 

The chiral central charge also implies a 'local', or 'perturbative'
gravitational anomaly, which, at least in the context of relativistic
QFT in curved space-time, physically corresponds to the non-conservation
of energy-momentum in the presence of curvature gradients \citep{alvarez1984gravitational,bertlmann2000anomalies,bastianelli2006path}.
Through the anomaly inflow mechanism, or in more physical terms, through
bulk+boundary energy-momentum conservation, this boundary anomaly
implies a gravitational Chern-Simons (gCS) term in the effective action
describing the 2+1D bulk of a chiral topological phase \citep{Chern-Simons,jackiw2003chern,kraus2006holographic,witten2007three,stone2012gravitational}\footnote{Whether the gCS term matches the boundary\textit{ global} gravitational
anomaly as well is, to the best of my knowledge, an open problem.}. In turn, the gCS term implies a quantized energy-momentum-stress
response to curvature gradients, in the bulk \footnote{In fact, the gCS contribution to the energy-momentum-stress tensor
is proportional to the mathematically important Cotton tensor of the
metric \citep{perez2010conserved}. }. 

Though the gCS term is relatively well understood in the context of
relativistic QFT, its concrete physical content in the non-relativistic
setting of condensed matter physics is quite subtle, due to the following
reasons: 
\begin{enumerate}
\item Physically, the actual gravitational field of the earth is usually
negligible in condensed matter experiments. It is therefore clear
that the adjective 'gravitational' used above cannot be taken literally,
and requires further interpretation. Namely, one must find a physical
probe relevant in condensed matter experiments, which will somehow
mimic the effects of a strong gravitational field. This scenario is
often referred to as 'analog gravity' or 'emergent gravity' \citep{volovik2009universe}.
Mathematically, this corresponds to a physically accessible geometric
structure on the space-time occupied by the system of interest, in
the spirit of general relativity\footnote{We use the words geometry and gravity interchangeably from here on.}.
The most straight forward example is given by strain - a physical
deformation of the sample on which the system resides \citep{avron1995viscosity}.
An additional set of examples is given by spin-2 inhomogeneities \citep{gromov2017geometric}
and collective excitations \citep{volovik1990gravitational,volovik2009universe,haldane2009hall,haldane2011geometrical,you2014theory,gromov2017bimetric}.
Finally, Luttinger's trick relates temperature gradients to an applied
gravitational field \citep{luttinger1964theory,cooper1997thermoelectric,shitade2014heat,gromov2015thermal,bradlyn2015low,nakai2016finite,nakai2017laughlin}.
\item Fundamentally, the coupling of a system to gravity generally depends
on its global space-time symmetries in the absence of gravity. For
example, relativistic systems will couple differently from Galilean
invariant systems. Even when the spatial symmetries are fixed, the
gravitational background may vary, e.g Riemannian vs. Riemann-Cartan
geometry, which are both relativistic.  Moreover, for systems defined
on a lattice, there is no definite, or universal, prescription for
a coupling to gravity at all, as opposed to lattice gauge fields which
are very well understood. The coupling of a system to gravity therefore
relies on more refined information than that used to classify topological
phases of matter. In particular, known results in relativistic QFT
do not directly apply to the non-relativistic condensed matter systems
we are interested in. 
\item Technically, when describing gravity in terms of a metric, the gCS
term is third order in derivatives, so obtaining effective actions
that contain it \textit{consistently}, i.e account for \textit{all}
possible terms up to the same order, is nontrivial. 
\end{enumerate}
Naturally, the pioneering approaches to the above difficulties were
based on an adaptation of known results in relativistic QFT \citep{volovik1990gravitational,read2000paired,ryu2012electromagnetic}
(see also \cite{wang2011topological,palumbo2016holographic}), an
approach that we carefully and critically follow in Sec.\ref{sec:Main-section-1:}.
A much more advanced treatment developed over the past decade, primarily
in the context of quantum Hall states \citep{haldane2009hall,haldane2011geometrical,you2014theory,abanov2014electromagnetic,gromov2014density,PhysRevB.90.045123,gromov2015framing,gromov2015thermal,bradlyn2015low,bradlyn2015topological,klevtsov2015geometric,gromov2016boundary,gromov2017bimetric,gromov2017investigating,klevtsov2017laughlin,klevtsov2017lowest,klevtsov2017quantum,wiegmann2018inner,Cappelli_2018,schine2018measuring,kapustin2019thermal,hu2020microscopic}.
In particular, a \textit{non-relativistic} gCS term arises in quantum
Hall states, and produces corrections to the odd viscosity (introduced
below) at finite wave-vector, and in curved background \citep{abanov2014electromagnetic,bradlyn2015topological,klevtsov2015geometric,gromov2015framing,klevtsov2017quantum}.
We follow this observation in Sec.\ref{sec:Main-section-2:}. We note
a couple of additional central results from the literature: 
\begin{enumerate}
\item In quantum Hall states, the chiral central charge contributes to the
braiding statistics and angular momentum of conical defects \citep{can2016emergent,gromov2016geometric},
the latter was recently observed in an optical realization of integer
quantum Hall states \citep{schine2016synthetic,schine2018measuring}.
\item The gCS term is not directly related to $K_{H}=c\pi T/6$ through
Luttinger's trick, simply because it is too high in derivatives of
the background metric \citep{stone2012gravitational}. Moreover,
careful analysis in quantum Hall states shows that $K_{H}$ receives
no bulk contribution at all \citep{gromov2015thermal,bradlyn2015low},
and is therefore purely a boundary phenomenon, as explained below
Eq.\eqref{eq:1-2}. Nevertheless, derivatives of $K_{H}$ can be
computed from the bulk Hamiltonian a la Luttinger \citep{cooper1997thermoelectric,qin2011energy,bradlyn2015low},
resulting in a relative topological invariant for gapped lattice systems
\citep{kapustin2019thermal}. The latter gives a rigorous 2+1D lattice
definition for the chiral central charge.
\end{enumerate}

\subsection{Chiral superfluids and superconductors \label{subsec:Chiral-superfluids-and}}

\begin{figure}[!th]
\begin{centering}
\includegraphics[width=0.35\columnwidth]{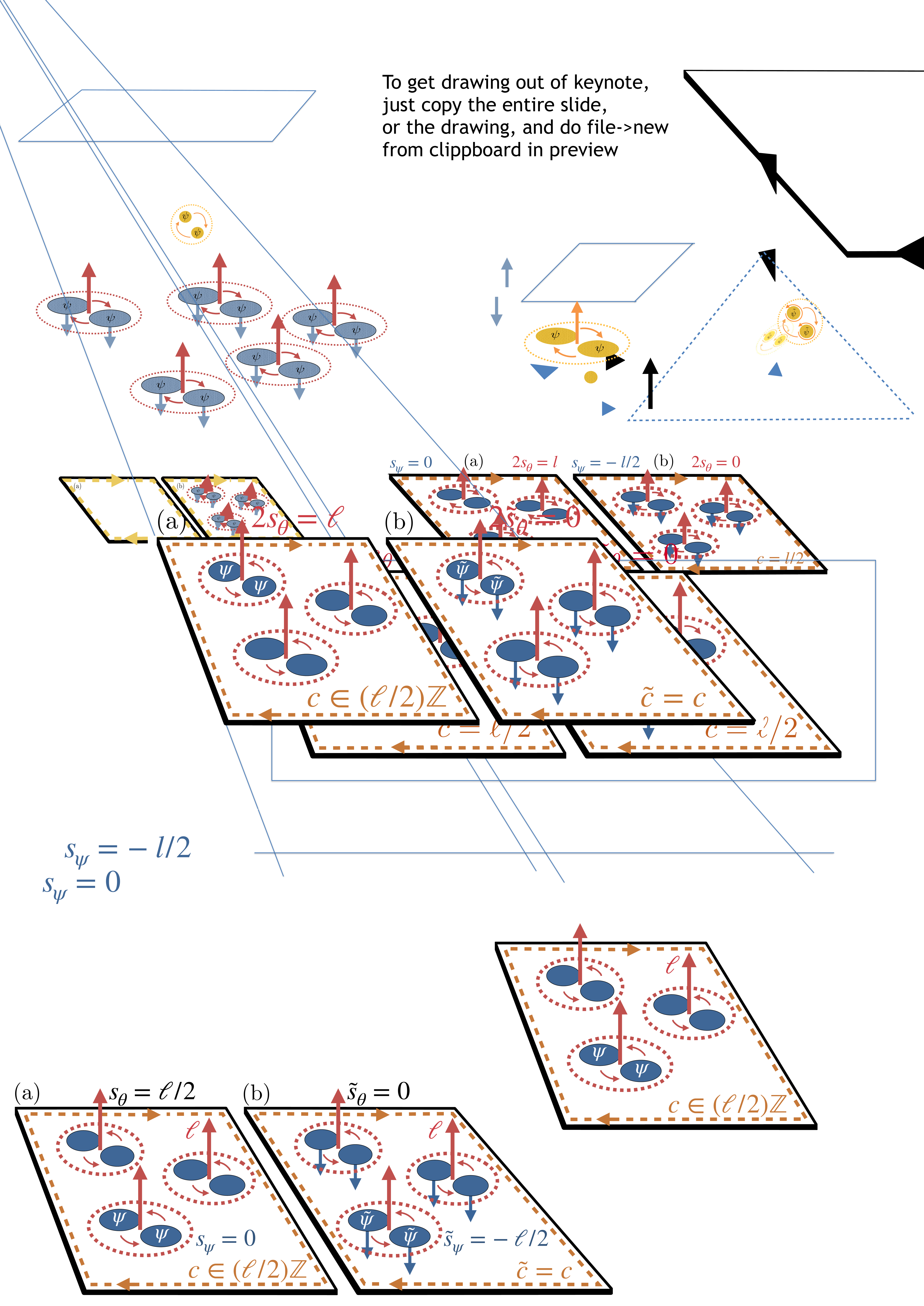}
\par\end{centering}
\caption{Chiral superfluids and superconductors (CSF/Cs) are comprised of fermions
$\psi$, which form Cooper pairs with non-vanishing relative angular
momentum $\ell\in\mathbb{Z}$ (red arrows), in units of $\hbar$.
As a result, CSF/Cs support boundary degrees of freedom (dashed orange)
with a chiral central charge $c\in\left(\ell/2\right)\mathbb{Z}$.
\label{fig:Chiral-Superfluid}}
\end{figure}

An important class of 2+1D chiral topological phases appears in chiral
superfluids and superconductors (CSFs and CSCs, or CSF/Cs), where
the ground state is a condensate of Cooper pairs of fermions, which
are spinning around their centre of mass with a non-vanishing angular
momentum $\ell\in\mathbb{Z}$ \citep{read2000paired,volovik2009universe},
see Fig.\ref{fig:Chiral-Superfluid}. As reviewed below, CSF/Cs appear
in a wide range of physical systems, all of which have been the subject
of extensive and continued research effort, going back to the classic
body of work on superfluid $^{3}\text{He}$ \citep{vollhardt2013superfluid}.
The interest in CSF/Cs comes from two fronts, a fermionic/topological
front, and a bosonic/symmetry-breaking front, both resulting directly
from the $\ell$-wave condensate. A central theme of this thesis is
the intricate interplay between these two facets of CSF/Cs.

On the fermionic/topological front, the $\ell$-wave condensate leads
to an energy gap for single fermion excitations, which form a chiral
SPT phase of matter: a topological superconductor \citep{Kallin_2016,Sato_2017}.
Topological superconductors are of interest since their chiral central
charge $c\in\left(\ell/2\right)\mathbb{Z}$ can be half-integer, indicating
the presence of chiral Majorana spinors on boundaries, each contributing
an additive $\pm1/2$ to $c$, where the sign depends on their chirality.
In turn, this implies the presence of Majorana bound states, or zero
modes, in the cores of vortices \citep{read2000paired}. The observation
of Majorana fermions, which are their own anti-particles, and may
not exist in nature as elementary particles, is of fundamental interest.
Moreover, Majorana bound states are closely related to non-abelian
Ising anyons \citep{moore1991nonabelions}, which have been proposed
as building blocks for topological quantum computers \citep{kitaev2003fault,nayak2008non}. 

On the bosonic/symmetry-breaking front, the $\ell$-wave condensate
implies an exotic symmetry breaking pattern, which leads to an unusual
spectrum of bosonic excitations, and as a result, an unusual hydrodynamic
description. In more detail, the condensation of $\ell$-wave Cooper
pairs corresponds to a non-vanishing ground-state expectation value
for the operator $\psi^{\dagger}\left(\partial_{x}\pm i\partial_{y}\right)^{\left|\ell\right|}\psi^{\dagger}$,
where $\psi^{\dagger}$ is a fermion creation operator\footnote{Due to Fermi statistics, $\ell$ must be odd if $\psi$ is spin-less.
An even $\ell$ requires spin-full fermions forming spin-less (singlet)
Cooper pairs, $\psi_{\uparrow}^{\dagger}\left(\partial_{x}\pm i\partial_{y}\right)^{\left|\ell\right|}\psi_{\downarrow}^{\dagger}$.
Spin-full fermions can also form spin-1 (triplet) Cooper pairs with
odd $\ell$, as is the case in $^{3}\text{He-A}$. Since the geometric
physics we are interested in is independent of the spin of the Cooper
pair, we restrict attention to spin-less fermions for odd $\ell$,
and write our expressions per spin component for even $\ell$.} and $\pm=\text{sgn}\left(\ell\right)$. An $\ell$-wave condensate
implies the breaking of time reversal symmetry $T$ and parity (spatial
reflection) $P$ down to $PT$, and of the symmetry groups generated
by particle number $N$ and angular momentum $L$ down to a diagonal
subgroup 
\begin{align}
 & U\left(1\right)_{N}\times SO\left(2\right)_{L}\rightarrow U\left(1\right)_{L-\left(\ell/2\right)N}.\label{eq:2-1-1}
\end{align}
In CSFs, this symmetry breaking occurs spontaneously, due to a symmetric
two-body attractive interaction between fermions. This phenomenon
is generic, at least from the perspective of perturbative Fermi-surface
renormalization group \citep{shankar1994renormalization}. Thin films
of $^{3}\text{He-A}$ are experimentally accessible $p$-wave CSFs
\citep{Levitin841,PhysRevLett.109.215301,Ikegami59,Zhelev_2017},
and there are many proposals for the realization of various $\ell$-wave
CSFs in cold atoms \citep{PhysRevLett.101.160401,PhysRevLett.103.020401,PhysRevA.86.013639,PhysRevA.87.053609,PhysRevLett.115.225301,BOUDJEMAA20171745,Hao_2017}.
The spontaneous symmetry breaking \eqref{eq:2-1-1} implies a single
Goldstone field, charged under the broken generator $N+\left(\ell/2\right)L$,
as well as massive Higgs fields, which are $U\left(1\right)_{N}$-neutral,
and carry angular momentum $0$ and $\pm2\ell$ \citep{brusov1981superfluidity,Volovik_2013,Sauls:2017aa,PhysRevB.98.064503,hsiao2018universal}.
In particular, $p$-wave ($\ell=\pm1$) superfluids support Higgs
fields of angular momentum $0$ and $\pm2$, which form a spatial
metric, including a non-relativistic analog of the graviton \citep{volovik1990gravitational,read2000paired,volovik2009universe}.
This observation will play a central role in Sec.\ref{sec:Main-section-1:}.
The angular momentum $\ell/2$ carried by the Goldstone field leads
to a $P,T$-odd hydrodynamic description, including an odd (or Hall)
viscosity, which is introduced below, and studied in Sec.\ref{sec:Main-section-2:}. 

An \textit{intrinsic} CSC is obtained if the $U\left(1\right)_{N}$
symmetry is gauged, by coupling to a dynamical gauge field. This gauge
field physically corresponds to the 3+1D electromagnetic interaction
between electrons, which are themselves confined to a 2+1D lattice
of ions. Experimental evidence for chiral superconductivity was recently
reported in Ref.\citep{Jiao:2020aa}. One may also consider an emergent
2+1D gauge field, with a Chern-Simons and/or Maxwell dynamics.
In particular, this leads to CSFs of 'composite fermions' \citep{read2000paired,PhysRevX.5.031027,Son_2018},
including field theoretic descriptions of the non-abelian candidates
for the fractional quantum Hall state observed at filling $5/2$ \citep{banerjee2018observation},
a subject of ongoing debate \citep{PhysRevB.98.045112,PhysRevLett.121.026801,PhysRevB.97.121406,PhysRevB.98.167401,PhysRevB.99.085309}.
The symmetry breaking pattern \eqref{eq:2-1-1} may also occur explicitly,
due the proximity of a conventional $s$-wave superconductor (SC)
to 2+1D spin-orbit coupled metal, in which case we speak of a \textit{proximity
induced} CSC, an observation of which was reported in Refs.\citep{PhysRevLett.114.017001,M_nard_2017}.
Note that in this case the Goldstone and Higgs fields can be viewed
as non-dynamical. 

Despite the large body of work on boundary Majorana fermions in CSFs
and CSCs (CSF/Cs), the bulk geometric physics corresponding to these
through anomaly inflow, and presumably captured by a gCS action, remains
poorly understood, due to the difficulties mentioned in Sec.\ref{subsec:Geometric-physics-in}.
In fact, most existing statements, though made in truly pioneering
and seminal work \citep{volovik1990gravitational,read2000paired,ryu2012electromagnetic},
are speculative, and are primarily based on an inaccurate adaptation
of known results in relativistic QFT to $p$-wave CSF/Cs. An understanding
of the bulk geometric physics is of particular importance since, in
the simplest case of spin-less fermions with no additional internal
symmetry, the only charge carried by the boundary Majorana fermions
is energy-momentum, and the only boundary anomalies and bulk topological
effective actions are therefore gravitational\footnote{For spin-full $p$-wave CSFs, one can exploit $SU\left(2\right)$
spin rotation symmetry, and does not have to resort to gravitational
probes \citep{volovik1989fractional,read2000paired,stone2004edge}.}. In particular, the boundary Majorana fermions are always $U\left(1\right)_{N}$-neutral,
and it follows that no $U\left(1\right)_{N}$ boundary anomaly or
bulk topological effective action occurs\footnote{An exception to this rule occurs in Galilean invariant systems, where
momentum and $U\left(1\right)_{N}$-current are identified, as we
will see in Sec.\ref{sec:Main-section-2:}.}. Motivated by this state of affairs, the goal of Sec.\ref{sec:Main-section-1:}-\ref{sec:Main-section-2:}
is to turn the insightful ideas of Refs.\citep{volovik1990gravitational,read2000paired,ryu2012electromagnetic}
into concrete physical predictions. 

As a first approach to the problem, in Sec.\ref{sec:Main-section-1:}
we follow Refs.\citep{volovik1990gravitational,read2000paired,ryu2012electromagnetic}
and utilize the low energy relativistic description of $p$-wave CSF/Cs,
which exists because the $p$-wave condensate $\psi^{\dagger}\left(\partial_{x}\pm i\partial_{y}\right)\psi^{\dagger}$
is first order in derivatives. The main questions we ask are:
\begin{quote}
\textit{What type of space-time geometry emerges in the low energy
relativistic description of $p$-wave superfluids and superconductors?
What are the physical implications of the emergent relativistic geometry
to these non-relativistic systems? }
\end{quote}
Our answer to the first question is that the fermionic excitations
in $p$-wave CSF/Cs correspond at low energy to a massive relativistic
Majorana spinor, which is minimally coupled to an emergent Riemann-Cartan
geometry. This geometry is described by the $p$-wave order parameter
$\Delta^{i}\sim\delta^{ij}\psi^{\dagger}\partial_{j}\psi^{\dagger}$,
made up of the Goldstone and Higgs fields, as well as a $U\left(1\right)_{N}$
gauge field. As opposed to the Riemannian geometry previously believed
to emerge \citep{volovik1990gravitational,read2000paired,ryu2012electromagnetic},
Riemann-Cartan space-times are characterized by a non-vanishing torsion
tensor, in addition to the curvature tensor \citep{ortin2004gravity}.
In condensed matter physics (or elasticity theory),  torsion is well
known to describe the density of lattice dislocations \citep{hughes2011torsional,hughes2013torsional,parrikar2014torsion,geracie2014hall}\footnote{Similarly, curvature traditionally describes the density of lattice
disclinations, as well a the curving of a two-dimensional material
in three dimensional space. It is also known that temperature gradients
correspond to time-like torsion via Luttinger's trick \citep{shitade2014heat,bradlyn2015low}. }, and our results provide a new mechanism by which  torsion can emerge
- due to the symmetry breaking pattern \eqref{eq:2-1-1} at $\ell=\pm1$
. The above statements are relevant if one aims at studying relativistic
fermions in nontrivial space-times using table-top experiments \citep{Kim2017},
or if one hopes that emergent relativistic geometry in condensed matter
can answer fundamental questions about the seemingly relativistic
geometry of our universe \citep{volovik2009universe}. Here, however,
we are interested in answering the second question posed above. 

As expected, a gCS term appears in the low energy effective action
of $p$-wave CSF/Cs, and we find that it produces a precisely quantized
bulk energy-momentum-stress response to the $p$-wave Higgs fields.
Accordingly, a (perturbative) gravitational anomaly that depends on
the Higgs fields appears on the boundary, implying a $c$-dependent
transfer of energy-momentum between bulk and boundary. The emergence
of torsion leads to additional interesting terms in the bulk effective
action. In particular, a non-topological 'gravitational \textit{pseudo}
Chern-Simons' term produces an energy-momentum-stress response closely
mimicking that of the gCS term, and we show how to disentangle the
two responses in order to extract $c$ from bulk measurements. In
lattice models, the low energy description consists of an even number
of relativistic Majorana spinors - a fermion doubling phenomena. Surprisingly,
we find that these spinors experience slightly different emergent
geometries. As a result, additional 'gravitational Chern-Simons difference'
terms are possible, which are again not of topological origin, but
nevertheless imply responses which must be carefully distinguished
from those of the gCS term. All other terms in the bulk effective
action are either higher in derivatives, or are lower in derivatives
but naively diverge within the relativistic description. The latter
'UV-sensitive' terms cannot be reliably interpreted based on the relativistic
description, and require a non-relativistic treatment. In particular,
the relativistic, UV-sensitive, and somewhat controversial 'torsional
Hall viscosity' \citep{hughes2011torsional,hughes2013torsional,parrikar2014torsion,geracie2014hall,hoyos2014hall,bradlyn2015low},
is found in Sec.\ref{sec:Main-section-2:} to correspond to the non-relativistic
and well understood odd (or Hall) viscosity of CSF/Cs \citep{Read:2009aa,read2011hall,hoyos2014effective,shitade2014bulk,moroz2015effective},
which is introduced below. 

Before continuing, we note that there has been considerable recent
interest in torsional physics in condensed matter, in systems described
at low energy by 3+1D Weyl (or Majorana-Weyl) spinors, namely Weyl
semi-metals and $^{3}$He-A \citep{PhysRevResearch.2.033269,PhysRevLett.124.117002,PhysRevB.101.125201,PhysRevB.101.165201,laurila2020torsional,huang2020torsional},
and in Kitaev's honeycomb model \citep{PhysRevB.101.245116,PhysRevB.102.125152}.

\subsection{Odd viscosity\label{subsec:Odd-viscosity}}

The odd (or Hall) viscosity $\eta_{\text{o}}$ is a non-dissipative,
time reversal odd, stress response to strain-rate \citep{PhysRevLett.75.697,Avron1998,PhysRevB.86.245309,hoyos2014hall,PhysRevE.89.043019},
which can appear even in superfluids (SFs) and incompressible (or
gapped) fluids,  where the more familiar and intuitive dissipative
viscosity vanishes. The observable signatures of $\eta_{\text{o}}$
are actively studied in a variety of systems \citep{PhysRevE.90.063005,PhysRevB.94.125427,PhysRevLett.119.226602,PhysRevLett.118.226601,PhysRevB.96.174524,PhysRevFluids.2.094101,banerjee2017odd,bogatskiy2018edge,holder2019unified,PhysRevLett.122.128001},
and recently led to its measurement in a colloidal chiral fluid \citep{soni2018free}
and in graphene's electron fluid under a magnetic field \citep{Berdyugineaau0685}.

 In isotropic 2+1 dimensional fluids, the odd viscosity tensor at
zero wave-vector ($\mathbf{q}=0$) reduces to a single component .
In analogy with the celebrated quantization of the odd (or Hall) conductivity
in the quantum Hall (QH) effect \citep{thouless1982quantized,avron1983homotopy,golterman1993chern,qi2008topological,Nobel-2016,mera2017topological},
this component obeys a quantization condition 
\begin{align}
 & \eta_{\text{o}}^{\left(1\right)}=-\left(\hbar/2\right)s\cdot n_{0},\;s\in\mathbb{Q},\label{eq:1-1}
\end{align}
in incompressible quantum fluids, such as integer and fractional QH
states \citep{PhysRevLett.75.697,Read:2009aa,read2011hall}. Here
$n_{0}$ is the ground state density, and $s$ is a rational-valued topological
invariant\footnote{In fact, an $SO\left(2\right)_{L}$-symmetry-protected topological
invariant.}, labeling the many-body ground state, which can be interpreted as
the average angular momentum per particle (in units of $\hbar$, which
is henceforth set to 1).  

Remarkably, Eq.\eqref{eq:1-1} also holds in CSFs, though they are
compressible. Computing $\eta_{\text{o}}^{\left(1\right)}$ in an
$\ell$-wave CSF, one finds Eq.\eqref{eq:1-1} with the intuitive
angular momentum per fermion, $s=\ell/2$ \citep{Read:2009aa,read2011hall,hoyos2014effective,shitade2014bulk,moroz2015effective}.
Thus, a measurement of $\eta_{\text{o}}^{\left(1\right)}$ at $\mathbf{q}=\mathbf{0}$
can be used to obtain the angular momentum $\ell$ of the Cooper pair,
but carries no additional information. It is therefore clear that
the symmetry breaking pattern \eqref{eq:2-1-1} which defines $\ell$,
rather than ground-state topology, is the origin of the quantization
$s=\ell/2$ in CSFs \footnote{Accordingly, the quantization of $s$ is broken in a mixture of CSFs
with different $\ell$s, where $U\left(1\right)_{N}\times SO\left(2\right)_{L}$
is completely broken. In the mixture $s\equiv-2\eta_{\text{o}}^{\left(1\right)}/n=\sum_{i}n_{i}\left(\ell_{i}/2\right)/\sum_{i}n_{i}$
retains its meaning as an average angular momentum per particle, but
is no longer quantized. This should be contrasted with multicomponent
QH states \citep{bradlyn2015topological}, where all $n_{i}$s are
proportional to the same applied magnetic field through the filling
factors $\nu_{i}\in\mathbb{Q}$, and $s$ remains quantized.}. 

Nevertheless, the gapped fermions in a CSF do carry non-trivial ground-state
topology labeled by the central charge $c\in\left(\ell/2\right)\mathbb{Z}$,
and, based on results in quantum Hall states \citep{abanov2014electromagnetic,klevtsov2015geometric,bradlyn2015topological},
a $c$-dependent correction to $\eta_{\text{o}}^{\left(1\right)}$
of Eq.\eqref{eq:1-1} is therefore expected to appear at small non-zero
wave-vector, 
\begin{align}
 & \delta\eta_{\text{o}}^{\left(1\right)}\left(\mathbf{q}\right)=-\frac{c}{24}\frac{1}{4\pi}q^{2}.\label{eq:2-1-2}
\end{align}
This raises the questions:
\begin{quote}
\textit{In chiral superfluids, can the boundary chiral central charge
be extracted from a measurement of the bulk odd viscosity? Can it
be extracted from any bulk measurement?}
\end{quote}
Providing a definite answer to these questions is the main goal of
Sec.\ref{sec:Main-section-2:}, and requires a fully non-relativistic
treatment of CSFs. The main reason for this is that the relativistic
low energy description misses most of the physics of the Goldstone
field. Analysis of Goldstone physics in CSFs was undertaken in Refs.\citep{volovik1988quantized,goryo1998abelian,goryo1999observation,furusaki2001spontaneous,stone2004edge,roy2008collective,lutchyn2008gauge,ariad2015effective},
most of which revolving around the non-vanishing, yet non-quantized,
Hall (or odd) conductivity in CSFs. More recently, Refs.\citep{hoyos2014effective,moroz2015effective}
considered CSFs in curved (or strained) space, following the pioneering
work \citep{son2006general} on $s$-wave ($\ell=0$) SFs. These works
demonstrated that the Goldstone field, owing to its charge $L+\left(\ell/2\right)N$,
produces the $\mathbf{q}=\mathbf{0}$ odd viscosity \eqref{eq:1-1},
and it is therefore natural to expect that a $q^{2}$ correction similar
to \eqref{eq:2-1-2} will also be produced. Nevertheless, Refs.\citep{hoyos2014effective,moroz2015effective}
did not consider the derivative expansion to the high order at which
$q^{2}$ corrections to $\eta_{\text{o}}$ would appear, nor did
they detect any bulk signature of $c$ at lower orders. In Sec.\ref{sec:Main-section-2:}
we obtain a low energy effective field theory that consistently captures
both the chiral Goldstone mode and the gCS term, thus unifying and
extending the seemingly unrelated analysis of Refs.\citep{son2006general,hoyos2014effective,moroz2015effective}
and Sec.\ref{sec:Main-section-1:}. Using the theory we show that
$c$ cannot be extracted for a measurement of the odd viscosity alone,
as suggested by Eq.\eqref{eq:2-1-2}. Nevertheless, a related observable,
termed 'improved odd viscosity', does allow for the bulk measurement
of $c$. Additional results of the same spirit are found in Galilean
invariant CSFs. 

\subsection{Quantum Monte Carlo sign problems in chiral topological matter \label{subsec:Complexity-of-simulating}}

Utilizing a random sampling of phase-space according to the Boltzmann
probability distribution, Monte Carlo simulations are arguably the
most powerful tools for numerically evaluating thermal averages in
classical many-body physics \citep{doi:10.1080/01621459.1949.10483310}.
 Though the phase-space of an $N$-body system scales exponentially
with $N$, a Monte-Carlo approximation with a fixed desired error is usually
obtained in polynomial time \citep{troyer2005computational,barahona1982computational}.
In \textit{Quantum} Monte Carlo (QMC), one attempts to perform Monte-Carlo
computations of thermal averages in quantum many-body systems, by
following the heuristic idea that quantum systems in $d$ dimensions
are equivalent to classical systems in $d+1$ dimensions \citep{Assaad,li2019sign}.

The difficulty with any such quantum to classical mapping, henceforth
referred to as a \textit{method}, is the infamous \textit{sign problem},
where the mapping can produce complex, rather than non-negative, Boltzmann
weights $p$, which do not correspond to a probability distribution.
  Faced with a sign problem, one can try to change the method used
and obtain $p\geq0$, thus \textit{curing the sign problem} \citep{marvian2018computational,klassen2019hardness}.
Alternatively, one can perform QMC using the weights $\left|p\right|$,
which is often done but generically leads to an exponential computational
complexity in evaluating physical observables, limiting ones ability
to simulate large systems at low temperatures \citep{troyer2005computational}.

Conceptually, the sign problem can be understood as an obstruction
to mapping quantum systems to classical systems, and accordingly,
from a number of complexity theoretic perspectives, a generic curing
algorithm in polynomial time is not believed to exist  \citep{troyer2005computational,bravyi2006complexity,hastings2016quantum,marvian2018computational,hangleiter2019easing,klassen2019hardness}.
 In many-body physics, however, one is mostly interested
in universal phenomena, i.e phases of matter and the transitions between
them, and therefore representative Hamiltonians which are free of the sing problem (henceforth 'sign-free')
often suffice \citep{kaul2013bridging}. In fact, QMC simulations
continue to produce unparalleled results, in all branches of many-body
quantum physics, precisely because new sign-free models are constantly
being discovered \citep{RevModPhys.67.279,PhysRevLett.83.3116,PhysRevX.3.031010,PhysRevD.88.021701,kaul2013bridging,GazitE6987,berg2019monte,li2019sign}.
Designing sign-free models requires \textit{design principles} (or
``de-sign'' principles) \citep{kaul2013bridging,wang2015split}
- easily verifiable properties that, if satisfied by a Hamiltonian
and method, lead to a sign-free representation of the corresponding
partition function.   An important example is
the condition $\bra iH\ket j\leq0$ where $i\neq j$ label a local
basis, which implies non-negative weights $p$ in a wide range of
methods  \citep{kaul2013bridging,hangleiter2019easing}.
Hamiltonians satisfying this condition in a given basis are known
as \textit{stoquastic} \citep{bravyi2006complexity}, and have
proven very useful in both application and theory of QMC in bosonic
(or spin, or 'qudit') systems \citep{kaul2013bridging,troyer2005computational,bravyi2006complexity,hastings2016quantum,marvian2018computational,hangleiter2019easing,klassen2019hardness}.

Fermionic Hamiltonians are not expected to be stoquastic in any local
basis \citep{troyer2005computational,li2019sign}, and alternative
methods, collectively known as determinantal quantum Monte-Carlo (DQMC),
are therefore used \citep{PhysRevD.24.2278,Assaad,Santos_2003,li2019sign,berg2019monte}.
 The search for design principles that apply to DQMC, and applications
thereof, has naturally played the dominant role in tackling the sign
problem in fermionic systems, and has seen a lot of progress in recent
years \citep{chandrasekharan2013fermion,wang2015split,li2016majorana,wei2016majorana,wei2017semigroup,berg2019monte,li2019sign}.
Nevertheless, long standing open problems in quantum many-body physics
continue to defy solution, and remain inaccessible for QMC. These
include the nature of high temperature superconductivity and the associated
repulsive Hubbard model  \citep{Santos_2003,PhysRevB.80.075116,PhysRevX.5.041041,kantian2019understanding},
dense nuclear matter and the associated lattice QCD at finite baryon
density \citep{Hands_2000,PhysRevD.66.074507,10.1093/ptep/ptx018},
and the enigmatic fractional quantum Hall state at filling $5/2$
and its associated Coulomb Hamiltonian \citep{banerjee2018observation,PhysRevB.98.045112,PhysRevLett.121.026801,PhysRevB.97.121406,PhysRevB.99.085309,hu2020microscopic},
all of which are fermionic. \textcolor{red}{} 

One may wonder if there is a fundamental reason that no design principle
applying to the above open problems has so far been found, despite
intense research efforts. More generally, 
\begin{quote}
\textit{Are there phases of matter that do not admit a sign-free representative?
Are there physical properties that cannot be exhibited by sign-free
models?}
\end{quote}
We refer to such phases of matter, where the sign problem simply cannot
be cured, as having an \textit{intrinsic sign problem} \citep{hastings2016quantum}.
From a practical perspective, intrinsic sign problems may prove useful
in directing research efforts and computational resources. From a
fundamental perspective, intrinsic sign problems identify certain
phases of matter as inherently quantum - their physical properties
cannot be reproduced by a partition function with positive Boltzmann
weights.

To the best of our knowledge, the first intrinsic sign problem was
discovered by Hastings \citep{hastings2016quantum}, who proved that
no stoquastic, commuting projector, Hamiltonians exist for the 'doubled
semion' phase \citep{PhysRevB.71.045110}, which is bosonic and topologically
ordered. In Ref.\citep{PhysRevResearch.2.033515}, we generalize this result
considerably - excluding the possibility of stoquastic Hamiltonians
in a broad class of bosonic non-chiral topological phases of matter.
Additionally, Ref.\citep{ringel2017quantized} demonstrated,
based on the algebraic structure of edge excitations, that no translationally
invariant stoquastic Hamiltonians exist for bosonic chiral topological
phases.

In Sec.\ref{sec:Main-section-3:}, we will establish a new criterion
for intrinsic sign problems in chiral topological matter, and take
the first step in analyzing intrinsic sign problems in fermionic systems.
First, based on the well established 'momentum polarization' method
for characterizing chiral topological matter \citep{PhysRevB.88.195412,PhysRevLett.110.236801,PhysRevB.90.045123,PhysRevB.92.165127},
we obtain a variant of the result of Ref.\citep{ringel2017quantized}
- excluding the possibility of stoquastic Hamiltonians in a broad
class of bosonic chiral topological phases. We then develop a formalism
with which we obtain analogous results for systems comprised of both
bosons \textit{and} fermions - excluding the possibility of sign-free
DQMC simulations. 

\begin{table*}[h]

  \renewcommand*{\arraystretch}{1.3}
\resizebox{\textwidth}{!}{

\begin{tabular}{lllll} 
\hline
\hline

\textbf{Phase of matter} & \textbf{Parameterization} & $\boldsymbol{c}$  & $\boldsymbol{\left\{h_{a}\right\}}$ & \textbf{Intrinsic sign problem?}

\tabularnewline 

\hline

Laughlin (B) \citep{hu2020microscopic} & Filling $1/q,\;(q\in2\mathbb{N})$ & $1$ & $\left\{ a^{2}/2q\right\} _{a=0}^{q-1}$ & In $98.5\%$ of first $10^3$
\tabularnewline Laughlin (F) \citep{hu2020microscopic} & Filling $1/q,\;(q\in2\mathbb{N}-1)$ & $1$ & $\left\{ \left(a+1/2\right)^{2}/2q\right\} _{a=0}^{q-1}$ & In $96.7\%$ of first $10^3$
\tabularnewline Chern insulator (F) \citep{PhysRevResearch.2.043032} & Chern number $\nu\in\mathbb{Z}$ & $\nu$ & $\left\{ \nu/8\right\} $ & For $\nu\notin 12\mathbb{Z}$
\tabularnewline $\ell$-wave superconductor (F) \citep{PhysRevB.100.104512} & Pairing channel $\ell\in2\mathbb{Z}-1$ & $-\ell/2$ & $\left\{ -\ell/16\right\} $ & Yes
\tabularnewline Kitaev spin liquid (B) \citep{kitaev2006anyons} & Chern number $\nu\in2\mathbb{Z}-1$ & $\nu/2$ & $\left\{ 0,1/2,\nu/16\right\} $ & Yes
\tabularnewline $SU\left(2\right)_{k}$ Chern-Simons (B) \citep{bonderson2007non} & Level $k\in\mathbb{N}$ & $3k/\left(k+2\right)$ & $\left\{ a\left(a+2\right)/4\left(k+2\right)\right\} _{a=0}^{k}$ & In  $91.6\%$ of first $10^3$
\tabularnewline $E_{8}$ $K$-matrix (B) \citep{PhysRevB.94.155113} & Stack of $n\in\mathbb{N}$ copies& $8n$ & $\left\{ 0\right\} $ & For $n\notin 3\mathbb{N}$
\tabularnewline Fibonacci anyon model (B) \citep{bonderson2007non} &  & $14/5$ (mod $8$)& $\left\{ 0,2/5\right\} $ & Yes
\tabularnewline Pfaffian (F) \citep{hsin2020effective} &  & $3/2$ & $\left\{   0,1/2,1/4,3/4,1/8,5/8     \right\}$ & Yes
\tabularnewline PH-Pfaffian (F) \citep{hsin2020effective} &  & $1/2$ & $\left\{   0,0,1/2,1/2,1/4,3/4     \right\}$ & Yes
\tabularnewline Anti-Pfaffian (F) \citep{hsin2020effective} &  & $-1/2$ & $\left\{   0,1/2,1/4,3/4,3/8,7/8     \right\}$ & Yes
\tabularnewline

\hline
\hline
\end{tabular}

}

\caption{Examples of intrinsic sign problems based on the criterion $e^{2\pi ic/24}\protect\notin\left\{ \theta_{a}\right\} $, in terms of the chiral central charge $c$ and the topological spins $\theta_{a}=e^{2\pi ih_{a}}$. The number of spins $h_a$ is equal to the dimension of the ground state subspace on the torus. We mark bosonic/fermionic phases by (B/F). The quantum Hall Laughlin phases corresond to $U(1)_q$ Chern-Simons theories. The $\ell$-wave superconductor is chiral, e.g $p+ip$ for $\ell=1$, and comprised of a single flavour of spinless fermions. The data shown  refers only to the SPT phase formed by gapped fermionic excitations, see Sec.\ref{subsec:Chiral-superfluids-and}. Data for the spinfull case is identical to that of the Chern insulator, with $-\ell$ odd (even) in place of $\nu$, for triplet (singlet) pairing. The modulo 8 ambiguity in the central charge of the Fibonacci anyon model corresponds to the stacking of a given realization  with copies of the $E_{8}$ $K$-matrix phase. Data for the three quantum Hall Pfaffian phases is given at the minimal filling 1/2. The physical filling 5/2 is obtained by stacking with a $\nu=2$ Chern insulator, and an intrinsic sign problem appears in this case as well. \label{tab:1} }

\end{table*}

All of the above mentioned topological phases are gapped, 2+1 dimensional,
and described at low energy by a topological field theory \citep{doi:10.1142/S0129055X90000107,kitaev2006anyons,freed2016reflection}.
The class of such phases in which we find an intrinsic sign problem
is defined in terms of robust data characterizing them: the chiral
central charge $c$, a rational number, as well as the set $\left\{ \theta_{a}\right\} $
of topological spins of anyons, a subset of roots of unity. Namely,
we find that 

\begin{quote}
 \nopagebreak[0]
\textit{An intrinsic sign problem exists if $e^{2\pi ic/24}$ is not
the topological spin of some anyon, i.e $e^{2\pi ic/24}\notin\left\{ \theta_{a}\right\} $.
}
\end{quote}
The above criterion applies to most chiral topological phases, see
Table \ref{tab:1} for examples.  In particular, we identify an intrinsic
sign problem in $96.7\%$ of the first one-thousand fermionic Laughlin
phases, and in all non-abelian Kitaev spin liquids.  We also find
intrinsic sign problems in $91.6\%$ of the first one-thousand $SU\left(2\right)_{k}$
Chern-Simons theories. Since, for $k\neq1,2,4$,  these allow for
universal quantum computation by manipulation of anyons \citep{Freedman:2002aa,nayak2008non},
our results support the strong belief that quantum computation cannot
be simulated with classical resources, in polynomial time \citep{Arute:2019aa}.
This conclusion is strengthened by examining the Fibonacci anyon model,
which is known to be universal for quantum computation \citep{nayak2008non},
and is found to be intrinsically sign-problematic. 

We stress that both $c$ and $\left\{ \theta_{a}\right\} $ have clear
observable signatures in both the bulk and boundary of chiral topological
matter, some of which have been experimentally observed. The chiral
central charge was extensively discussed in previous section, including
its observation in Refs.\citep{jezouin2013quantum,banerjee2017observed,banerjee2018observation,Kasahara:2018aa,schine2018measuring}.
The topological spins determine the exchange statistics of anyons,
predicted to appear in interferometry experiments \citep{nayak2008non}.
Experimental observation remained elusive \citep{PhysRevLett.122.246801}
until it was recently reported in the Laughlin 1/3 quantum Hall state
\citep{Nakamura:2020aa}. Additionally, a measurement of anyonic statistics
via current correlations \citep{PhysRevLett.116.156802} was recently
reported in the Laughlin 1/3 state \citep{Bartolomei173}.

\pagebreak{}

\section{Probing topological superconductors with emergent relativistic gravity\label{sec:Main-section-1:}}

In this section, we restrict our attention to spin-less $p$-wave
chiral superfluids and superconductors (CSF/Cs), and analyze the relativistic
geometry, or gravity, that emerges at low energy. We seek answers
to the questions posed in Sec.\ref{subsec:Chiral-superfluids-and}. 


\subsection{Approach and main results\label{subsec:Results0}}

\subsubsection{Model and approach\label{subsec:Results1}}

As a starting point, we consider a simple model for spin-less $p$-wave
CSF/Cs \citep{Volovik:1988aa}. The action is given by 
\begin{align}
S\left[\psi;\Delta\right]= & \int\text{d}^{2+1}x\left[\psi^{\dagger}\left(i\partial_{t}+\frac{\delta^{ij}\partial_{i}\partial_{j}}{2m^{*}}-m\right)\psi+\left(\frac{1}{2}\psi^{\dagger}\Delta^{j}\partial_{j}\psi^{\dagger}+h.c\right)\right],\label{eq:13}
\end{align}
and describes the coupling of a spin-less fermion $\psi$, with effective
mass $m^{*}$ and chemical potential $-m$, to a $p$-wave order parameter
$\Delta=\left(\Delta^{x},\Delta^{y}\right)\in\mathbb{C}^{2}$. The
order parameter corresponds to the condensate of Cooper pairs described
in Sec.\ref{subsec:Chiral-superfluids-and}. In a proximity induced
CSC, the order parameter can be thought of as a non-dynamical background
field, as it appears in Eq.\eqref{eq:13}. In intrinsic CSCs, and
in CSFs, the order parameter is a quantum field which mediates an
attractive interaction between fermions, and a treatment of the dynamics
of $\Delta$ is deferred to Sec.\ref{sec:Conclusion-and-discussion}
and \ref{sec:Main-section-2:}. The ground-state, or unperturbed,
configuration of $\Delta$ is the $p_{x}\pm ip_{y}$ configuration
$\Delta=\Delta_{0}e^{i\theta}\left(1,\pm i\right)$, where $\Delta_{0}>0$
and $\theta$ are constants. The phase $\theta$ corresponds to the
Goldstone field implied by Eq.\eqref{eq:2-1-1}, while the orientation
(or chirality) $o=\pm$ corresponds to the breaking of reflection
and time reversal symmetries to their product, $P\times T\rightarrow PT$. 

One may view the model \eqref{eq:13} as 'microscopic', as will be
done in Sec.\ref{sec:Main-section-2:}, but here we will think of
it as the low energy description of a lattice model, introduced and analyzed
in Sec.\ref{sec:Lattice-model}. In the 'relativistic regime'
where the order parameter is much larger than the single particle
scales, the lattice model is essentially a lattice regularization
of four, generically massive, relativistic Majorana spinors, centered
at the particle-hole invariant points $k=-k$ in the Brillouin zone.
Around each of these four points, the low energy description is given
by an action of the form \eqref{eq:13}. In the relativistic regime
the effective mass $m^{*}$ is large, and in the limit $m^{*}\rightarrow\infty$
Eq.\eqref{eq:13} reduces to a relativistic action, with mass $m$
and speed of light $c_{\text{light}}=\Delta_{0}/\hbar$, for the Nambu
spinor $\Psi^{\dagger}=\left(\psi^{\dagger},\psi\right)$, which is
a Majorana spinor. The different Majorana spinors, associated with
the four particle-hole invariant points, have different orientations
$o_{n}$ and masses $m_{n}$, where $1\leq n\leq4$. 

The chiral central charge of the lattice model can be deduced from
its Chern number \citep{read2000paired,kitaev2006anyons,volovik2009universe,ryu2010topological}.
The $n$th Majorana spinor contributes $c_{n}=o_{n}\text{sgn}\left(m_{n}\right)/4$,
and summing over $n$ one obtains the central charge of the lattice
model $c=\sum_{n=1}^{4}c_{n}=\sum_{n=1}^{4}o_{n}\text{sgn}\left(m_{n}\right)/4$,
which gives the topological phase diagram purely in terms of the low
energy relativistic data $o_{n},m_{n}$, see Sec.\ref{sec:Lattice-model}.
This formula motivates a study of the geometric physics associated
with $c$, purely within the low energy relativistic description,
which we now pursue. 

In order to access the physics associated with $c$, we perturb the
order parameter out of the $p_{x}\pm ip_{y}$ configuration, and treat
$\Delta=\left(\Delta^{x},\Delta^{y}\right)\in\mathbb{C}^{2}$ as a
general space-time dependent field. This is analogous to applying
an electromagnetic field in order to measure a quantized Hall conductivity
in the quantum Hall effect. Following the observations of Refs.\cite{volovik2009universe,read2000paired}, we show in Sec.\ref{sec:Emergent-Riemann-Cartan-geometry}
that, in the relativistic limit, the Majorana spinor $\Psi$ experiences
such a general order parameter (along with a general $U\left(1\right)_{N}$
gauge field) as a non trivial gravitational background, namely Riemann-Cartan
geometry. See also Sec.\ref{subsec:The-order-parameter} for
the basics of this statement. Some of the emergent geometry is described
by the (inverse) spatial metric 
\begin{eqnarray}
 &  & g^{ij}=-\Delta^{(i}\Delta^{j)*},\label{eq:emergent metric}
\end{eqnarray}
where brackets denote the symmetrization, and the sign is a matter
of convention. The metric $g^{ij}$ corresponds to the Higgs field
included in the order parameter. Parameterizing $\Delta=e^{i\theta}\left(\left|\Delta^{x}\right|,e^{i\phi}\left|\Delta^{y}\right|\right)$
with the overall phase $\theta$ and relative phase $\phi\in\left(-\pi,\pi\right]$,
the metric is independent of $\theta$ and of the orientation $o=\text{sgn}\phi=\pm$,
which splits order parameters into $p_{x}+ip_{y}$-like and $p_{x}-ip_{y}$-like.
Note that in the $p_{x}\pm ip_{y}$ configuration the metric is euclidian,
$g^{ij}=-\Delta_{0}^{2}\delta^{ij}$. For our purposes it is important
that the metric be perturbed out of this form, and in particular it
is not enough to take the $p_{x}\pm ip_{y}$ configuration with a
space-time dependent Goldstone field $\theta$.
\begin{figure}[!th]
\begin{centering}
\includegraphics[scale=0.4]{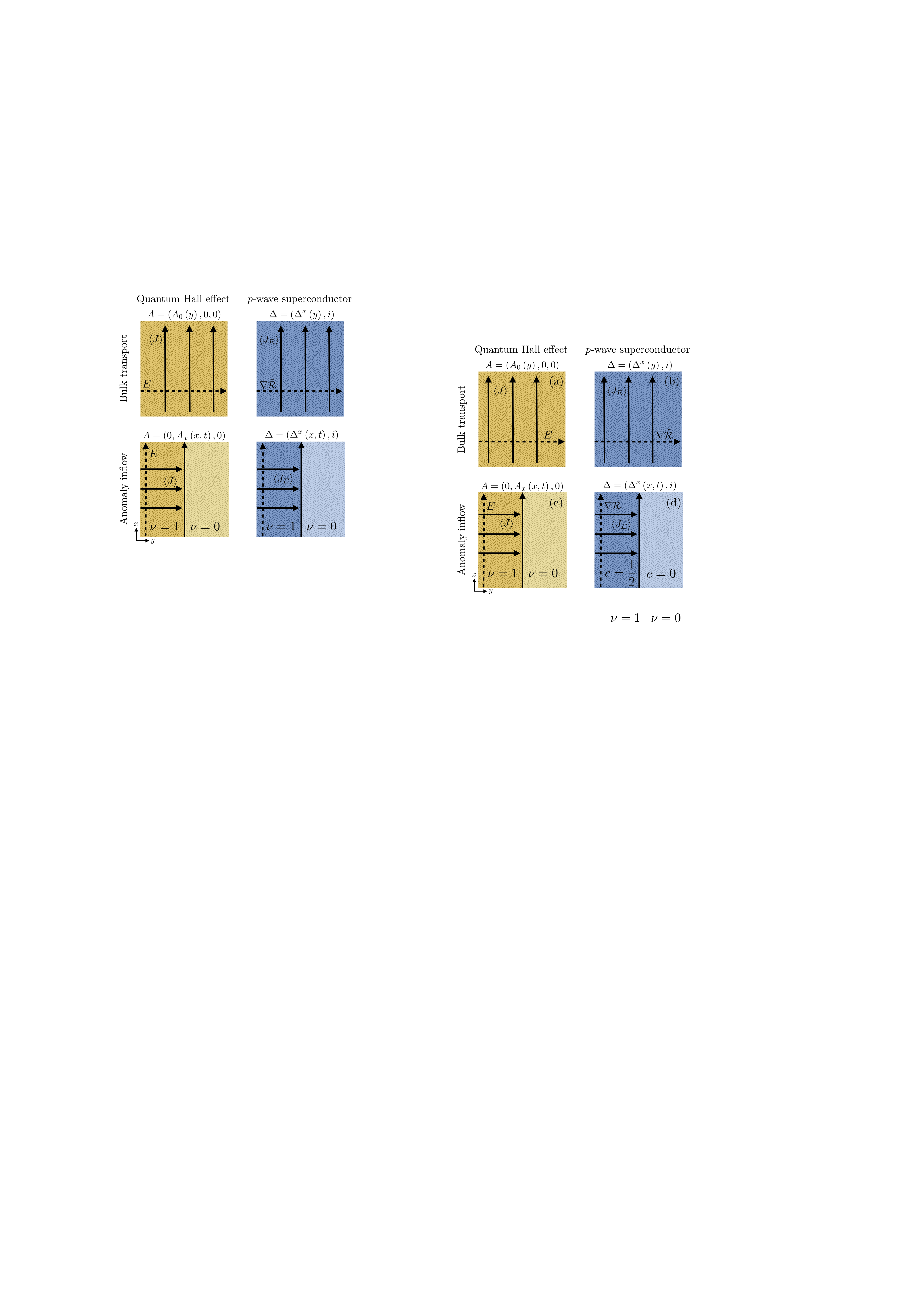}
\par\end{centering}
\caption{A comparison of the
integer quantum Hall effect (IQHE) and its energy-momentum analog in
$p$-wave CSF/Cs. (a) In the IQHE there is a perpendicular
electric current $\left\langle J\right\rangle $ in response to an
applied electric field $E$, with a quantized Hall conductivity, proportional
to the Chern number $\nu\in \mathbb{Z}$, as encoded in a $U(1)_N$ Chern-Simons term. (b)
In $p$-wave CSF/Cs, an energy current $\left\langle J_{E}\right\rangle $
flows in response to a space dependent order parameter $\Delta$,
as encoded in a gravitational Chern-Simons term. Derivatives of the
curvature $\tilde{\mathcal{R}}$ associated with $\Delta$ play the
role of the electric field in the IQHE, and $\left\langle J_{E}\right\rangle $
is perpendicular to the curvature gradient $\nabla\tilde{\mathcal{R}}$.
The ratio between the magnitudes of $\left\langle J_{E}\right\rangle $
and $\nabla\tilde{\mathcal{R}}$ is quantized, and proportional to
 the chiral central charge $c\in (1/2)\mathbb{Z}$. As described in the text, the spontaneous breaking
of $U(1)_N$ symmetry in $p$-wave CSF/Cs allows for a gravitational
\textit{pseudo} Chern-Simons term, encoding closely related
bulk responses, which are not topological in nature. (c) The quantized
Hall conductivity implies the existence of a chiral boundary fermion
with a $U\left(1\right)_N$ anomaly, which can be described as a Weyl
fermion at low energy. (d) The analogous response in  $p$-wave CSF/Cs
 implies the existence of a boundary chiral Majorana fermion with
a gravitational anomaly, which can be described as a Majorana-Weyl
fermion at low energy.\label{fig:A-comparison-of-1}}
\end{figure}

\subsubsection{Topological bulk responses from a gravitational Chern-Simons term \label{subsubsec:Topological bulk responses from a gravitational Chern-Simons term}}

Using the mapping of $p$-wave CSF/Cs to relativistic Majorana fermions
in Riemann-Cartan space-time, we compute and analyze in Sec.\ref{sec:Bulk-response} the effective
action obtained by integrating over the bulk fermions in the presence
of a general order parameter $\Delta$, and $U\left(1\right)_{N}$
gauge field. Here we discuss the main physical implications of this
effective action. As already explained in Sec.\ref{subsec:Chiral-superfluids-and},
we only describe UV-insensitive physics, which can be reliably understood
within the low energy relativistic description. This physics is controlled
by dimensionless coefficients, including the chiral central charge
$c$ in which we are primarily interested. 

As expected, the effective action contains a gCS term  \citep{Chern-Simons,jackiw2003chern,kraus2006holographic,witten2007three,perez2010conserved,stone2012gravitational}

\begin{align}
S_{\text{gCS}}= & \alpha\int\text{tr}\left(\tilde{\Gamma}\wedge\text{d}\tilde{\Gamma}+\frac{2}{3}\tilde{\Gamma}\wedge\tilde{\Gamma}\wedge\tilde{\Gamma}\right),
\end{align}
with  coefficient $\alpha=\frac{c}{96\pi}\in\frac{1}{192\pi}\mathbb{Z}$,
and where $\tilde{\Gamma}$ is the Christoffel symbol of the space-time
metric obtained from Eq.\eqref{eq:emergent metric}, see Sec.\ref{subsec:Effective-action-for}. Although we obtain
this result in the limit $m^{*}\rightarrow\infty$, we expect it to
hold throughout the phase diagram. This is based on known arguments
for the quantization of the coefficient $\alpha$ due to symmetry,
and on the relation with the boundary gravitational anomaly described
below.

The gCS term implies a topological bulk response, where energy-momentum currents and densities appear due to a space-time dependent order parameter. To see this in the simplest setting, assume that the order parameter is time independent, and that the relative phase is $\phi=\pm\frac{\pi}{2}$, as in the $p_{x}\pm ip_{y}$ configuration, so that $\Delta=e^{i\theta}\left(\left|\Delta^{x}\right|,\pm i\left|\Delta^{y}\right|\right)$, $o=\pm$. Then the metric is time independent, and takes the simple form  \begin{eqnarray}  &  & g^{ij}=-\begin{pmatrix}\left|\Delta^{x}\right|^{2} & 0\\ 0 & \left|\Delta^{y}\right|^{2} \end{pmatrix}.\label{eq:3-3} \end{eqnarray} On this background, we find the following contributions to the expectation values of the  fermionic energy current $J_{E}^{i}$, and momentum density $P_{i}$ \footnote{$P_{x}$ ($P_{y}$) is the density of the $x$ ($y$) component of momentum.},  \begin{eqnarray}   \left\langle J_{E}^{i}\right\rangle _{\text{gCS}}&=&-\frac{c}{96\pi}\frac{1}{\hbar}\varepsilon^{ij}\partial_{j}\tilde{\mathcal{R}},\label{eq:4}\\   \left\langle P_{i}\right\rangle _{\text{gCS}}&=&-\frac{c}{96\pi}\hbar g_{ik}\varepsilon^{kj}\partial_{j}\tilde{\mathcal{R}}.\nonumber  \end{eqnarray} Here
$\tilde{\mathcal{R}}$ is the curvature, or Ricci scalar, of the metric
$g_{ij}$, which is the inverse of $g^{ij}$, and $\varepsilon^{xy}=-\varepsilon^{yx}=1$.
 The curvature for the above metric is given explicitly by  
\begin{eqnarray}
 & \tilde{\mathcal{R}} & =-2\left|\Delta^{x}\right|\left|\Delta^{y}\right|\left(\partial_{y}\left(\frac{\left|\Delta^{y}\right|\partial_{y}\left|\ensuremath{\Delta}^{x}\right|}{\left|\ensuremath{\Delta}^{x}\right|^{2}}\right)+\partial_{x}\left(\frac{\left|\Delta^{x}\right|\partial_{x}\left|\ensuremath{\Delta}^{y}\right|}{\left|\ensuremath{\Delta}^{y}\right|^{2}}\right)\right).\label{eq:3-8-8}
\end{eqnarray}
It is a nonlinear expression in the order parameter, which is second
order in derivatives. Thus the responses \eqref{eq:4} are third order
in derivatives, and start at linear order but include nonlinear contributions
as well. The first equation in \eqref{eq:4} is analogous to the response
$\left\langle J^{i}\right\rangle =\frac{\nu}{2\pi}\varepsilon^{ij}E_{j}$
of the IQHE, see Fig.\ref{fig:A-comparison-of-1}(a),(b). The second
equation is analogous to the dual response $\left\langle \rho\right\rangle =\frac{\nu}{2\pi}B$.

\subsubsection{Additional bulk responses from a gravitational pseudo Chern-Simons term\label{subsubsec:Additional bulk responses from a gravitational pseudo}}

Apart from the gCS term, the effective action obtained by integrating
over the bulk fermions also contains an additional term of interest,
which we refer to as a gravitational \textit{pseudo} Chern-Simons
term (gpCS). To the best of our knowledge, the gpCS term has not appeared previously in the context of  $p$-wave CSF/Cs. It is possible because $U\left(1\right)_N$ symmetry is spontaneously broken in  $p$-wave CSF/Cs. In the geometric point of view, this translates to the emergent geometry in  $p$-wave CSF/Cs being not only curved but also torsion-full.
The gpCS term  produces bulk responses which are closely related to those of gCS, despite it being fully gauge invariant. This gauge invariance implies that it is not associated with a boundary anomaly, nor does its coefficient $\beta$ need to be quantized. Hence, gpCS does not encode \textit{topological} bulk responses. Remarkably, we find that $\beta$ is  quantized and identical to the coefficient $\alpha=\frac{c}{96\pi}$ of the gCS term in the limit of $m^*\rightarrow\infty$, but we do not expect this value to hold outside of this limit.  
Let us now describe the bulk responses from gpCS, setting $\beta=\frac{c}{96\pi}$. First, we find the following contributions to the fermionic energy current and momentum density, \begin{eqnarray}   \left\langle J_{E}^{i}\right\rangle _{\text{gpCS}}&=&\frac{c}{96\pi}\varepsilon^{ij}\partial_{j}\tilde{\mathcal{R}},\label{eq:11-1}\\   \left\langle P_{i}\right\rangle _{\text{gpCS}}&=&-\frac{c}{96\pi}g_{ik}\varepsilon^{kj}\partial_{j}\tilde{\mathcal{R}}.\nonumber  \end{eqnarray} Up to the sign difference in the first equation, these responses are the same as those from gCS \eqref{eq:4}. 
As opposed to gCS, the gpCS term also contributes to the fermionic charge density $\rho=-\psi^{\dagger}\psi$. For the bulk responses we have written thus far, every Majorana spinor contributed $c_{n}=\frac{o_{n}}{4}\text{sgn}\left(m_{n}\right)$, and summing over $n$ produced the central charge $c$. For the density response this is not the case. Here, the $n$th Majorana spinor contributes \begin{eqnarray}   \left\langle \rho\right\rangle _{\text{gpCS}}=\frac{o_{n}c_{n}}{24\pi}\sqrt{g}\tilde{\mathcal{R}},\label{eq:9-1} \end{eqnarray} where $\sqrt{g}=\sqrt{\text{det}g_{ij}}$ is the emergent volume element. The orientation $o_{n}$ in Eq. (\ref{eq:9-1}) makes the sum over the four Majorana spinors different from the central charge, $\sum_{n=1}^{4}o_{n}c_{n}=\sum_{n=1}^{4}\frac{1}{2}\text{sgn}\left(m_{n}\right)\neq c$.  The appearance of $o_{n}$ can be understood by considering the effect of time reversal, since both the density and curvature are time reversal even. The response \eqref{eq:9-1} also holds when the order parameter is time dependent, in which case $\tilde{\mathcal{R}}$ will also contain time derivatives. One then finds a time dependent density, but there is no corresponding current response, which is due to the breaking of $U(1)_N symmetry$.

To gain some insight into the expressions we have written thus far, we write the operators $P,J_{E}$ more explicitly. For each Majorana spinor (suppressing the index $n$),  \begin{eqnarray}   P_{j}&=&\frac{i}{2}\psi^{\dagger}\overleftrightarrow{D_{j}}\psi,\label{eq:10-5}\\   J_{E}^{j}&=&g^{jk}P_{k}+\frac{o}{2}\partial_{k}\left(\frac{1}{\sqrt{g}}\varepsilon^{jk}\rho\right)+O\left(\frac{1}{m^{*}}\right).\nonumber   \end{eqnarray}   The momentum density is the familiar expression for free fermions, but in the energy current we have only written explicitly contributions that survive the limit $m^{*}\rightarrow\infty$. These contributions are only possible due to the $p$-wave pairing, and are of order $\Delta^{2}$.
From the relation \eqref{eq:10-5} between $J_{E}$, $P$ and $\rho$ we can understand that the equality $\left\langle J_{E}^{j}\right\rangle _{\text{gCS}}=g^{jk}\left\langle P_{k}\right\rangle _{\text{gCS}}$ expressed in equation \eqref{eq:4} is a result of the vanishing contribution of gCS to the density $\rho$. We can also understand the sign difference between the first and second line of \eqref{eq:11-1} as a result of \eqref{eq:9-1}. The important point is that a measurement of the charge density $\rho$ can be used to fix the value of the coefficient $\beta$, which is generically unquantized,  and thus separate the contributions of gpCS to $P,J_{E}$, from those of gCS. In this manner, one can overcome the obscuring of gCS by gpCS.  

\subsubsection{Bulk-boundary correspondence from gravitational anomaly }

Among the two terms in the bulk effective action which we described
in Sec.\ref{subsubsec:Topological bulk responses from a gravitational Chern-Simons term}-\ref{subsubsec:Additional bulk responses from a gravitational pseudo}, only gCS is related to the boundary gravitational anomaly.
This relation can be explicitly analyzed in the case where $\Delta=\Delta_{0}e^{i\theta\left(t,x\right)}\left(1+f\left(x,t\right),i\right)$
is a perturbation of the $p_{x}+ip_{y}$ configuration with small
$f$, and there is a domain wall (or boundary) at $y=0$ where the
value of $c$ jumps. For simplicity, assume $c=1/2$ for $y<0$
and $c=0$ for $y>0$. This situation is illustrated in Fig.\ref{fig:A-comparison-of-1}(d).
In Appendix \ref{sec:Boundary-fermions-and} we derive the action for the boundary, or edge mode, 
\begin{align}
  S_{\text{e}}=\frac{i}{2}\int\mbox{d}t\text{d}x\tilde{\xi}\left(\partial_{t}-\left|\Delta^{x}\left(t,x\right)\right|\partial_{x}\right)\tilde{\xi},
\end{align}
which describes a chiral $D=1+1$ Majorana fermion $\tilde{\xi}$
localized on the boundary, with a space-time dependent velocity $\left|\Delta^{x}\left(x,t\right)\right|=\Delta_{0}\left|1+f\left(x,t\right)\right|$.
Classically, the edge fermion $\tilde{\xi}$ conserves energy-momentum
in the following sense, 
\begin{eqnarray}
  \partial_{\beta}t_{\mbox{e}\;\alpha}^{\beta}+\partial_{\alpha}\mathcal{L}_{\mbox{e}}=0.\label{eq:12-1}
\end{eqnarray}
Here $t_{\mbox{e}}$ is the canonical energy-momentum tensor for $\tilde{\xi}$,
with indices $\alpha,\beta=t,x$, and $\mathcal{L}_{\mbox{e}}$ is
the edge Lagrangian, $S_{\text{e}}=\int\text{d}t\mathcal{L}_{\mbox{e}}$, see Sec.\ref{subsec:Energy-momentum}.
For $\alpha=t$ ($\alpha=x$), Eq.\eqref{eq:12-1} describes
the sense in which the edge fermion conserves energy (momentum) classically.
The source term $\partial_{\alpha}\mathcal{L}_{\mbox{e}}$ 
follows from the space-time dependence of $\mathcal{L}_{\text{e}}$
through $\Delta^{x}$. Quantum mechanically, the action $S_{\text{e}}$
is known to have a gravitational anomaly, which means that energy-momentum
is not conserved at the quantum level \cite{bertlmann2000anomalies}.
In the context of emergent gravity, this implies that Eq.\eqref{eq:12-1}
is violated for the expectation values,  
\begin{align}
  \partial_{\beta}\left\langle t_{\mbox{e}\;\alpha}^{\beta}\right\rangle +\partial_{\alpha}\left\langle \mathcal{L}_{\mbox{e}}\right\rangle =-\frac{c}{96\pi}g_{\alpha\gamma}\varepsilon^{\gamma\beta y}\partial_{\beta}\tilde{\mathcal{R}}.\label{eq:12}
\end{align}
This equation is written with $\hbar=1$ and $c_{\mbox{light}}=\Delta_{0}/\hbar=1$
for simplicity. Since $\Delta^x$ depends on time, $\tilde{\mathcal{R}}$ is not the curvature
of the spatial metric $g_{ij}$, but of a corresponding space-time
metric $g_{\mu\nu}$ \eqref{eq:10-1}, and is given by $\tilde{\mathcal{R}}=\ddot{f}-2\dot{f}^{2}+O(f\ddot{f},f\dot{f}^{2})$
in this case. Note that time dependence in this example is crucial.
 From gCS we find for $\Delta=\Delta_{0}e^{i\theta\left(t,x\right)}\left(1+f\left(x,t\right),i\right)$
the bulk energy-momentum tensor
\begin{eqnarray}
 &  & \left\langle t_{\;\alpha}^{y}\right\rangle _{\text{gCS}}=-\frac{c}{96\pi}g_{\alpha\gamma}\varepsilon^{\gamma\beta y}\partial_{\beta}\tilde{\mathcal{R}},\label{eq:13b}
\end{eqnarray}
which explains the anomaly as the inflow of energy-momentum from the
bulk to the boundary,
\begin{eqnarray}
 &  & \partial_{\beta}\left\langle t_{\mbox{e}\;\alpha}^{\beta}\right\rangle +\partial_{\alpha}\left\langle \mathcal{L}_{\mbox{e}}\right\rangle =\left\langle t_{\;\alpha}^{y}\right\rangle _{\text{gCS}}.
\end{eqnarray}
Since $c$ jumps from 1/2 to 0 at $y=0$ the energy-momentum
current \eqref{eq:13b}  stops at the boundary and does not extend to the $y>0$ region. The gravitationally anomalous boundary
mode is then essential for the conservation of total energy-momentum to hold. As this example shows, 
bulk-boundary correspondence follows from bulk+boundary conservation
of energy-momentum in the presence of a space-time dependent order
parameter.

\subsection{Lattice model\label{sec:Lattice-model} }

In this section  we review and slightly generalize a simple lattice
model for a $p$-wave SC \cite{bernevig2013topological}, which will
serve as our microscopic starting point.   We describe its band structure
and its symmetry protected topological phases, and also  explain
some of the basics of the emergent geometry which can be seen in this
setting.

 The hamiltonian is given in real space by 
\begin{eqnarray}
  H=-\frac{1}{2}\sum_{\boldsymbol{l}}\left[t\psi_{\boldsymbol{l}}^{\dagger}\psi_{\boldsymbol{l}+x}+t\psi_{\boldsymbol{l}}^{\dagger}\psi_{\boldsymbol{l}+y}+\mu\psi_{\boldsymbol{l}}^{\dagger}\psi_{\boldsymbol{l}}\right.
  +\left.\delta^{x}\psi_{\boldsymbol{l}}^{\dagger}\psi_{\boldsymbol{l}+x}^{\dagger} 
  + \delta^{y}\psi_{\boldsymbol{l}}^{\dagger}\psi_{\boldsymbol{l}+y}^{\dagger}+h.c\right].\label{eq:2-1}
\end{eqnarray}
Here the sum is over all lattice sites $\boldsymbol{l}\in L$ of
a 2 dimensional square lattice $L=a\mathbb{Z}\times a\mathbb{Z}$,
with a lattice spacing $a$. $\psi_{\boldsymbol{l}}^{\dagger},\psi_{\boldsymbol{l}}$
are creation and annihilation operators for spin-less fermions on
the lattice, with the canonical anti commutators $\left\{ \psi_{\boldsymbol{l}}^{\dagger},\psi_{\boldsymbol{l}'}\right\} =\delta_{\boldsymbol{l}\boldsymbol{l}'}$.
$\boldsymbol{l}+x$ denotes the nearest neighboring site to $\boldsymbol{l}$
in the $x$ direction. The hopping amplitude $t$ is real and $\mu$
is the chemical potential. Apart from the single particle terms $t\psi_{\boldsymbol{l}}^{\dagger}\psi_{\boldsymbol{l}+x}+t\psi_{\boldsymbol{l}}^{\dagger}\psi_{\boldsymbol{l}+y}+\mu$,
there is also the pairing term $\delta^{x}\psi_{\boldsymbol{l}}^{\dagger}\psi_{\boldsymbol{l}+x}^{\dagger}+\delta^{y}\psi_{\boldsymbol{l}}^{\dagger}\psi_{\boldsymbol{l}+y}^{\dagger}$
, with the order parameter $\delta=\left(\delta^{x},\delta^{y}\right)\in\mathbb{\mathbb{C}}^{2}$.
We think of $\delta$ as resulting from a Hubbard-Stratonovich decoupling
of interactions, in which case we refer to it as intrinsic, or as
being induced by proximity to an $s$-wave SC. In both cases we treat
$\delta$ as a bosonic background field that couples to the fermions.

The generic order parameter is charged under a few  symmetries of
the single particle terms. The order parameter has charge 2 under
the global $U\left(1\right)$ group generated by $Q=-\sum_{\boldsymbol{l}}\psi_{\boldsymbol{l}}^{\dagger}\psi_{\boldsymbol{l}}$,
in the sense that $e^{-i\alpha Q}H\left(e^{2i\alpha}\delta\right)e^{i\alpha Q}=H\left(\delta\right)$,
which physically represents the electromagnetic charge $-2$ of Cooper
pairs\footnote{Since $\delta$ has charge 2, $H$ commutes with the fermion parity
$\left(-1\right)^{Q}$.  The Ground state of $H$ will therefore
be labelled by a fermion parity eigenvalue $\pm1$, in addition to
the topological label which is the Chern number \cite{read2000paired,kitaev2009periodic}.
Fermion parity is a subtle quantity in the thermodynamic limit, and
will not be important in the following.}.  The order parameter is also charged under time reversal $T$,
which is an anti unitary transformation satisfying $T^{2}=1$, that
acts as the complex conjugation of coefficients in the Fock basis
corresponding to $\psi_{\boldsymbol{l}},\psi_{\boldsymbol{l}}^{\dagger}$.
The equation $T^{-1}H\left(\delta^{*}\right)T=H\left(\delta\right)$
shows $\delta\mapsto\delta^{*}$ under time reversal. Finally, $\delta$
is also charged under the point group symmetry of the lattice, which
for the square lattice is the Dihedral group $D_{4}$. The continuum
analog of this is that the order parameter is charged under spatial
rotations and reflections, and more generally, under space-time transformations
(diffeomorphisms), which is due to the orbital angular momentum 1
of Cooper pairs in a $p$-wave SC. This observation will be important
for our analysis, and will be discussed further below. 

 In an intrinsic $p_{x}\pm ip_{y}$ SC, the configuration of $\delta$
which minimizes the ground state energy is given by $\delta=\delta_{0}e^{i\theta}\left(1,\pm i\right)$,
where $\delta_{0}>0$ is determined by the minimization, but the
sign $o=\pm1$ and the phase $\theta$ (which dynamically corresponds
to a goldstone mode) are left undetermined. See \cite{volovik2009universe}
for a pedagogical discussion of a closely related model within mean
field theory. A choice of $\theta$ and $o$ corresponds to a spontaneous
symmetry breaking of the group $U\left(1\right)\rtimes\left\{ 1,T\right\} $
including both the $U\left(1\right)$ and time reversal transformations.
More accurately, in the $p_{x}\pm ip_{y}$ SC, the group $\left(U\left(1\right)\rtimes\left\{ 1,T\right\} \right)\times D_{4}$
is spontaneously broken down to a certain diagonal subgroup. We discuss
the continuum analog of this and its implications in section \ref{subsec:Energy-momentum}. 

Crucially, we do not restrict $\delta$ to the $p_{x}\pm ip_{y}$
configuration, and treat it as a general two component complex vector
$\delta=\left(\delta^{x},\delta^{y}\right)\in\mathbb{\mathbb{C}}^{2}$.
In the following  we will take $\delta$ to be space time dependent,
$\delta\mapsto\delta_{\boldsymbol{l}}\left(t\right)$, and show that
this space time dependence can be thought of as a perturbation to
which there is a topological response, but for now we assume $\delta$
is constant. 

\subsubsection{\label{subsec:Band-structure-and}Band structure and phase diagram }

Writing the Hamiltonian \eqref{eq:2-1} in Fourier space, and in the
BdG form in terms of the Nambu spinor $\Psi_{\boldsymbol{q}}=\left(\psi_{\boldsymbol{q}},\psi_{-\boldsymbol{q}}^{\dagger}\right)^{T}$
we find
\begin{align}
  H & =\frac{1}{2}\int_{BZ}\frac{\mbox{d}^{2}\boldsymbol{q}}{\left(2\pi\right)^{2}}\Psi_{\boldsymbol{q}}^{\dagger}\begin{pmatrix}h_{\boldsymbol{q}} &  \delta_{\boldsymbol{q}}\\
\delta_{\boldsymbol{q}}^{*}  & -h_{\boldsymbol{q}}
\end{pmatrix}\Psi_{\boldsymbol{q}}+const\nonumber \\
   & =\frac{1}{2}\int_{BZ}\frac{\mbox{d}^{2}\boldsymbol{q}}{\left(2\pi\right)^{2}}\Psi_{\boldsymbol{q}}^{\dagger}\left(\boldsymbol{d}_{\boldsymbol{q}}\cdot\boldsymbol{\sigma}\right)\Psi_{\boldsymbol{q}}+const,\label{eq:3}
\end{align}
with $h_{\boldsymbol{q}}=-t\cos\left(aq_{x}\right)-t\cos\left(aq_{y}\right)-\mu$
real and symmetric, and $\delta_{\boldsymbol{q}}=-i\delta^{x}\sin\left(aq_{x}\right)-i\delta^{y}\sin\left(aq_{y}\right)$
complex and anti-symmetric.  Here $\boldsymbol{\sigma}=\left(\sigma^{x},\sigma^{y},\sigma^{z}\right)$
is the vector of Pauli matrices and $BZ$ is the Brillouin zone $BZ=\left(\mathbb{R}/\frac{2\pi}{a}\mathbb{Z}\right)^{2}$.
By definition, the Nambu spinor obeys the reality condition $\Psi_{\boldsymbol{q}}^{\dagger}=\left(\sigma^{x}\Psi_{-\boldsymbol{q}}\right)^{T}$,
and is therefore a Majorana spinor, see appendix \ref{subsec:Charge-conjugation-(Appendix)}.
Accordingly, the BdG Hamiltonian is particle-hole (or charge conjugation)
symmetric, $\sigma^{x}H\left(\boldsymbol{q}\right)^{*}\sigma^{x}=-H\left(-\boldsymbol{q}\right)$,
and therefore belongs to symmetry class D of the Altland-Zirnbauer
classification of free fermion Hamiltonians \cite{ryu2010topological}.
The constant in \eqref{eq:3} is $\frac{1}{2}\text{tr}h=\frac{V}{2}\int\frac{\text{d}^{2}\boldsymbol{q}}{\left(2\pi\right)^{2}}h_{\boldsymbol{q}}$
where $V$ is the infinite volume. This operator ordering correction
is important as it contributes to physical quantities such as the
energy density and charge density, but we will mostly keep it implicit
in the following. The BdG band structure is given by $E_{\boldsymbol{q},\pm}=\pm\frac{1}{2}E_{\boldsymbol{q}}$
where 
\begin{eqnarray}
 E_{\boldsymbol{q}}=\left|\boldsymbol{d}_{\boldsymbol{q}}\right|=\sqrt{h_{\boldsymbol{q}}^{2}+\left|\delta_{\boldsymbol{q}}\right|^{2}}.
\end{eqnarray}

\begin{figure}[!th]

\begin{centering}

\subfloat[]{
\includegraphics[width=0.35\columnwidth]{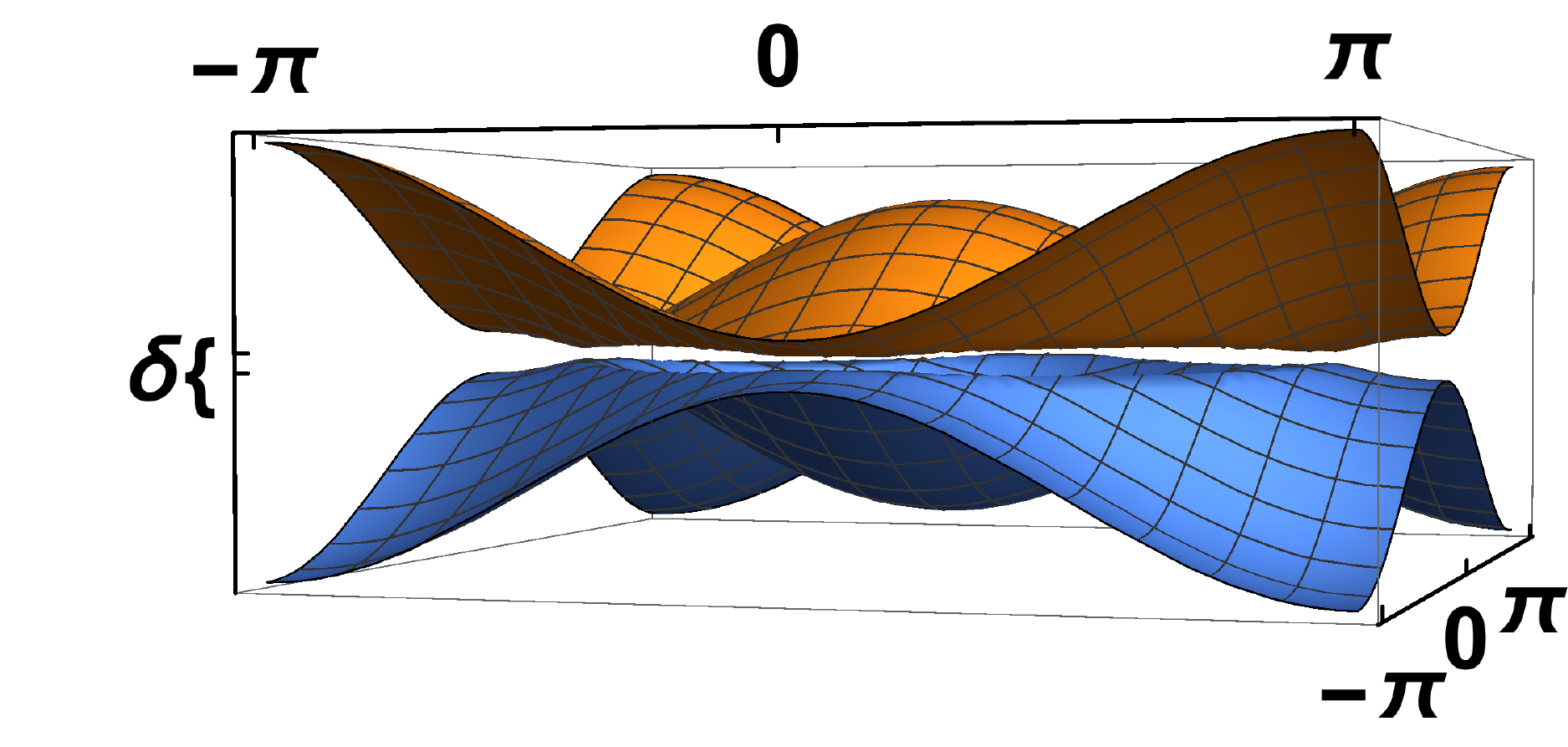}
} \subfloat[]{
\includegraphics[width=0.35\columnwidth]{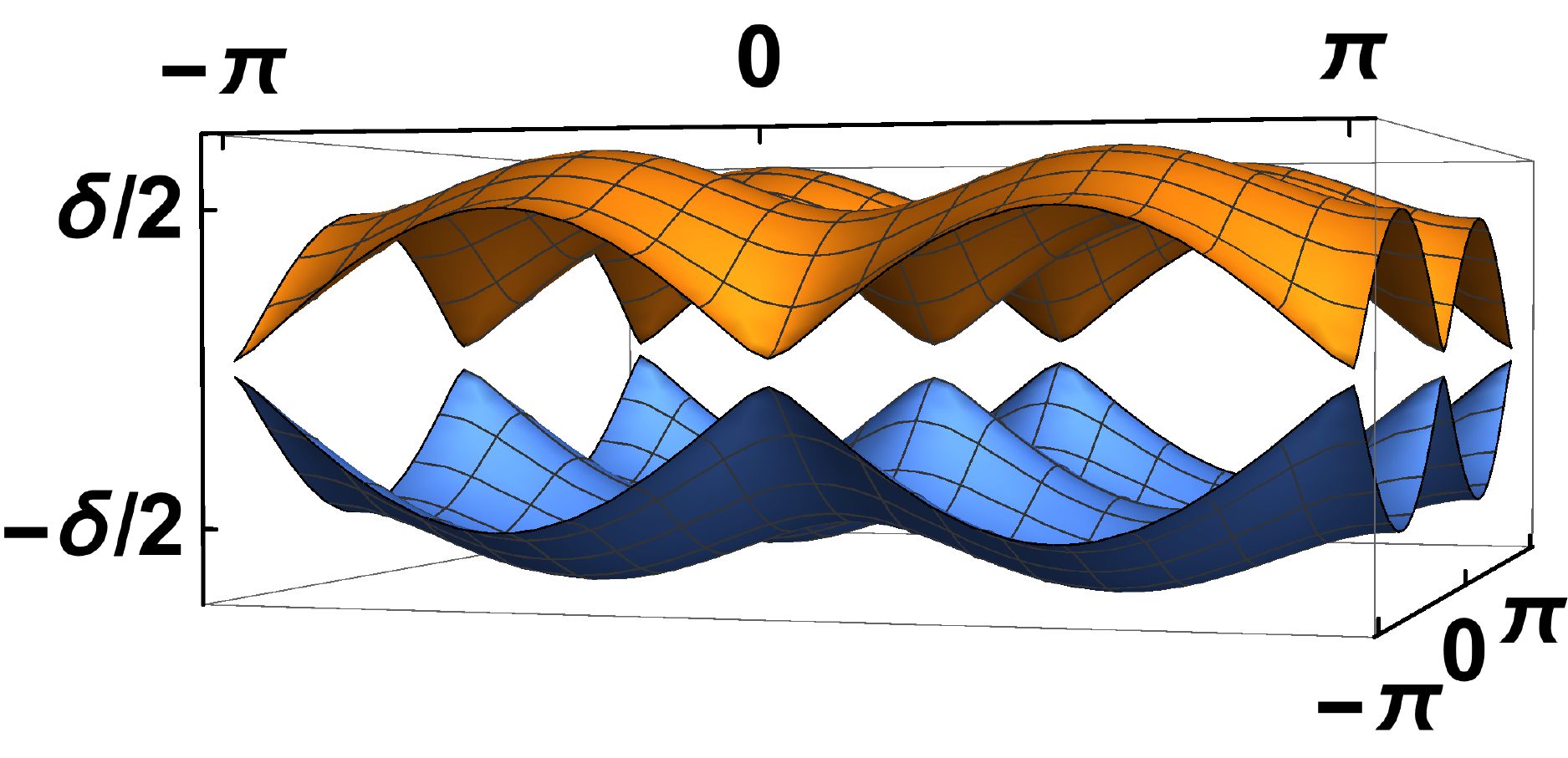}
}
\par\end{centering}
\caption{Generic band structure of the lattice model. (a) When the
order parameter is much smaller than the single particle bandwidth
$\delta\ll t$, the spectrum takes the form of a gapped single particle
Fermi surface with gap $\sim\delta$. This regime describes the onset
of superconductivity, and it is appropriate to refer to $\delta$
as the ``gap function''. (b) When the order parameter is
much larger than the single particle scales $\delta\gg t,\mu$, the
spectrum takes the form of four regulated relativistic fermions centered
at the particle-hole invariant points $\left(0,0\right),\left(0,\pi\right),\left(\pi,0\right),\left(\pi,\pi\right)$,
in units of the inverse lattice spacing $a^{-1}$. We will be working
in this regime. \label{fig:Generic-band-structure}}
\end{figure}

For the $p_{x}\pm ip_{y}$ configuration $\left|\delta_{\boldsymbol{q}}\right|^{2}=\delta_{0}^{2}\left(\sin^{2}aq_{x}+\sin^{2}aq_{y}\right)$,
and therefore $E_{\boldsymbol{q}}$ can only vanish at the particle-hole
invariant points $a\boldsymbol{K}^{\left(1\right)}=\left(0,0\right),a\boldsymbol{K}^{\left(2\right)}=\left(0,\pi\right),a\boldsymbol{K}^{\left(3\right)}=\left(\pi,\pi\right),a\boldsymbol{K}^{\left(4\right)}=\left(\pi,0\right)$,
which happens when $\mu=-2t,0,2t,0$. Representative band structures
are plotted in Fig.\ref{fig:Generic-band-structure}. For $\delta_{0}\ll t$
the spectrum takes the form of a gapped single particle Fermi surface
with gap $\sim\delta_{0}$, while for $\delta_{0}\gg t$ one obtains
Four regulated relativistic fermions centered at the points $\boldsymbol{K}^{\left(n\right)},\;1\leq n\leq4$
with masses $m_{n}=-2t-\mu,-\mu,2t-\mu,-\mu$, speed of light $c_{\text{light}}=a\delta_{0}/\hbar$,
bandwidth $\sim\delta_{0}$ and momentum cutoff $\sim a^{-1}$. 

\begin{figure}[!th]

\begin{centering}

 \subfloat[]{
\includegraphics[width=0.35\columnwidth]{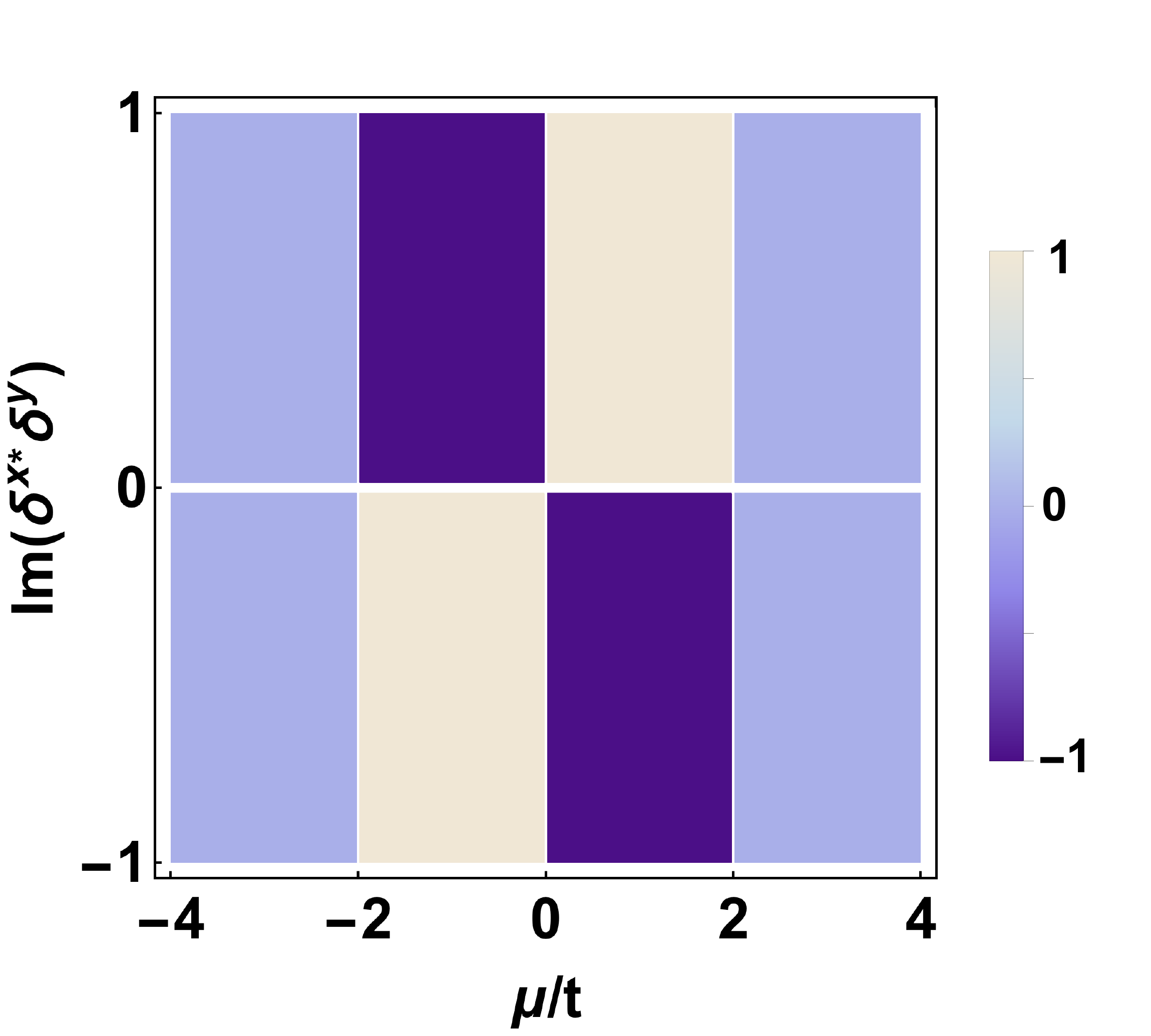}
} \subfloat[]{
\includegraphics[width=0.35\columnwidth]{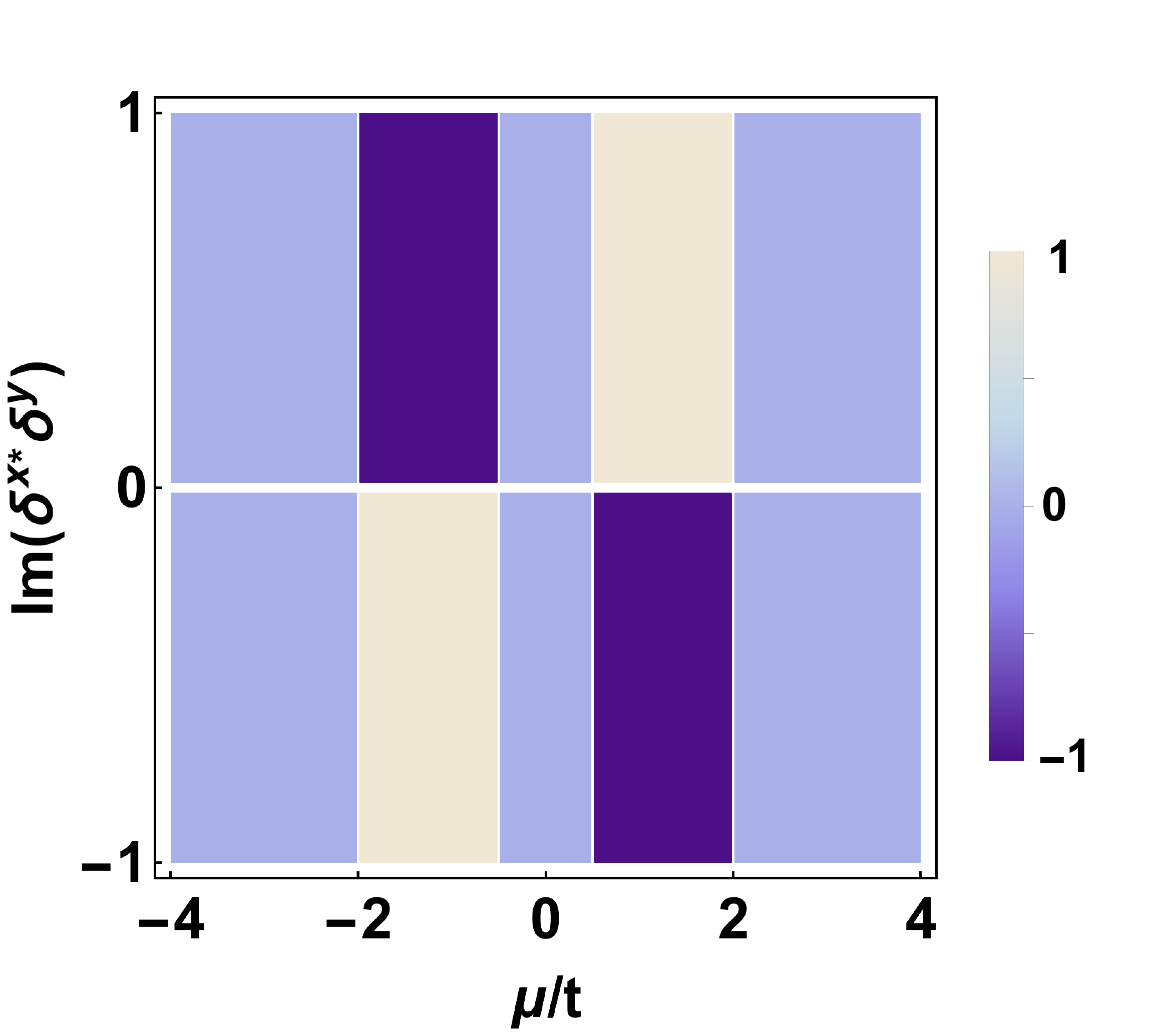}
}
\par\end{centering}

\caption{The topological phase diagram of the lattice model is simplest to
understand from the formula $\nu=\frac{1}{2}\sum_{n=1}^{4}o_{n}\text{sgn}\left(m_{n}\right)$
for the Chern number in terms of the masses and orientations of low energy relativistic
fermions. (a) Topological phase diagram for isotropic hopping
$t$. Units on the vertical axis are arbitrary, the topological phase
diagram only depends  on the orientation $o=\text{sgn}\left(\text{Im}\left(\delta^{x*}\delta^{y}\right)\right)$.
(b) Topological phase diagram for anisotropic hopping $t^{x}\protect\neq t^{y}$,
 additional trivial phases exist around $\mu=0$. Here $t=\frac{t^{x}+t^{y}}{2}$.
\label{fig:Phase-Diagram}}
\end{figure}

With generic $\mu,\delta_{0}$ the spectrum is gapped, and the Chern
number $\nu$ labeling the different topological phases is well defined. It can be calculated by $\nu=\int_{BZ}\frac{\text{d}^{2}k}{2\pi}\text{tr}\left(\mathcal{F}\right)$
where $\mathcal{F}$ is the Berry curvature on the Brillouin zone $BZ$ \cite{ryu2010topological}.
A more general definition is $\nu=\frac{1}{24\pi^{2}}\int_{\mathbb{R}\times BZ}\text{tr}\left(G\text{d}G^{-1}\right)^{3}$\footnote{More explicitly, $\nu=\frac{1}{24\pi^{2}}\mbox{tr}\int_{\mathbb{R}\times BZ}\mbox{d}^{3}k\varepsilon^{\alpha\beta\gamma}\left(G\partial_{\alpha}G^{-1}\right)\left(G\partial_{\beta}G^{-1}\right)\left(G\partial_{\gamma}G^{-1}\right)$.}, where 
$G\left(k_{0},k_{x},k_{y}\right)$ is the single particle propagator
\cite{volovik2009universe}, which remains valid in the presence of weak interactions, as long as the gap does not close. For two band Hamiltonians such as \eqref{eq:3}, $\nu$ reduces to the homotopy type of the map $\hat{\boldsymbol{d}_{\boldsymbol{q}}}=\boldsymbol{d}_{\boldsymbol{q}}/\left|\boldsymbol{d}_{\boldsymbol{q}}\right|$
from $BZ$ (which is a flat torus) to the sphere, 
\begin{align}
\nu=\frac{1}{\left(2\pi\right)^{2}}\int_{BZ}\text{d}^{2}\boldsymbol{q}\hat{\boldsymbol{d}_{k}}\cdot\left(\partial_{q_{y}}\hat{\boldsymbol{d}_{\boldsymbol{q}}}\times\partial_{q_{y}}\hat{\boldsymbol{d}_{\boldsymbol{q}}}\right)\in\mathbb{Z}.
\end{align}
One obtains $\nu=0$ for $\left|\mu\right|>2t$, $\nu=\pm1$ for $\mu\in\left(0,2t\right)$
and $\nu=\mp1$ for $\mu\in\left(-2t,0\right)$. The topological phase
diagram is plotted in Fig.\ref{fig:Phase-Diagram}(a). 

Away from the $p_{x}\pm ip_{y}$ configuration, the topological phase
diagram is essentially unchanged. For $\text{Im}\left(\delta^{x*}\delta^{y}\right)\neq0$,
 gap closings happen at the same points $\boldsymbol{K}^{\left(n\right)}$
and the same values of $\mu$ described above. $\nu$ takes the same
values, with the orientation $o=\text{sgn}\left(\text{Im}\left(\delta^{x*}\delta^{y}\right)\right)$,
described below, generalizing the sign $\pm1$ that characterizes
the $p_{x}\pm ip_{y}$ configuration. For $\text{Im}\left(\delta^{x*}\delta^{y}\right)=0$
the spectrum is always gapless. The topological phase diagram is most easily understood from the
formula $\nu=\frac{1}{2}\sum_{n=1}^{4}o_{n}\text{sgn}\left(m_{n}\right)$
where $o_{n}=\pm1$ are orientations associated with the relativistic
fermions which we describe below \cite{sticlet2012edge}. 

It will also be useful consider a slight generalization of the single
particle part of the lattice model, with un-isotropic hopping $t^{x}\psi_{\boldsymbol{l}}^{\dagger}\psi_{\boldsymbol{l}+x}+t^{y}\psi_{\boldsymbol{l}}^{\dagger}\psi_{\boldsymbol{l}+y}$.
This changes the masses to $m_{1}=-\left(t_{1}+t_{2}\right)-\mu,m_{2}=t_{1}-t_{2}-\mu,m_{3}=t_{1}+t_{2}-\mu,m_{4}=-\left(t_{1}-t_{2}\right)-\mu$.
In particular, the degeneracy between the masses $m_{2},m_{4}$ breaks,
and additional trivial phases appear around $\mu=0$. See Fig.\ref{fig:Phase-Diagram}(b).

\subsubsection{Basics of the emergent geometry \label{subsec:The-order-parameter}}

A key insight which we will extensively use, originally due to Volovik,
is that the order parameter is in fact a \textit{vielbein}. In the
present space-time independent situation, this vielbein is just a
$2\times2$ matrix which generically will be invertible
\begin{eqnarray}
 &  & e_{A}^{\;\;j}=\left(\begin{array}{cc}
\mbox{Re}(\delta^{x}) & \mbox{Re}(\delta^{y})\\
\mbox{Im}(\delta^{x}) & \mbox{Im}(\delta^{y})
\end{array}\right)\in GL\left(2\right),\label{eq:5-1}
\end{eqnarray}
where $A=1,2,\;j=x,y$. More accurately, $e_{A}^{\;\;j}$ is invertible if $\text{det}\left(e_{A}^{\;\;i}\right)=\text{Im}\left(\delta^{x*}\delta^{y}\right)\neq0$.
We refer to an order parameter as singular if $\text{Im}\left(\delta^{x*}\delta^{y}\right)=0$.
From the vielbein one can calculate a metric, which in the present
situation is a general symmetric positive semidefinite matrix 
\begin{align}
 g^{ij}=e_{A}^{\;\;i}\delta^{AB}e_{B}^{\;\;j}=\delta^{(i}\delta^{j)*}\label{eq:6-2}
 =\left(\begin{array}{cc}
\left|\delta^{x}\right|^{2}  \mbox{Re}\left(\delta^{x}\delta^{y*}\right)\\
\mbox{Re}\left(\delta^{x}\delta^{y*}\right) & \left|\delta^{y}\right|^{2}
\end{array}\right).\nonumber
\end{align}
 Every vielbein determines a metric uniquely, but the converse is
not true. Vielbeins $e,\tilde{e}$ that are related by an internal
reflection and rotation $e_{A}^{\;j}=\tilde{e}{}_{B}^{\;\;j}L_{\;A}^{B}$
with $L\in O\left(2\right)$ give rise to the same metric. By diagonalization,
it is also clear that any metric can be written in terms of a vielbein.
Therefore the set of (constant) metrics can be parameterized by the
coset $GL\left(2\right)/O\left(2\right)$. To see this explicitly
we parameterize $\delta=e^{i\theta}\left(\left|\delta^{x}\right|,e^{i\phi}\left|\delta^{y}\right|\right)$
with the overall phase $\theta$ and relative phase $\phi\in\left(-\pi,\pi\right]$.
Then 
\begin{align}
  g^{ij}=\left(\begin{array}{cc}
\left|\delta^{x}\right|^{2} & \left|\delta^{x}\right|\left|\delta^{y}\right|\cos\phi\\
\left|\delta^{x}\right|\left|\delta^{y}\right|\cos\phi & \left|\delta^{y}\right|^{2}
\end{array}\right)
\end{align}
is independent of $\theta$ which parametrizes $SO\left(2\right)$
and $\text{sgn}\phi$ which parametrizes $O\left(2\right)/SO\left(2\right)$.
Note that the group $O\left(2\right)$ of internal rotations and reflections
is just $U\left(1\right)\rtimes\left\{ 1,T\right\} $ acting on $e_{A}^{\;\;j}$.
In more detail, $\delta\mapsto e^{2i\alpha}\delta$ (or $\delta\mapsto\delta^{*}$)
corresponds to $e_{A}^{\;\;i}\mapsto L_{\;A}^{B}e_{B}^{\;\;i}$ with

\begin{align} 
L=\begin{pmatrix}\cos2\alpha & \sin2\alpha\\
-\sin2\alpha & \cos2\alpha
\end{pmatrix} 
\left(\text{or } 
L=\begin{pmatrix}1 & 0\\
0 & -1 
\end{pmatrix}\right).
\end{align}

The internal reflections, corresponding to a reversal of time,
flip the \textit{orientation} of the vielbein $o=\text{sgn}\left(\text{det}\left(e_{A}^{\;\;i}\right)\right)=\text{sgn}\left(\text{Im}\left(\delta^{x*}\delta^{y}\right)\right)$\textit{,
}and therefore every quantity that depends on $o$ is time reversal
odd\textit{. }We will also refer to $o$ as the orientation of the
order parameter. An order parameter with a positive (negative) orientation
can be thought of as $p_{x}+ip_{y}$-like ($p_{x}-ip_{y}$-like). 

For the $p_{x}\pm ip_{y}$ configuration, $\delta=e^{i\theta}\delta_{0}\left(1,\pm i\right)$,
one obtains a scalar metric $g^{ij}=\delta_{0}\delta^{ij}$, independent
of the phase $\theta$ and the orientation $o=\pm$. We see that $\theta,o$
correspond precisely to the $O\left(2\right)=U\left(1\right)\rtimes\left\{ 1,T\right\} $
degrees of freedom of the vielbein to which the metric is blind to.
Thus the metric $g^{ij}$ corresponds to the Higgs part of the order
parameter, by which we mean the part of the order parameter on which
the ground state energy depends, in the intrinsic case. 

The fact that $U\left(1\right)$ transformations map to internal rotations
also appears naturally in the BdG formalism which we will use in the
following. Consider the Nambu spinor $\Psi=\left(\psi,\psi^{\dagger}\right)^{T}$.
It follows from the $U\left(1\right)$ action $\psi\mapsto e^{i\alpha}\psi$
that $\Psi\mapsto e^{i\alpha\sigma^{z}}\Psi$ where $\sigma^{z}$
is the Pauli matrix. We see that $U\left(1\right)$ acts on $\Psi$
as a spin rotation. Moreover, the fact that $\delta$ has charge $2$
while $\psi$ has charge 1 implies $e$ is an $SO\left(2\right)$
vector while $\Psi$ is a spinor. 


\subsubsection{Non-relativistic continuum limit \label{subsec:Coupling-the-Lattice}}

Consider the lattice model \eqref{eq:2-1}, with a general space time
dependent order parameter $\delta_{\boldsymbol{l}}=\left(\delta_{\boldsymbol{l}}^{x}\left(t\right),\delta_{\boldsymbol{l}}^{y}\left(t\right)\right)$,
and minimally coupled to electromagnetism, 
\begin{eqnarray}
  H=-\frac{1}{2}\sum_{\boldsymbol{l}}\left[t\psi_{\boldsymbol{l}}^{\dagger}e^{iA_{\boldsymbol{l},\boldsymbol{l}+x}}\psi_{\boldsymbol{l}+x}+\left(\mu_{\boldsymbol{l}}+A_{t,\boldsymbol{l}}\right)\psi_{\boldsymbol{l}}^{\dagger}\psi_{\boldsymbol{l}}\right.
 +\left.\delta_{\boldsymbol{l}}^{x}\psi_{\boldsymbol{l}}^{\dagger}e^{iA_{\boldsymbol{l},\boldsymbol{l}+x}}\psi_{\boldsymbol{l}+x}^{\dagger}+\left(x\leftrightarrow y\right)+h.c\right].\label{eq:8}
\end{eqnarray}

Here $A_{\boldsymbol{l},\boldsymbol{l}'},A_{t,\boldsymbol{l}}$ are the
components of a $U\left(1\right)$ gauge field describing background
electromagnetism, on the discrete space and continuous time. We will
work in the relativistic regime $\delta_{0}\gg t,\mu$ where $\delta_{0}$
is a characteristic scale for $\delta$. To obtain a continuum description,
we split $BZ$ into four quadrants $BZ=\cup_{n=1}^{4}BZ^{\left(n\right)}$
centered around the four points $\boldsymbol{K}^{\left(n\right)}$,
 and decompose the fermion operator $\psi_{\boldsymbol{l}}$ as a
sum $\psi_{\boldsymbol{l}}=\sum_{n=1}^{4}\psi_{\boldsymbol{l}}^{\left(n\right)}e^{i\boldsymbol{K}^{\left(n\right)}\cdot\boldsymbol{l}}$,
where $\psi_{\boldsymbol{l}}^{\left(n\right)}e^{i\boldsymbol{K}^{\left(n\right)}\cdot\boldsymbol{l}}$
has non zero Fourier modes only in $BZ^{\left(n\right)}$. Thus the
fermions $\psi^{\left(n\right)}$ all have non zero Fourier modes
only in $BZ^{\left(n\right)}-\boldsymbol{K}^{\left(n\right)}=\left[-\frac{\pi}{2a},\frac{\pi}{2a}\right]^{2}$.
This restriction of the quasi momenta provides the fermions $\psi^{\left(n\right)}$
with a \textit{physical} cutoff $\sim a^{-1}$, which will be important
when we compare results from the continuum description to the lattice
model. Assuming $\mu,\delta,A$ have small derivatives relative to
$a^{-1}$, the inter fermion terms in $H$ can be neglected and $H$
splits into a sum $H\approx\sum_{n=1}^{4}H^{\left(n\right)}$, with
$H^{\left(n\right)}$ a Hamiltonian for $\psi^{\left(n\right)}$.
We then expand the Hamiltonians $H^{\left(n\right)}$ in small $\psi^{\left(n\right)}$
derivatives relative to $a^{-1}$. The resulting Hamiltonian, focusing
on the point $\boldsymbol{K}^{\left(1\right)}=\left(0,0\right)$,
is the $p$-wave superfluid (SF) Hamiltonian 
\begin{eqnarray}
  H_{\text{SF}}=\int\text{d}^{2}x\left[\psi^{\dagger}\left( - \frac{D^{2}}{2m^{*}}+m-A_{t}\right)\psi\right.\label{eq:9}
  -\left.\left(\frac{1}{2}\psi^{\dagger}\Delta^{j}\partial_{j}\psi^{\dagger}+h.c\right)\right],
\end{eqnarray}
where the fermion field has been redefined such that $\left\{ \psi^{\dagger}\left(x\right),\psi\left(x'\right)\right\} =\delta^{\left(2\right)}\left(x-x'\right)$.
Here $D_{\mu}=\partial_{\mu}-iA_{\mu}$ is the $U\left(1\right)$-covariant
derivative, with the connection $A=A_{j}\text{d}x^{j}$ related to
$A_{\boldsymbol{l},\boldsymbol{l}'}$ by $A_{\boldsymbol{l},\boldsymbol{l}'}=\int_{\boldsymbol{l}}^{\boldsymbol{l}'}A$,
and $D^{2}=\delta^{ij}D_{i}D_{j}$ with $i,j=x,y$. Note the appearance
of the flat background spatial metric $\delta^{ij}$. The effective
mass is related to the hopping amplitude $1/m^{*}=a^{2}t$, and the
order parameter is $\Delta=a\delta$, so it is essentially the lattice
order parameter. The chemical potential for the $p$-wave SF is $-m$. The coupling to $A$ in the pairing term is lost,
since $\psi^{\dagger}\psi^{\dagger}=0$. For this reason it is a derivative
and not a covariant derivative that appears in $\psi^{\dagger}\Delta^{j}\partial_{j}\psi^{\dagger}$,
and one can verify that this term is gauge invariant. Moreover, due
to the anti-commutator $\left\{ \psi^{\dagger}\left(x\right),\psi^{\dagger}\left(y\right)\right\} =0$
any operator put between two $\psi^{\dagger}$s is anti-symmetrized,
and in particular $\psi^{\dagger}\Delta^{j}\partial_{j}\psi^{\dagger}=\psi^{\dagger}\frac{1}{2}\left\{ \Delta^{j},\partial_{j}\right\} \psi^{\dagger}$
where $\left\{ \Delta^{j},\partial_{j}\right\} $ is the anti-commutator
of differential operators. This Hamiltonian is essentially the one
considered in \cite{read2000paired} for the $p$-wave SF. The corresponding
action is the $p$-wave SF action 
\begin{eqnarray}
 S_{\text{SF}}\left[\psi,\Delta,A\right]=\int\text{d}^{2+1}x\left[\psi^{\dagger}\left(iD_{t}+\frac{D^{2}}{2m^{*}}-m\right)\psi\right.
 +\left.\left(\frac{1}{2}\psi^{\dagger}\Delta^{j}\partial_{j}\psi^{\dagger}+h.c\right)\right],\label{eq:10}
\end{eqnarray}
in which $\psi,\psi^{\dagger}$ are no longer fermion operators, but
independent Grassmann valued fields, $\left\{ \psi\left(x\right),\psi^{\dagger}\left(x'\right)\right\} =0$.
This action comes equipped with a momentum cutoff $\Lambda_{UV}\sim a^{-1}$
inherited from the lattice model. 

For the other points $\boldsymbol{K}^{\left(2\right)},\boldsymbol{K}^{\left(3\right)},\boldsymbol{K}^{\left(4\right)}$
the SF action obtained is slightly different. The chemical potential for the $n$th fermion is $-m_{n}$.The order parameter
for the $n$th fermion is $\Delta_{\left(n\right)}^{x}=a\delta^{x}e^{iK_{x}^{\left(n\right)}},\;\Delta_{\left(n\right)}^{y}=a\delta^{y}e^{iK_{y}^{\left(n\right)}}$,
and we note that $e^{iK_{j}^{\left(n\right)}}=\pm1$. The order parameters
for $\boldsymbol{K}^{\left(1\right)}=\left(0,0\right),\;\boldsymbol{K}^{\left(3\right)}=\left(\pi,\pi\right)$
are related by an overall sign, which is a $U\left(1\right)$ transformation,
and so are the order parameters for $\boldsymbol{K}^{\left(2\right)}=\left(0,\pi\right),\boldsymbol{K}^{\left(4\right)}=\left(\pi,0\right)$.
Thus the order parameters for $n=1,3$ are physically indistinguishable,
and so are order parameters for $n=2,4$. The order parameters for
$n=1$ and $n=2$ are however physically distinct. First, the orientations
$o_{n}=\text{sgn}\left(\text{Im}\left(\Delta_{\left(n\right)}^{x*}\Delta_{\left(n\right)}^{y}\right)\right)$
are different, with $o_{1}=-o_{2}$. Second, the metrics $g_{\left(n\right)}^{ij}=\Delta^{(i}\Delta^{j)*}$
are generically different, with the same diagonal components, but
$g_{\left(1\right)}^{xy}=-g_{\left(2\right)}^{xy}$. We note that
if the relative phase between $\delta^{x}$ and $\delta^{y}$ is $\pm\pi/2$,
as in the $p_{x}\pm ip_{y}$ configuration, then all metrics $g_{\left(n\right)}^{ij}$
are diagonal and therefore equal. These differences between the orientations
and metrics of the different lattice fermions will be important later
on. 

Similarly, the effective mass tensor which in \eqref{eq:9}, for $n=1$,
is $\left(M^{-1}\right)^{ij}=\frac{\delta^{ij}}{m^{*}}$, has different
signatures for different $n$, but this will not be important for our analysis. For now we continue working with the action \eqref{eq:10}
for the $n=1$ fermion, keeping the other lattice fermions implicit
until section \ref{subsec:Summing-over-lattice}.

\subsubsection{Relativistic  continuum limit \label{subsec:Relativistic-limit-of}}

Since we work in the relativistic regime $\delta\gg t,\mu$ we can
treat the term $\psi^{\dagger}\frac{D^{2}}{2m^{*}}\psi$ as a perturbation
and compute quantities to zeroth order in $1/m^{*}$. Then $S_{\text{SF}}$
reduces to what we refer to as the relativistic limit of the $p$-wave
SF action, given in BdG form by 
\begin{eqnarray}
 S_{\text{rSF}}\left[\psi,\Delta,A\right]\label{eq:14-0}
 =\frac{1}{2}\int\mbox{d}^{2+1}x\Psi^{\dagger}\begin{pmatrix}i\partial_{t}+A_{t}-m  \frac{1}{2}\left\{ \Delta^{j},\partial_{j}\right\} \\
-\frac{1}{2}\left\{ \Delta^{*j},\partial_{j}\right\}   i\partial_{t}-A_{t}+m
\end{pmatrix}\Psi.
\end{eqnarray}
It is well known that when $\Delta$ takes the $p_{x}\pm ip_{y}$
configuration $\Delta=\Delta_{0}e^{i\theta}\left(1,\pm i\right)$
and $A=0$ this action is that of a relativistic Majorana spinor in
Minkowski space-time, with mass $m$ and speed of light $c_{\text{light}}=\frac{\Delta_{0}}{\hbar}$.
In the following, we will see that for general $\Delta$ and $A$, 
\eqref{eq:14-0} is the action of a relativistic Majorana spinor in
curved and torsion-full space-time. We wil sometimes refer to the relativistic
limit as $m^{*}\rightarrow\infty$, though this is somewhat loose,
because in the relativistic regime both $m^{*}$ is large and $m$
is small.

Before we go on to analyze the $p$-wave SF in the relativistic limit,
it is worth considering what of the physics of the $p$-wave SF is
captured by the relativistic limit, and what is not. First, the coupling
to $A_{x},A_{y}$ is lost, so the relativistic limit is blind to the
magnetic field. Since superconductors are usually defined by their
interaction with the magnetic field, the relativistic limit is actually
insufficient to describe the properties of the $p$-wave SF as a superconductor.
Of course, a treatment of superconductivity also requires the dynamics
of $\Delta$. Likewise, the term $\frac{1}{2m^{*}}\psi^{\dagger}D^{2}\psi=\frac{1}{2m^{*}}\psi^{\dagger}\delta^{ij}D_{i}D_{j}\psi$
seems to be the only term in $S_{\text{SF}}$ that includes the flat
background metric $\delta^{ij}$, describing the real geometry of
space. It appears that the relativistic limit is insufficient to describe
the response of the system to a change in the real geometry of space\footnote{In fact, some of the response to the real geometry can be obtained,
see our discussion, section \ref{sec:Conclusion-and-discussion}.}. Nevertheless, as is well known, the relativistic limit does suffice
to determine the topological phases of the $p$-wave SC as a free
(and weakly interacting) fermion system. Indeed, the Chern number
labeling the different topological phases can be calculated by the
formula $\nu=\frac{1}{2}\sum_{n=1}^{4}o_{n}\text{sgn}\left(m_{n}\right)$,
which only uses data from the relativistic limit. Here the sum is
over the four particle-hole invariant points of the lattice model,
with orientations $o_{n}$ and masses $m_{n}$. This suggests that
at least some physical properties characterizing the different free
fermion topological phases can be obtained from the relativistic limit.
Indeed, in the following we will see how a topological bulk response
and a corresponding boundary anomaly can be obtained within
the relativistic limit.

\subsection{Emergent Riemann-Cartan geometry\label{sec:Emergent-Riemann-Cartan-geometry}}

We argue that \eqref{eq:14-0} is precisely the action which describes
a relativistic massive Majorana spinor in a curved and torsion-full
background known as Riemann-Cartan (RC) geometry, with a particular
form of background fields. We refer the reader to \cite{ortin2004gravity}
parts I.1 and I.4.4, for a review of RC geometry and the coupling
of fermions to it, and provide only the necessary details here, focusing
on the implications for the $p$-wave SF. For simplicity we work locally
and in coordinates, and we differ the treatment of global aspects
to appendix \ref{subsec:Global-structures-and}. 

The action describing the dynamics of a Majorana spinor on RC background
in 2+1 dimensional space-time can be written as 
\begin{eqnarray}
 S_{\text{RC}}\left[\chi,e,\omega\right]\label{eq:43-1}
 =\frac{1}{2}\int\mbox{d}^{2+1}x\left|e\right|\overline{\chi}\left[\frac{i}{2}e_{a}^{\;\mu}\left(\gamma^{a}D_{\mu}-\overleftarrow{D_{\mu}}\gamma^{a}\right)-m\right]\chi.
\end{eqnarray}
 Here $\chi$ is a Majorana spinor with mass $m$  obeying, as a
field operator, the canonical anti-commutation relation $\left\{ \chi\left(x\right),\chi\left(y\right)\right\} =\frac{\delta^{\left(2\right)}\left(x-y\right)}{\left|e\left(x\right)\right|}$,
where we suppressed spinor indices. As a Grassmann field $\left\{ \chi\left(x\right),\chi\left(y\right)\right\} =0$.
The field $e_{a}^{\;\mu}$ is an inverse vielbein which is an invertible
matrix at each point in space-time. The indices $a,b,\dots\in\left\{ 0,1,2\right\} $
are $SO\left(1,2\right)$ (Lorentz) indices which we refer to as
internal indices, while $\mu,\nu,\dots\in\left\{ t,x,y\right\} $
are coordinate indices.

We will also use $A,B,\dots\in\left\{ 1,2\right\} $ for spatial internal
indices and $i,j,\dots\in\left\{ x,y\right\} $ for spatial coordinate
indices

The vielbein $e_{\;\mu}^{a}$, is the inverse of $e_{a}^{\;\mu}$,
such that $e_{\;\mu}^{a}e_{a}^{\;\nu}=\delta_{\mu}^{\nu},\;e_{\;\mu}^{a}e_{b}^{\;\mu}=\delta_{b}^{a}$.
It is often useful to view the vielbein as a set of linearly independent
(local) one-forms $e^{a}=e_{\;\mu}^{a}\text{d}x^{\mu}$. The metric
corresponding to the vielbein is $g_{\mu\nu}=e_{\;\mu}^{a}\eta_{ab}e_{\;\nu}^{b}$
and the inverse metric is $g^{\mu\nu}=e_{a}^{\;\mu}\eta^{ab}e_{b}^{\;\nu}$,
where $\eta_{ab}=\eta^{ab}=\text{diag}\left[1,-1,-1\right]$ is the
flat Minkowski metric. Internal indices are raised and lowered using
$\eta$, while coordinate indices are raised and lowered using $g$
and its inverse. Using $e$ one can replace internal indices with
coordinate indices and vice versa, e.g $v^{a}=e_{\;\mu}^{a}v^{\mu}$.
The volume element is defined by $\left|e\right|=\left|\text{det}e_{\;\mu}^{a}\right|=\sqrt{g}$.
$\left\{ \gamma^{a}\right\} _{a=0}^{2}$ are gamma matrices obeying
$\left\{ \gamma^{a},\gamma^{b}\right\} =2\eta^{ab}$, and we will
work with $\gamma^{0}=\sigma^{z},\;\gamma^{1}=-i\sigma^{x},\;\gamma^{2}=i\sigma^{y}$\footnote{The gamma matrices form a basis for the Clifford algebra  associated
with $\eta$. The above choice of basis is a matter of convention.
}. The covariant derivative $D_{\mu}=\partial_{\mu}+\omega_{\mu}$\footnote{We use the notation $D$ for spin, Lorentz, and $U\left(1\right)$
covariant derivatives in any representation, and the exact meaning
should be clear from the field $D$ acts on. } contains the spin connection $\omega_{\mu}=\frac{1}{2}\omega_{ab\mu}\Sigma^{ab}$,
where $\Sigma^{ab}=\frac{1}{4}\left[\gamma^{a},\gamma^{b}\right]$
generate the spin group $Spin\left(1,2\right)$ which is the double
cover of the Lorentz group $SO\left(1,2\right)$. Note that $\omega_{ab\mu}=-\omega_{ba\mu}$
and therefore $\omega_{\;b\mu}^{a}$ is an $SO\left(1,2\right)$ connection.
It follows that $\omega$ is metric compatible, $D_{\mu}\eta_{ab}=0$.
It is often useful to work (locally) with a connection one-form $\omega=\omega_{\mu}\text{d}x^{\mu}$.
$\overline{\chi}$ is the Dirac conjugate defined as in Minkowski
space-time $\overline{\chi}=\chi^{\dagger}\gamma^{0}$. The derivative
$\overleftarrow{D_{\mu}}$ acts only on $\overline{\chi}$ and is
explicitly given by $\chi\overleftarrow{D_{\mu}}=\partial_{\mu}\overline{\chi}-\overline{\chi}\omega_{\mu}$.

Our statement is that $S_{\text{RC}}\left[\chi,e,\omega\right]$ evaluated
on the fields 
\begin{eqnarray}
 \chi=\left|e\right|^{-1/2}\Psi,\hspace{7bp}\label{17}
 e_{a}^{\;\mu}=\left(\begin{array}{ccc}
1 & 0 & 0\\
0 & \mbox{Re}(\Delta^{x}) & \mbox{Re}(\Delta^{y})\\
0 & \mbox{Im}(\Delta^{x}) & \text{Im}(\Delta^{y})
\end{array}\right),\hspace{7bp}\omega_{\mu}=-2A_{\mu}\Sigma^{12},
\end{eqnarray}
reduces precisely to $S_{\text{rSF}}\left[\psi,\Delta,A\right]$ of
equation \eqref{eq:14-0}, where one must keep in mind that $S_{\text{RC}}$
is written in relativistic units where $\hbar=1$ and $c_{\text{light}}=\Delta_{0}/\hbar=1$,
which we will use in the following. Moreover, the functional integral
over $\chi$ is equal to the functional integral over $\Psi$. This
refines the original statement by Volovik  and
subsequent work by Read and Green \cite{read2000paired}. We defer
the proof to appendices \ref{subsec:Equivalent-forms-of} and \ref{subsec:Equality-of-path},
where we also address certain subtleties that arise. Here we describe
the particular RC geometry that follows from \eqref{17}, and attempt
to provide some intuition for this geometric description of the $p$-wave
SF.  Starting with the vielbein, note that the only nontrivial part
of $e_{a}^{\;\mu}$ is the spatial part $e_{A}^{\;j}$, which is just
the order parameter $\Delta$, as in \eqref{eq:5-1}. The inverse
metric we obtain from our vielbein is 
\begin{eqnarray}
 g^{\mu\nu}=e_{a}^{\;\mu}\eta^{ab}e_{b}^{\;\nu}\label{eq:10-1}
 =\left(\begin{array}{ccc}
1 & 0 & 0\\
0 & -\left|\Delta^{x}\right|^{2} & -\mbox{Re}\left(\Delta^{x}\Delta^{*y}\right)\\
0 & -\mbox{Re}\left(\Delta^{x}\Delta^{*y}\right) & -\left|\Delta^{y}\right|^{2}
\end{array}\right),
\end{eqnarray}
where the spatial part $g^{ij}=-\Delta^{(i}\Delta^{j)*}$
is the Higgs part of the order parameter, as in \eqref{eq:6-2}. For
the $p_{x}\pm ip_{y}$ configuration the metric reduces to the Minkowski
metric. If $\Delta$ is time independent $g^{\mu\nu}$ describes a
Riemannian geometry which is trivial in the time direction, but we
allow for a time dependent $\Delta$. A metric of the form \eqref{eq:10-1}
is said to be in gaussian normal coordinates with respect to space
\cite{carroll2004spacetime}. 

The $U\left(1\right)$ connection $A_{\mu}$ maps to a $Spin\left(2\right)$
connection $\omega_{\mu}=-2A_{\mu}\Sigma^{12}=-iA_{\mu}\sigma^{z}$
 which corresponds to spatial spin rotations. This is a special case
of the general $Spin\left(1,2\right)$ connection which appears in
RC geometry. The fact that $U\left(1\right)$ transformations map
to spin rotations when acting on the Nambu spinor $\Psi$ is a general
feature of the BdG formalism as was already discussed in section \ref{subsec:The-order-parameter}.
 From the spin connection $\omega$ it is natural to construct a
curvature, which is a matrix valued two-form defined by $R_{\;b}^{a}=\text{d}\omega_{\;b}^{a}+\omega_{\;c}^{a}\wedge\omega_{\;b}^{c}$.
In local coordinates $x^{\mu}$ it can be written as $R_{\;b}^{a}=\frac{1}{2}R_{\;b\mu\nu}^{a}\text{d}x^{\mu}\wedge\text{d}x^{\nu}$,
where the components are given explicitly by $R_{\;b\mu\nu}^{a}=\partial_{\mu}\omega_{\;b\nu}^{a}-\partial_{\nu}\omega_{\;b\mu}^{a}+\omega_{\;c\mu}^{a}\omega_{\;b\nu}^{c}-\omega_{\;c\nu}^{a}\omega_{\;b\mu}^{c}$.
It follows from \eqref{17} that in our case the only non zero components
are  
\begin{eqnarray}
 R_{12}=-R_{21}=-2F,
\end{eqnarray}
 where the two form $F=\text{d}A$ is the $U\left(1\right)$ field
strength, or curvature, comprised of the electric and magnetic fields.

\subsubsection{Torsion and additional geometric quantities}

Since we treat $A$ and $\Delta$ as independent background fields,
so are the spin connection $\omega$ and vielbein $e$. This situation
is referred to as the first order vielbein formalism for gravity \cite{ortin2004gravity}.
 Apart from the metric $g$ and the curvature $R$ which we already
described, there are a few more geometric quantities which can be
constructed from $e,\omega$, and that will be used in the following.
These additional quantities revolve around the notion of torsion.

The torsion tensor $T$ is an important geometrical quantity, but
a pragmatic way to view it is as a useful parameterization for the
set of all spin connections $\omega$, for a fixed vielbein $e$.
Thus one can work with the variables $e,T$ instead of $e,\omega$.
We will see later on that the bulk responses in the $p$-wave SC are
easier to describe using $e,T$. This is analogous to, and as we will
see, generalizes, the situation in $s$-wave SC, where the independent
degrees of freedom are $A$ and $\Delta=\left|\Delta\right|e^{i\theta}$,
but it is natural to change variables and work with $\Delta$ and
$D_{\mu}\theta=\partial_{\mu}\theta-2A_{\mu}$ instead. We now provide
the details. 

The torsion tensor, or two-form, is defined in terms of $e,\omega$
as $T^{a}=De^{a}$, or in coordinates $T_{\mu\nu}^{a}=2D_{[\mu}e_{\nu]}^{a}$.
Since our temporal vielbein $e^{0}=\text{d}t$ is trivial and the
connection $\omega$ is only an $SO\left(2\right)$ connection, $T^{0}=0$
for all $A$ and $\Delta$. All other components of the torsion are
in general non trivial, and are given by $T_{ij}^{A}=D_{i}e_{\;j}^{A}-D_{j}e_{\;i}^{A},\;T_{ti}^{A}=-T_{it}^{A}=D_{t}e_{\;i}^{A}$.
This describes the simple change of variables from $\omega$ to $T$. 

Going from $T$ back to $\omega$ is slightly more complicated, and
is done as follows. One starts by finding the $\omega$ that corresponds
to $T=0$. The solution is the unique torsion free spin connection
$\tilde{\omega}=\tilde{\omega}\left(e\right)$ which we refer to as
the Levi Civita (LC) spin connection\footnote{The unique torsion free spin connection $\tilde{\omega}$ is also
referred to as the Cartan connection is the literature.}. This connection is given explicitly by $\tilde{\omega}_{abc}=\frac{1}{2}\left(\xi_{abc}+\xi_{bca}-\xi_{cab}\right)$
where $\xi_{\;bc}^{a}=2e_{b}^{\;\mu}e_{c}^{\;\nu}\partial_{[\mu}e_{\;\nu]}^{a}$.
Now, for a general $\omega$ the difference $C_{\;b\mu}^{a}=\omega_{\;b\mu}^{a}-\tilde{\omega}_{\;b\mu}^{a}$
is referred to as the contorsion tensor, or one-form. It carries the
same information as $T$ and the two are related by $T^{a}=C_{\;b}^{a}\wedge e^{b}$
($T_{\mu\nu}^{a}=2C_{\;b[\mu}^{a}e_{\;\nu]}^{b}$) and $C_{\mu\alpha\nu}=\frac{1}{2}\left(T_{\alpha\mu\nu}+T_{\mu\nu\alpha}-T_{\nu\alpha\mu}\right)$.
One can then reconstruct $\omega$ from $e,T$ as $\omega=\tilde{\omega}\left(e\right)+C\left(e,T\right)$.
Note that $\omega,\tilde{\omega}$ are both connections, but $C,T$
are tensors. 

For the $p_{x}\pm ip_{y}$ configuration $\Delta=\Delta_{0}e^{i\theta}\left(1,\pm i\right)$
one finds $\tilde{\omega}_{12\mu}=-\tilde{\omega}_{21\mu}=-\partial_{\mu}\theta$
(with all other components vanishing), and it follows that $C_{12\mu}=D_{\mu}\theta$.
These are familiar quantities in the theory of superconductivity,
and one can view $\tilde{\omega}$ and $C$ as generalizations of
these. General formulas are given in appendix \ref{subsec:Explicit-formulas-for}.

Using $\tilde{\omega}$ one can define a covariant derivative $\tilde{D}$
and curvature $\tilde{R}$ just as $D$ and $R$ are constructed from
$\omega$. The quantity $\tilde{R}_{\;\nu\rho\sigma}^{\mu}$ is the
usual Riemann tensor of Riemannian geometry and general relativity.
Note that $\tilde{R}_{\;\nu\rho\sigma}^{\mu}$ depends solely on $g$
which is the Higgs part of the order parameter $\Delta$. Since $g$
is flat in the $p_{x}\pm ip_{y}$ configuration, we conclude that
a non vanishing Riemann tensor requires a deviation of the Higgs part
of $\Delta$ from the $p_{x}\pm ip_{y}$ configuration. As in Riemannian
geometry we can define the Ricci tensor $\tilde{\mathcal{R}}_{\nu\sigma}=\tilde{R}_{\;\nu\mu\sigma}^{\mu}$
and Ricci scalar $\tilde{\mathcal{R}}=\tilde{\mathcal{R}}_{\;\nu}^{\nu}$.
Examples for the calculation of $\tilde{\mathcal{R}}$ in terms of
$\Delta$ were given in section \ref{subsubsec:Topological bulk responses from a gravitational Chern-Simons term}.

Another important quantity which can be constructed from $e,\omega$
is the affine connection $\Gamma_{\;\beta\mu}^{\alpha}=e_{a}^{\;\alpha}\left(\partial_{\mu}e_{\;\beta}^{a}+\omega_{\;b\mu}^{a}e_{\;\beta}^{b}\right)=e_{a}^{\;\alpha}D_{\mu}e_{\;\beta}^{a}$,
or affine connection (local) one-form $\Gamma_{\;\beta}^{\alpha}=\Gamma_{\;\beta\mu}^{\alpha}\text{d}x^{\mu}$.
It is not difficult to see that $T$ is the anti symmetric part of
$\Gamma$, $T_{\;\mu\nu}^{\rho}=\Gamma_{\mu\nu}^{\rho}-\Gamma_{\nu\mu}^{\rho}$,
and it follows that the LC affine connection $\tilde{\Gamma}_{\;\beta\mu}^{\alpha}=e_{a}^{\;\alpha}\tilde{D}_{\mu}e_{\;\beta}^{a}$,
for which $T=0$, is symmetric in its the two lower indices. This
is the usual metric compatible and torsion free connection of Riemannian
geometry, given by the Christoffel symbol $\tilde{\Gamma}_{\alpha\beta\mu}=\frac{1}{2}\left(\partial_{\mu}g_{\beta\alpha}+\partial_{\beta}g_{\alpha\mu}-\partial_{\alpha}g_{\mu\beta}\right)$.
$\Gamma$ appears in covariant derivatives of tensors with coordinate
indices, for example $\nabla_{\mu}v^{\alpha}=\partial_{\mu}v^{\alpha}+\Gamma_{\;\beta\mu}^{\alpha}v^{\beta}$,
$\nabla_{\mu}v_{\alpha}=\partial_{\mu}v_{\alpha}-v_{\beta}\Gamma_{\;\alpha\mu}^{\beta}$,
and so on. We also denote by $\nabla$ the total covariant derivative
of tensors with both coordinate and internal indices, which includes
both $\omega$ and $\Gamma$. Thus, for example, $\nabla_{\mu}v_{\;\nu}^{a}=\partial_{\mu}v_{\;\nu}^{a}+\omega_{\;b\mu}^{a}v_{\;\nu}^{b}-v_{\;\nu}^{a}\Gamma_{\;\mu\alpha}^{\nu}=D_{\mu}v_{\;\nu}^{a}-v_{\;\nu}^{a}\Gamma_{\;\mu\alpha}^{\nu}$.
The most important occurrence of $\nabla$ is in the identity $\nabla_{\nu}e_{\;\mu}^{a}=0$,
which follows from the definition of $\Gamma$ in this formalism,
and is sometimes called the first vielbein postulate. It means that
the covariant derivative $\nabla$ commutes with index manipulation
preformed using $e,\eta$ and $g$. To obtain more intuition for what
$\Gamma$ is from the $p$-wave SC point of view, we can write it
as $\Gamma_{\;a\mu}^{\alpha}=-D_{\mu}e_{a}^{\;\alpha}$. Then it is
clear that the non vanishing components of $\Gamma_{\;a\mu}^{\alpha}$
are given by $\Gamma_{\;1\mu}^{j}+i\Gamma_{\;2\mu}^{j}=-D_{\mu}\Delta^{j}$

\subsection{Symmetries, currents, and conservation laws \label{sec:Symmetries,-currents,-and}}

In order to map fermionic observables in the $p$-wave SF to those
of a Majorana fermion in RC space-time, it is usefull is to map the symmetries and the
corresponding conservation laws between the two. We start with $S_{\text{SF}}$,
and then review the analysis of $S_{\text{RC}}$ and show how it maps
to that of $S_{\text{SF}}$, in the relativistic limit. The bottom
line is that there is a sense in which electric charge and energy-momentum
are conserved in a $p$-wave SC, and this maps to the sense in
which spin and energy-momentum are conserved for a Majorana spinor
in RC space-time.

\subsubsection{Symmetries, currents, and conservation laws of the $p$-wave superfluid action \label{subsec:Symmetries,-currents,-and}}

\paragraph{Electric charge }

$U\left(1\right)$ gauge transformations act on $\psi,\Delta,A$ by
\begin{align}
  \psi\mapsto e^{i\alpha}\psi,\;\Delta\mapsto e^{2i\alpha}\Delta,\;A_{\mu}\mapsto A_{\mu}+\partial_{\mu}\alpha.\label{eq:6.1}
\end{align}
This symmetry of  $S_{\text{SF}}\left[\psi,\Delta,A\right]$
implies a conservation law for electric charge, 
\begin{eqnarray}
  \partial_{\mu}J^{\mu}=-i\psi^{\dagger}\Delta^{j}\partial_{j}\psi^{\dagger}+h.c,\label{eq:15}
\end{eqnarray}
where $J^{\mu}=-\frac{\delta S}{\delta A_{\mu}}$ is the fermion
electric current. Since $A_{\mu}$ does not enter the pairing term,
$J^{\mu}$ is the same as in the normal state where $\Delta=0$, 
\begin{eqnarray}
  J^{t}=-\psi^{\dagger}\psi,\;J^{j}=-\frac{\delta^{jk}}{m^{*}}\frac{i}{2}\psi^{\dagger}\overleftrightarrow{D_{k}}\psi.\label{16}
\end{eqnarray}
Here $\psi^{\dagger}\overleftrightarrow{D_{k}}\psi=\psi^{\dagger}D_{k}\psi-\left(D_{k}\psi^{\dagger}\right)\psi$.
The conservation law \eqref{eq:15} shows that the fermionic charge alone is not conserved
due to the exchange of charge between the fermions $\psi$ and Cooper
pairs $\Delta$. If one adds a ($U\left(1\right)$ gauge invariant)
term $S'\left[\Delta,A\right]$ to the action and  considers $\Delta$
as a dynamical field, then it is possible to use the equation of motion
 $\frac{\delta\left(S'+S\right)}{\delta\Delta}=0$ for $\Delta$ and
the definition $J_{\Delta}^{\mu}=-\frac{\delta S'}{\delta A_{\mu}}$
of the Cooper pair current in order to rewrite \eqref{eq:15} as $\partial_{\mu}\left(J^{\mu}+J_{\Delta}^{\mu}\right)=0$.
This expresses the conservation of total charge in the $p$-wave SC.

\paragraph{Energy-momentum \label{subsec:Energy-momentum}}

Energy and momentum are at the heart of our analysis, and obtaining
the correct expressions for these quantities, as well as interpreting
correctly the conservation laws they satisfy, will be crucial. 

In flat space, one usually starts with the canonical energy-momentum
tensor. For a Lagrangian $\mathcal{L}\left(\phi,\partial\phi,x\right)$,
where $\phi$ is any fermionic of bosonic field, it is given by 
\begin{eqnarray}
 &  & t_{\;\nu}^{\mu}=\frac{\partial\mathcal{L}}{\partial\partial_{\mu}\phi}\partial_{\nu}\phi-\delta_{\nu}^{\mu}\mathcal{L},
\end{eqnarray}
and satisfies, on the equation of motion for $\phi$, 
\begin{eqnarray}
 &  & \partial_{\mu}t_{\;\nu}^{\mu}=-\partial_{\nu}\mathcal{L},\label{eq:18}
\end{eqnarray}
which can be obtained from Noether's first theorem for space-time
translations. Thus $t_{\;\nu}^{\mu}$ is conserved if and only if
the Lagrangian is independent of the coordinate $x^{\nu}$. This motivates
the identification of $t_{\;t}^{\mu}$ as the energy current, and
of $t_{\;j}^{\mu}$ as the current of the $j$th component of momentum
($j$-momentum). $t_{\;t}^{t}$ is just the Hamiltonian density, or
energy density, and $t_{\;j}^{t}$ is the $j$-momentum density. 

It is well known however, that the canonical energy-momentum tensor
may fail to be gauge invariant, symmetric in its indices, or traceless,
in situations where these properties are physically required, and
it is also sensitive to the addition of total derivatives to the Lagrangian.
To obtain the physical energy-momentum tensor one can either ``improve``
$t_{\;\nu}^{\mu}$ or appeal to a geometric (gravitational) definition
which directly provides the physical energy-momentum tensor \cite{ortin2004gravity,forger2004currents}.

We will comment on the coupling of the $p$-wave SF to a real background
geometry our discussion, section \ref{sec:Conclusion-and-discussion},
but here we fix the background geometry to be flat, and instead continue
by introducing the $U\left(1\right)$-covariant canonical energy-momentum
tensor. It can be shown to coincide with the physical energy-momentum
tensor obtained by coupling the $p$-wave SF to a real background
geometry. Since we work with a fixed flat background geometry in this
section, we will only consider space-time transformations which are
symmetries of this background, and it will suffice to consider space-time
translations and spatial rotations. 

The $U\left(1\right)$-covariant canonical energy-momentum tensor
is relevant in the following situation. Assume that the $x$ dependence
in $\mathcal{L}$ is only through a $U\left(1\right)$ gauge field
to which $\phi$ is minimally coupled, $\mathcal{L}\left(\phi,\partial\phi,x\right)=\mathcal{L}\left(\phi,D\phi\right)$.
Then, $t_{\;\nu}^{\mu}$ is not gauge invariant, and therefore physically
ambiguous. This is reflected in the conservation law \eqref{eq:18}
which takes the non covariant form 
\begin{eqnarray}
  \partial_{\mu}t_{\;\nu}^{\mu}=J^{\mu}\partial_{\nu}A_{\mu},\label{19}
\end{eqnarray}
where $J^{\mu}=-\frac{\partial\mathcal{L}}{\partial A_{\mu}}$ is
the $U\left(1\right)$ current. This lack of gauge invariance is to
be expected, as this conservation law follows from translational symmetry,
and translations do not commute with gauge transformations. Instead,
one should use $U\left(1\right)$-covariant space-time translations,
which are translations from $x$ to $x+a$ followed by a $U\left(1\right)$
parallel transport from $x+a$ back to $x$, $\phi\left(x\right)\mapsto e^{iq\int_{x-a}^{x}A}\phi\left(x-a\right)$
where $\phi\mapsto e^{iq\alpha}\phi$ under $U\left(1\right)$ and
the integral is over the straight line from $x-a$ to $a$. This is
still a symmetry because the additional $e^{iq\int_{x-a}^{x}A}$ is
just a gauge transformation. The conservation law that follows from
this modified action of translations is 
\begin{eqnarray}
  \partial_{\mu}t_{\text{cov}\;\nu}^{\mu}=F_{\nu\mu}J^{\mu},\label{eq:21}
\end{eqnarray}
where $F_{\mu\nu}=\partial_{\mu}A_{\nu}-\partial_{\nu}A_{\mu}$ is
the electromagnetic field strength, and 
\begin{eqnarray}
  t_{\text{cov}\;\nu}^{\mu}=\frac{\partial\mathcal{L}}{\partial D_{\mu}\phi}D_{\nu}\phi-\delta_{\nu}^{\mu}\mathcal{L}=t_{\;\nu}^{\mu}-J^{\mu}A_{\nu}\label{eq:32-00}
\end{eqnarray}
is the $U\left(1\right)$-covariant  version of $t_{\;\nu}^{\mu}$,
which we refer to as the $U\left(1\right)$-covariant canonical energy-momentum
tensor. The right hand side of \eqref{eq:21} is just the usual
Lorentz force, which acts as a source of $U\left(1\right)$-covariant
energy-momentum. We stress that the covariant and non covariant conservation
laws are equivalent, as can be verified by using the fact that $\partial_{\mu}J^{\mu}=0$
in this case. Both hold in any gauge, but in \eqref{eq:21} all quantities
are gauge invariant. 

 For the $p$-wave SF one obtains the $U\left(1\right)$-covariant
energy-momentum tensor 
\begin{eqnarray}
  t_{\text{cov}\;t}^{t}&=&\frac{i}{2}\psi^{\dagger}\overleftrightarrow{D_{t}}\psi-\mathcal{L}\label{25-2}\\
   &=&
  \frac{\delta^{ij}D_{i}\psi^{\dagger}D_{j}\psi}{2m^{*}}+m\psi^{\dagger}\psi-\left(\frac{1}{2}\psi^{\dagger}\Delta^{j}\partial_{j}\psi^{\dagger}+h.c\right),\nonumber\\
  t_{\text{cov}\;j}^{t}&=&\frac{i}{2}\psi^{\dagger}\overleftrightarrow{D_{j}}\psi,\nonumber \\
  t_{\text{cov}\;t}^{i}&=&-\frac{\delta^{ik}\left(D_{k}\psi\right)^{\dagger}D_{t}\psi}{2m^{*}}+\frac{1}{2}\psi^{\dagger}\Delta^{i}\partial_{t}\psi^{\dagger}+h.c,\nonumber \\
  t_{\text{cov}\;j}^{i}&=&-\frac{\delta^{ik}\left(D_{k}\psi\right)^{\dagger}D_{j}\psi}{2m^{*}}+\frac{1}{2}\psi^{\dagger}\Delta^{i}\partial_{j}\psi^{\dagger}+h.c-\delta_{j}^{i}\mathcal{L}.\nonumber 
\end{eqnarray}
 The $U\left(1\right)$-covariant conservation law is slightly
more complicated than \eqref{eq:21} due the additional background
field $\Delta$, 
\begin{align}
  \partial_{\mu}t_{\text{cov}\;\nu}^{\mu}=\frac{1}{2}\psi^{\dagger}\partial_{j}\psi^{\dagger}D_{\nu}\Delta^{j}+h.c+F_{\nu\mu}J^{\mu},\label{32}
\end{align}
where we have used the $U\left(1\right)$ conservation law \eqref{eq:15},
and $D_{\mu}\Delta^{j}=\left(\partial_{\mu}-2iA_{\mu}\right)\Delta^{j}$.
This conservation law shows that ($U\left(1\right)$-covariant) fermionic
energy-momentum is not conserved due to the exchange of energy-momentum
with the background fields $A,\Delta$. Apart from the Lorentz force
there is an additional source term due to the space-time dependence
of $\Delta$. 

 As in the case of the electric charge, if one considers $\Delta$
as a dynamical field and uses its equation of motion, \eqref{32}
can be written as\footnote{$t_{\Delta\;\text{cov}\;\nu}^{\mu}$ is the $U\left(1\right)$-covariant
energy-momentum tensor of Cooper pairs. It is defined by \eqref{eq:32-00}
with $\phi=\Delta$ and $\mathcal{L}=\mathcal{L}'\left(\Delta,\Delta^{*},D\Delta,D\Delta^{*}\right)$
being the (gauge invariant) term added to the $p$-wave SF Lagrangian.
Here it is important that the coupling of $\Delta$ to $\psi$ in
\eqref{eq:10} can be written without derivatives of $\Delta$. } 
\begin{align}
  \partial_{\mu}\left(t_{\text{cov}\;\nu}^{\mu}+t_{\Delta\;\text{cov}\;\nu}^{\mu}\right)=F_{\nu\mu}\left(J^{\mu}+J_{\Delta}^{\mu}\right),\label{33}
\end{align}
which is of the general form \eqref{eq:21}.

Note that the spatial part $t_{\text{cov}\;j}^{i}$ is not symmetric,
\begin{align}
  t_{\text{cov}\;y}^{x}-t_{\text{cov}\;x}^{y}=\frac{1}{2}\psi^{\dagger}\left(\Delta^{x}\partial_{y}-\Delta^{y}\partial_{x}\right)\psi^{\dagger}+h.c, \label{eq:28-0}
\end{align}
which physically represents an exchange of angular momentum between
$\Delta$ and $\psi$, possible because of the intrinsic angular momentum
of Cooper pairs in a $p$-wave SC. Explicitly, the ($U\left(1\right)$-covariant)
angular momentum current is given by $J_{\varphi}^{\mu}=t_{\text{cov}\;\varphi}^{\mu}=t_{\text{cov}\;\nu}^{\mu}\zeta^{\nu}$
where $\zeta=x\partial_{y}-y\partial_{x}=\partial_{\varphi}$ is the
generator of spatial rotations around $x=y=0$, and $\varphi$ is
the polar angle. From \eqref{32} and \eqref{eq:28-0} we find its conservation
law
\begin{align}
  \partial_{\mu}J_{\varphi}^{\mu}=\left(\frac{1}{2}\psi^{\dagger}\partial_{j}\psi^{\dagger}D_{\varphi}\Delta^{j}+h.c+F_{\varphi\mu}J^{\mu}\right)\label{eq:36}\\
  +\frac{1}{2}\psi^{\dagger}\left(\Delta^{x}\partial_{y}-\Delta^{y}\partial_{x}\right)\psi^{\dagger}+h.c,\nonumber 
\end{align}
which shows that even when the Lorentz force in the $\varphi$ direction
vanishes and $\Delta$ is ($U\left(1\right)$-covariantly) constant
in the $\varphi$ direction, $\Delta$ still acts a source for fermionic
angular momentum, due to the last term. 

Even though fermionic angular momentum is never strictly conserved
in a $p$-wave SF, it is well known that a certain combination of
fermionic charge and fermionic angular momentum can be strictly conserved
\cite{shitade2014bulk,tada2015orbital,volovik2015orbital,shitade2015orbital}.
Indeed, using \eqref{eq:36} and \eqref{eq:15},
\begin{align}
  &\partial_{\mu}\left(J_{\varphi}^{\mu}\mp\frac{1}{2}J^{\mu}\right)=\left(\frac{1}{2}\psi^{\dagger}\partial_{j}\psi^{\dagger}D_{\varphi}\Delta^{j}+h.c+F_{\varphi\mu}J^{\mu}\right)\nonumber \\
 & \pm\frac{i}{2}\psi^{\dagger}\left(\Delta^{x}\pm i\Delta^{y}\right)\left(\partial_{x}\mp i\partial_{y}\right)\psi^{\dagger}+h.c.
\end{align}
We see that when $F_{\varphi\mu}=0$, $D_{\varphi}\Delta=0$ and
$\Delta^{y}=\pm i\Delta^{x}$, the above current is strictly conserved
\begin{eqnarray}
 &  & \partial_{\mu}\left(J_{\varphi}^{\mu}\mp\frac{1}{2}J^{\mu}\right)=0,
\end{eqnarray}
which occurs in the generalized $p_{x}\pm ip_{y}$ configuration $\Delta=e^{i\theta\left(r,t\right)}\Delta_{0}\left(r,t\right)\left(1,\pm i\right)$,
written in the gauge $A_{\varphi}=0$, and where $r=\sqrt{x^{2}+y^{2}}$.
This conservation law follows from the symmetry of the generalized
$p_{x}\pm ip_{y}$ configuration under the combination of a spatial
rotation by an angle $\alpha$ and a $U\left(1\right)$ transformation
by a phase $\mp\alpha/2$.

\subsubsection{Symmetries, currents, and conservation laws in the geometric description\label{subsec:Currents,-symmetries,-and}}

The symmetries and conservation laws for Dirac fermions have been
described recently in \cite{hughes2013torsional}. Here we review
the essential details (for Majorana fermions) and focus on the mapping
to the symmetries and conservation laws of the $p$-wave SF action
\eqref{eq:14-0}, which were described in section \ref{subsec:Symmetries,-currents,-and}. 

\paragraph{Currents in the geometric description \label{currents}}

The natural currents in the geometric description are defined by the
functional derivatives of the action $S_{\text{RC}}$ with respect
to the background fields $e,\omega$,
\begin{eqnarray}
  \mathsf{J}_{\;a}^{\mu}=\frac{1}{\left|e\right|}\frac{\delta S_{\text{RC}}}{\delta e_{\;\mu}^{a}},\;\mathsf{J}^{ab\mu}=\frac{1}{\left|e\right|}\frac{\delta S_{\text{RC}}}{\delta\omega_{ab\mu}}.\label{eq:54}
\end{eqnarray}
$\mathsf{J}_{\;a}^{\mu}$ is the energy momentum (energy-momentum)
tensor and $\mathsf{J}^{ab\mu}$ is the spin current. 
Note that we use $\mathsf{J}$ as opposed to $J$ to distinguish the
geometric currents from the $p$-wave SF currents described in the
previous section, though the two are related as shown below. 

Calculating the geometric currents for the action \eqref{eq:43-1}
one obtains 
\begin{eqnarray}
  2\mathsf{J}_{\;a}^{\mu}&=&\mathcal{L}_{\text{RC}}e_{a}^{\;\mu}-\frac{i}{2}\overline{\chi}\left(\gamma^{\mu}D_{a}-\overleftarrow{D_{a}}\gamma^{\mu}\right)\chi,\label{eq:55}\\
  2\mathsf{J}^{ab\mu}&=&-\frac{1}{4}\overline{\chi}\chi e_{c}^{\;\mu}\varepsilon^{abc},\nonumber 
\end{eqnarray}
where $\mathcal{L}_{\text{RC}}=\overline{\chi}\left[\frac{i}{2}e_{a}^{\;\mu}\left(\gamma^{a}D_{\mu}-\overleftarrow{D_{\mu}}\gamma^{a}\right)-m\right]\chi$
is (twice) the Lagrangian, which vanishes on the $\chi$ equation
of motion. We see that $\mathsf{J}_{\;a}^{\mu}$ is essentially the
$SO\left(1,2\right)$-covariant version of the canonical energy-momentum
tensor of the spinor $\chi$. We also see that the spin current $\mathsf{J}^{ab\mu}$
has a particularly simple form in $D=2+1$, it is just the spin density
$\frac{1}{2}\overline{\chi}\chi$ times a tensor $-\frac{1}{2}e_{c}^{\;\mu}\varepsilon^{abc}$
that only depends on the background field $e$. Using the expressions
\eqref{17} for the geometric  fields we find that $\mathsf{J}_{\;a}^{\mu},\;\mathsf{J}^{ab\mu}$
are related simply to the electric current and the ($U\left(1\right)$-covariant)
canonical energy-momentum tensor described in section \ref{subsec:Symmetries,-currents,-and},
in the limit $m^{*}\rightarrow\infty$,
\begin{eqnarray}
 &  & J^{\mu}=4\left|e\right|\mathsf{J}^{12\mu}=-\psi^{\dagger}\psi\delta_{t}^{\mu},\label{eq:56}\\
 &  & t_{\text{cov}\;\nu}^{\mu}=-\left|e\right|\mathsf{J}_{\;\nu}^{\mu}=\begin{cases}
\frac{i}{2}\psi^{\dagger}\overleftrightarrow{D_{\nu}}\psi & \mu=t\\
\frac{1}{2}\psi^{\dagger}\Delta^{j}\partial_{\nu}\psi^{\dagger}+h.c & \mu=j
\end{cases}.\nonumber 
\end{eqnarray}
Here we have simplified $t_{\text{cov}}$ using the equation of motion
for $\psi$, and one can also use the equation of motion to remove
time derivatives and obtain Schrodinger picture operators. For example,
$t_{\text{cov}\;t}^{t}=\frac{i}{2}\psi^{\dagger}\overleftrightarrow{D_{t}}\psi=m\psi^{\dagger}\psi-\left(\frac{1}{2}\psi^{\dagger}\Delta^{j}\partial_{j}\psi^{\dagger}+h.c\right)$
is just the ($U\left(1\right)$-covariant) Hamiltonian density in
the relativistic limit. The expression for the energy current $t_{\text{cov}\;t}^{i}$
is more complicated, and it is convenient to write it using some of
the geometric quantities introduced above
\begin{align}
  t_{\text{cov }t}^{j}=g^{jk}\frac{i}{2}\psi^{\dagger}\overleftrightarrow{D_{k}}\psi-&\frac{o}{2}\partial_{k}\left(\frac{1}{\left|e\right|}\varepsilon^{jk}\psi^{\dagger}\psi\right)\label{eq:49}\\
  -&\left(\psi^{\dagger}\psi\right)g^{jk}C_{12k}.\nonumber
\end{align}
This is an expression for the energy current in terms of the momentum
and charge densities, and it will be obtained below as a consequence
of Lorentz symmetry in the relativistic limit. We now describe the
symmetries of the action \eqref{eq:43-1} and the conservation laws
they imply for these currents. As expected, these conservation laws
turn out to be essentially the ones derived in section \eqref{subsec:Symmetries,-currents,-and},
in the relativistic limit. 

\paragraph{Spin \label{spin}}

The Lorentz Lie algebra $so\left(1,2\right)$ is comprised of matrices
$\theta\in\mathbb{R}^{3\times3}$ with entries $\theta_{\;b}^{a}$
such that $\theta_{ab}=-\theta_{ba}$. These can be spanned as $\theta=\frac{1}{2}\theta_{ab}L^{ab}$
where the generators $L^{ab}=-L^{ba}$ are defined such that $\eta L^{ab}$
is the antisymmetric matrix with $1$ ($-1$) at position $a,b$ ($b,a$)
and zero elsewhere. The spinor representation of $\theta$ is 
\begin{eqnarray}
  \hat{\theta}=\frac{1}{2}\theta_{ab}\Sigma^{ab},\;\Sigma^{ab}=\frac{1}{4}\left[\gamma^{a},\gamma^{b}\right].
\end{eqnarray}
Local Lorentz transformations act on $\chi,e,\omega$ by 
\begin{eqnarray}
  \chi&\mapsto& e^{-\hat{\theta}}\chi,\;e_{a}^{\;\mu}\mapsto e_{b}^{\;\mu}\left(e^{\theta}\right)_{\;a}^{b},\nonumber\\
 \omega_{\mu}&\mapsto& e^{-\hat{\theta}}\left(\partial_{\mu}+\omega_{\mu}\right)e^{\hat{\theta}}.\label{eq:20-00}
\end{eqnarray}
The subgroup of $SO\left(1,2\right)$ that is physical in the $p$-wave
SC is $SO\left(2\right)$ generated by $L^{12}$. Using the relations
\eqref{17} between the $p$-wave SC fields and the geometric fields,
and choosing $\theta=\theta_{12}L^{12}=-2\alpha L^{12}$, the transformation
law \eqref{eq:20-00} reduces to the $U\left(1\right)$ transformation \eqref{eq:6.1},
\begin{align}
  \psi\mapsto e^{i\alpha}\psi,\;\Delta\mapsto e^{2i\alpha}\Delta,\;A_{\mu}\mapsto A_{\mu}+\partial_{\mu}\alpha.
\end{align} 
The factor of 2 in $\theta_{12}=-2\alpha$ shows that $U(1)$ actually maps to $Spin(2)$, the double cover of $SO(2)$. Moreover, the fact that $\Delta$ has $U\left(1\right)$ charge
2 while $\psi$ has $U\left(1\right)$ charge 1 corresponds to $e_{a}^{\;\mu}$
being an $SO\left(1,2\right)$ vector while $\chi$ is an $SO\left(1,2\right)$
spinor. 
The Lie algebra version of \eqref{eq:20-00} is 
\begin{align}
  \delta\chi=-\frac{1}{2}\theta_{ab}\Sigma^{ab}\chi,\;\delta e_{\;\mu}^{a}=-\theta_{\;b}^{a}e_{\;\mu}^{b},\;\delta\omega_{\;b\mu}^{a}=D_{\mu}\theta_{\;b}^{a}.
\end{align}
Invariance of $S_{\text{RC}}$ under this variation implies the conservation law 
\begin{eqnarray}
  \nabla_{\mu}\mathsf{J}^{ab\mu}-\mathsf{J}^{ab\rho}T_{\mu\rho}^{\mu}=\mathsf{J}^{[ab]},\label{eq:57}
\end{eqnarray}
valid on the equations of motion for $\chi$ \cite{hughes2013torsional,bradlyn2015low}. This conservation
law relates the anti symmetric part of the energy-momentum tensor
to the divergence of spin current. Essentially, the energy-momentum
tensor isn't symmetric due to the presence of the background field
$\omega$ which transforms under $SO\left(1,2\right)$. From a different
point of view, the vielbein $e$ acts as a source for the fermionic
spin current since it is charged under $SO\left(1,2\right)$. 
Inserting the expressions \eqref{17} into the $\left(a,b\right)=\left(1,2\right)$ component of
\eqref{eq:57} we obtain \eqref{eq:15}, 
\begin{eqnarray}
 &  & \partial_{\mu}J^{\mu}=-i\psi^{\dagger}\Delta^{j}\partial_{j}\psi^{\dagger}+h.c.\label{eq:15-1}
\end{eqnarray}
The other components of \eqref{eq:57} follow from the symmetry under
local boosts, which is only a symmetry of $S_{\text{SF}}$
when $m^{*}\rightarrow\infty$. These can be used to obtain the formula \eqref{eq:49} for the energy current of the $p$-wave
SF, in the limit $m^{*}\rightarrow\infty$, in terms of the momentum
and charge densities.

\paragraph{Energy-momentum\label{subsec:Diffeomorphism-symmetry}}

A diffeomorphism is a smooth invertible map between manifolds. We
consider only diffeomorphisms from space-time to itself and denote
the group of such maps by $Diff$. Since the flat background metric
$\delta^{ij}$ decouples in the relativistic limit, it makes sense
to consider all diffeomorphisms, and not restrict to symmetries of
$\delta^{ij}$ as we did in section \ref{subsec:Energy-momentum}. 

Locally, diffeomorphisms can be described by coordinate transformations
$x\mapsto x'=f\left(x\right)$. The lie algebra is that of vector
fields $\zeta^{\nu}\left(x\right)$, which means diffeomorphisms in
the connected component of the identity $Diff_{0}$ can be written
as $f\left(x\right)=f_{1}\left(x\right)$ where $f_{\varepsilon}\left(x\right)=\exp_{x}\left(\varepsilon\zeta\right)=x+\varepsilon\zeta\left(x\right)+O\left(\varepsilon^{2}\right)$
is the flow of $\zeta$ \cite{nakahara2003geometry}. $Diff$ acts on
the geometric fields by the pullback
\begin{eqnarray}
  \chi\left(x\right)&\mapsto&\chi\left(f\left(x\right)\right),\;e_{\;\mu}^{a}\left(x\right)\mapsto\partial_{\mu}f^{\nu}e_{\;\nu}^{a}\left(f\left(x\right)\right),\nonumber\\
  \omega_{\mu}\left(x\right)&\mapsto&\partial_{\mu}f^{\nu}\omega_{\nu}\left(f\left(x\right)\right).\label{eq:51}
\end{eqnarray}
The action of $Diff$ on the $p$-wave SF fields is similar, and follows from \eqref{eq:51}
supplemented by the dictionary \eqref{17}. For $f\in Diff_{0}$ generated
by $\zeta$, the Lie algebra version of \eqref{eq:51} is given by the Lie derivative, 
\begin{eqnarray}
 \delta\chi&=&\mathcal{L}_{\zeta}\chi=\zeta^{\mu}\partial_{\mu}\chi,\label{eq:53-0}\\
  \delta e_{\;\mu}^{a}&=&\mathcal{L}_{\zeta}e_{\;\mu}^{a}=\partial_{\mu}\zeta^{\nu}e_{\;\nu}^{a}+\zeta^{\nu}\partial_{\nu}e_{\;\mu}^{a},\nonumber \\
  \delta\omega_{\mu}&=&\mathcal{L}_{\zeta}\omega_{\mu}=\partial_{\mu}\zeta^{\nu}\omega_{\nu}+\zeta^{\nu}\partial_{\nu}\omega_{\mu}.\nonumber 
\end{eqnarray}
Since these variations are not Lorentz covariant, they will give rise
to a conservation law which is not Lorentz covariant.  This follows
from the fact that the naive $Diff$ action \eqref{eq:51} does not
commute with Lorentz gauge transformations, as was described for the
simpler case of translations and $U\left(1\right)$ gauge transformations
in section \ref{subsec:Energy-momentum}. Instead, one should use
the Lorentz-covariant $Diff$ action, which is the pull back from
$f\left(x\right)$ to $x$ followed by a Lorentz parallel transport
from $f\left(x\right)$ to $x$ along the integral curve $\gamma_{x,\zeta}\left(\varepsilon\right)=\exp_{x}\left(\varepsilon\zeta\right)=f_{\varepsilon}\left(x\right)$,
\begin{eqnarray}
 \chi\left(x\right)&\mapsto& P\chi\left(f\left(x\right)\right),\label{eq:62}\\ e_{\;\mu}^{a}\left(x\right)&\mapsto& P_{\;b}^{a}\partial_{\mu}f^{\nu}e_{\;\nu}^{b}\left(f\left(x\right)\right),\nonumber\\
 \omega_{\mu}\left(x\right)&\mapsto& P\left[\partial_{\mu}f^{\nu}\omega_{\mu}\left(f\left(x\right)\right)+\partial_{\mu}\right]P^{-1},\nonumber
\end{eqnarray}
 where $P=\frac{1}{2}P_{ab}\Sigma^{ab}$ and $P=\mathcal{P}\exp\left(-\int_{\gamma_{x,\zeta}}\omega\right)$
is the spin parallel transport given by the path ordered exponential.
At the Lie algebra level, this modification of \eqref{eq:51} amounts
to an infinitesimal Lorentz gauge transformation generated by $\theta_{ab}=-\zeta^{\rho}\omega_{ab\rho}$,
which modifies \eqref{eq:53-0} to the covariant expressions 
\begin{eqnarray}
 \delta\chi&=&\zeta^{\mu}\nabla_{\mu}\chi,\label{eq:53-1}\\
  \delta e_{\;\mu}^{a}&=&\nabla_{\mu}\zeta^{a}-T_{\mu\nu}^{a}\zeta^{\nu},\nonumber \\
  \delta\omega_{\mu}&=&\zeta^{\nu}R_{ab\nu\mu}.\nonumber 
\end{eqnarray}
Since the usual $Diff$ and Lorentz actions on the fields are both
symmetries of $S_{\text{RC}}$, so is the Lorenz-covariant $Diff$
action. This leads directly to the conservation law 
\begin{eqnarray}
 &  & \nabla_{\mu}\mathsf{J}_{\;\nu}^{\mu}-\mathsf{J}_{\;\nu}^{\rho}T_{\mu\rho}^{\mu}=T_{\nu\mu}^{b}\mathsf{J}_{\;b}^{\mu}+R_{bc\nu\mu}\mathsf{J}^{bc\mu},\label{eq:71}
\end{eqnarray}
valid on the equations of motion for $\chi$ \cite{bradlyn2015low,hughes2013torsional}. We find it useful
to rewrite \eqref{eq:71} in a way which isolates the effect
of torsion, 
\begin{eqnarray}
 &  & \tilde{\nabla}_{\mu}\mathsf{J}_{\;\nu}^{\mu}=C_{ab\nu}\mathsf{J}^{[ab]}+R_{ab\nu\mu}\mathsf{J}^{ab\mu},\label{eq:60}
\end{eqnarray}
where we note that the curvature also depends on the torsion, $R=\tilde{R}+\tilde{D}C+C\wedge C$.
Equation \eqref{eq:71} can also be massaged to the non-covariant
form 
\begin{align}
  \partial_{\mu}\left(\left|e\right|\mathsf{J}_{\;\nu}^{\mu}\right)=\left(e_{a}^{\;\rho}D_{\nu}e_{\;\mu}^{a}\right)\left|e\right|\mathsf{J}_{\;\rho}^{\mu}+R_{\nu\mu ab}\left|e\right|\mathsf{J}^{ab\mu}.\label{eq:66}
\end{align}
Using the dictionary \eqref{17} and the subsequent paragraph, and
\eqref{eq:56}, this reduces to 
\begin{align}
  \partial_{\mu}t_{\text{cov}\;\nu}^{\mu}=\left(D_{\nu}\Delta^{j}\right)\frac{1}{2}\psi^{\dagger}\partial_{j}\psi^{\dagger}+h.c+F_{\nu\mu}J^{\mu},
\end{align}
which is just the energy-momentum conservation law \eqref{32} for
the $p$-wave SF (with $m^{*}\rightarrow\infty$). 

Writing the conservation law in the form \eqref{eq:66} may not seem
natural from the geometric point of view because it uses the partial
derivative as opposed to a covariant derivative. It is however natural
from the $p$-wave SC point of view, where space-time is actually
flat and $e$ is viewed as a bosonic field with no geometric role,
which is the order parameter $\Delta$. This point is important
in the context of the gravitational anomaly in the $p$-wave SC, see Appendix  \ref{subsec:Boundary-gravitational-anomaly}. 

Similar statements hold for other mechanisms for emergent/analogue
gravity, see section I.6 of \cite{volovik2009universe} and \cite{keser2016analogue},
and were also made in the gravitational context without reference
to emergent phenomena \cite{leclerc2006canonical}.

\subsection{Bulk response\label{sec:Bulk-response} }

\subsubsection{Currents from effective action\label{subsec:Bulk-response-from}}

The effective action for the background fields is obtained by integrating
over the spin-less fermion $\psi$,
\begin{eqnarray}
  e^{iW_{\text{SF}}\left[\Delta,A\right]}=\int\text{D}\psi^{\dagger}\text{D}\psi e^{iS_{\text{SF}}\left[\psi,\psi^{\dagger},\Delta,A\right]}.
\end{eqnarray}

The integral is a fermionic coherent state functional integral, over
the Grassmann valued fields $\psi,\psi^{\dagger}$, and the action
$S_{\text{SF}}$ is given in \eqref{eq:10}. 

As described in section \ref{sec:Emergent-Riemann-Cartan-geometry},
in the relativistic limit $W_{\text{SF}}$ is equal to the effective
action obtained by integrating over a Majorana fermion coupled to
RC geometry, 
\begin{eqnarray}
  e^{iW_{\text{SF}}\left[\Delta,A\right]}&=&e^{iW_{\text{RC}}\left[e,\omega\right]}\label{eq:73-1}\\
  &=&\int\text{D}\left(\left|e\right|^{1/2}\chi\right)e^{iS_{\text{RC}}\left[\chi,e,\omega\right]},\nonumber
\end{eqnarray}
where $e,\omega$ are given in terms of $\Delta,A$ by \eqref{17}. 

It follows from the definition \eqref{eq:54} of the spin current
$\mathsf{J}^{ab\mu}$ and the energy-momentum tensor $\mathsf{J}_{\;a}^{\mu}$
as functional derivatives of $S_{\text{RC}}$ that their ground state
expectation values are given by 
\begin{eqnarray}
 &  & \left\langle \mathsf{J}_{\;a}^{\mu}\right\rangle =\frac{1}{\left|e\right|}\frac{\delta W_{\text{RC}}}{\delta e_{\;\mu}^{a}},\;\left\langle \mathsf{J}^{ab\mu}\right\rangle =\frac{1}{\left|e\right|}\frac{\delta W_{\text{RC}}}{\delta\omega_{ab\mu}}.\label{eq:54-1}
\end{eqnarray}
Using the mapping \eqref{eq:56} between $\mathsf{J}_{\;a}^{\mu},\;\mathsf{J}^{ab\mu}$
and $t_{\text{cov}\;\nu}^{\mu},\;J^{\mu}$ 
we see that 
\begin{eqnarray}
 &  & \left\langle J^{\mu}\right\rangle =4\left|e\right|\left\langle \mathsf{J}^{12\mu}\right\rangle =4\frac{\delta W_{\text{RC}}\left[e,\omega\right]}{\delta\omega_{12\mu}},\label{eq:71-1}\\
 &  & \left\langle t_{\text{cov}\;\nu}^{\mu}\right\rangle =-\left|e\right|e_{\;\nu}^{a}\left\langle \mathsf{J}_{\;a}^{\mu}\right\rangle =-e_{\;\nu}^{a}\frac{\delta W_{\text{RC}}\left[e,\omega\right]}{\delta e_{\;\mu}^{a}}.\nonumber 
\end{eqnarray}
This is the recipe we will use to obtain the expectation values $\left\langle J^{\mu}\right\rangle ,\left\langle t_{\text{cov}\;\nu}^{\mu}\right\rangle $
from the effective action $W_{\text{RC}}$
for a Majorana spinor in RC space-time. 

Note that in \eqref{eq:71-1} there are derivatives with respect to
all components of the vielbein, not just the spatial ones which we
can physically obtain from $\Delta$. For this reason, to get all
components of $\left\langle t_{\text{cov}\;\nu}^{\mu}\right\rangle $,
we should obtain $W_{\text{RC}}$ for general $e$, take the functional
derivative in \eqref{eq:71-1}, and only then set $e$ to the configuration
obtained from $\Delta$ according to \eqref{17}. From the $p$-wave
SF point of view, this corresponds to the introduction of a fictitious
background field $e_{0}^{\;\mu}$ which enters $S_{\text{SF}}$ by
generalizing $\psi^{\dagger}iD_{t}\psi$ to $\psi^{\dagger}\frac{i}{2}e_{0}^{\;\mu}\overleftrightarrow{D_{\mu}}\psi$,
and setting $e_{0}^{\;\mu}=\delta_{t}^{\mu}$ at the end of the calculation, as in
\cite{bradlyn2015low}. 

Before we move on, we offer some intuition for the expressions \eqref{eq:71-1}.
The first equation in \eqref{eq:71-1} follows from the definition
$J^{\mu}=-\frac{\delta S_{\text{SF}}}{\delta A_{\mu}}$ of the electric
current and the simple relation $\omega_{12\mu}=-\omega_{21\mu}=-2A_{\mu}$
between the spin connection and the $U\left(1\right)$ connection.
The second equation in \eqref{eq:71-1} is slightly trickier. It implies
that the (relativistic part of the) energy-momentum tensor $t_{\text{cov}\;\nu}^{\mu}$
is given by a functional derivative with respect to the order parameter
$\Delta$, because $\Delta$ is essentially the vielbein $e$. This
may seem strange, and it is certainly not the case in an $s$-wave
SC, where $\frac{\delta H}{\delta\Delta}\sim\psi_{\uparrow}^{\dagger}\psi_{\downarrow}^{\dagger}$
has nothing to do with energy-momentum. In a $p$-wave SC, the operator
$\frac{\delta H}{\delta\Delta^{j}}\sim\psi^{\dagger}\partial_{j}\psi^{\dagger}$
contains a spatial derivative which hints that it is related to fermionic
momentum. More accurately, we see from \eqref{25-2} that the operator
$\psi^{\dagger}\partial_{j}\psi^{\dagger}$ enters the energy-momentum
tensor in a $p$-wave SC. 

\subsubsection{Effective action from perturbation theory }

\paragraph{Setup and generalities}

We consider the effective action for a $p$-wave SF on the plane $\mathbb{R}^{2}$,
with the corresponding space-time manifold $M_{3}=\mathbb{R}_{t}\times\mathbb{R}^{2}$,
by using perturbation theory around the $p_{x}\pm ip_{y}$ configuration
$\Delta=\Delta_{0}e^{i\theta}\left(1,\pm i\right)$ with no electromagnetic
fields $\partial_{\mu}\theta-2A_{\mu}=0$. After $U\left(1\right)$
gauge fixing $\theta=0$\footnote{In doing so we are ignoring the possibility of vortices, see \cite{ariad2015effective}.},
we obtain $\Delta=\Delta_{0}\left(1,\pm i\right),\;A=0$. Let us start
with the $p_{x}+ip_{y}$ configuration, which has a positive orientation,
in which case the corresponding (gauge fixed) vielbein and spin connection
are just $e_{a}^{\;\mu}=\delta_{a}^{\mu}$ and $\omega_{ab\mu}=0$.
A perturbation of the $p_{x}+ip_{y}$ configuration corresponds to
$e_{a}^{\;\mu}=\delta_{a}^{\mu}+h_{a}^{\;\mu}$ with a small $h$
and to a small spin connection $\omega_{ab\mu}$. In other words,
a perturbation of the $p_{x}+ip_{y}$ configuration without electromagnetic
fields corresponds to a perturbation of flat and torsion-less space-time. 

The effective action for a Dirac spinor in a background RC geometry
was recently calculated perturbatively around flat and torsionless
space-time, with a positive orientation, in the context of geometric
responses of Chern insulators \cite{hughes2013torsional,parrikar2014torsion}.
This is equal to $2W_{\text{RC}}$ where $W_{\text{RC}}$ is the effective
action for a Majorana spinor in RC geometry. 

At this point is seems that we can apply these results in order to
obtain the effective action for the $p$-wave SC, in the relativistic
limit. There is however, an additional ingredient in the perturbative
calculation of the effective action which we did not yet discuss,
which is the renormalization scheme  used to handle diverging integrals.
We refer to terms in the effective action that involve diverging integrals
as \textit{UV sensitive}. The values one obtains for such terms depend
on the details of the renormalization scheme, or in other words, on
microscopic details that are not included in the continuum action.

For us, the continuum description is simply an approximation to the
lattice model, where space is a lattice but time is continuous. This
implies a physical cutoff $\Lambda_{UV}$ for wave-vectors, but not
for frequencies. In particular, such a scheme is not Lorentz invariant,
even though the action in the relativistic limit is. Lorentz symmetry
is in any case broken down to spatial $SO\left(2\right)$ for finite
$m^{*}$. For these reasons, UV sensitive terms in the effective action
$W_{\text{RC}}$ for the $p$-wave SC will be assigned different values
than those obtained before, using a fully relativistic scheme. 

The perturbative calculation within the renormalization scheme outlined
above is described in appendix \ref{subsec:Perturbative-calculation-of},
where we also demonstrate that it produces physical quantities that
approximate those of the lattice model, and compare to the fully relativistic
schemes used in previous works. In the following we will focus on
the UV \textit{insensitive} part of the effective action, and in doing
so we will obtain results which are essentially\footnote{See the discussion of $O\left(\frac{m}{\Lambda_{UV}}\right)$ corrections
below.} independent of microscopic details that do not appear in the continuum
action. We start by quoting the fully relativistic results of \cite{hughes2013torsional,parrikar2014torsion},
and then restrict our attention to the UV insensitive part of the
effective action, and describe the physics of the $p$-wave SC it
encodes.

\paragraph{Effective action for a single Majorana spinor \label{subsec:Effective-action-for}}

The results of \cite{hughes2013torsional,parrikar2014torsion} can
be written as 
\begin{eqnarray}
  2W_{\text{RC}}\left[e,\omega\right]&=&\frac{\kappa_{H}}{2}\int_{M_{3}}Q_{3}\left(\tilde{\omega}\right)\label{eq:72}\\
 &+&\frac{\zeta_{H}}{2}\int_{M_{3}}e^{a}De_{a}-\frac{\kappa_{H}}{2}\int_{M_{3}}\tilde{\mathcal{R}}e^{a}De_{a}\nonumber\\
  &+&\frac{1}{2\kappa_{N}}\int_{M_{3}}\left(\tilde{\mathcal{R}}-2\Lambda+\frac{3}{2}c^{2}\right)\left|e\right|\mbox{d}^{3}x+\cdots\nonumber 
\end{eqnarray}
where 
\begin{eqnarray}
Q_{3}\left(\tilde{\omega}\right)=\text{tr}\left(\tilde{\omega}\text{d}\tilde{\omega}+\frac{2}{3}\tilde{\omega}^{3}\right)
\end{eqnarray}
is the Chern-Simons (local) 3-form,  $c=C_{abc}\varepsilon^{abc}$ is the totally antisymmetric
piece of the contorsion tensor, and $\kappa_{H},\zeta_{H},1/\kappa_{N},\Lambda/\kappa_{N}$
are coefficients that will be discussed further below. The first two
lines of \eqref{eq:72} are written in terms of differential forms, and the third line is written in terms of scalars. By scalars we mean $Diff$ invariant
objects.  In the differential forms the wedge product
is implicit, as it will be from now on, so $\tilde{\omega}\wedge\text{d}\tilde{\omega}$
is written as $\tilde{\omega}\text{d}\tilde{\omega}$ and so on. The
integrals over differential forms can be written as integrals over
pseudo-scalars,
\begin{eqnarray}
 &  & e^{a}De_{a}=\left(e_{\;\alpha}^{a}D_{\beta}e_{\;\gamma}^{b}\frac{1}{\left|e\right|}\varepsilon^{\alpha\beta\gamma}\right)\left|e\right|\text{d}^{3}x=-oc\left|e\right|\text{d}^{3}x,\label{eq:76}\\
 &  & Q_{3}\left(\tilde{\omega}\right)=\left(\tilde{\omega}_{\;b\alpha}^{a}\partial_{\beta}\tilde{\omega}_{\;a\gamma}^{b}+\frac{2}{3}\tilde{\omega}_{\;b\gamma}^{a}\tilde{\omega}_{\;c\beta}^{b}\tilde{\omega}_{\;a\gamma}^{c}\right)\frac{1}{\left|e\right|}\varepsilon^{\alpha\beta\gamma}\left|e\right|\text{d}^{3}x,\nonumber 
\end{eqnarray}
which are only invariant under the orientation preserving
subgroup of $Diff$ which we denote $Diff_{+}$. Here $o=\text{sgn}\left(\text{det}\left(e\right)\right)$ is the
orientation of $e$. These expressions are odd under orientation reversing
diffeomorphisms because so are $o$ and the pseudo-tensor
$\frac{1}{\left|e\right|}\varepsilon^{\alpha\beta\gamma}$ \footnote{In this section $\varepsilon$ always stands for the usual totally anti
symmetric symbol, normalized to 1. Thus $\varepsilon^{123}=\varepsilon^{xyt}=\varepsilon_{xyt}=1$.
Note that $\varepsilon^{abc}$ is an $SO\left(1,2\right)$ tensor,
and an $O\left(1,2\right)$ pseudo-tensor, while $\varepsilon^{\mu\nu\rho}=\text{det}\left(e\right)e_{a}^{\;\mu}e_{b}^{\;\nu}e_{c}^{\;\rho}\varepsilon^{abc}$
is a (coordinate) tensor density, $\frac{1}{\text{det}\left(e\right)}\varepsilon^{\mu\nu\rho}$
is a tensor and $\frac{1}{\left|e\right|}\varepsilon^{\mu\nu\rho}=\frac{1}{\left|\text{det}\left(e\right)\right|}\varepsilon^{\mu\nu\rho}$
is a pseudo-tensor. }.

Equation \eqref{eq:72} can be expanded in the perturbations $h_{a}^{\;\mu}$
and $\omega_{ab\mu}$ to reveal the order in perturbation theory at
which the different terms arise, see appendix \ref{subsec:Perturbative-calculation-of}. Additionally, at every order in the perturbations
the effective action can be expanded in powers of derivatives of the
perturbations over the mass $m$. The terms written explicitly above
show up at first and second order in $h,\omega$ and at up to third order
in their derivatives. They also include higher order corrections that
make them $Diff_{+}$ and Lorentz gauge invariant, or invariant up
to total derivatives. 

All other contributions denoted by $+\cdots$ are at least third order
in the perturbations or fourth order in derivatives. Such
a splitting is not unique \cite{hughes2013torsional}, but the form
\eqref{eq:72} has been chosen because it is well suited for the study
of the bulk responses. 

Let us now describe the different terms in \eqref{eq:72}. The first
term is the gravitational Chern-Simons (gCS) term. It has a similar structure to the more familiar $U\left(1\right)$
CS term $\int A\text{d}A$, and is in fact an $SO\left(1,2\right)$
CS term, but note that the LC spin connection $\tilde{\omega}$ is a functional
of the vielbein $e$. It is important that the spin connection in
gCS is not $\omega$, since through $\omega_{\mu}=-2A_{\mu}\Sigma^{12}$
this would imply a quantized Hall conductivity in a $p$-wave SC,
which does not exist \cite{read2000paired,stone2004edge}. As it is
written in \eqref{eq:72}, gCS is invariant under $Diff_{+}$, but
not under $SO\left(1,2\right)$ if $M_{3}$ has a boundary. This is
the boundary $SO\left(1,2\right)$ anomaly, which is discussed further
in section \ref{subsec:Gauge-symmetry-of}. Using the relation $\tilde{\Gamma}_{\;\beta\mu}^{\alpha}=e_{a}^{\;\alpha}\left(\delta_{b}^{a}\partial_{\mu}+\tilde{\omega}_{\;b\mu}^{a}\right)e_{\;\beta}^{b}$
between $\tilde{\Gamma}$ and $\tilde{\omega}$, one can derive an
important formula,
\begin{align}
 Q_{3}\left(\tilde{\Gamma}\right)-Q_{3}\left(\tilde{\omega}\right)=\text{tr}\left[\frac{1}{3}\left(e\text{d}e^{-1}\right)^{3}+\text{d}\left(\text{d}e^{-1}e\tilde{\Gamma}\right)\right],\nonumber\\\label{eq:70}
\end{align}
where unusually, $e=\left(e_{\;\mu}^{a}\right)$ is treated in this
expression as a matrix valued function \cite{kraus2006holographic,stone2012gravitational}.
The variation with respect to $e$ of the two terms on the right hand
side is a total derivative, which means that they are irrelevant for
the purpose of calculating bulk responses. One can therefore use $Q_{3}\left(\tilde{\Gamma}\right)$,
which only depends on the metric $g_{\mu\nu}$, instead of $Q_{3}\left(\tilde{\omega}\right)$.
The form $\int_{M_{3}}Q_{3}\left(\tilde{\Gamma}\right)$ of gCS is
invariant under $SO\left(1,2\right)$ but not under $Diff_{+}$, as
opposed to $\int_{M_{3}}Q_{3}\left(\tilde{\omega}\right)$. Thus the
right hand side of \eqref{eq:70} has the effect of shifting the boundary
anomaly from $SO\left(1,2\right)$ to $Diff$. 

The second term in \eqref{eq:72} has a structure similar to a CS
term with $e^{a}$ playing the role of a connection, and indeed some
authors refer to it as such \cite{zanelli2012chern}. Nevertheless,
it is $SO\left(1,2\right)$ and $Diff_{+}$ invariant, as can be seen
from \eqref{eq:76}. This term was related to the torsional Hall
viscosity in \cite{hughes2013torsional}, where it was discussed
extensively. The third term in \eqref{eq:72} is also $SO\left(1,2\right)$
and $Diff_{+}$ invariant. We refer to this term as \textit{gravitational
pseudo Chern-Simons} (gpCS), to indicate its similarity to gCS, and
the fact that it is not a Chern-Simons term. The similarity between
gCS and gpCS is demonstrated and put in a broader context in the discussion,
section \ref{sec:Conclusion-and-discussion}. In section \ref{subsec:Calculation-of-bulk} we will see that gCS and gpCS produce similar contributions to bulk responses. For now, we simply note that both terms are second order in $h$ and third order in derivatives
of $h$. 

The third line in \eqref{eq:72} contains the Einstein-Hilbert action
with a cosmological constant $\Lambda$ familiar from general relativity,
and an additional torsional contribution $\propto c^{2}$. The coefficient $1/\kappa_{N}$ of
the Einstein-Hilbert term is usually related to a Newton's constant
$G_{N}=\kappa_{N}/8\pi$. Note that in Riemannian geometry,
where torsion vanishes and $\omega=\tilde{\omega}$, only the gCS
term, the Einstein-Hilbert term, and the cosmological constant survive. 

The coefficients $\kappa_{H},\zeta_{H},1/\kappa_{N},\Lambda/\kappa_{N}$
are given by frequency and wave-vector integrals that arise
within the perturbative calculation, and are described in appendix
\ref{subsec:Perturbative-calculation-of}. In particular $\zeta_{H},1/\kappa_{N},\Lambda/\kappa_{N}$
are dimension-full, with mass dimensions $2,1,3$, and naively diverge. In other words, they are UV sensitive. On the other hand, $\kappa_{H}$ is dimensionless and UV insensitive. With no regularization, one finds
\begin{eqnarray}
 \kappa_{H}=\frac{1}{48\pi}\frac{\text{sgn}\left(m\right)o}{2}.
\end{eqnarray}
Thus, the effective action for a single Majorana spinor can be written
as
\begin{align}
  W_{\text{RC}}\left[e,\omega\right]=\frac{1/2}{96\pi}\frac{\text{sgn}\left(m\right)o}{2}W\left[e,\omega\right]+\cdots\label{eq:77}
\end{align}
where 
\begin{align}
  W\left[e,\omega\right]=\int_{M_{3}}Q_{3}\left(\tilde{\omega}\right)-\int_{M_{3}}\tilde{\mathcal{R}}e^{a}De_{a}\label{eq:RelEffAction}
\end{align}
is the sum of gCS and gpCS, and the dots include UV sensitive terms, or terms of a higher order in derivatives or perturbations, as described above. 

Since the lattice model implies a finite physical cutoff $\Lambda_{UV}$
for wave-vectors, \eqref{eq:77} is exact only for $m/\Lambda_{UV}\rightarrow0$.
For non-zero $m$ there are small $O\left(m/\Lambda_{UV}\right)$
corrections\footnote{All expressions here are with $\hbar=c_{\text{light}}=1$. Restoring
units one finds $\frac{m}{\Lambda_{\text{UV}}}\sim\frac{\text{max}\left(t,\mu\right)}{\delta}$
and so $\frac{m}{\Lambda_{\text{UV}}}\ll1$ in the relativistic regime.} to \eqref{eq:77}. We will keep these corrections implicit for now,
and come back to them in section \ref{subsec:Gauge-symmetry-of}.

\paragraph{Summing over Majorana spinors \label{subsec:Summing-over-lattice}}

As discussed in Appendix \ref{sec:Lattice-model},
the continuum description of the $p$-wave SC includes four Majorana
spinors labeled by $1\leq n\leq4$, with masses $m_{n}$, which are
coupled to vielbeins $e_{\left(n\right)}$. Let us repeat the necessary details. The vielbein $e_{\left(1\right)}$
is associated with the order parameter $\delta$ of the underlying
lattice model, as in \eqref{17}, up to an unimportant rescaling by
the lattice spacing $a$. For this reason we treat it as a fundamental
vielbein and write $e=e_{\left(1\right)}$ in some expressions. The
other vielbeins $\left(e_{\left(n\right)}\right)_{a}^{\;\mu}$ are
obtained from $e$ by multiplying one of the columns $\mu=x,y$ or
both by $-1$. This implies that $o=o_{1}=o_{3}=-o_{2}=-o_{4}$, and
that the metrics $g_{\left(n\right)}^{\mu\nu}$ are identical apart
from $g^{xy}=g_{\left(1\right)}^{xy}=g_{\left(3\right)}^{xy}=-g_{\left(2\right)}^{xy}=-g_{\left(4\right)}^{xy}$.
With this in mind, we can sum over the four Majorana spinors and obtain
and effective action for the $p$-wave SC, 
\begin{eqnarray}
 W_{\text{SC}}\left[e,\omega\right]&=&\sum_{n=1}^{4}W_{\text{RC}}\left[e_{\left(n\right)},\omega\right]\label{eq:77-1-1}\\
 &=&\frac{1/2}{96\pi}\sum_{n=1}^{4}\frac{\text{sgn}\left(m_{n}\right)o_{n}}{2}W\left[e_{\left(n\right)},\omega\right]+\cdots\nonumber
\end{eqnarray}
Note that the Chern number of the lattice
model is given by $\nu=\sum_{n=1}^{4}\text{sgn}\left(m_{n}\right)o_{n}/2$,
but since $W$ also depends on the different vielbeins $e_{\left(n\right)}$,
\eqref{eq:77-1-1} does not only depend on $\nu$ in the general case. 

Some simplification is possible however. Since $e_{\left(1\right)}=e_{\left(3\right)}$
and $e_{\left(2\right)}=e_{\left(4\right)}$ up to a space-time independent
$SO\left(2\right)$ ($U\left(1\right)$) transformation, 
\begin{eqnarray}
  W_{\text{SC}}\left[e,\omega\right]&=&\sum_{l=1}^{2}\frac{\nu_{l}/2}{96\pi}W\left[e_{\left(l\right)},\omega\right]+\cdots\label{eq:77-3}\\
  &=&\sum_{l=1}^{2}\frac{\nu_{l}/2}{96\pi}\int_{M_{3}}Q_{3}\left(\tilde{\omega}{}_{\left(l\right)}\right)+\cdots\nonumber 
\end{eqnarray}
where in the second line, we have only written explicitly gCS terms.
Here we defined
\begin{align}
\nu_{1}&=\frac{o_{1}}{2}\left(\text{sgn}\left(m_{1}\right)+\text{sgn}\left(m_{3}\right)\right),\\
\nu_{2}&=\frac{o_{2}}{2}\left(\text{sgn}\left(m_{2}\right)+\text{sgn}\left(m_{4}\right)\right),\nonumber
\end{align}
which are both integers,  $\nu_{1},\nu_{2}\in\mathbb{Z}.$ The Chern
number of the lattice model is given by the sum $\nu=\nu_{1}+\nu_{2}$.
Thus the lattice model seems to behave like a bi-layer, with layer
index $l=1,2$. In the topological phases of the model $\nu_{1}=0$, $\nu=\nu_{2}=\pm1$, and so 
\begin{eqnarray}
 & W_{\text{SC}}\left[e,\omega\right] & =\frac{\nu/2}{96\pi}W\left[e_{\left(2\right)},\omega\right]+\cdots\label{eq:77-4}\\
 &  & =\frac{\nu/2}{96\pi}\int_{M_{3}}Q_{3}\left(\tilde{\omega}_{\left(2\right)}\right)+\cdots\nonumber 
\end{eqnarray}
where again, in the second line we have only written explicitly the
gCS term. This result is close to what one may have guessed. In the
topological phases with Chern number $\nu\neq0$, the effective action
contains a single gCS term, with coefficient $\frac{\nu/2}{96\pi}$.
A result of this form has been anticipated in \cite{volovik1990gravitational,read2000paired,wang2011topological,ryu2012electromagnetic,palumbo2016holographic},
but there are a few details which are important to note. First, apart
from gCS, $W$ also contains the a gpCS term of the form $\int_{M_{3}}\tilde{\mathcal{R}}e^{a}De_{a}$,
which is possible due to the emergent torsion. Second, the connection
that appears in the CS form $Q_{3}$ is a LC connection, and not the torsion-full connection $\omega$. Moreover, this LC connection is not $\tilde{\omega}$, but
a modification of it $\tilde{\omega}{}_{\left(2\right)}$, where the
subscript $\left(2\right)$ indicates the effect of the multiple Majorana
spinors in the continuum description of the lattice model.  Third, the geometric fields $e,\omega$ are given by $\Delta,A$.

In the trivial phases $\nu_{1}=-\nu_{2}\in\left\{ -1,0,1\right\} $,
$\nu=0$, and we find
\begin{eqnarray}
 & W_{\text{SC}}\left[e,\omega\right] & =\frac{\nu_{1}/2}{96\pi}\left[W\left[e_{\left(1\right)},\omega\right]-W\left[e_{\left(2\right)},\omega\right]\right]+\cdots\label{eq:77-5}\\
 &  & =\frac{\nu_{1}/2}{96\pi}\left[\int_{M_{3}}Q_{3}\left(\tilde{\omega}{}_{\left(1\right)}\right)-\int_{M_{3}}Q_{3}\left(\tilde{\omega}{}_{\left(2\right)}\right)\right]+\cdots\nonumber 
\end{eqnarray}
 This result is quite surprising. Instead of containing no gCS terms,
some trivial phases contain the difference of two such terms, with
slightly different spin connections. One may wonder if these trivial
phases are really trivial after all. This is part of a larger issue
which we now address. 

\subsubsection{Symmetries of the effective action \label{subsec:Gauge-symmetry-of}}

By considering the gauge symmetry of the effective action we can reconstruct
the topological phase diagram appearing in Fig.\ref{fig:Phase-Diagram}
from \eqref{eq:77-3}. This will also help us understand which of
our results are special to the relativistic limit, and which should
hold throughout the phase diagram. By gauge symmetry we refer in this
section to the $SO\left(2\right)$ subgroup of $SO\left(1,2\right)$,
which corresponds to the physical $U\left(1\right)$ symmetry of the
$p$-wave SC. Equation \eqref{eq:70} shows that we can equivalently
consider $Diff$ symmetry. The physical reason for this equivalence
is that the $p$-wave order parameter is charged under both symmetries,
and therefore maps them to one another. 

The effective action was calculated within perturbation theory on
the space-time manifold $M_{3}=\mathbb{R}_{t}\times\mathbb{R}^{2}$,
but for this discussion, we use its locality to assume it remains
locally valid on more general $M_{3}$, which may be closed (compact
and without a boundary) or have a boundary. A closed space-time is
most simply obtained by working on $M_{3}=\mathbb{R}_{t}\times M_{2}$
with $M_{2}$ closed, and with background fields $\Delta,A$ which
are periodic in time, such that $\mathbb{R}_{t}$ can be  compactified
to a circle. 

As described in appendix \ref{subsec:Global-structures-and}, a non
singular order parameter endows $M_{3}$ with an orientation and a
spin structure, and in particular requires that $M_{2}$ be orientable
\cite{quelle2016edge}, which we assume. Thus, for example, we exclude
the possibility of $M_{2}$ being the Mobius strip. Moreover, a non
singular order parameter on a closed $M_{2}$ requires that $M_{2}$
contain $\left(g-1\right)o$ magnetic monopoles \cite{read2000paired},
where $g$ is the genus of $M_{2}$, and we assume that this condition
is satisfied. For example, if $M_{2}$ is the sphere then it must
contain a single monopole or anti-monopole depending on the orientation
$o$ \cite{kraus2009majorana,moroz2016chiral}. 

\paragraph{Quantization of coefficients\label{subsec:quantization}}

The first fact about the gCS term that we will need, is that gauge
symmetry of $\alpha\int_{M_{3}}Q_{3}\left(\tilde{\omega}\right)$
for all closed $M_{3}$ requires that $\alpha$ be quantized such
that $\alpha\in\frac{1}{192\pi}\mathbb{Z}$, see equation (2.27) of
\cite{witten2007three}. 
In order to understand how generic is our result \eqref{eq:77-3}, we
will check what quantization condition on $\alpha_{1},\alpha_{2}$
is required for gauge symmetry of $\alpha_{1}\int_{M_{3}}Q_{3}\left(\tilde{\omega}_{\left(1\right)}\right)+\alpha_{2}\int_{M_{3}}Q_{3}\left(\tilde{\omega}_{\left(2\right)}\right)$
on all closed $M_{3}$. Following the arguments of \cite{witten2007three}
we find that $\alpha_{1}+\alpha_{2}\in\frac{1}{192\pi}\mathbb{Z}$,
but $\alpha_{1},\alpha_{2}\in\mathbb{R}$ are not separately restricted, see appendix \ref{subsec:quntization-of-coefficients}.  It is
therefore natural to define $\alpha=\alpha_{1}+\alpha_{2}$ and rewrite
\begin{eqnarray}
 && \alpha_{1}\int_{M_{3}}Q_{3}\left(\tilde{\omega}_{\left(1\right)}\right)+\alpha_{2}\int_{M_{3}}Q_{3}\left(\tilde{\omega}_{\left(2\right)}\right)\\
  &&=\alpha\int_{M_{3}}Q_{3}\left(\tilde{\omega}_{\left(2\right)}\right)+\alpha_{1}\int_{M_{3}}\left[Q_{3}\left(\tilde{\omega}_{\left(1\right)}\right)-Q_{3}\left(\tilde{\omega}_{\left(2\right)}\right)\right],\nonumber 
\end{eqnarray}
where $\alpha\in\frac{1}{192\pi}\mathbb{Z}$ but $\alpha_{1}\in\mathbb{R}$.
Comparing with the result \eqref{eq:77-3}, we identify $\alpha=\frac{\nu/2}{96\pi}$, $\alpha_{1}=\frac{\nu_{1}/2}{96\pi}$,
and we conclude that $\nu$ must be precisely an integer and equal
to the Chern number, while $\nu_{1}$ need not be quantized. We therefore
interpret the $O\left(m/\Lambda_{\text{UV}}\right)$ corrections to
$\alpha=\frac{\nu/2}{96\pi}$ produced in our computation as artifacts
of our approximations\footnote{Specifically, in obtaining the relativistic continuum approximation
we split the Brillouin zone $BZ$ into four quadrants and linearized
the lattice Hamiltonian \eqref{eq:3} in every quadrant. Applying
any integral formula for the Chern number to the approximate Hamiltonian
will give a result $\nu_{\text{apprx}}=\frac{1}{2}\sum_{n=1}^{4}o_{n}\text{sgn}\left(m_{n}\right)+O\left(m/\Lambda_{\text{UV}}\right)$
which is only approximately quantized in the relativistic regime,
simply because the approximate Hamiltonian is discontinuous on $BZ$.
Nevertheless, the known quantization $\nu\in\mathbb{Z}$ and the fact
that $\nu_{\text{apprx}}\approx\nu$ are enough to obtain the exact
result $\nu=\frac{1}{2}\sum_{n=1}^{4}o_{n}\text{sgn}\left(m_{n}\right)$.}, which must vanish due to gauge invariance. On the other hand, we
interpret the quantization $\alpha_{1}=\frac{\nu_{1}/2}{96\pi}$ as
a special property of the relativistic limit with both $m^{*}\rightarrow\infty$ and $m\rightarrow0$,
which should not hold throughout the phase diagram. 

So far we have only considered gCS terms. As already explained, the
gpCS term is gauge invariant on any $M_{3}$, and we therefore see
no reason for the quantization of its coefficient. Explicitly, $-\beta\int_{M_{3}}\tilde{\mathcal{R}}e^{a}De_{a}$
is gauge invariant for all $\beta\in\mathbb{R}$. Thus we interpret
the approximate quantization of the coefficients of gpCS terms as
a special property of the relativistic limit, which should not hold
throughout the phase diagram. We note that even for a relativistic spinor any $\beta\in\mathbb{R}$ can be obtained, by adding a non minimal coupling to torsion \cite{hughes2013torsional}.

In light of the above, it is natural to interpret \eqref{eq:77-3}
as a special case of
\begin{eqnarray}
 && W_{\text{SC}}\left[e,\omega\right]=\frac{\nu/2}{96\pi}\int_{M_{3}}Q_{3}\left(\tilde{\omega}_{\left(2\right)}\right)\label{eq:80-1}\\
  &&+\alpha_{1}\int_{M_{3}}\left[Q_{3}\left(\tilde{\omega}_{\left(1\right)}\right)-Q_{3}\left(\tilde{\omega}_{\left(2\right)}\right)\right]\nonumber\\
  &&-\beta_{1}\int_{M_{3}}\tilde{\mathcal{R}}_{\left(1\right)}e_{\left(1\right)}^{a}De_{\left(1\right)a}-\beta_{2}\int_{M_{3}}\tilde{\mathcal{R}}e_{\left(2\right)}^{a}De_{\left(2\right)a}+\cdots\nonumber 
\end{eqnarray}
where $\nu\in\mathbb{Z}$ is the Chern number and $\alpha_{1},\beta_{1},\beta_{2}$
are additional, non quantized, yet dimensionless, response coefficients. In the relativistic
limit $\alpha_{1},\beta_{1},\beta_{2}$ happen to be quantized, but
this is not generic. Only the first gCS term encodes topological bulk
responses, proportional to the Chern number $\nu$, and below we will
see that only this term is related to an edge anomaly. We can also
write \eqref{eq:80-1} more symmetrically, 
\begin{eqnarray}
 && W_{\text{SC}}\left[e,\omega\right]\label{eq:81-1}\\
 &&=\sum_{l=1}^{2}\left[\alpha_{l}\int_{M_{3}}Q_{3}\left(\tilde{\omega}_{\left(l\right)}\right)-\beta_{l}\int_{M_{3}}\tilde{\mathcal{R}}_{\left(l\right)}e_{\left(l\right)}^{a}De_{\left(l\right)a}\right]+\cdots\nonumber
\end{eqnarray}
but here we must keep in mind the quantization condition $\alpha_{1}+\alpha_{2}=\frac{\nu/2}{96\pi}\in\frac{1}{192\pi}\mathbb{Z}$.

This equation should be compared with the result in the relativistic
limit \eqref{eq:77-3}, where $\alpha_{l},\beta_{l}$ are all quantized,
and $\alpha_{l}=\beta_{l}$. We note that the quantization of $\alpha_{l},\beta_{l}$
in the relativistic limit can be understood on dimensional grounds: in this limit there are simply not enough dimension-full
quantities which can be used to construct dimensionless quantities,
beyond $\text{sgn}\left(m_{n}\right)$ and $o_{n}$. Of course, this
does not explain why $\alpha_{l}=\beta_{l}$ in the relativistic limit.

\paragraph{Boundary anomalies\label{subsec:Boundary-anomalies}}

We can strengthen the above conclusions by considering space-times
$M_{3}$ with a boundary. The second fact about the gCS term that
we will need is that it is not gauge invariant when $M_{3}$ has a
boundary, even with a properly quantized coefficient. In more detail,
the $SO\left(2\right)$ variation of gCS is given by 
\begin{align}
  \delta_{\theta}\int_{M_{3}}Q_{3}\left(\tilde{\omega}\right)=-\text{tr}\int_{\partial M_{3}}\mbox{d}\theta\tilde{\omega}.
\end{align}
Up to normalization, the boundary term above is called the consistent
Lorentz anomaly, which is one of the forms in which the gravitational
anomaly manifests itself \cite{bertlmann2000anomalies}\footnote{Generally speaking, \textit{consistent} anomalies are given by symmetry
variations of functionals. We will also discuss below the more physical
\textit{covariant} anomalies, which correspond to the actual inflow
of some charge from bulk to boundary}. The anomaly $\text{tr}\int_{\partial M_{3}}\mbox{d}\theta\tilde{\omega}$
is a local functional that can either be written as the gauge variation
of a local bulk functional, as it is written above, or as the gauge
variation of a \textit{nonlocal} boundary functional $F\left[\tilde{\omega}\right]$,
such that $\delta_{\theta}F\left[\tilde{\omega}\right]=\int_{\partial M_{3}}\mbox{d}\theta\tilde{\omega}$,
but cannot be written as the gauge variation of a local boundary functional
\cite{manes1985algebraic}. The difference of two gCS terms is also
not gauge invariant,
\begin{align}
 & \delta_{\theta}\left[\int_{M_{3}}Q_{3}\left(\tilde{\omega}_{\left(1\right)}\right)-\int_{M_{3}}Q_{3}\left(\tilde{\omega}_{\left(2\right)}\right)\right]\\
 &=-\text{tr}\int_{\partial M_{3}}\mbox{d}\theta\left(\tilde{\omega}_{\left(1\right)}-\tilde{\omega}_{\left(2\right)}\right),\nonumber
\end{align}
but here there is a local boundary term that can produce the same
variation, given by $\text{tr}\left(\tilde{\omega}_{(1)}\tilde{\omega}_{(2)}\right)$.

The physical interpretation is as follows. Since $F\left[\tilde{\omega}\right]$
is non local it can be interpreted as the effective action obtained
by integrating over a gapless, or massless, boundary field coupled
to $e$. These are the boundary chiral Majorana fermions of the $p$-wave
SC. The statement that $F$ cannot be local implies that this boundary
field cannot be gapped. In this manner the existence of the gCS term
in the bulk effective action, with a coefficient that is fixed within
a topological phase, implies the existence of gapless degrees of freedom
that cannot be gapped within a topological phase. We  study this
bulk-boundary correspondence in more detail in Appendix \ref{sec:Boundary-fermions-and}.
Naively, the difference of two gCS terms implies the existence of
two boundary fermions with opposite chiralities, one of which is coupled
to $e_{\left(1\right)}$ and the other coupled to $e_{\left(2\right)}$.
The boundary term $\int_{\partial M_{3}}\text{tr}\left(\tilde{\omega}_{\left(1\right)}\tilde{\omega}_{\left(2\right)}\right)$
can only be generated if the two counter propagating fermions are
coupled, and its locality indicates that this coupling can open a
gap. Thus the term $\int_{\partial M_{3}}\text{tr}\left(\tilde{\omega}_{\left(1\right)}\tilde{\omega}_{\left(2\right)}\right)$
represents the effect of a generic interaction between two counter
propagating chiral Majorana fermions.

Again, as opposed to the gCS term, the gpCS term is gauge invariant
on any $M_{3}$, and is therefore unrelated to edge anomalies. Thus,
in the effective action \eqref{eq:80-1}, only the first gCS term
is related to an edge anomaly.

\paragraph{Time reversal and reflection symmetry of the effective action}

Time reversal $T$ and reflection $R$ are discussed in appendices
\ref{subsec:Spatial-reflections-and} and \ref{subsec:relativisitc Spatial-reflection-and}.
The orientation $o$ of the order parameter is odd under both $T,R$,
and it follows that so are the coefficients $\nu_{l}$. Therefore
$\nu_{l}$ are $T,R$-odd response coefficients. More generally, $\alpha_{l},\beta_{l}$
in \eqref{eq:81-1} are $T,R$-odd response coefficients. As described
in section \ref{subsec:Effective-action-for}, integrals over differential
forms are also odd under the orientation reversing diffeomorphisms
$T,R$, and therefore $W_{\text{SC}}$ is invariant under $T,R$.

\subsubsection{Calculation of currents\label{subsec:Calculation-of-bulk}}

To derive the currents we start with the expression
\begin{align}
  \alpha_{1}\int_{M_{3}}Q_{3}\left(\tilde{\omega}\right)-\beta_{1}\int_{M_{3}}\tilde{\mathcal{R}}e^{a}De_{a}+\cdots\label{eq:72-1}
\end{align}
which is the effective action for the layer $l=1$. We then sum the results over $l=1,2$, as in
\eqref{eq:81-1}, to get the full low energy response of the lattice
model, keeping in mind that $\alpha_{1}+\alpha_{2}=\frac{\nu/2}{96\pi}\in\frac{1}{192\pi}\mathbb{Z}$. 

\paragraph{Bulk response from gravitational Chern-Simons terms\label{subsec:Currents-from-the}}

    For the purpose of calculating the contribution of gCS
to the bulk energy-momentum tensor it is easier to use $Q_{3}\left(\tilde{\Gamma}\right)$
instead of $Q_{3}\left(\tilde{\omega}\right)$. The result is \cite{jackiw2003chern,perez2010conserved,stone2012gravitational}
\begin{align}
  \left\langle \mathsf{J}_{\;a}^{\mu}\right\rangle _{\text{gCS}}=\frac{1}{\left|e\right|}\frac{\delta}{\delta e_{\;\mu}^{a}}\left[\alpha_{1}\int_{M_{3}}Q_{3}\left(\tilde{\Gamma}\right)\right]=4\alpha_{1}\tilde{C}_{\;a}^{\mu},\label{eq:74}
\end{align}
where $\tilde{C}$ is the Cotton tensor, which can be written as 
\begin{eqnarray}
  \tilde{C}^{\mu\nu}=-\frac{1}{\sqrt{g}}\varepsilon^{\rho\sigma(\mu}\tilde{\nabla}_{\rho}\tilde{\mathcal{R}}_{\sigma}^{\nu)}.
\end{eqnarray}
Relevant properties of the Cotton tensor are $\tilde{\nabla}_{\mu}\tilde{C}^{\mu\nu}=0$,
$\tilde{C}_{\;\mu}^{\mu}=0$, and $C^{[\mu\nu]}=0$. It follows from \eqref{eq:74} that 
\begin{align}
 \left\langle t_{\text{cov}\;\nu}^{\mu}\right\rangle _{\text{gCS}}=-\left|e\right|\left\langle \mathsf{J}_{\;\nu}^{\mu}\right\rangle _{\text{gCS}}=-4\alpha_{1}\left|e\right|\tilde{C}_{\;\nu}^{\mu}.\label{eq:110}
\end{align}
For order parameters of the form 
\begin{eqnarray}
 &  & \Delta=e^{i\theta}\left(\left|\Delta^{x}\right|,\pm i\left|\Delta^{y}\right|\right)\label{eq:110-10}
\end{eqnarray}
the metrics for both layers $l=1,2$ are identical. Since $\tilde{C}$
only depends on the metric it follows that for such order parameters
the summation over $l=1,2$ gives 

\begin{align}
  \left\langle t_{\text{cov}\;\nu}^{\mu}\right\rangle _{\text{gCS}}=-\left|e\right|\left\langle \mathsf{J}_{\;\nu}^{\mu}\right\rangle _{\text{gCS}}=-\frac{\nu/2}{96\pi}4\left|e\right|\tilde{C}_{\;\nu}^{\mu}.\label{eq:110-1}
\end{align}
Put differently, the difference of gCS terms in \eqref{eq:81-1},
with coefficient $\alpha_{1}$, does not produce a bulk response for
such order parameters. This provides a simple way to separate the
topological invariant $\nu$ from the non quantized $\alpha_{1}$. 

The Cotton tensor takes a simpler form if the geometry is a product
geometry, where the metric is of the form $\text{d}s^{2}=g_{\alpha\beta}\left(x^{\alpha}\right)\text{d}x^{\alpha}\text{d}x^{\beta}+\sigma\text{d}z^{2}$.
Here $\sigma=\pm1$ depends on whether $z$ is a space-like or time-like
coordinate, and we will use both in the following. The two coordinates
$x^{\alpha}$ are space-like if $z$ is time-like and mixed if $z$
is space-like. In this case the curvature is determined by the curvature
scalar, which corresponds to the curvature scalar of the two dimensional
metric $g_{\alpha\beta}$. In particular $\mathcal{R}_{\;\beta}^{\alpha}=\frac{1}{2}\mathcal{R}\delta_{\beta}^{\alpha}$
and the other components of $\mathcal{R}_{\;\nu}^{\mu}$ vanish. Then
\begin{align}
  \left\langle \mathsf{J}^{\alpha z}\right\rangle _{\text{gCS}}=\left\langle \mathsf{J}^{z\alpha}\right\rangle _{\text{gCS}}=\alpha_{1}\frac{1}{\left|e\right|}\varepsilon^{z\alpha\beta}\partial_{\beta}\tilde{\mathcal{R}},\label{eq:111}
\end{align}
and the other components vanish. In terms of $t_{\text{cov}\;\nu}^{\mu}$,
\begin{eqnarray}
 &  & \left\langle t_{\text{cov}\;z}^{\alpha}\right\rangle _{\text{gCS}}=-\alpha_{1}\sigma\varepsilon^{z\alpha\beta}\partial_{\beta}\tilde{\mathcal{R}},\\
 &  & \left\langle t_{\text{cov}\;\alpha}^{z}\right\rangle _{\text{gCS}}=-\alpha_{1}g_{\alpha\beta}\varepsilon^{z\beta\gamma}\partial_{\gamma}\tilde{\mathcal{R}}.\nonumber 
\end{eqnarray}

Taking $z=t$ is natural in the context of the $p$-wave SC, since
the emergent metric \eqref{eq:10-1} is always a product metric if
$\Delta$ is time independent. Then, with a general time independent
order parameter, 
\begin{eqnarray}
 &  & \left\langle J_{E}^{i}\right\rangle _{\text{gCS}}=\left\langle t_{\text{cov}\;t}^{i}\right\rangle _{\text{gCS}}=-\alpha_{1}\varepsilon^{ij}\partial_{j}\tilde{\mathcal{R}}\label{eq:87-2},\\
 &  & \left\langle P_{i}\right\rangle _{\text{gCS}}=\left\langle t_{\text{cov}\;i}^{t}\right\rangle _{\text{gCS}}=-\alpha_{1}g_{ik}\varepsilon^{kj}\partial_{j}\tilde{\mathcal{R}},\nonumber 
\end{eqnarray}
where $\tilde{\mathcal{R}}$ is the curvature associated with the
spatial metric $g^{ij}=-\Delta^{(i}\Delta^{j)*}$. Again, for order
parameters of the form \eqref{eq:110-10} the metrics for both layers $l=1,2$ are identical, and the summation
over $l=1,2$ produces
\begin{eqnarray}
 &  & \left\langle J_{E}^{i}\right\rangle _{\text{gCS}}=-\frac{\nu/2}{96\pi}\varepsilon^{ij}\partial_{j}\tilde{\mathcal{R}}\label{eq:87-2-2},\\
 &  & \left\langle P_{i}\right\rangle _{\text{gCS}}=-\frac{\nu/2}{96\pi}g_{ik}\varepsilon^{kj}\partial_{j}\tilde{\mathcal{R}}.\nonumber 
\end{eqnarray}
These are the topological bulk responses described in section \ref{subsubsec:Topological bulk responses from a gravitational Chern-Simons term}.
It is also usefull to consider order parameters of the form
\begin{eqnarray}
 &  & \Delta=\Delta_{0}e^{i\theta}\left(1,e^{i\phi}\right),\label{eq:87-9}
\end{eqnarray}
where $\phi$ is space dependent. Here the metrics satisfy $g^{xy}=g_{\left(1\right)}^{xy}=-g_{\left(2\right)}^{xy}=\Delta_{0}^{2}\cos\phi$,
with the other components constant, and therefore the Ricci scalars
satisfy $\mathcal{R}=\mathcal{R}_{\left(1\right)}=-\mathcal{R}_{\left(2\right)}$.
The summation over $l=1,2$ for such order parameters then gives
\begin{eqnarray}
 &  & \left\langle J_{E}^{i}\right\rangle _{\text{gCS}}=-\left(\alpha_{1}-\alpha_{2}\right)\varepsilon^{ij}\partial_{j}\tilde{\mathcal{R}}.\label{eq:87-2-1}
\end{eqnarray}
Unlike the sum $\alpha_{1}+\alpha_{2}=\frac{\nu/2}{96\pi}$, the difference
$\alpha_{1}-\alpha_{2}=2\alpha_{1}-\frac{\nu/2}{96\pi}$ is not quantized.
The response \eqref{eq:87-2-1} is therefore not a topological bulk
response. Measuring $\left\langle J_{E}\right\rangle $ for an order
parameter such that $\mathcal{R}=\mathcal{R}_{\left(1\right)}=\mathcal{R}_{\left(2\right)}$,
and then for an order parameter such that $\mathcal{R}=\mathcal{R}_{\left(1\right)}=-\mathcal{R}_{\left(2\right)}$,
allows one to fix both $\alpha_{1},\alpha_{2}$, or both $\nu$ and
$\alpha_{1}$. 

To demonstrate how closely \eqref{eq:87-2-1} can resemble a topological
bulk response, we go back to the lattice model. In the relativistic
limit we found that some trivial phases, where $\nu=0$, have $\alpha_{1}=\frac{\nu_{1}/2}{192\pi}\neq0$.
It follows that these trivial phases have \textit{in the relativistic
limit} a quantized response 
\begin{align}
  \left\langle J_{E}^{i}\right\rangle _{\text{gCS}}=-2\alpha_{1}\varepsilon^{ij}\partial_{j}\tilde{\mathcal{R}}=-\frac{\nu_{1}}{96\pi}\varepsilon^{ij}\partial_{j}\tilde{\mathcal{R}},
\end{align}
for order parameters $\Delta=\Delta_{0}e^{i\theta}\left(1,e^{i\phi}\right)$.

Another case of interest is when $z$ is a spatial coordinate. As
an example, we take $z=y$. This decomposition is less natural in
the $p$-wave SC, as can be seen from \eqref{eq:10-1}. It allows
for time dependence, but restricts the configuration the order parameter
can take at any given time. A simple example for an order parameter
that gives rise to a product metric with respect to $y$ is $\Delta=\Delta_{0}e^{i\theta\left(t,x\right)}\left(1+f\left(t,x\right),\pm i\right)$,
which is a perturbation of the $p_{x}\pm ip_{y}$ configuration with
a small real function $f$. Then 
\begin{eqnarray}
 &  & \left\langle t_{\text{cov}\;\alpha}^{y}\right\rangle _{\text{gCS}}=-\frac{\nu/2}{96\pi}g_{\alpha\beta}\varepsilon^{\beta\gamma y}\partial_{\gamma}\tilde{\mathcal{R}},\label{eq:104}
\end{eqnarray}
where we have summed over $l=1,2$. This an interesting contribution
to the $x$-momentum current and energy current in the $y$ direction.
If we consider, as in Fig.\ref{fig:A-comparison-of-1}, a boundary or
domain wall at $y=0$, between a topological phase and a trivial phase
where $\nu=0$, we see that there is an inflow of energy and $x$-momentum
into the boundary from the topological phase. This shows that energy
and $x$-momentum are accumulated on the boundary, at least locally,
which corresponds to the boundary gravitational anomaly. We complete
the analysis of this situation from the boundary point of view in
Appendix \ref{subsec:Implication-for-the}. 

\paragraph{Bulk response from the gravitational pseudo Chern-Simons term \label{subsec:Additional-contributions}}

 The gpCS term $-\beta_{1}\int_{M_{3}}\tilde{\mathcal{R}}e^{a}De_{a}$
contributes to the energy-momentum tensor, and also provides a contribution
to the spin density, 
\begin{align}
  \left\langle \mathsf{J}^{\mu\nu}\right\rangle _{\text{gpCS}}=&\beta_{1}\left\{\frac{1}{\left|e\right|}\varepsilon^{\mu\nu\rho}\partial_{\rho}\tilde{\mathcal{R}}-\frac{1}{\left|e\right|}\varepsilon^{\mu\rho\sigma}\tilde{\mathcal{R}}T_{\rho\sigma}^{\nu}\right.\label{eq:92-1}+\left.2o\left[\left(\tilde{\nabla}^{\mu}\tilde{\nabla}^{\nu}-g^{\mu\nu}\tilde{\nabla}^{2}\right)-\tilde{\mathcal{R}}^{\mu\nu}\right]c \vphantom{\frac{1}{\left|e\right|}} \right\}.\\
 \left\langle \mathsf{J}^{ab\mu}\right\rangle _{\text{gpCS}}=&\beta_{1}o\tilde{\mathcal{R}}\varepsilon^{abc}e_{c}^{\;\mu}\nonumber 
\end{align}
These are calculated in appendix \ref{subsec:Calculation-of-certain}.

 Using \eqref{eq:71-1}, the above contribution to the spin density
corresponds to a contribution to the charge density, 
\begin{eqnarray}
 &  & \left\langle J^{t}\right\rangle _{\text{gpCS}}=4\beta_{1}o\left|e\right|\tilde{\mathcal{R}}.\label{eq:98}
\end{eqnarray}
The most notable feature of this density is that it is not accompanied
by a current, even for time dependent background fields, where $\partial_{\mu}\left\langle J^{\mu}\right\rangle =\partial_{t}\left\langle J^{t}\right\rangle \neq0$.
This represents the non conservation of fermionic charge in a $p$-wave
SC \eqref{eq:15}. The appearance of $o$ can be understood from \eqref{eq:76}.
One can also understand the appearance of $o$ based on time reversal
symmetry. Since both $J^{t}$ and $\tilde{\mathcal{R}}$ are time
reversal even, the coefficient of the above response cannot be $\beta_{1}$,
which is time reversal odd. 

We now discuss the energy-momentum contributions $\left\langle \mathsf{J}^{\mu\nu}\right\rangle _{\text{gpCS}}$
in \eqref{eq:92-1}, with the purpose of comparing them to the gCS
contributions $\left\langle \mathsf{J}^{\mu\nu}\right\rangle _{\text{gCS}}$.
To do this in the simplest setting, we restrict to a product geometry
with respect to the coordinate $z$ as described in the previous section.
We will also assume for simplicity that torsion vanishes, and generalize
to non-zero torsion in appendix \ref{subsec:Calculation-of-certain}.
For a torsion-less product geometry $\left\langle \mathsf{J}^{\mu\nu}\right\rangle _{\text{gpCS}}$
reduces to 
\begin{align}
  -\left\langle \mathsf{J}^{\alpha z}\right\rangle _{\text{gpCS}}=\left\langle \mathsf{J}^{z\alpha}\right\rangle _{\text{gpCS}}=\beta_{1}\frac{1}{\left|e\right|}\varepsilon^{z\alpha\beta}\partial_{\beta}\tilde{\mathcal{R}}.\label{eq:94-1}
\end{align} 
Note that while the gpCS term vanishes in a torsion-less geometry, the currents it produces, given by its functional derivatives, do not. Comparing with \eqref{eq:111}, we see that $\left\langle \mathsf{J}^{z\alpha}\right\rangle _{\text{gpCS}}\propto\left\langle \mathsf{J}^{z\alpha}\right\rangle _{\text{gCS}}$,
while $\left\langle \mathsf{J}^{\alpha z}\right\rangle _{\text{gpCS}}\propto-\left\langle \mathsf{J}^{\alpha z}\right\rangle _{\text{gCS}}$,
with the proportionality constant $\alpha_{1}/\beta_{1}$, that goes
to 1 in the relativistic limit.  This demonstrates the similarity
between the gpCS and gCS terms. 

In particular, we find in a time independent situation the following
contributions to the energy current and momentum density,
\begin{eqnarray}
 &  & \left\langle J_{E}^{i}\right\rangle _{\text{gpCS}}=\left\langle t_{\text{cov}\;t}^{i}\right\rangle _{\text{gpCS}}=\beta_{1}\varepsilon^{ij}\partial_{j}\tilde{\mathcal{R}},\label{eq:92-2}\\
 &  & \left\langle P_{i}\right\rangle _{\text{gpCS}}=\left\langle t_{\text{cov}\;i}^{t}\right\rangle _{\text{gpCS}}=-\beta_{1}g_{ik}\varepsilon^{kj}\partial_{j}\tilde{\mathcal{R}}.\nonumber 
\end{eqnarray}
Comparing with \eqref{eq:87-2}, we see that $\left\langle P_{i}\right\rangle _{\text{gpCS}}\propto\left\langle P_{i}\right\rangle _{\text{gCS}}$,
while $\left\langle J_{E}^{i}\right\rangle _{\text{gpCS}}\propto-\left\langle J_{E}^{i}\right\rangle _{\text{gCS}}$.
This sign difference can be understood from the density response \eqref{eq:98},
and the relation \eqref{eq:49} between the operators $J_{E}$ and
$P$, in the relativistic limit. With vanishing torsion it reduces
to
\begin{eqnarray}
 &  & J_{E}^{j}-g^{jk}P_{k}=\frac{o}{2}\varepsilon^{jk}\partial_{k}\left(\frac{1}{\left|e\right|}J^{t}\right).
\end{eqnarray}
 Thus the gCS contributions \eqref{eq:87-2} satisfy $\left\langle J_{E}^{j}\right\rangle _{\text{gCS}}-g^{jk}\left\langle P_{k}\right\rangle _{\text{gCS}}=0$
because gCS does not contribute to the density. On the other hand,
the gpCS does contribute to the density, which is why $\left\langle J_{E}^{j}\right\rangle _{\text{gpCS}}-g^{jk}\left\langle P_{k}\right\rangle _{\text{gpCS}}\neq0$.
This conclusion holds regardless of the value of the coefficient $\beta_{1}$
of gpCS. One can therefore fix the value of $\beta_{1}$ by a measurement
of the density, and thus separate the topological bulk responses (gCS)
from the non-topological bulk responses (gpCS). 

More accurately, we have seen that the lattice model behaves as a
bi-layer with layer index $l=1,2$, and there are actually two coefficients
$\beta_{1},\beta_{2}$. As in the previous section, one can extract
both $\beta_{1},\beta_{2}$ by first considering an order parameter \eqref{eq:110-10}
such that $\mathcal{R}=\mathcal{R}_{\left(1\right)}=\mathcal{R}_{\left(2\right)}$,
and then considering an order parameter \eqref{eq:87-9} such that $\mathcal{R}=\mathcal{R}_{\left(1\right)}=-\mathcal{R}_{\left(2\right)}$.

Another case of interest is when $z$ is a spatial coordinate, and
as in the previous section we take $z=y$, $\Delta=\Delta_{0}e^{i\theta\left(t,x\right)}\left(1+f\left(t,x\right),\pm i\right)$.
We then find from \eqref{eq:94-1}, $\left\langle \mathsf{J}^{y\alpha}\right\rangle _{\text{gpCS}}=\beta_{1}\frac{1}{\left|e\right|}\varepsilon^{z\alpha\beta}\partial_{\beta}\tilde{\mathcal{R}}$,
or
\begin{eqnarray}
 &  & \left\langle t_{\text{cov}\;\alpha}^{y}\right\rangle _{\text{gpCS}}=-\beta_{1}g_{\alpha\beta}\varepsilon^{\beta\gamma y}\partial_{\gamma}\tilde{\mathcal{R}}.\label{eq:104-1}
\end{eqnarray}
In the presence of a boundary (or domain wall) at $y=0$, this describes
an inflow of energy and $x$-momentum from the bulk to the boundary,
such that $\left\langle t_{\text{cov}\;\alpha}^{y}\right\rangle _{\text{gpCS}}\propto\left\langle t_{\text{cov}\;\alpha}^{y}\right\rangle _{\text{gCS}}$.
After summing over $l=1,2$ one finds the proportionality constant
$\frac{\alpha_{1}+\alpha_{2}}{\beta_{1}+\beta_{2}}$, that goes to
1 in the relativistic limit. Nevertheless, we argue that $\left\langle t_{\text{cov}\;\alpha}^{y}\right\rangle _{\text{gCS}}$
corresponds to a boundary gravitational anomaly while $\left\langle t_{\text{cov}\;\alpha}^{y}\right\rangle _{\text{gpCS}}$
does not, in accordance with section \ref{subsec:Boundary-anomalies}.
The relation between gCS and the boundary gravitational anomaly is
well known within the gravitational description, and is described
from the $p$-wave SC point of view in Appendix \ref{subsec:Implication-for-the}.
The fact that $\left\langle t_{\text{cov}\;\alpha}^{y}\right\rangle _{\text{gpCS}}$
is unrelated to any boundary anomaly follows from the fact that it
is $SO\left(1,2\right)$ and $Diff$ invariant. Due to this invariance
the bulk gpCS term produces not only the bulk currents \eqref{eq:92-1},
but also boundary currents, such that bulk+boundary energy-momentum
is conserved. In a product geometry with $z=y$ we find the boundary currents
\begin{align}
  \left\langle \mathsf{j}^{\alpha\beta}\right\rangle _{\text{gpCS}}=&-\beta_{1}\frac{1}{\left|e\right|}\varepsilon^{\alpha\beta y}\tilde{\mathcal{R}},\label{eq:98-3}\\
  \left\langle \mathsf{j}^{ab\mu}\right\rangle _{\text{gpCS}}=&0,\nonumber 
\end{align}
which are calculated in appendix \ref{subsec:Calculation-of-certain}.
We see that 
\begin{align}
  \tilde{\nabla}_{\alpha}\left\langle \mathsf{j}^{\alpha\beta}\right\rangle _{\text{gpCS}}=\left\langle \mathsf{J}^{y\beta}\right\rangle _{\text{gpCS}}.\label{eq:99}
\end{align}
This conservation law is the statement of bulk+boundary conservation
of energy-momentum within the gravitational description. It can be
understood from \eqref{eq:60}, by noting that the source terms in
\eqref{eq:60} vanish because $\left\langle \mathsf{j}^{ab\mu}\right\rangle _{\text{gpCS}}=0$,
and because we assumed torsion vanishes. The additional source term
$\left\langle \mathsf{J}^{y\beta}\right\rangle _{\text{gpCS}}$, absent
in \eqref{eq:60}, represents the inflow from the bulk. In Appendix
\ref{subsec:Implication-for-the} we translate \eqref{eq:99} to the
language of the $p$-wave SC.

\subsection{Discussion\label{sec:Conclusion-and-discussion}}

\paragraph*{Beyond the relativistic limit}

In this section we have shown that there is a topological bulk response
of $p$-wave CSF/Cs to a perturbation of their order parameter, which
follows from a gCS term, and we described a corresponding gravitational
anomaly of the edge states. The coefficient of gCS was found to be
$\alpha=\frac{c}{96\pi}$ where $c$ is the chiral central charge,
as anticipated. We provided arguments, based on symmetry and topology,
for the validity of these results beyond the relativistic limit in
which they were computed. The appearance of torsion in the emergent
geometry brought about a surprise: an additional term, closely related
but distinct from gCS, which we referred to as gravitational pseudo
Chern-Simons (gpCS), with a dimensionless coefficient $\beta=\frac{c}{96\pi}=\alpha$.
The gpCS term is completely invariant under the symmetries we considered,
and is therefore unrelated to edge anomalies. Therefore, the quantization
of $\beta$ seems to be a property of the relativistic limit\footnote{The quantization of $\beta$ can be understood on dimensional grounds
- in the relativistic limit there are simply not enough dimension-full
quantities which can be used to construct dimensionless quantities,
beyond $\text{sgn}\left(m\right)$ and $o$. Of course, this does
not explain why $\beta=\alpha$ in the relativistic limit. }, which will not hold throughout the phase diagram. 

The above results are based on a careful mapping of $p$-wave CSF/Cs,
in the regime where the order parameter is large, to relativistic
Majorana spinors in a curved and torsion-full space-time. Though the
relativistic limit captures a rich geometric physics in CSF/Cs, it
is in fact very limited in its ability to describe the full non-relativistic
system, as discussed in Appendix \ref{subsec:Relativistic-limit-of}.
This raises the question of whether the results obtained in this section
truly characterize $p$-wave CSF/Cs, especially in the physically
important weak coupling limit where $\Delta$ is small, which is opposite
to the relativistic limit. Moreover, it is natural to ask whether
the results obtained here apply to $\ell$-wave CSF/Cs with $\left|\ell\right|>1$,
which do not admit a relativistic low energy description. These questions
will be answered, in part, in Sec.\ref{sec:Main-section-2:}, where
we perform a general non-relativistic analysis of continuum models
for $\ell$-wave CSF/Cs. In order to address these questions in the
context of lattice models, numerical studies seem to be required.
Recently, such studies where initiated in the context of emergent
gravity in Kitaev's honeycomb model \citep{PhysRevB.101.245116,PhysRevB.102.125152},
which is closely related to the $p$-wave CSF/Cs discussed in this
section. For non-topological physical properties, the results show
that the lattice model is well described by relativistic predictions,
but only in the relativistic regime, where the masses of low energy
relativistic fermions are small and the correlation length is large,
as might be expected. An interesting question is whether the topological
physics associated with the gCS term is more robustly captured by
the relativistic predictions made in this section.

\paragraph*{Real background geometry and manipulation of the order parameter}

In this section we considered $p$-wave CSF/Cs in flat space, and
focused on the emergent geometry described by a general $p$-wave
order parameter. It is also natural to consider the effect of a real
background geometry, obtained by deforming, or straining, the 2-dimensional
sample in 3-dimensional space, possibly in a time dependent manner.
Treating this at the level of the lattice model is beyond the scope
of this thesis, but we can take the non-relativistic action \eqref{eq:13}
as a starting point.   On a general deformed sample it generalizes
to 
\begin{align}
 & S\left[\psi;\Delta,A,G\right]=\int\mbox{d}^{2+1}x\sqrt{G}\left[\psi^{\dagger}\frac{i}{2}\overleftrightarrow{D_{t}}\psi-\frac{1}{2m^{*}}G^{ij}D_{i}\psi^{\dagger}D_{j}\psi+\left(\frac{1}{2}\Delta^{j}\psi^{\dagger}\partial_{j}\psi^{\dagger}+h.c\right)\right],\label{eq:120}
\end{align}
which now depends on three background fields: the order parameter
$\Delta^{j}$, the $U\left(1\right)$ connection $A_{\mu}$, which
enters $D_{\mu}=\partial_{\mu}-iA_{\mu}$, and includes a chemical
potential, and the real background metric $G$, coming from the embedding
of the 2-dimensional sample in 3-dimensional space, which corresponds
to the strain tensor $u^{ij}=\left(G^{ij}-\delta^{ij}\right)/2$.
This action is written for the fermion $\psi$, which satisfies
$\left\{ \psi^{\dagger}\left(x\right),\psi\left(y\right)\right\} =\delta^{\left(2\right)}\left(x-y\right)/\sqrt{G\left(x\right)}$
as an operator. In this problem there are two (inverse) metrics, the
real $G^{ij}$ and emergent $g^{ij}=\Delta^{(i}\Delta^{j)*}$, and
it is interesting to study their interplay. In our analysis we focused
on the relativistic limit, where $m^{*}\rightarrow\infty$. In this
limit the metric $G$ completely decouples from the action, when written
in terms of the fundamental fermion density $\tilde{\psi}=G^{1/4}\psi$
\citep{hawking1977zeta,fujikawa1980comment,abanov2014electromagnetic}
. Thus, results obtained within the relativistic limit, are essentially
unaffected by the background metric $G$. This conclusion is appropriate
as long as the order parameter is treated as an independent background
field, which is always suitable for the purpose of integrating out
the gapped fermion density $\psi$. One then obtains the bulk currents
and densities that we have described, which depend on the configuration
of $\Delta$, and the question that remains is what this configuration
physically is. Two scenarios are of importance.

The first scenario is that of a proximity induced CSC, where the order
parameter is induced by proximity to a conventional $s$-wave SC.
In this case both $G$ and $g$ are background metrics, a scenario
similar to a bi-metric description of anisotropic quantum Hall states
\citep{gromov2017investigating}. In this case the magnitude of the
order parameter depends on the distance between the sample and the
$s$-wave SC, so if the position of the $s$-wave SC is fixed but
the sample is deformed, a space-time dependent order parameter is
obtained. One can also obtain the same effect by considering a flat
sample, where $G$ is euclidian, and an $s$-wave SC with a curved
surface, leading to non-Euclidian $g$. This provides one rout to
a manipulation of the order parameter that will result in the bulk
effects we have described. 

The second scenario is that of intrinsic CSCs, and CSFs, where the
order parameter is dynamical. The order parameter splits into a massive
Higgs field, which is the emergent metric $g^{ij}$, and a massless
Goldstone field which is the overall phase $\theta$. The quantum
theory of the emergent metric $g^{ij}$ is on its own an interesting
problem, which will be discussed in Sec.\ref{sec:Discussion-and-outlook}.
Nevertheless, as long as the probes $A,G$ are slow compared to the
Higgs mass, $g^{ij}$ can be treated as fixed to its instantaneous
ground state configuration, which in general will depend on the details
of the attractive fermionic interaction. For an interaction that depends
only on the geodesic distance, the ground state configuration is expected
to be the curved space $p_{x}\pm ip_{y}$ configuration, where the
pairing term is $\Delta_{0}e^{i\theta}\psi^{\dagger}\left(E_{1}^{\;j}\partial_{j}\pm iE_{2}^{\;j}\partial_{j}\right)\psi^{\dagger}$,
see Appendix \ref{sec:microscopic model} and Refs.\citep{read2000paired,hoyos2014effective,moroz2015effective,quelle2016edge,moroz2016chiral}.
Here $\Delta_{0}$ is a constant, $\theta$ is the Goldstone phase,
and $E$ is a vielbein for the real metric $G$, such that $G^{ij}=E_{A}^{\;i}\delta^{AB}E_{B}^{\;j}$,
which is a fixed background field\footnote{The $SO\left(2\right)_{L}$ ambiguity in choosing $E$ can be incorporated
into $\theta$, which has $SO\left(2\right)_{L}$ charge 1}. 
What this means, in the language of this section, is that the emergent
metric is proportional to the real metric, $g^{ij}=\Delta_{0}^{2}G^{ij}$.
It follows that the responses to the emergent metric $g$ that we
have described are, in this case, responses to the real metric $G$.
This suggests a second rout to the manipulation of the order parameter
that will result in the bulk effects we have described. 

Of course, in the intrinsic case one cannot ignore the dynamics of
the Goldstone phase $\theta$, which will be gapless as long as $A$
is treated as a background field, as appropriate in CSFs. It is then
interesting to see how the results obtained in this section are modified
by the dynamics of $\theta$. The interplay between the topological
gCS term and the $\theta$-dependent gpCS term suggests a non-trivial
modification, which will be explored in Sec.\ref{sec:Main-section-2:}.

\paragraph*{Towards experimental observation }

There are a few basic questions that arise when trying to make contact
between the phenomena described in this section and a possible experimental
observation. Here we take as granted that one has at one's disposal
either a $p$-wave CSF/C, or a candidate material. The first question
is how to manipulate the Higgs part of the order parameter, which
is the emergent metric, and was discussed above. 

The second natural question is how to measure energy currents and
momentum densities. Also relevant, though not accentuated in this
section, is a measurement of the stress tensor, comprised of the spatial
components of the energy-momentum tensor. One possible approach, which
provides both a means to manipulate the order parameter, and a measurement
of energy-momentum-stress is a measurement of the phonon spectrum
a la \citep{barkeshli2012dissipationless,schmeltzer2014propagation,schmeltzer2017detecting}.
 For the gpCS term, apart from energy-momentum-stress, there is also
the density response \eqref{eq:9-1}, which is a simpler quantity
for measurement, though not a \textit{topological} bulk response.
A possible way to avoid the need to measure energy-momentum-stress
may be possible in a Galilean invariant system, where electric current
and momentum density are closely related. The simplest scenario is
that of a $p$-wave CSF on a curved sample \eqref{eq:120}, where
one assumes that the emergent metric follows the real metric, $g^{ij}=\Delta_{0}^{2}G^{ij}$.
Here the electric current is related to the momentum density by 
\begin{align}
 & J^{i}=-\frac{G^{ij}}{m^{*}}P_{j}.
\end{align}
Our result \eqref{eq:4} then implies that the expectation value $\left\langle J^{i}\right\rangle $
has a contribution related to the gCS term, 
\begin{align*}
 & \left\langle J^{i}\right\rangle _{\text{gCS}}=-\frac{G^{ij}}{m^{*}}\left\langle P_{j}\right\rangle _{\text{gCS}}=\frac{1}{m^{*}}\frac{c}{96\pi}\hbar\varepsilon^{ij}\partial_{j}\tilde{\mathcal{R}}.
\end{align*}
This is not a topological response per se, due to the appearance of
$m^{*}$. But, if $m^{*}$ is known, then the central charge $c$
can be extracted from a measurement of the electric current, which
may be simpler to measure than energy-momentum-stress. This motivates
the study of Galilean invariant CSFs, which will pursued in Sec.\ref{sec:Main-section-2:}. 

\paragraph*{Hall viscosity and torsional Hall viscosity}

In the effective action \eqref{eq:72} for a Majorana spinor in Riemann-Cartan
space-time, there is a term $\left(\zeta_{H}/2\right)\int e_{a}De^{a}$,
which describes an energy-momentum-stress response termed \textit{torsional
Hall viscosity} \citep{hughes2011torsional,hughes2013torsional,parrikar2014torsion},
due to its similarities with the odd (or Hall) viscosity $\eta_{\text{o}}$
that occurs in quantum Hall states and CSF/Cs \citep{avron1995viscosity,read2009non,abanov2014electromagnetic,hoyos2014effective,moroz2015effective}.
As opposed to the well understood $\eta_{\text{o}}$, the torsional
Hall viscosity $\zeta_{H}$ remains somewhat controversial \citep{geracie2014hall,hoyos2014hall,bradlyn2015low},
because it only appears if a non-symmetric stress tensor is used,
and because it is UV-sensitive. The latter is also the reason that,
in this section, we avoided from interpreting $\zeta_{H}$ in the
context of $p$-wave CSF/Cs. However, the mapping between $p$-wave
CSF/Cs and relativistic Majorana spinors in Riemann-Cartan space-time
developed in this section, along with the existence of a Hall viscosity
on one side of the mapping, and of a torsional Hall viscosity on the
other, strongly suggests that the two types of Hall viscosity should
map to one another, in this context. Moreover, it is known that the
gCS term implies universal corrections to $\eta_{\text{o}}$ at non-zero
wave-vector or in curved background \citep{abanov2014electromagnetic,bradlyn2015topological,klevtsov2015geometric},
and that the gpCS term implies curvature corrections to $\zeta_{H}$
\citep{hughes2013torsional}, which we expect to be non-universal.
These observations provide us with a strong motivation to study the
interplay between torsional Hall viscosity, Hall viscosity, and chiral
central charge in CSFs, which we do in Sec.\ref{sec:Main-section-2:}.

\pagebreak{}

\section{Boundary central charge from bulk odd viscosity: chiral superfluids\label{sec:Main-section-2:}}

As discussed in Sec.\ref{sec:Conclusion-and-discussion}, our analysis
of CSF/Cs within the relativistic limit raises many questions which
require a fully non-relativistic treatment. These questions revolve
around the quantization of coefficients, application of strain (or
a background metric), odd viscosity, dynamics of the Goldstone mode,
and Galilean symmetry, and will be addressed in this section. In particular,
we will answer the questions posed in Sec.\ref{subsec:Odd-viscosity},
by deriving a low energy effective field theory that consistently
captures both the chiral Goldstone mode implied the symmetry breaking pattern \eqref{eq:2-1-1},
and the gCS term discussed in Sec.\ref{sec:Main-section-1:}. This
theory unifies and extends the seemingly unrelated analysis of Refs.\citep{son2006general,hoyos2014effective,moroz2015effective}
and Sec.\ref{sec:Main-section-1:}.

\subsection{Building blocks for the effective field theory\label{sec: building blocks}}

In order to probe a CSF, we minimally couple it to two background
fields - a time-dependent spatial metric $G_{ij}$, which we use to
apply strain $u_{ij}=\left(G_{ij}-\delta_{ij}\right)/2$ and strain-rate
$\partial_{t}u_{ij}$, and a $U\left(1\right)_{N}$-gauge field $A_{\mu}=\left(A_{t},A_{i}\right)$,
where we absorb a chemical potential $A_{t}=-\mu+\cdots$. The microscopic
action $S$ is then invariant under $U\left(1\right)_{N}$ gauge transformations,
implying the number conservation $\partial_{\mu}(\sqrt{G}J^{\mu})=0$,
where $\sqrt{G}J^{\mu}=-\delta S/\delta A_{\mu}$. It is also clear
that $S$ is invariant under \textit{spatial} diffeomorphisms generated
by $\delta x^{i}=\xi^{i}\left(\mathbf{x}\right)$, if $G_{ij}$ transforms
as a tensor and $A_{\mu}$ as a 1-form.  Less obvious is the fact
that a Galilean invariant fluid is additionally symmetric under $\delta x^{i}=\xi^{i}\left(t,\mathbf{x}\right)$,
provided one adds to the transformation rule of $A_{i}$ a non-standard
 piece that dependens on the non-relativistic mass $m$  \citep{son2006general,hoyos2012hall,hoyos2014effective,gromov2014density,Andreev:2014aa,geracie2015spacetime,Andreev:2015aa,Geracie:2017aa},
\begin{align}
 & \delta A_{i}=-\xi^{k}\partial_{k}A_{i}-A_{k}\partial_{i}\xi^{k}+mG_{ij}\partial_{t}\xi^{j}.\label{eq:4-3-1-1}
\end{align}
We refer to $\delta x^{i}=\xi^{i}\left(\mathbf{x},t\right)$ as \textit{local
Galilean symmetry} (LGS), as it can be viewed as a local version of
the Galilean transformation $\delta x^{i}=v^{i}t$. The LGS implies
the momentum conservation law\footnote{We use the notation $\varepsilon^{\mu\nu\rho}$ for the totally anti-symmetric
(pseudo) tensor, normalized such that $\varepsilon^{xyt}=1/\sqrt{G}$,
as well as $\varepsilon^{ij}=\varepsilon^{ijt}$.} 
\begin{align}
 & \frac{1}{\sqrt{G}}\partial_{t}\left(\sqrt{G}mJ^{i}\right)+\nabla_{j}T^{ji}=nE_{i}+\varepsilon^{ij}J_{j}B,\label{eq:5-3-1}
\end{align}
where $\sqrt{G}T^{ij}=2\delta S/\delta G^{ij}$ is the stress tensor
and the right hand side is the Lorentz force. This identifies the
momentum density $P^{i}=mJ^{i}$ - a familiar Galilean relation.

Since CSFs spontaneously break the rotation symmetry in flat space,  in order to describe them in curved, or strained, space, it is necessary to introduce a background vielbein. This is a field $E_{\;j}^{A}$ valued in $GL\left(2\right)$, such that $G_{ij}=E_{\;i}^{A}\delta_{AB}E_{\;j}^{B}$, where  $A,B\in\left\{ 1,2\right\} $. For a given metric $G$ the vielbein $E$ is not unique - there is an internal $O\left(2\right)_{P,L}=\mathbb{Z}_{2,P}\ltimes SO\left(2\right)_{L}$ ambiguity, or symmetry, acting by $E_{\;j}^{A}\mapsto O_{\;B}^{A}E_{\;j}^{B}$, $O\in O\left(2\right)_{P,L}$. The generators $L,P$ correspond to \textit{internal} spatial rotations and reflections, and are analogs of angular momentum and spatial reflection (parity), acting  on the tangent space rather than on space itself. The inverse vielbein $E_{B}^{\;j}$ is defined by $E_{\;j}^{A}E_{B}^{\;j}=\delta_{B}^{A}$.

\begin{figure}[!th]
\begin{centering}
\par\end{centering}
\begin{centering}
\par\end{centering}
\begin{centering}
\par\end{centering}
\begin{centering}
\includegraphics[width=0.6\linewidth]{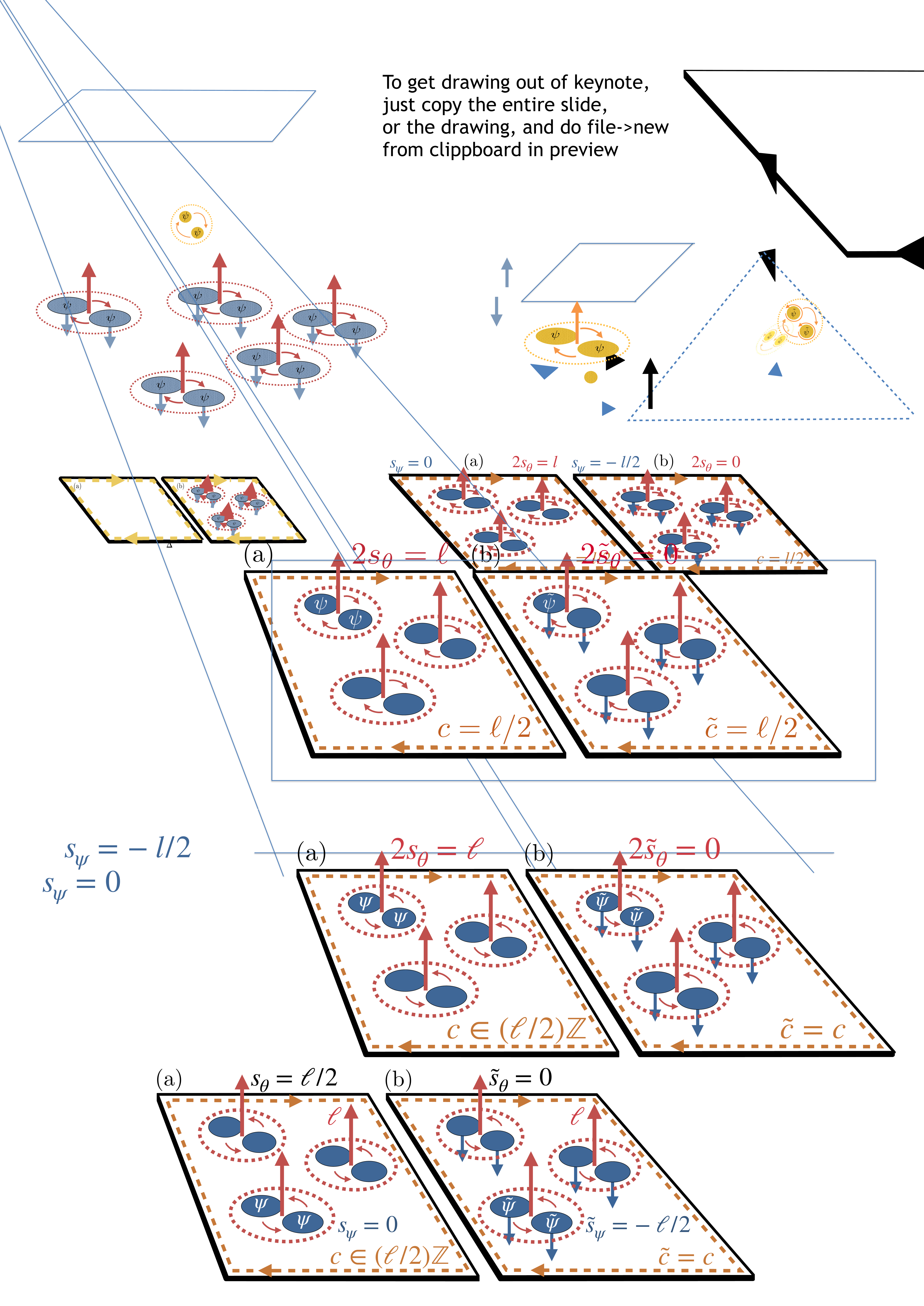}
\par\end{centering}
\caption{(a) A CSF is comprised of fermions $\psi$ which carry no \textit{geometric
}spin, $s_{\psi}=0$, which form Cooper pairs with relative angular
momentum $\ell\in\mathbb{Z}$ (red arrows). The geometric spin $s_{\theta}=\ell/2$
of the Cooper pair gives rise to the $\mathbf{q}=\mathbf{0}$ odd
viscosity \eqref{eq:1-1}, with $s=s_{\theta}$. The CSF supports
boundary degrees of freedom (dashed orange) with a chiral central
charge $c\in\left(\ell/2\right)\mathbb{Z}$, which cannot be extracted
from the odd viscosity $\eta_{\text{o}}\left(\mathbf{q}\right)$ alone
\eqref{eq:14-1}. (b) In an auxiliary CSF the fermion $\tilde{\psi}$
is assigned a geometric spin $\tilde{s}_{\psi}=-\ell/2$ (blue arrows).
The geometric spin of the Cooper pair therefore vanishes, $\tilde{s}_{\theta}=\ell/2+\tilde{s}_{\psi}=0$,
as in an $s$-wave superfluid, but the central charge is unchanged,
$\tilde{c}=c$. As a result, the small $\mathbf{q}$ behavior of the
odd viscosity $\tilde{\eta}_{\text{o}}\left(\mathbf{q}\right)$ depends
only on $c$ \eqref{eq:18-1-1}. The improved odd viscosity of the
CSF is defined as the odd viscosity of the auxiliary CSF, and is given
explicitly by \eqref{eq:16-2-1}. \label{fig:Comparison-of-the-1-1}}
\end{figure}

 The charge $N+\left(\ell/2\right)L$ of the Goldstone field $\theta$
implies the covariant derivative
\begin{align}
 & \nabla_{\mu}\theta=\partial_{\mu}\theta-A_{\mu}-s_{\theta}\omega_{\mu},\label{eq:6}
\end{align}
with a \textit{geometric spin} $s_{\theta}=\ell/2$. Here $\omega_{\mu}$
is the non-relativistic spin connection, an $SO\left(2\right)_{L}$-gauge
field which is $\ensuremath{E_{\;j}^{A}}$-compatible, see Appendix \ref{sec:Geometric quantities}.
So far we assumed that the microscopic fermion $\psi$ does not carry
a geometric spin, $s_{\psi}=0$, which defines the physical system
of interest, see Fig.\ref{fig:Comparison-of-the-1-1}(a). It will be useful, however, to generalize to $s_{\psi}\in\left(1/2\right)\mathbb{Z}$.
A non-zero $s_{\psi}$ changes the SSB pattern \eqref{eq:2-1-1},
by modifying the geometric spin of the Goldstone field to $s_{\theta}=s_{\psi}+\ell/2$
and the unbroken generator to $L-s_{\theta}N$. In the special case
$s_{\psi}=-\ell/2$ the Cooper pair is geometrically spin-less and
$L$ is unbroken, as in an $s$-wave SF, see Fig.\ref{fig:Comparison-of-the-1-1}(b).
This $s_{\theta}=0$ CSF is, however, distinct from a conventional
$s$-wave SF, because $P$ and $T$ are still broken down to $PT$,
and we therefore refer to it as a \textit{geometric} $s$-wave (g$s$-wave)
CSF, to distinguish the two. In particular, a central charge $c\neq0$,
which is $P,T$-odd, is not forbidden by symmetry, and is in fact
independent of $s_{\psi}$. This makes the g$s$-wave CSF particularly
useful for our purposes. 

We note that $\omega_{\mu}$ transforms as a 1-form under LGS only
if $B/2m$ is added to $\omega_{t}$ \citep{hoyos2014effective,moroz2015effective},
which we do implicitly throughout. For $\psi$, this is equivalent
to adding  a g-factor $g_{\psi}=2s_{\psi}$ \citep{geracie2015spacetime}. 

\subsection{Effective field theory\label{sec: effective field theory}}

 Based on the above characterization of CSFs, the low energy, long
wave-length, behavior of the system can be captured by an effective
action $S_{\text{eff}}\left[\theta;A,G;\ell,c\right]$, formally obtained
by integrating out all massive degrees of freedom - the single fermion
excitations and the Higgs fields. In this Section we describe a general
expression for $S_{\text{eff}}$, compatible with the symmetries,
SSB pattern, and ground state topology of CSFs. 

The effective action can be written order by order
in a derivative expansion, with the power counting scheme $\partial_{\mu}=O\left(p\right),\;A_{\mu},G_{ij}=O\left(1\right),\;\theta=O\left(p^{-1}\right)$
\citep{son2006general,hoyos2014effective}. The spin connection
is a functional of $G_{ij}$ that involves a single derivative (see Appendix \ref{sec:Geometric quantities}),
so $\omega_{\mu}=O\left(p\right)$. Denoting by $\mathcal{L}_{n}$
the term in the Lagrangian which is $O\left(p^{n}\right)$ and invariant
under all symmetries, we have $S_{\text{eff}}=\sum_{n=0}^{\infty}\int\text{d}^{2}x\text{d}t\sqrt{G}\mathcal{L}_{n}$.
The desired $q^{2}$ corrections to $\eta_{\text{o}}$ are $O\left(p^{3}\right)$,
which poses the main technical difficulty. 

The leading order Lagrangian
\begin{align}
 & \mathcal{L}_{0}=P\left(X\right),\;X=\nabla_{t}\theta-\frac{1}{2m}G^{ij}\nabla_{i}\theta\nabla_{j}\theta,\label{eq:9-0}
\end{align}
was studied in detail in \citep{hoyos2014effective}, and contains
the earlier results of \citep{volovik1988quantized,goryo1998abelian,goryo1999observation,furusaki2001spontaneous,stone2004edge,roy2008collective,lutchyn2008gauge}.
Here $X$ is the unique $O\left(1\right)$ scalar, which reduces
to the chemical potential $\mu$ in the ground state(s) $\partial_{\mu}\theta=0$,
and $P$ is an arbitrary function of $X$  that physically corresponds
to the ground state pressure $P_{0}=P\left(\mu\right)$. The function
$P$ also determines the ground state density $n_{0}=P'\left(\mu\right)$,
and the leading dispersion of the Goldstone mode $\omega^{2}=c_{s}^{2}q^{2}$,
where $c_{s}^{2}=\partial_{n_{0}}P_{0}/m=P'(\mu)/P''(\mu)m$ is the speed of
sound. For $\ell\neq0$, the spin connection appears in each $\nabla\theta$, see
Eq.\eqref{eq:6}, and so $\mathcal{L}_{0}$ includes $O\left(p\right)$
contributions, which produce the leading odd viscosity and conductivity,
discussed below. There are no additional terms at $O\left(p\right)$,
so that $\mathcal{L}_{1}=0$ \citep{hoyos2014effective}. 

At $O\left(p^{2}\right)$ one has  
\begin{align}
\mathcal{L}_{2}= & F_{1}\left(X\right)R+F_{2}\left(X\right)\left[mK_{\;i}^{i}-\nabla^{2}\theta\right]^{2}+F_{3}\left(X\right)\left[2m\left(\nabla_{i}K_{\;j}^{j}-\nabla^{j}K_{ji}\right)\nabla^{i}\theta\right]+\cdots,\label{eq:10-1-0}
\end{align}
where $K_{ij}=\partial_{t}G_{ij}/2$ and $R$ are the extrinsic curvature
and Ricci scalar of the spatial slice at time $t$ \citep{carroll2004spacetime},
the $F$s are arbitrary functions of $X$, and dots indicate additional
terms which do not contribute to $\eta_{\text{o}}$ up to $O\left(p^{2}\right)$,
see Appendix \ref{subsec: Second order effective action} for the full expression. The Lagrangian $\mathcal{L}_{2}$ was obtained
in \citep{son2006general} for $s$-wave SFs. For $\ell\neq0$ the
spin connection in $\nabla\theta$ produces $O\left(p^{3}\right)$
contributions to $\mathcal{L}_{2}$, and, in turn, non-universal $q^{2}$
corrections to $\eta_{\text{o}}$.

The term $\mathcal{L}_{3}$ is the last ingredient required for reliable
results at $O\left(p^{3}\right)$. Most importantly, it includes the
non-relativistic gCS term  \citep{bradlyn2015topological,gromov2016boundary},
$\mathcal{L}_{3}=\text{\ensuremath{\mathcal{L}}}_{\text{gCS}}+\cdots$,
where  
\begin{align}
\text{\ensuremath{\mathcal{L}}}_{\text{gCS}} & =-\frac{c}{48\pi}\omega\text{d}\omega.\label{eq:10-0}
\end{align}


Here $\omega\text{d}\omega=\varepsilon^{\mu\nu\rho}\omega_{\mu}\partial_{\nu}\omega_{\rho}$,
and $c$ is the chiral central charge of the boundary degrees of freedom,
as required to match the boundary gravitational anomaly \citep{kraus2006holographic,stone2012gravitational,PhysRevB.98.064503}.
 Unlike the lower order terms, $\mathcal{L}_{\text{gCS}}$ is independent
of $\theta$, and encodes only the response of the fermionic topological
phase to the background metric. The direct confirmation of \eqref{eq:10-0}
within a non-relativistic microscopic model has been anticipated for
some time \citep{volovik1990gravitational,read2000paired},
and is a main result of Appendix \ref{sec:microscopic model}.
         In Appendices \ref{subsec:Odd-viscosity-from} and \ref{subsec:Additional terms at third order} we argue that additional terms in $\mathcal{L}_{3}$ do
not produce $q^{2}$ corrections to $\eta_{\text{o}}$.

There are three symmetry-allowed topological terms that can be added
to $S_{\text{eff}}$ \citep{ferrari2014fqhe,can2014fractional,abanov2014electromagnetic,gromov2014density,gromov2015framing,can2015geometry,klevtsov2015geometric,bradlyn2015topological,Klevtsov_2016,klevtsov2017laughlin,Cappelli_2018}.
These are the $U\left(1\right)$ Chern-Simons (CS) and first and second
Wen-Zee (WZ1, WZ2) terms, which can be added to $\mathcal{L}_{1}$,
$\mathcal{L}_{2}$, $\mathcal{L}_{3}$, respectively \footnote{These are not LGS invariant, and would need to be modified along the
lines of \citep{hoyos2012hall,Andreev:2014aa,Andreev:2015aa}. }, 
\begin{align}
 & \frac{\nu}{4\pi}\left(A\text{d}A-2\overline{s}\omega\text{d}A+\overline{s^{2}}\omega\text{d}\omega\right).\label{eq:11}
\end{align}
As our notation suggests, WZ2 and gCS are identical for the purpose
of local bulk responses, of interest here, but the two are globally
distinct \citep{bradlyn2015topological,gromov2016boundary,Cappelli_2018}.
Based on symmetry, and ignoring boundary physics, the independent
coefficients $\nu$, $\nu\overline{s}$, and $\nu\overline{s^{2}}$
obey certain quantization conditions \citep{witten2007three}, but
are otherwise unconstrained. The absence of a boundary $U\left(1\right)_{N}$-anomaly
then fixes $\nu=0$ \citep{PhysRevB.98.064503}, but leaves $\nu\overline{s},\nu\overline{s^{2}}$
undetermined \citep{bradlyn2015topological,gromov2016boundary,Cappelli_2018}.
One can argue that a Chern-Simons term can only appear for the unbroken
generator $L-s_{\theta}N$, so that $\nu=0$ implies $\nu\overline{s}=\nu\overline{s^{2}}=0$.
Moreover, in the following section we will see that a perturbative
computation within a canonical model for $\ell=\pm1$ shows that $\nu\overline{s}=\nu\overline{s^{2}}=0$,
which applies to any deformation of the model (which preserves the
symmetries, SSB pattern, and single fermion gap), due to the quantization
of $\nu\overline{s},\nu\overline{s^{2}}$. Accordingly, we set $\nu\overline{s}=\nu\overline{s^{2}}=0$
in the following.

\subsection{Benchmarking the effective theory against a microscopic model\label{subsec:Benchmarking-the-effective}}

In this section we take a complementary approach and compute $S_{\text{eff}}$
perturbatively, starting from a canonical microscopic model for a
$p$-wave CSF. The perturbative computation verifies the general expression
in a particular example, and determines the coefficients of topological
terms which are not completely fixed by symmetry. It also gives one
a sense of the behavior of the coefficients of non-topological terms
as a function of microscopic parameters such as the chemical potential $\mu$ and mass $m$.  In particular, the results of Sec.\ref{sec:Main-section-1:} are reproduced in the relativistic limit $m\rightarrow \infty$.

Here we will outline the computation and describe its results, omitting many technical details which can be found in Appendix \ref{sec:microscopic model}.
The microscopic model is given by 
\begin{align}
S_{\text{m}}=\int\mbox{d}^{2}x & \text{d}t\sqrt{G}\left[\frac{i}{2}\psi^{\dagger}\overleftrightarrow{\nabla_{t}}\psi-\frac{1}{2m}G^{ij}\nabla_{i}\psi^{\dagger}\nabla_{j}\psi+\left(\frac{1}{2}\Delta^{j}\psi^{\dagger}\nabla_{j}\psi^{\dagger}+h.c\right)-\frac{1}{2\lambda}G_{ij}\Delta^{i*}\Delta^{j}\right],\label{eq:3-1-1}
\end{align}
where the covariant derivative of the spin-less (or single component) fermion $\psi$ is
$\nabla_{\mu}=\partial_{\mu}+iA_{\mu}+is_{\psi}\omega_{\mu}$. Note that  we allow for a  \textit{geometric} spin $s_{\psi}$, as discussed in Sec.\ref{sec: building blocks}. Apart
from the standard non-relativistic kinetic term, the action includes
the simplest attractive two-body interaction \citep{Volovik:1988aa,quelle2016edge},
mediated by the complex vector $\Delta^{i}$, the order parameter,
with coupling constant $\lambda>0$. A simplified version of this
action was already discussed in Sec.\ref{sec:Conclusion-and-discussion}. 

For a given $\Delta^{j}$, the fermion $\psi$ is gapped, unless the
chemical potential $\mu$ or chirality $\ell=\text{sgn}\left(\text{Im}\left(\Delta^{x}\Delta^{y*}\right)\right)$
 are tuned to 0, and forms a fermionic topological phase characterized
by the boundary chiral central charge \citep{read2000paired,volovik2009universe,ryu2010topological}
\begin{align}
 & c=-\left(\ell/2\right)\Theta\left(\mu\right)\in\left\{ 0,\pm1/2\right\} .\label{eq:12-1-2}
\end{align}

An effective action $S_{\text{eff},\text{m}}\left[\Delta;A,G\right]$
for $\Delta^{j}$ in the background $A_{\mu},G_{ij}$ is then obtained
by integrating over the fermion. The subscript 'm' indicates that
this is obtained from the particular microscopic model $S_{\text{m}}$.
Since Eq.\eqref{eq:3-1-1} is quadratic in $\psi,\psi^{\dagger}$,
obtaining $S_{\text{eff},\text{m}}$ is formally straight forward,
and leads to a functional Pfaffian. 

To zeroth order in derivatives $S_{\text{eff},\text{m}}\left[\Delta;A,G\right]=-V\left[\Delta;G\right]$,
where the potential $V$ is minimized by the $p_{x}\pm ip_{y}$ order
parameter. In flat space this is given by the familiar $\Delta^{j}\partial_{j}=\Delta_{0}e^{-2i\theta}\left(\partial_{x}\pm i\partial_{y}\right)$.
Here $\Delta_{0}$ is a fixed function of $m,\mu$ and $\lambda$,
determined by the minimization, while the phase $\theta$ and chirality
$\ell=\pm1$ are undetermined. In order to write down the $p_{x}\pm ip_{y}$
configuration in curved space it is necessary to use the background
vielbein \citep{hoyos2014effective,quelle2016edge}
\begin{align}
\Delta^{j} & =\Delta_{0}e^{-2i\theta}\left(E_{1}^{\;j}\pm iE_{2}^{\;j}\right).\label{eq:12-3}
\end{align}

Fluctuations of $\Delta$ away from these configurations correspond
to massive Higgs modes, which should in principle be integrated out
to obtain a low energy action $S_{\text{eff},\text{m}}\left[\theta;A,G\right]$
that can be compared with the general $S_{\text{eff}}$ of the previous
section. We will simply ignore these fluctuations, and obtain $S_{\text{eff},\text{m}}\left[\theta;A,G\right]$
by plugging Eq.\eqref{eq:12-3} into $S_{\text{eff},\text{m}}\left[\Delta;A,G\right]$.
This will suffice as a derivation of $S_{\text{eff}}$ from a microscopic
model. A proper treatment of the massive Higgs modes will only further
renormalize the coefficients we find, apart from the central charge
$c$.

To practically compare the actions $S_{\text{eff}}$ and $S_{\text{eff},\text{m}}$
we expand them in fields, to second order around $\theta=0,A_{\nu}=-\mu\delta_{\nu}^{t},G_{ij}=\delta_{ij}$,
and in derivatives, to third order, see appendices \ref{subsec:Effective-action-and} and \ref{subsec:Perturbative-expansion}. Equating
these two double expansions leads to an overdetermined system of equations
for the phenomenological parameters in $S_{\text{eff}}$ in terms
of the microscopic parameters in $S_{\text{m}}$, with a unique solution.
In particular, we find the dimensionless parameters
\begin{align}
\frac{P''}{m}= & \frac{1}{2\pi}\begin{cases}
1\\
\frac{1}{1+2\kappa}
\end{cases},\;\;\;\;\;F_{1}'=\frac{1}{96\pi}\begin{cases}
1\\
\frac{3}{1+2\kappa}
\end{cases},\;\;\;\;\;mF_{2}=-\frac{1}{128\pi}\begin{cases}
1+2\kappa\\
\frac{1}{1+2\kappa}
\end{cases},\label{eq:c2}\\
mF_{3}= & \frac{1}{48\pi}\begin{cases}
1+\kappa\\
\frac{1}{1+2\kappa}
\end{cases},\;\;\;\;\;c=\begin{cases}
-\ell/2\\
0
\end{cases},\nonumber 
\end{align}
where $\kappa=\left|\mu\right|/m\Delta_{0}^{2}>0$, primes denote derivatives with respect to $\mu$, and the cases
refer to $\mu>0$ and $\mu<0$. We note that for $\mu>0$ there is
a single particle Fermi surface, with energy $\varepsilon_{F}=\mu$
and wave-vector $k_{F}=\sqrt{2m\mu}$, which for small $\lambda$
will acquire an energy gap $\varepsilon_{\Delta}=\Delta_{0}k_{F}\ll\varepsilon_{F}$.
In this  weak-coupling regime, it is natural to parametrize the
coefficients in \eqref{eq:c2} using the small parameter $\varepsilon_{\Delta}/\varepsilon_{F}=\sqrt{2/\kappa}$.

The coefficient $P''$ determines the leading odd (or Hall) conductivity
and has been computed previously in the literature \citep{volovik1988quantized,goryo1998abelian,goryo1999observation,furusaki2001spontaneous,stone2004edge,roy2008collective,lutchyn2008gauge,hoyos2014effective,ariad2015effective},
while $F_{1},F_{2}$ and $F_{3}$, to the best of our knowledge, have
not been computed previously, even for an $s$-wave SF. Crucially,
Eq.\eqref{eq:c2} shows that the coefficient $c$ of the bulk gCS
term matches the known boundary central charge \eqref{eq:12-1-2},
which is a main result of the perturbative computation. It follows
that there is no WZ2 term in $S_{\text{eff},\text{m}}$, so $\nu\overline{s^{2}}=0$,
in accordance with the general discussion of the previous section.
We additionally confirm that $\nu=\nu\overline{s}=0$.

A few additional comments regarding Eq.\eqref{eq:c2} are in order:
\begin{enumerate}
\item The seeming quantization of $P''/m$ and $F_{1}'$ for $\mu>0$ is
a non-generic result, as was shown explicitly for $P''/m$ \citep{ariad2015effective}.
\item The free fermion limit $\kappa\rightarrow\infty$, or $\Delta_{0}\rightarrow0$,
of certain coefficients in \eqref{eq:c2} diverges for $\mu>0$
but not for $\mu<0$. This signals the breakdown of the gradient expansion
for a gapless Fermi surface, but not for gapped free fermions.
\item   The opposite limit, $\kappa\rightarrow0$, or $m\rightarrow\infty$,
is the relativistic limit studied in Sec.\ref{sec:Main-section-1:}, in which the fermionic
part of the model reduces to a 2+1 dimensional Majorana spinor with
mass $\mu$ and speed of light $\Delta_{0}$, coupled to a Riemann-Cartan
geometry described by $\Delta^{i},\;A_{\mu}$.  Accordingly, there is a sense in which the limit $\kappa\rightarrow\infty$ of $S_{\text{eff,m}}$ reproduces the  effective action of a massive Majorana spinor in Riemann-Cartan space-time (see Sec.\ref{sec:Bulk-response}). In particular, in the limit $\kappa\rightarrow0$ the dimensionless coefficients \eqref{eq:c2} are all quantized, as expected based on dimensional analysis. Apart from $c$, only the coefficient $F_{1}'$ is discontinuous at $\mu=0$ within this limit, with a quantized discontinuity $-\left(\ell/4\right)\left[F'_{1}\left(0^{+}\right)-F'_{1}\left(0^{-}\right)\right]=\left(\ell/2\right)/96\pi$ that matches the coefficient $\beta$  of the \textit{gravitational pseudo Chern-Simons} term of Sec.\ref{sec:Main-section-1:}. As anticipated in Sec.\ref{sec:Main-section-1:}, the coefficient $c$ remains quantized away from the relativistic limit, while  $\beta$, or $F_{1}'$, does not. Finally, we note that our perturbative computation of the gCS term generalizes the computations of Refs.\cite{goni1986massless,van1986topological,vuorio1986parity,vuorio1986parityErr,kurkov2018gravitational} and Appendix \ref{subsec:Perturbative-calculation-of} for relativistic fermions, and reduces to these as $\kappa\rightarrow0$.

\end{enumerate}

\subsection{Induced action and linear response\label{sec: induced action and linear response}}

By expanding $S_{\text{eff}}$ to second order in the fields $\theta,A_{t}-\mu,A_{i},u_{ij}$,
and performing Gaussian integration over $\theta$, we obtain an induced
action $S_{\text{ind}}\left[A_{\mu},u_{ij}\right]$ that captures
the linear response of CSFs to the background fields, see Appendix \ref{subsec:Obtaining--from} for explicit expressions. 
Taking functional derivatives one obtains the expectation values $J^{\mu}=G^{-1/2}\delta S_{\text{ind}}/\delta A_{\mu}$,
$T^{ij}=G^{-1/2}\delta S_{\text{ind}}/\delta u_{ij}$ of the current
and stress, and from them the conductivity $\sigma^{ij}=\delta J^{i}/\delta E_{j}$,
the viscosity $\eta^{ij,kl}=\delta T^{ij}/\delta\partial_{t}u_{kl}$,
 and the mixed response function $\kappa^{ij,k}=\delta T^{ij}/\delta E_{k}=\delta J^{k}/\delta\partial_{t}u_{ij}$.
We will also need the static susceptibilities $\chi_{JJ}^{\mu,\nu},\;\chi_{TJ}^{ij,\nu}$,
defined by restricting to time independent $A_{\mu},u_{ij}$, and
computing $\delta J^{\mu}/\delta A_{\nu}$ and $\delta J^{\nu}/\delta u_{ij}$,
respectively. 

Before we compute $\eta_{\text{o}}$, it is useful to restrict its
form based on dimensionality and symmetries: space-time translations
and spatial rotations, as well as $PT$. The analysis is performed in Appendices \ref{subsec: B.1}-\ref{subsec: B.4}, and results in the expression
\begin{align}
\eta_{\text{o}}\left(\omega,\mathbf{q}\right)= & \eta_{\text{o}}^{\left(1\right)}\sigma^{xz}+\eta_{\text{o}}^{\left(2\right)}\left[\left(q_{x}^{2}-q_{y}^{2}\right)\sigma^{0x}-2q_{x}q_{y}\sigma^{0z}\right],\label{eq:3-3-2-1-1}
\end{align}
which is written in a basis of anti-symmetrized tensor products
of the symmetric Pauli matrices, $\sigma^{ab}=2\sigma^{[a}\otimes\sigma^{b]}$
\citep{Avron1998}. As components of the strain tensor, the matrices
$\sigma^{x},\sigma^{z}$ correspond to shears, while the identity
matrix $\sigma^{0}$ corresponds to a dilatation. The details of the
system are encoded in two independent coefficients $\eta_{\text{o}}^{\left(1\right)},\eta_{\text{o}}^{\left(2\right)}\in\mathbb{C}$,
which are themselves arbitrary functions of $\omega,q^{2}$. In the
case of uniform strain ($\mathbf{q}=\mathbf{0}$), the odd viscosity
tensor reduces to a single component, $\eta_{\text{o}}\left(\omega,\mathbf{0}\right)=\eta_{\text{o}}^{\left(1\right)}\left(\omega\right)\sigma^{[x}\otimes\sigma^{z]}$,
as is well known \citep{PhysRevLett.75.697,Avron1998,PhysRevB.86.245309,hoyos2014hall,PhysRevE.89.043019}.
The additional component $\eta_{\text{o}}^{\left(2\right)}$ has not
been discussed much in the literature \citep{abanov2014electromagnetic,hoeller2018second},
and also appears in the presence of vector, or pseudo-vector, anisotropy
\citep{PhysRevB.99.045141,PhysRevB.99.035427}, in which case $\mathbf{q}$
should be replaced by a background vector $\mathbf{b}$. The expression
\eqref{eq:3-3-2-1-1} applies at finite temperature, out of equilibrium,
and in the presence of disorder that preserves the symmetries on average.
For clean systems at zero temperature, $\eta_{\text{o}}^{\left(1\right)},\eta_{\text{o}}^{\left(2\right)}$
are both real, even functions of $\omega$. In gapped systems $\eta_{\text{o}}^{\left(1\right)},\eta_{\text{o}}^{\left(2\right)}$
will usually be regular at $\omega=0=q^{2}$, though an exception
to this rule was recently found in Ref.\citep{10.21468/SciPostPhys.9.1.006}.

For the CSF, we find the $\omega=0$ coefficients 
\begin{align}
\eta_{\text{o}}^{\left(1\right)}\left(q^{2}\right)= & -\frac{1}{2}s_{\theta}n_{0}-\left(\frac{c}{24}\frac{1}{4\pi}+s_{\theta}C^{\left(1\right)}\right)q^{2}+O\left(q^{4}\right),\nonumber \\
\eta_{\text{o}}^{\left(2\right)}\left(q^{2}\right)= & \frac{1}{2}s_{\theta}n_{0}q^{-2}+\left(\frac{c}{24}\frac{1}{4\pi}+s_{\theta}C^{\left(2\right)}\right)+O\left(q^{2}\right),\label{eq:14-1}
\end{align}
where $C^{\left(1\right)},C^{\left(2\right)}\in\mathbb{R}$ are generically
non-zero, and are given by particular linear combinations of the dimensionless
coefficients $F_{1}'\left(\mu\right),mF_{2}\left(\mu\right),mF_{3}\left(\mu\right)$,
defined in \eqref{eq:10-1-0}, see Appendix \ref{subsec:Obtaining--from} for more details. 

The leading term in $\eta_{\text{o}}^{\left(1\right)}$ is the familiar
\eqref{eq:1-1}, which also appears in gapped states, while the non-analytic
leading term in $\eta_{\text{o}}^{\left(2\right)}$ is possible because
the superfluid is gapless, and does not contribute to the viscosity
tensor when $q\rightarrow0$ at $\omega\neq0$ \citep{hoyos2014effective}.
Both leading terms obey the same quantization condition due to SSB,
can be used to extract $s_{\theta}$, and are independent of $c$.
The sub-leading corrections to both $\eta_{\text{o}}^{\left(1\right)},\eta_{\text{o}}^{\left(2\right)}$
contain the quantized gCS contributions proportional to $c$, but
also the non-universal coefficients $C^{\left(1\right)},C^{\left(2\right)}$.
Thus the central charge cannot be extracted from a measurement of
$\eta_{\text{o}}$ alone. 

Noting that the non-universal sub-leading corrections to $\eta_{\text{o}}$
originate from the geometric spin $s_{\theta}=\ell/2$ of the Goldstone
field, one is naturally led to consider the g$s$-wave CSF, where
$s_{\theta}=0$, and the odd viscosity is, to leading order in $q$,
purely due to the gCS term
\begin{align}
\tilde{\eta}_{\text{o}}^{\left(1\right)}\left(q^{2}\right)= & -\frac{c}{24}\frac{1}{4\pi}q^{2}+O\left(q^{4}\right),\label{eq:18-1-1}\\
\tilde{\eta}_{\text{o}}^{\left(2\right)}\left(q^{2}\right)= & \frac{c}{24}\frac{1}{4\pi}+O\left(q^{2}\right).\nonumber 
\end{align}
Here and below we use $O$ and $\tilde{O}$, for the quantity $O$
in the CSF and in the corresponding $g$s-wave CSF, respectively.
Equation \eqref{eq:18-1-1} follows from \eqref{eq:14-1} by setting
$s_{\theta}=0$, but can be understood directly from $S_{\text{eff}}$.
Indeed, for the g$s$-wave CSF, $S_{\text{eff}}$ is identical to
that of the conventional $s$-wave SF up to $O\left(p^{2}\right)$,
but contains the additional $\mathcal{L}_{\text{gCS}}$ at $O\left(\omega q^{2}\right)$,
which is the leading $P,T$-odd term, and produces the leading odd
viscosity \eqref{eq:18-1-1}. 

 Due to the LGS \eqref{eq:4-3-1-1}-\eqref{eq:5-3-1}, the viscosity
\eqref{eq:18-1-1} implies also
\begin{align}
 & \tilde{\chi}_{TJ,\text{o}}^{ij,k}=-\frac{i}{m}\frac{c}{48\pi}q_{\perp}^{i\vphantom{j}}q_{\perp}^{j}q_{\bot}^{k\vphantom{j}}+O\left(q^{4}\right),\label{eq:17-1}
\end{align}
where $q_{\perp}^{i}=\varepsilon^{ij}q_{j}$, and the subscript ``o''
(``e'') refers to the $P,T$-odd (even) part of an object, which
is odd (even) in $\ell$. In particular, a steady $P,T$-odd current
$\tilde{J}_{\text{o}}^{k}=-\frac{1}{m}\frac{c}{96\pi}\partial_{\perp}^{k}R+O\left(q^{4}\right)$
flows perpendicularly to gradients of curvature, which has the linearized
form $R=-2\partial_{\perp}^{i}\partial_{\perp}^{j}u_{ij}$. We stress
that the fundamental relation is the momentum density $\tilde{P}_{\text{o}}^{k}=-\frac{c}{96\pi}\partial_{\perp}^{k}R+O\left(q^{4}\right)$,
which follows from the odd viscosity \eqref{eq:18-1-1} along with
momentum conservation, irrespective of Galilean symmetry. This relation
was predicted for $p$-wave CSFs based on the relativistic limit,
see Sec.\ref{subsubsec:Topological bulk responses from a gravitational Chern-Simons term}
and \ref{sec:Conclusion-and-discussion}. 
We conclude that, in the g$s$-wave CSF, $c$ can be extracted
from a measurement of $\tilde{\eta}_{\text{o}}$, and in the Galilean
invariant case, also from a measurement of the current $\tilde{J}$
in response to (time-independent) strain. 

Though the simple results above do not apply to the physical system
of interest, the CSF, there is a relation between the observables
of the CSF and the corresponding $g$s-wave CSF, which we can utilize.
At the level of induced actions it is given by the simple relation 
\begin{align}
\tilde{S}_{\text{ind}}\left[A_{\mu},u_{ij}\right]=S_{\text{ind}}\left[A_{\mu}-\left(\ell/2\right)\omega_{\mu},u_{ij}\right],\label{eq:induced}
\end{align}
where $\omega_{\mu}$ is expressed through $u_{ij}$ as in Appendix \ref{sec:Geometric quantities}, and by taking functional derivatives one finds relations between response functions \citep{geracie2015spacetime}. In particular, 
\begin{align}
\tilde{\eta}_{\text{o}}^{ij,kl}= & \eta_{\text{o}}^{ij,kl}-\frac{\ell}{4}n_{0}\left(\sigma^{xz}\right)^{ij,kl}+\frac{i\ell}{4}\left(\kappa_{\text{e}\vphantom{\bot}}^{ij,(k}q_{\perp}^{l)}-\kappa_{\text{e}\vphantom{\bot}}^{kl,(i}q_{\perp}^{j)}\right)+\frac{\ell^{2}}{16}\sigma_{\text{o}}q_{\perp}^{(i}\varepsilon_{\vphantom{\bot}}^{j)(k}q_{\perp}^{l)},\label{eq:16-2-1}
\end{align}
where the response functions $\eta_{\text{o}},\sigma_{\text{o}},\kappa_{\text{e}}$
depend on $\omega,\mathbf{q}$.  In a Galilean
invariant system one further has 
\begin{align}
\tilde{\chi}_{TJ,\text{o}}^{ij,k}= & \chi_{TJ,\text{o}}^{ij,k}-\frac{\ell}{4m}\chi_{TJ,\text{e}}^{ij,t}iq_{\bot}^{k}+\frac{\ell}{2}iq_{\bot}^{(i}\chi_{JJ,\text{e}}^{j),k}+\frac{\ell^{2}}{8m}q_{\bot}^{(i}\chi_{JJ,\text{o}}^{j),t}q_{\bot}^{k},\label{eq:21-1}
\end{align}
and we note the relations $\chi_{TJ,\text{e}}^{ij,t}=\kappa_{\text{e}}^{ij,k}iq_{k},\;\chi_{JJ,\text{o}}^{j,t}=\sigma_{\text{o}}q_{\bot}^{j},\;\chi_{JJ,\text{e}}^{j,k}=\rho_{\text{e}}q_{\bot}^{j}q_{\bot}^{k}$,
between the above susceptibilities, the response functions $\kappa_{\text{e}},\sigma_{\text{o}}$,
and the London diamagnetic response $\rho_{\text{e}}$. Though the above expressions are a mouthful, they correspond to the simple subtraction of an angular momentum $\ell/2$ per fermion, as expressed by Eq.\eqref{eq:induced}.


\subsection{Discussion\label{sec:discussion}}

Equations \eqref{eq:18-1-1} and \eqref{eq:16-2-1} are the main results
of this Section. They rely on the SSB pattern \eqref{eq:2-1-1}, but
not on Galilean symmetry. Equation \eqref{eq:16-2-1} expresses $\tilde{\eta}_{\text{o}}$
as a bulk observable of CSFs, which we refer to as the \textit{improved
odd viscosity}. According to \eqref{eq:18-1-1}, the leading term
in the expansion of $\tilde{\eta}_{\text{o}}\left(0,\mathbf{q}\right)$
around $\mathbf{q}=\mathbf{0}$ is fixed by $c$. Since this leading
term occurs at second order in $\mathbf{q}$, in order to extract
$c$ one needs to measure $\sigma_{\text{o}},\chi_{\text{e}}$, and
$\eta_{\text{o}}$, at zeroth, first, and second order, respectively.
In a Galilean invariant system, \eqref{eq:18-1-1} and \eqref{eq:16-2-1}
imply \eqref{eq:17-1} and \eqref{eq:21-1} respectively, which, in
turn, show that $c$ can be extracted in an experiment where $U\left(1\right)_{N}$
fields and strain are applied, and the resulting number current and
density are measured. In particular, a measurement of the stress tensor
is not required. Since $U\left(1\right)_{N}$ fields can be applied
in Galilean invariant fluids by tilting and rotating the sample \citep{Viefers_2008},
we believe that a bulk measurement of the boundary central charge,
through \eqref{eq:17-1} and \eqref{eq:21-1}, is within reach of
existing experimental techniques \citep{PhysRevLett.109.215301,Levitin841,Ikegami59,Zhelev_2017}.
\textcolor{red}{}

The problem of obtaining $c$ from a bulk observable has been previously
studied in QH states, described by \eqref{eq:10-0}-\eqref{eq:11} \citep{ferrari2014fqhe,can2014fractional,abanov2014electromagnetic,gromov2014density,gromov2015framing,can2015geometry,klevtsov2015geometric,bradlyn2015topological,Klevtsov_2016,gromov2016boundary,klevtsov2017laughlin,Cappelli_2018}.
 It was found that $c$ can only be extracted if $\text{var}s=\overline{s^{2}}-\overline{s}^{2}=0$,
as in Laughlin states, or in a single filled Landau level. Under this
condition, the response to strain, at fixed $A_{\mu}-\overline{s}\omega_{\mu}$,
depends purely on $c$ \citep{bradlyn2015topological,gromov2016boundary}
- a useful theoretical characterization, which seems challenging experimentally.
  However, the improved odd viscosity \eqref{eq:16-2-1}, constructed
here, applies also to $\text{var}s=0$ QH states, with $\ell$ replaced
by $-2\overline{s}$, and defines a concrete bulk observable which
is precisely quantized, and determined by $c$.

\pagebreak{}

\section{Intrinsic sign problem in chiral topological matter\label{sec:Main-section-3:}}

The question of intrinsic Monte Carlo sign problems was motivated
in Sec.\ref{subsec:Complexity-of-simulating}, where we also stated
and discussed the criterion $e^{2\pi ic/24}\notin\left\{ \theta_{a}\right\} $,
which we obtain for intrinsic sign problems in chiral topological phases. Here we precisely state and derive the results that hold under this criterion. 

In Sec.\ref{sec:Signs-from-geometric} we discuss the universal finite-size correction to the boundary momentum density  
in chiral topological phases,
arriving at the 'momentum polarization' Eq.\eqref{eq:12-3-1}. Relying on the above, Sec.\ref{sec:No-stoquastic-Hamiltonians}
then obtains Result \hyperref[Result 1]{1} - an intrinsic sign problem
in bosonic chiral topological matter. In Sec.\ref{sec:Spontaneous-chirality}
we perform a similar analysis for the case where chirality (or time
reversal symmetry breaking), appears spontaneously rather than explicitly,
arriving at Result \hyperref[Result 2]{2}. We then turn to fermionic
systems. In Sec.\ref{sec:Determinantal-quantum-Monte} we develop
a formalism which unifies and generalizes the currently used DQMC
algorithms, and the corresponding design principles. In Sec.\ref{sec:No-sign-free-DQMC},
we obtain within this formalism Result \hyperref[Result 1F]{1F} and Result
\hyperref[Result 2F]{2F}, the fermionic analogs of Results \hyperref[Result 1]{1}
and \hyperref[Result 2]{2}. Section \ref{sec:Generalization-and-extension}
describes a conjectured extension of our results that applies beyond
chiral phases, and unifies them with the intrinsic sign problems found
in our parallel work \citep{PhysRevResearch.2.033515}. In Sec.\ref{sec:Discussion-and-outlook}
we discuss our results and provide an outlook for future work.

\subsection{Signs from geometric manipulations \label{sec:Signs-from-geometric}}

\subsubsection{Momentum finite-size correction\label{subsec:Boundary-finite-size}}

In analogy with the $T>0$ correction to the energy current in Eq.\eqref{eq:1-2},
the boundary of a chiral topological phase, described by a chiral
CFT, also supports a non-vanishing ground state (or $T=0$) momentum
density $p$, which receives a universal correction on a cylinder
with finite circumference $L<\infty$, 
\begin{align}
 & p\left(L\right)=p\left(\infty\right)+\frac{2\pi}{L^{2}}\left(h_{0}-\frac{c}{24}\right).\label{eq:2-2}
\end{align}

Equation \eqref{eq:2-2} is the main property of chiral topological
matter that we use below, so we discuss it in detail. The appearance
of the chiral central charge is a manifestation of the global gravitational
anomaly, as explained in Sec.\ref{subsec:Geometric-physics-in}.
The rational number $h_{0}$ is a \textit{chiral} conformal weight
from the boundary CFT. Like the chiral central charge, the two boundary
components of the cylinder have opposite $h_{0}$s, see Fig.\ref{fig:Chiral-topological-phases}.
 From the bulk TFT perspective, $h_{0}$ corresponds to the topological
spin of an anyon quasi-particle, defined by the phase $\theta_{0}=e^{2\pi ih_{0}}$
accumulated as the anyon undergoes a $2\pi$ rotation \citep{kitaev2006anyons}.
The set $\left\{ \theta_{a}\right\} _{a=1}^{N}$ of topological spins
of anyons is associated with the $N$-dimensional ground state subspace
on the torus, and the unique $\theta_{0}=e^{2\pi ih_{0}}$ defined
by \eqref{eq:2-2} corresponds to the generically unique ground state
on the cylinder, with a finite-size energy separation $\sim1/L$ from
the low lying excited states, see Appendix \ref{subsec:Further-details-regarding}. 

As the equation $\theta_{0}=e^{2\pi ih_{0}}$ suggests, only $h_{0}\mod1$
is universal for a topological phase. The integer part of $h_{0}$
can change as the Hamiltonian is deformed on the cylinder, while maintaining
the bulk gap, and even as a function of $L$ for a fixed Hamiltonian.
Additionally, the choice of $\theta_{0}$ from the set $\left\{ \theta_{a}\right\} $
is non-universal, and can change due to bulk gap preserving deformations,
or as a function of $L$. Both types of discontinuous jumps in $h_{0}$
may be accompanied by an accidental degeneracy of the ground state
on the cylinder. Therefore, the universal and $L$-independent statement
regarding $h_{0}$ is that, apart from accidental degeneracies, $e^{2\pi ih_{0}}=\theta_{0}\in\left\{ \theta_{a}\right\} $
- a fact that will be important in our analysis. 

The non-trivial behavior of $h_{0}$ described above appears when
the boundary corresponds to a non-conformal deformation of a CFT,
by e.g a chemical potential. As demonstrated analytically and numerically
in Appendix \ref{subsec:Beyond-the-assumption}, such behavior appears
already in the simple context of Chern insulators with non-zero Fermi
momenta, as would be the case in Fig.\ref{fig:Chiral-topological-phases}(b)
if the chemical potential $\mu$ is either raised or lowered. 

\subsubsection{Momentum polarization\label{subsec:Momentum-polarization}}

In this section we describe a procedure for the extraction of $h_{0}-c/24$
in Eq.\eqref{eq:2-2}, given a lattice Hamiltonian on the cylinder.
Since the two boundary components carry opposite momentum densities,
the ground state on the cylinder does not carry a total momentum,
only a 'momentum polarization'. It is therefore clear that some sort
of one-sided translation will be required.

\begin{figure}[!th]
\begin{centering}
\includegraphics[width=0.6\columnwidth]{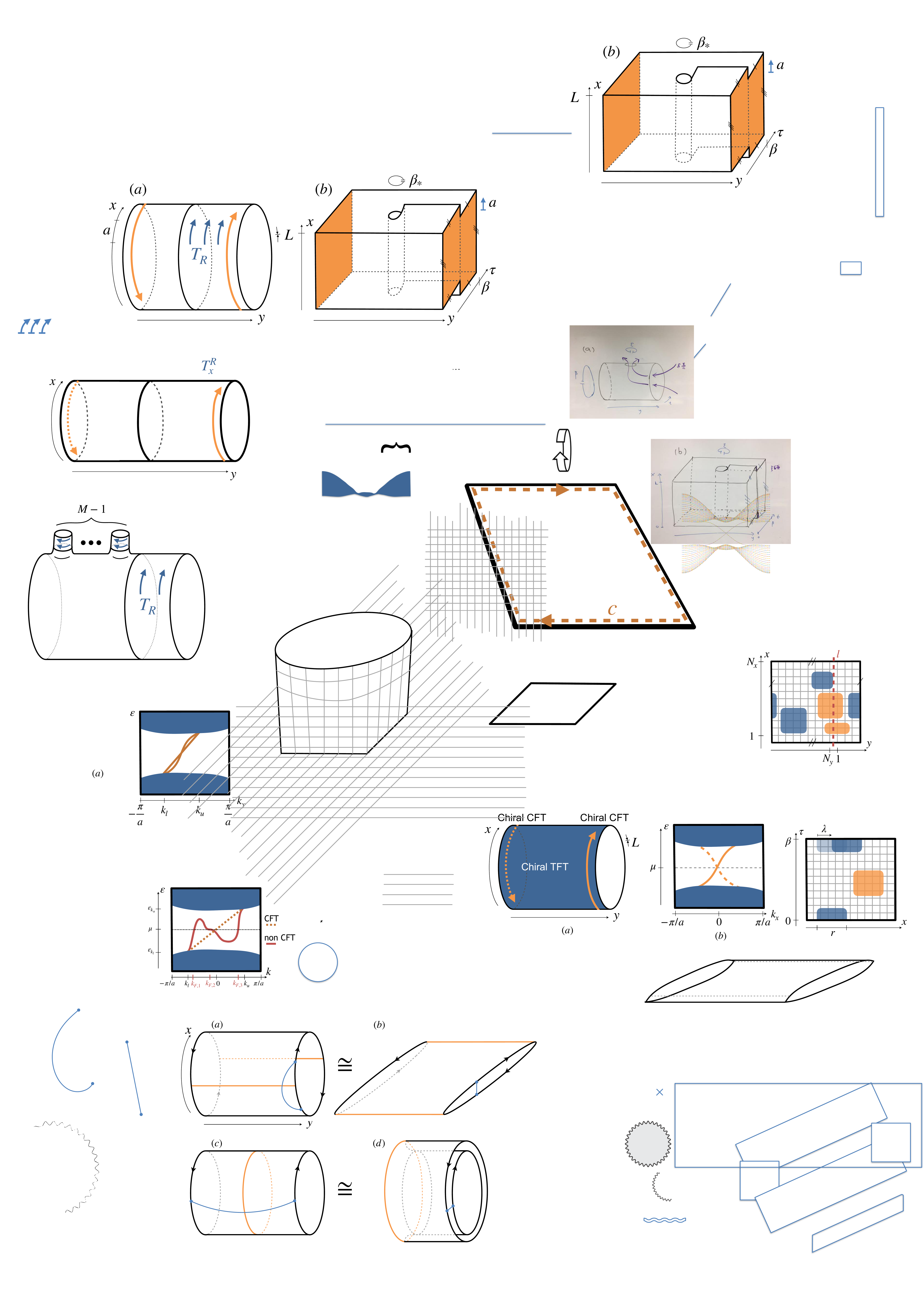}
\par\end{centering}
\caption{Momentum polarization. (a) Hamiltonian, or spatial, point of view.
The operator $T_{R}$ translates the right half of the cylinder by
one unit cell, a distance $a$, in the $x$ direction. It acts as
the identity on the left boundary component, and as a translation
on the right boundary component. The object $\tilde{Z}/Z$ is the
thermal expectation value of $T_{R}$. (b) Field theory, or space-time,
point of view. The object $\tilde{Z}$ is the partition function on
a space-time carrying a screw dislocation. The space-time region occupied
by the boundary components of the spatial cylinder is colored in orange.
The screw dislocation can be described as an additional boundary component,
on which $T_{R}$ acts as a translation, with a high effective temperature
$1/\beta_{*}$. \label{fig:Defect} }
\end{figure}
 Following Ref.\citep{PhysRevB.88.195412}, we define $\tilde{Z}:=\text{Tr}\left(T_{R}e^{-\beta H}\right)$,
which is related to the usual partition function $Z=\text{Tr}\left(e^{-\beta H}\right)$
($\beta=1/T$), by the insertion of the operator $T_{R}$, which translates
the right half of the cylinder by one unit cell in the periodic $x$
direction, see Fig.\ref{fig:Defect}(a). The object $\tilde{Z}$ satisfies
\begin{align}
\tilde{Z} & =Z\exp\left[\alpha N_{x}+\frac{2\pi i}{N_{x}}\left(h_{0}-\frac{c}{24}\right)+o\left(N_{x}^{-1}\right)\right],\label{eq:12-3-1}
\end{align}
where $N_{x}$ is the number of sites in the $x$ direction, $\alpha\in\mathbb{C}$
is non-universal and has a negative real part, and $o\left(N_{x}^{-1}\right)$
indicates corrections that decay faster than $N_{x}^{-1}$ as $N_{x}\rightarrow\infty$.
\textcolor{red}{}The above expression is valid at temperatures low
compared to the finite-size energy differences on the boundary, $\beta^{-1}=o\left(N_{x}^{-1}\right)$,
 see Appendix \ref{subsec:Further-details-regarding}.

Equation \eqref{eq:12-3-1} follows analytically from the low energy
description of chiral topological matter in terms of chiral TFT and
CFT \citep{PhysRevB.88.195412}, and was numerically scrutinized
in a large number of examples in Refs.\citep{PhysRevB.88.195412,PhysRevLett.110.236801,PhysRevB.90.045123,PhysRevB.90.115133,PhysRevB.92.165127},
as well as in Appendix \ref{subsec:Beyond-the-assumption}. Nevertheless,
we are not aware of a rigorous proof of Eq.\eqref{eq:12-3-1} for
gapped lattice Hamiltonians. Therefore, in stating our results we
will use the assumption 'the Hamiltonian $H$ is in a chiral topological
phase of matter', the content of which is that $H$ admits a low energy
description in terms of a chiral TFT with chiral central charge $c$
and topological spins $\left\{ \theta_{a}\right\} $, and in particular,
Eq.\eqref{eq:12-3-1} holds for any bulk-gap preserving deformation
of $H$, with $e^{2\pi ih_{0}}\in\left\{ \theta_{a}\right\} $ (apart
from accidental degeneracies on the cylinder, as explained below Eq.\eqref{eq:2-2}).
In the remainder of this section we further discuss the content of
Eq.\eqref{eq:12-3-1} and its expected range of validity, in light
of the Hamiltonian and space-time interpretations of $\tilde{Z}$. 

From a Hamiltonian perspective, $\tilde{Z}/Z$ is the thermal expectation
value of $T_{R}$, evaluated at a temperature $\beta^{-1}$ low enough
to isolate the ground state. The exponential decay expressed in Eq.\eqref{eq:12-3-1}
appears because $T_{R}$ is not a symmetry of $H$, and $-\text{Re}\left(\alpha\right)$
can be understood as the energy density of the line defect where $T_{R}$
is discontinuous, see Fig.\ref{fig:Defect}(a).  In fact, we expect Eq.\eqref{eq:12-3-1} to hold irrespective of whether the \textit{uniform} translation is a symmetry of $H$, or of the underlying 'lattice' on which $H$ is defined, which may be any polygonalization of the cylinder (see Ref.\citep{PhysRevLett.115.036802} for a similar scenario). The only expected requirement is that the low energy description of $H$ is homogeneous. Furthermore, if Eq.\eqref{eq:12-3-1} only holds after a disorder averaging of $\tilde{Z}/Z$, our results and derivations in the following sections remain unchanged. 

There is also a simple space-time interpretation of $\tilde{Z}$,
which will be useful in the context of DQMC. The usual partition function
$Z=\text{Tr}\left(e^{-\beta H}\right)$ has a functional integral
representation in terms of bosonic fields $\phi$ (fermionic fields
$\psi$) defined on space, the cylinder $C$ in our case, and the
imaginary time circle $S_{\beta}^{1}=\mathbb{R}/\beta\mathbb{Z}$,
with periodic (anti-periodic) boundary conditions \citep{altland2010condensed}.
In $\tilde{Z}=\text{Tr}\left(T_{R}e^{-\beta H}\right)$, the insertion
of $T_{R}$ produces a twisting of the boundary conditions of $\phi,\psi$
in the time direction, such that $\tilde{Z}$ is the partition function
on a space-time carrying a screw dislocation, see Fig.\ref{fig:Defect}(b).

The above interpretation of $\tilde{Z}$, supplemented by Eq.\eqref{eq:2-2},
allows for an intuitive explanation of Eq.\eqref{eq:12-3-1}, which
loosely follows its analytic derivation \citep{PhysRevB.88.195412}.
As seen in Fig.\ref{fig:Defect}(b), the line where $T_{R}$ is discontinuous
can be interpreted as an additional boundary component at a high effective
temperature, $\beta_{*}\ll L/v$. Since the effective temperature
is much larger than the finite size energy-differences $2\pi v/L$
on the boundary CFT, the screw dislocation contributes no finite size
corrections to $\tilde{Z}$. This leaves only the contribution of
the boundary component on the right side of the cylinder, where $T_{R}$
produces the phase $e^{iaLp\left(L\right)}$, assuming $\beta_{*}\ll L/v\ll\beta$.
Equation \eqref{eq:2-2} then leads to the universal finite size correction
$\left(2\pi i/N_{x}\right)\left(h_{0}-c/24\right)$.

\subsection{Excluding stoquastic Hamiltonians for chiral topological matter\label{sec:No-stoquastic-Hamiltonians}}

In this section we consider bosonic (or 'qudit', or spin) systems,
and a single design principle  - existence of a local basis in which
the many-body Hamiltonian is stoquastic. A sketch of the derivation
of Result \hyperref[Result 1]{1} is that the momentum polarization
$\tilde{Z}$ is positive for Hamiltonians $H'$ which are stoquastic
in an on-site and homogenous basis, and this implies that $\theta_{0}=e^{2\pi ic/24}$
for any Hamiltonian $H$ obtained from $H'$ by conjugation with a
local unitary. 

\subsubsection{Setup\label{subsec:Setup}}

The many body Hilbert space is given by $\mathcal{H}=\otimes_{\mathbf{x}\in X}\mathcal{H}_{\mathbf{x}}$,
where the tensor product runs over the sites $\mathbf{x}=\left(x,y\right)$
of a 2-dimensional lattice $X$, and $\mathcal{H}_{\mathbf{x}}$ are
on-site 'qudit' Hilbert spaces of finite dimension $\mathsf{d}\in\mathbb{N}$.
With finite-size QMC simulations in mind, we consider a square lattice
with spacing 1, $N_{x}\times N_{y}$ sites, and periodic boundary
conditions, so that $X=\mathbb{Z}_{N_{x}}\times\mathbb{Z}_{N_{y}}$
is a discretization of the flat torus $\left(\mathbb{R}/N_{x}\mathbb{Z}\right)\times\left(\mathbb{R}/N_{y}\mathbb{Z}\right)$.
Generalization to other 2-dimensional lattices is straight forward.
On this Hilbert space a gapped $r$-local Hamiltonian $H=\sum_{\mathbf{x}}H_{\mathbf{x}}$
is assumed to be given. Here the terms $H_{\mathbf{x}}$ are supported
within a range $r$ of $\mathbf{x}$ - they are defined on $\otimes_{\left|\mathbf{y}-\mathbf{x}\right|\leq r}\mathcal{H}_{\mathbf{y}}$
and act as the identity on all other qudits.

Fix an tensor product basis $\ket s=\otimes_{\mathbf{x}\in X}\ket{s_{\mathbf{x}}}$,
labeled by strings $s=\left(s_{\mathbf{x}}\right)_{\mathbf{x}\in X}$,
where $s_{\mathbf{x}}\in\left\{ 1,\cdots,\mathsf{d}\right\} $ labels
a basis $\ket{s_{\mathbf{x}}}$ for $\mathcal{H}_{\mathbf{x}}$. For
any vector $\mathbf{d}\in X$, the corresponding translation operator
$T^{\mathbf{d}}$ is defined in this basis, $T^{\mathbf{d}}\ket s=\ket{t^{\mathbf{d}}s}$,
with $\left(t^{\mathbf{d}}s\right)_{\mathbf{x}}=s_{\mathbf{x}+\mathbf{d}}$.
These statements assert that $\ket s$ is both an on-site and a homogeneous
basis, or \textit{on-site homogenous }for short. Note that $T^{\mathbf{d}}$
acts as a permutation matrix on the $\ket s$s, and in particular,
has non-negative matrix elements in this basis.

In accordance with Sec.\ref{subsec:Momentum-polarization}, we assume
that the low energy description of $H$ is invariant under $T^{\mathbf{d}}$,
as defined above. In doing so, we exclude the possibility of generic
background gauge fields for any on-site symmetry that $H$ may posses,
which is beyond the scope of this thesis. Nevertheless, commonly used
background gauge fields, such as those corresponding to uniform magnetic
fields with rational flux per plaquette, can easily be incorporated
into our analysis, by restricting to translation vectors $\mathbf{d}$
in a sub-lattice of $X$. A restriction to sub-lattice translations
can also be used to guarantee that $T^{\mathbf{d}}$ acts purely as
a translation in the low energy TQFT description. In particular, a
lattice translation may permute the anyon types $a$ \footnote{We thank Michael Levin for pointing out this phenomenon.}.
Since the number of anyons is finite, restricting to large enough
translations will eliminate this effect. An example is given by Wen's
plaquette model, where different anyons are localized on the even/odd
sites of a bipartite lattice \citep{PhysRevB.87.184402}, and a restriction
to translations that maps the even (odd) sites to themselves will
be made.

Finally, we assume that $H$ is \textit{locally stoquastic}: it is
term-wise stoquastic in a local basis. This means that a local unitary
operator $U$ exists, such that the conjugated Hamiltonian $H'=UHU^{\dagger}$
 is a sum of local terms $H_{\mathbf{x}}'=UH_{\mathbf{x}}U^{\dagger}$,
which have non-positive matrix elements in the on-site homogeneous
basis, $\bra sH_{\mathbf{x}}'\ket{\tilde{s}}\leq0$ for all basis
states $\ket s,\ket{\tilde{s}}$. Note that we include the diagonal
matrix elements in the definition, without loss of generality. 

The term \textit{local unitary} used above refers to a depth-$D$
quantum circuit, a product $U=U_{D}\cdots U_{1}$ where each $U_{i}$
is itself a product of unitary operators with non-overlapping supports of diameter $w$. It follows that $H'$ has a range $r'=r+2r_{U}$,
where $r_{U}=Dw$ is the range of $U$. Equivalently, we may take
$U$ to be a finite-time evolution with respect to an $\tilde{r}$-local,
smoothly time-dependent, Hamiltonian $\tilde{H}\left(t\right)$, given
by the time-ordered exponential $U=\text{TO}e^{-i\int_{0}^{1}\tilde{H}\left(t\right)dt}$.
The two types of locality requirements are equivalent, as finite-time
evolutions can be efficiently approximated by finite-depth circuits,
while finite-depth circuits can be written as finite-time evolutions
over time $D$ with piecewise constant $w$-local Hamiltonians \citep{Lloyd1073,zeng2019quantum}. 

\subsubsection{Constraining $c$ and $\left\{ \theta_{a}\right\} $}

\begin{figure}[!th]
\begin{centering}
\includegraphics[width=0.3\columnwidth]{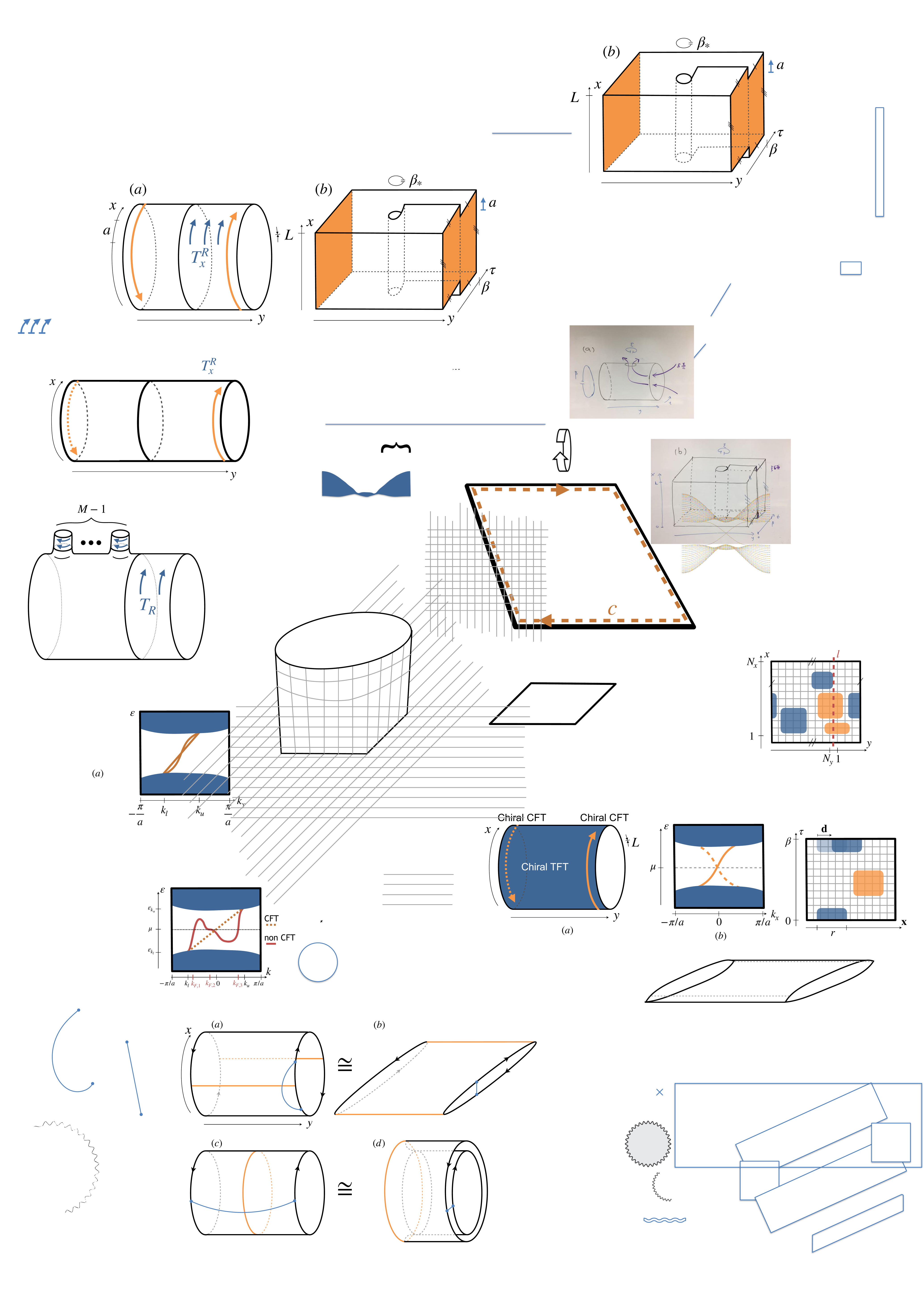}
\par\end{centering}
\caption{Cutting the torus to a cylinder along the line $l$. Orange areas
mark the supports of Hamiltonian terms $H_{\mathbf{x}}'$ which are
removed from $H'$, while blue areas mark the supports of terms which
are kept. \label{fig:cutting} }
\end{figure}

In order to discuss the momentum polarization, we need to map the
stoquastic Hamiltonian $H'$ from the torus $X$ to a cylinder $C$.
This is done by choosing a translation vector $\mathbf{d}\in X$,
and then cutting the torus $X$ along a line $l$ parallel to $\mathbf{d}$.
 To simplify the presentation we restrict attention to the case
$\mathbf{d}=\left(1,0\right)$, where $T^{\mathbf{d}}=T$ (and in
the following $T_{R}^{\mathbf{d}}=T_{R}$). All other cases amount
to a lattice-spacing redefinition, see Appendix \ref{subsec:Cutting-the-torus-2}. The cylinder $C=\mathbb{Z}_{N_{x}}\times\left\{ 1,\dots,N_{y}\right\} $
is then obtained from the torus $X=\mathbb{Z}_{N_{x}}\times\mathbb{Z}_{N_{y}}$
by cutting along the line $l=\left\{ \left(i,1/2\right):\;i\in\mathbb{Z}_{N_{x}}\right\} $.
A stoquastic Hamiltonian on the cylinder can be obtained from that
on the torus by removing all local terms $H'_{\mathbf{x}}$ whose
support overlaps $l$, see Fig.\ref{fig:cutting}. Note that this
procedure may render $H'$ acting as $0$ on certain qudits $\mathcal{H}_{\mathbf{x}}$
with $\mathbf{x}$ within a range $r'$ of $l$, but this does not
bother us. Since all terms $H_{\mathbf{x}}'$ are individually stoquastic,
this procedure leaves $H'$, now defined on the cylinder, stoquastic.
One can similarly map $H$ and $U$ to the cylinder $C$ such that
the relation $H'=UHU^{\dagger}$ remains valid on $C$.

Let us now make contact with the momentum polarization Eq.\eqref{eq:12-3-1}.
Having mapped $H'$ to the cylinder, we consider the 'partition function'
\begin{align}
\tilde{Z}' & :=\text{Tr}\left(e^{-\beta H'}T_{R}\right),\label{eq:4-1}
\end{align}
where $T_{R}$ is defined by $T_{R}\ket s=\ket{T_{R}s}$, $\left(T_{R}s\right)_{x,y}=s_{x+\Theta\left(y\right),y}$
, and $\Theta$ is a heavy side function supported on the right half
of the cylinder. Though $\tilde{Z}'$ is generally different from
$\tilde{Z}=\text{Tr}\left(e^{-\beta H}T_{R}\right)$ appearing in
Eq.\eqref{eq:12-3-1}, it satisfies two useful properties:
\begin{enumerate}
\item $\tilde{Z}'>0$. Both $-H'$ and $T_{R}$ have non-negative entries
in the on-site basis $\ket s$, and therefore so does $e^{-\beta H'}T_{R}$.
\item $H'=UHU^{\dagger}$ is in the same phase of matter as $H$, so $c'=c$
and $\left\{ \theta_{a}'\right\} =\left\{ \theta_{a}\right\} $. Moreover,
$h_{0}'=h_{0}$ for all $N_{x}$. Treating $U$ as a finite time evolution,
we have $H\left(\lambda\right)=U\left(\lambda\right)HU\left(\lambda\right)^{\dagger}$,
where $U\left(\lambda\right):=\text{TO}e^{-i\int_{0}^{\lambda}\tilde{H}\left(t\right)dt}$,
as a deformation from $H$ to $H'$ which maintains locality and preserves
the bulk-gap. Moreover, the full spectrum on the cylinder is $\lambda$-independent,
and therefore so is $h_{0}$. 
\end{enumerate}
Combining Eq.\eqref{eq:12-3-1}, for $H'$ instead of $H$, with the
two above properties leads to

\begin{align}
1= & \tilde{Z}'/\left|\tilde{Z}'\right|\label{eq:6-1-0}\\
= & \exp2\pi i\left[\epsilon'N_{x}+\frac{1}{N_{x}}\left(h_{0}-\frac{c}{24}\right)+o\left(N_{x}^{-1}\right)\right],\nonumber 
\end{align}
where $\epsilon':=\text{Im}\left(\alpha'\right)$ is generally different
from $\epsilon=\text{Im}\left(\alpha\right)$ of Eq.\eqref{eq:12-3-1}
since $H'\neq H$. The non-universal integer part of $h_{0}$ can
then be eliminated by raising Eq.\eqref{eq:6-1-0} to the $N_{x}$th
power,

\begin{align}
1 & =e^{2\pi i\epsilon'N_{x}^{2}}\theta_{0}\left(N_{x}\right)e^{-2\pi ic/24}+o\left(1\right),\label{eq:6-1}
\end{align}
where we used $\theta_{0}=e^{2\pi ih_{0}}$, and $o\left(1\right)\rightarrow0$
as $N_{x}\rightarrow\infty$. We also indicated explicitly the possible
$N_{x}$-dependence of $\theta_{0}$, as described in Sec.\ref{subsec:Boundary-finite-size}.
We proceed under the assumption that no accidental degeneracies occur
on the cylinder, so that $\theta_{0}\left(N_{x}\right)\in\left\{ \theta_{a}\right\} $
for all $N_{x}$, deferring the degenerate case to Appendix \ref{subsec:Dealing-with-accidental}.
Now, for rational $\epsilon'=n/m$, the series $e^{2\pi i\epsilon'N_{x}^{2}}$
($N_{x}\in\mathbb{N}$) covers periodically a subset $S$ of the $m$th
roots of unity, including $1\in S$. On the other hand, for irrational
$\epsilon'$ the series $e^{2\pi i\epsilon'N_{x}^{2}}$ is dense in
the unit circle. Combined with the fact that $\theta_{0}\left(N_{x}\right)$
is valued in the finite set $\left\{ \theta_{a}\right\} $, while
$c$ is $N_{x}$-independent, Equation \eqref{eq:6-1} implies that
$\epsilon'$ must be rational, and that the values attained by $\theta_{0}\left(N_{x}\right)e^{-2\pi ic/24}$
cover the set $S$ periodically, for large enough $N_{x}$. It follows
that $1\in S\subset\left\{ \theta_{a}e^{-2\pi ic/24}\right\} $. We
therefore have 

\begin{description}
\item [{Result$\;$1\label{Result 1}}] If a local bosonic Hamiltonian
$H$ is both locally stoquastic and in a chiral topological phase
of matter, then one of the corresponding topological spins satisfies
$\theta_{a}=e^{2\pi ic/24}$. Equivalently, a bosonic chiral topological
phase of matter where $e^{2\pi ic/24}$ is not the topological spin
of some anyon, i.e $e^{2\pi ic/24}\notin\left\{ \theta_{a}\right\} $,
admits no local Hamiltonians which are locally stoquastic.
\end{description}
The above result can be stated in terms of the topological $\mathbf{T}$-matrix,
which is the representation of a Dehn twist on the torus ground state
subspace, and has the spectrum $\text{Spec}\left(\mathbf{T}\right)=\left\{ \theta_{a}e^{-2\pi ic/24}\right\} _{a}$
\citep{doi:10.1142/S0129055X90000107,kitaev2006anyons,PhysRevLett.110.236801,PhysRevB.88.195412,PhysRevLett.110.067208,PhysRevB.91.125123}. 
\begin{description}
\item [{Result$\;$1'\label{Result 1'}}] If a local bosonic Hamiltonian
$H$ is is both locally stoquastic and in a chiral topological phase
of matter, then the corresponding $\mathbf{T}$-matrix satisfies $1\in\text{Spec}\left(\mathbf{T}\right)$.
Equivalently, a bosonic chiral topological phase of matter where $1\notin\text{Spec}\left(\mathbf{T}\right)$,
admits no local Hamiltonians which are locally stoquastic.
\end{description}
The above result is our main statement for bosonic phases of matter.
The logic used in its derivation is extended in Sec.\ref{sec:Spontaneous-chirality}-\ref{sec:No-sign-free-DQMC}, where we generalize Result \hyperref[Result 1]{1} to systems which
are fermionic, spontaneously-chiral, or both. 


\subsection{Spontaneous chirality\label{sec:Spontaneous-chirality}}

The invariants $h_{0}$ and $c$ change sign under both time reversal
$\mathcal{T}$ and parity (spatial reflection) $\mathcal{P}$, and
therefore require a breaking of $\mathcal{T}$ and $\mathcal{P}$
down to $\mathcal{PT}$ to be non-vanishing. The momentum polarization
Eq.\eqref{eq:12-3-1} is valid if this symmetry breaking is explicit,
i.e $H$ does not commute with $\mathcal{P}$ and $\mathcal{T}$ separately.
Here we consider the case where $H$ is $\mathcal{P},\mathcal{T}$-symmetric,
but these are broken down to $\mathcal{PT}$ spontaneously, as in
e.g intrinsic topological superfluids and superconductors \citep{volovik2009universe,PhysRevB.100.104512,rose2020hall}.
We first generalize Eq.\eqref{eq:12-3-1} to this setting, and then
use this generalization to obtain a spontaneously-chiral analog of
Result \hyperref[Result 1]{1}.

Note that the physical time-reversal $\mathcal{T}$ is an \textit{on-site}
anti-unitary operator acting \textit{identically} on all qudits, which
implies $\left[\mathcal{T},T_{R}\right]=0$, while $\mathcal{P}$
is a unitary operator that maps the qudit at $\mathbf{x}$ to that
at $P\mathbf{x}$, where $P$ is the nontrivial element in $O\left(2\right)/SO\left(2\right)$,
e.g $\left(x,y\right)\mapsto\left(-x,y\right)$. 

\subsubsection{Momentum polarization for spontaneously-chiral Hamiltonians \label{subsec:Momentum-polarization-for}}

For simplicity, we begin by assuming that $H$ is 'classically symmetry
breaking' - it has two exact ground states on the cylinder, already
at finite system sizes. We therefore have two ground states $\ket{\pm}$,
such that $\ket -$ is obtained from $\ket +$ by acting with either
$\mathcal{T}$ or $\mathcal{P}$. In particular, $\ket{\pm}$ have
opposite values of $h_{0}$ and $c$. The $\beta\rightarrow\infty$
density matrix is then $e^{-\beta H}/Z=\left(\rho_{+}+\rho_{-}\right)/2$,
where $\rho_{\pm}=\ket{\pm}\bra{\pm}$, and this modifies the right
hand side of Eq.\eqref{eq:12-3-1} to its real part,
\begin{align}
\tilde{Z}:= & \text{Tr}\left(T_{R}e^{-\beta H}\right)\label{eq:12-3-1-1}\\
= & Ze^{-\delta N_{x}}\cos2\pi\left[\epsilon N_{x}+\frac{2\pi}{N_{x}}\left(h_{0}-\frac{c}{24}\right)+o\left(N_{x}^{-1}\right)\right],\nonumber 
\end{align}
where $-\delta\pm2\pi i\epsilon$ are the values of the non-universal
$\alpha$ obtained from Eq.\eqref{eq:12-3-1}, by replacing $e^{-\beta H}$
by $\rho_{\pm}$ . Indeed, it follows from $\left[\mathcal{T},T_{R}\right]=0$
that if two density matrices are related by $\rho_{-}=\mathcal{T}\rho_{+}\mathcal{T}^{-1}$,
then $\tilde{Z}_{\pm}:=Z_{\pm}\text{Tr}\left(T_{R}\rho_{+}\right)$
are complex conjugates, $\tilde{Z}_{-}=\tilde{Z}_{+}^{*}$. 

Now, for a generic symmetry breaking Hamiltonian $H$, exact ground
state degeneracy happens only in the infinite volume limit \citep{sachdev_2011}.
At finite size, the two lowest lying eigenvalues of $H$ would be
separated by an exponentially small energy difference
$\Delta E=O\left(e^{-fN_{x}^{\lambda}}\right)$, with some $f>0,\lambda>0$.
The two corresponding eigenstates would be $\mathcal{T},\mathcal{P}$-even/odd,
of the form $W\left[\ket +\pm\ket -\right]$, where $W$ is a $\mathcal{T},\mathcal{P}$-invariant
local unitary \citep{zeng2019quantum}.  One can think of these statements as resulting from the existence
of a bulk-gap preserving and $\mathcal{T},\mathcal{P}$-symmetric
deformation of $H$ to a 'classically symmetry breaking' Hamiltonian\footnote{The canonical example is the transverse field Ising model $H\left(g\right)=-\sum_{i=1}^{N_{x}}\left(Z_{i}Z_{i+1}+gX_{i}\right)$
in 1+1d. Exact ground state degeneracy appears at finite $N_{x}$
only for $g=0$, though spontaneous symmetry breaking occurs for all
$\left|g\right|<1$, where a splitting $\sim\left|g\right|^{N_{x}}$
appears.}. 

In the generic setting, we have 
\begin{align}
e^{-\beta H}/Z & =W\left(\rho_{+}+\rho_{-}\right)W^{\dagger}/2+O\left(\beta\Delta E\right),
\end{align}
and, following our treatment of the local unitary $U$ in the previous
section, Equation \eqref{eq:12-3-1-1} remains valid, with modified
$\delta,\epsilon$, but unchanged $h_{0}-c/24$. This statement holds
for temperatures much higher than $\Delta E$ and much smaller that
the CFT energy spacing, $\Delta E\ll\beta^{-1}\ll N_{x}^{-1}$,
or more accurately $\beta^{-1}=o\left(N_{x}^{-1}\right)$ and $\beta\Delta E=o\left(N_{x}^{-1}\right)$
(cf. Sec.\ref{subsec:Momentum-polarization}).   Note that the
universal content of Eq.\eqref{eq:12-3-1-1} is the absolute value
$\left|h_{0}-c/24\right|$, since the cosine is even and $\text{sgn}\left(\epsilon\right)$
is non-universal. 

\subsubsection{Constraining $c$ and $\left\{ \theta_{a}\right\} $ }

Let us now assume that a gapped and local Hamiltonian $H$ is $\mathcal{T}$,$\mathcal{P}$-symmetric,
and is locally stoquastic, due to a unitary $U$. It follows that
$\tilde{Z}'=\text{Tr}\left(T_{R}e^{-\beta H'}\right)>0$, where $H'=UHU^{\dagger}$.
If $U$ happens to be $\mathcal{T}$,$\mathcal{P}$-symmetric, then
so is $H'$, and Eq.\eqref{eq:12-3-1-1} holds for $\tilde{Z}'$,
with $\delta',\epsilon'$ in place of $\delta,\epsilon$. For a general
$U$, we have 
\begin{align}
e^{-\beta H'}/Z & '=UW\left(\rho_{+}+\rho_{-}\right)W^{\dagger}U^{\dagger}/2+O\left(\beta\Delta E\right),
\end{align}
where $UW$ need not be $\mathcal{T}$,$\mathcal{P}$-symmetric. As
result, $\tilde{Z}'$ satisfies a weaker form of Eq.\eqref{eq:12-3-1-1},
\begin{align}
0<\tilde{Z}'=\left(Z'/2\right) & \sum_{\sigma=\pm}e^{-\delta_{\sigma}'N_{x}}e^{2\pi i\sigma\left[\epsilon_{\sigma}'N_{x}+\frac{1}{N_{x}}\left(h_{0}-\frac{c}{24}\right)+o\left(N_{x}^{-1}\right)\right]},\label{eq:10-00}
\end{align}
where $\delta_{+}',\epsilon_{+}'$ may differ from $\delta_{-}',\epsilon_{-}'$,
and we also indicated the positivity of $\tilde{Z}'$. Now, if $\delta_{+}'\neq\delta_{-}'$,
one of the chiral contributions is exponentially suppressed relative
to the other as $N_{x}\rightarrow\infty$, and we can apply the analysis
of Sec.\ref{sec:No-stoquastic-Hamiltonians}. If $\delta_{+}'=\delta_{-}'$,
we obtain
\begin{align}
0< & \sum_{\sigma=\pm}\exp2\pi i\sigma\left[\epsilon_{\sigma}'N_{x}+\frac{1}{N_{x}}\left(h_{0}-\frac{c}{24}\right)+o\left(N_{x}^{-1}\right)\right],\label{eq:11-1-0}
\end{align}
in analogy with Eq.\eqref{eq:6}.Unlike Eq.\eqref{eq:6}, taking
the $N_{x}$th power of this equation does not eliminate the mod 1
ambiguity in $h_{0}$. This corresponds to the fact that, as opposed
to explicitly chiral systems, stacking copies of a spontaneously chiral
system does not increase its net chirality. One can replace $T_{R}$
in $\tilde{Z}'$ with a larger half-translation $T_{R}^{m}$, which
would multiply the argument of the cosine by $m$. However, since
the largest translation on the cylinder is obtained for $m\approx N_{x}/2$,
this does not eliminate the mod 1 ambiguity in $h_{0}$. Moreover,
even if it so happens that $\epsilon'_{+}=\epsilon'_{-}=0$, Equation
\eqref{eq:11-1-0} does not imply $h_{0}-c/24=0$ (mod 1) since $N_{x}$
is large. 

\begin{figure}[th]
\begin{centering}
\includegraphics[width=0.3\columnwidth]{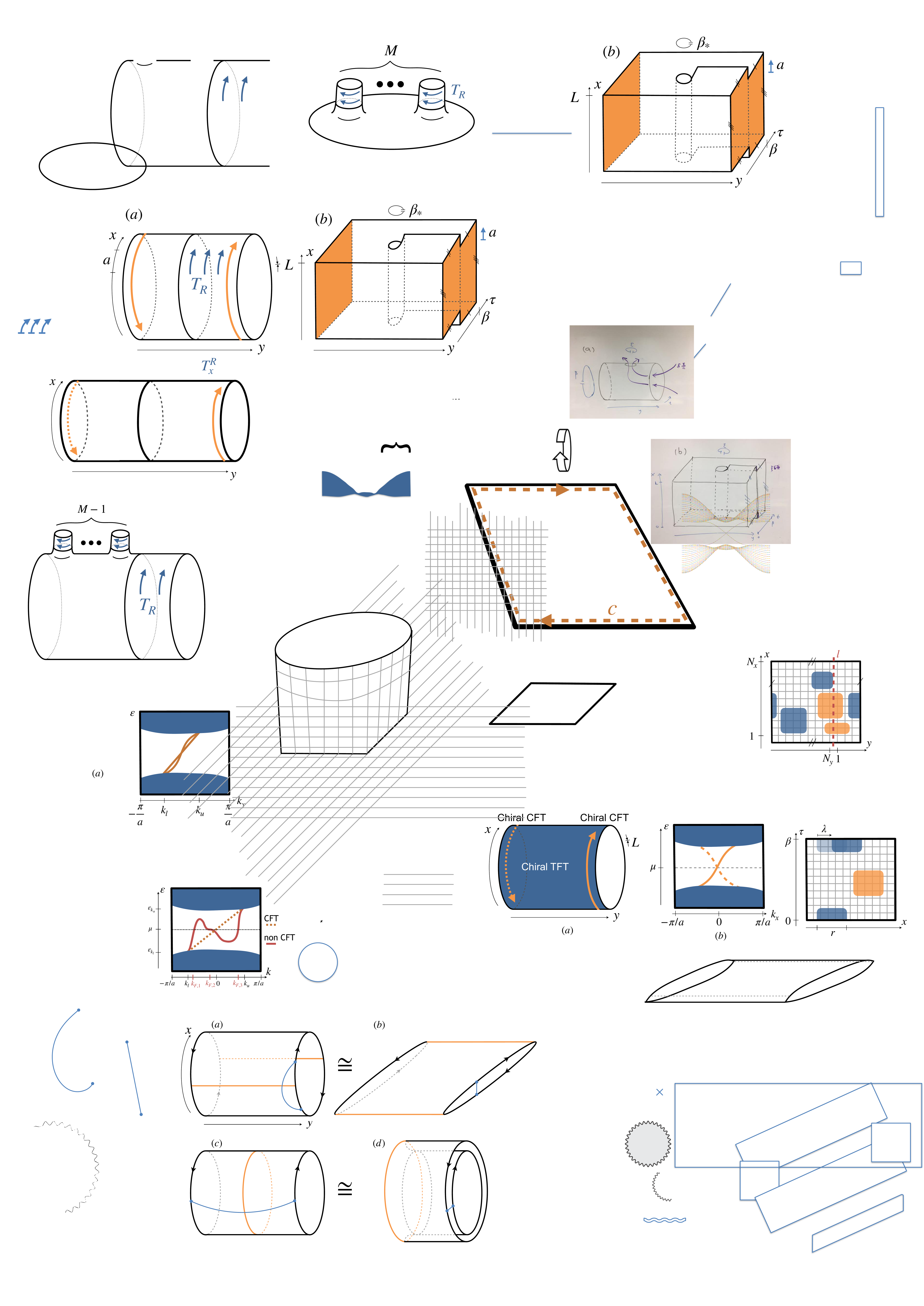}
\par\end{centering}
\caption{Bagpipes construction. We attach $M$ identical cylinders, or 'pipes',
to the given lattice, and define the half translation $T_{R}$ to
act on their top halves, as indicated by blue arrows. The contributions
of the pipes to the momentum polarization adds, producing the factor
$M$ in Eq.\eqref{eq:12-4}. \label{fig:bagpipes}}
\end{figure}

In order to make progress, we make use of the bagpipes construction
illustrated in Fig.\ref{fig:bagpipes}. We attach
$M$ identical cylinders, or 'pipes', to the given lattice, and act
with $T_{R}$ on these cylinders. The global topology of the given
lattice is unimportant - all that is needed is a large enough disk
in which the construction can be applied. The construction does require some form of homogeneity in order to have a unique extension of the Hamiltonian $H'$ to the pipes, and which will be identical for all pipes. We will assume a strict translation symmetry with respect to a sub-lattice, but we believe that this assumption can be relaxed.

The resulting surface, shown in Fig.\ref{fig:bagpipes}, has negative
curvature at the base of each pipe, which requires a finite number
of lattice disclinations in this region. In order to avoid any possible
ambiguity in the definition of $H'$ at a disclination, one can simply
remove any local term $H_{\mathbf{x}}'$ whose support contains a
disclination, which amounts to puncturing a hole around each disclination.
The resulting boundary components do not contribute to the momentum
polarization since $T_{R}$ acts on these as the identity. 

With the construction at hand, the identical contributions of all
cylinders to $\tilde{Z}'$ add, which implies
\begin{align}
0< & \sum_{\sigma=\pm}\exp2\pi i\sigma M\left[\epsilon_{\sigma}'N_{x}+\frac{1}{N_{x}}\left(h_{0}-\frac{c}{24}\right)+o\left(N_{x}^{-1}\right)\right].\label{eq:12-4}
\end{align}
Setting $M=N_{x}$ gives 
\begin{align}
0< & e^{2\pi i\epsilon_{+}'N_{x}^{2}}\theta_{0}\left(N_{x}\right)e^{-2\pi ic/24}\\
 & +e^{-2\pi i\epsilon_{-}'N_{x}^{2}}\theta_{0}^{*}\left(N_{x}\right)e^{2\pi ic/24}+o\left(1\right),\nonumber 
\end{align}
where we indicate explicitly the possible $N_{x}$-dependence of $\theta_{0}$.
This is the spontaneously chiral analog of Eq.\eqref{eq:6-1}, and
can be analyzed similarly. Since $\theta_{0}\left(N_{x}\right)$ is
valued in the finite set $\left\{ \theta_{a}\right\} $, both $\epsilon'_{\pm}$
must be rational, $\epsilon'_{\pm}=n_{\pm}/m_{\pm}$. Restricting
then to $N_{x}=n_{x}m_{+}m_{-}$, such that $e^{2\pi i\epsilon_{\pm}N_{x}^{2}}=1$,
and $\theta_{0}$ attains a constant value $\theta_{a}$ for large
enough $n_{x}$, we have 
\begin{align}
0<\text{Re}\left(\theta_{a}e^{-2\pi ic/24}\right),\label{eq:14}
\end{align}
for some anyon $a$. Repeating the analysis with $k$ times more pipes
$M=kN_{x}$, replaces $\theta_{a}e^{-2\pi ic/24}$ in Eq.\eqref{eq:14}
with its $k$th power, for all $k\in\mathbb{N}$. This infinite set
of equations then implies $\theta_{a}e^{-2\pi ic/24}=1$. To summarize,

\begin{description}
\item [{Result$\;$2\label{Result 2}}] If a local bosonic Hamiltonian
$H$ is both locally stoquastic and in a spontaneously-chiral topological
phase of matter, then one of the corresponding topological spins satisfies
$\theta_{a}=e^{2\pi ic/24}$. Equivalently, a bosonic spontaneously-chiral
topological phase of matter where $e^{2\pi ic/24}$ is not the topological
spin of some anyon, i.e $e^{2\pi ic/24}\notin\left\{ \theta_{a}\right\} $,
admits no local Hamiltonians which are locally stoquastic.
\end{description}
This extends Result \hyperref[Result 1]{1} beyond explicitly-chiral Hamiltonians,
and clarifies that the essence of the intrinsic sign problem we find
is the macroscopic, physically observable, condition $e^{2\pi ic/24}\notin\left\{ \theta_{a}\right\} $,
as opposed to the microscopic absence (or presence) of time reversal
and reflection symmetries. 

\subsection{DQMC: locality, homogeneity, and geometric manipulations\label{sec:Determinantal-quantum-Monte}}

In order to obtain fermionic analogs of the bosonic results of the
previous sections, we first need to establish a framework in which
such results can be obtained. In this section we develop a formalism
that unifies and generalizes the currently used DQMC algorithms and
design principles, and implement within it the geometric manipulations
used in previous sections, in a sign-free manner. Since we wish to
treat the wide range of currently known DQMC algorithms and design
principles on equal footing, the discussion will be more abstract
than the simple setting of locally stoquastic Hamiltonians used above.
In particular, Sections \ref{subsec:Local-determinantal-QMC}-\ref{subsec:Local-and--homogeneous}
lead up to the definition of \textit{locally sign-free DQMC}, which
is our fermionic analog of a locally stoquastic Hamiltonian. This
definition is used later on in Sec.\ref{sec:No-sign-free-DQMC} to
formulate Result \hyperref[Result 1F]{1F} and Result \hyperref[Result 2F]{2F},
the fermionic analogs of Results \hyperref[Result 1]{1} and \hyperref[Result 2]{2}.
The new tools needed to establish these results are the sign-free
geometric manipulations described in Sec.\ref{sec:Sign-free-geometric-manipulation}.

\subsubsection{Local DQMC\label{subsec:Local-determinantal-QMC}}

In the presence of bosons and fermions, the many-body Hilbert space
is given by $\mathcal{H}=\mathcal{H}_{\text{F}}\otimes\mathcal{H}_{\text{B}}$,
where $\mathcal{H}_{\text{F}}$ is a fermionic Fock space, equipped
with an on-site occupation basis $\ket{\nu}_{\text{F}}=\prod_{\mathbf{x},\alpha}\left(f_{\mathbf{x},\alpha}^{\dagger}\right)^{\nu_{\mathbf{x},\alpha}}\ket 0_{\text{F}}$,
$\nu_{\mathbf{x},\alpha}\in\left\{ 0,1\right\} $, generated by acting
with fermionic (anti-commuting) creation operators $f_{\mathbf{x},\alpha}^{\dagger}$
on the Fock vacuum $\ket 0_{\text{F}}$. The product is taken with
respect to a fixed ordering of fermion species $\alpha\in\left\{ 1,\cdots,\mathsf{d}_{\text{F}}\right\} $
and lattice sites $\mathbf{x}\in X$. We will also make use of the
single-fermion space $\mathcal{H}_{1\text{F}}\cong\mathbb{C}^{\left|X\right|}\otimes\mathbb{C}^{\mathsf{d}_{\text{F}}}$,
spanned by $\ket{\mathbf{x},\alpha}_{\text{F}}=f_{\mathbf{x},\alpha}^{\dagger}\ket 0_{\text{F}}$,
where $\left|X\right|=N_{x}N_{y}$ is the system size. As in Sec.\ref{sec:No-stoquastic-Hamiltonians},
$\mathcal{H}_{\text{B}}$ is a many-qudit Hilbert space with local
dimension $\mathsf{d}$. It can also be a bosonic Fock space where
$\mathsf{d}=\infty$.

We consider local fermion-boson Hamiltonians $H$, of the form 
\begin{align}
H=\sum_{\mathbf{x},\mathbf{y}}f_{\mathbf{x}}^{\dagger}h_{0}^{\mathbf{x},\mathbf{y}}f_{\mathbf{y}}+H_{I},\label{eq:11-0}
\end{align}
where the free-fermion Hermitian matrix $h_{0}^{\mathbf{x},\mathbf{y}}$
is $r_{0}$-local, it vanishes unless $\left|\mathbf{x}-\mathbf{y}\right|\leq r_{0}$,
and we suppress, here and in the following, the fermion species indices.
The Hamiltonian $H_{I}$ describes all possible $r_{0}$-local interactions
which preserve the fermion parity $\left(-1\right)^{N_{f}}$, where
$N_{f}=\sum_{\mathbf{x}}f_{\mathbf{x}}^{\dagger}f_{\mathbf{x}}$,
including fermion-independent terms $H_{\text{B}}$ as in Sec.\ref{sec:No-stoquastic-Hamiltonians}.
Thus $H_{I}$ is of the form 
\begin{align}
H_{I}= & H_{\text{B}}+\sum_{\mathbf{x},\mathbf{y}}f_{\mathbf{x}}^{\dagger}K_{\text{B}}^{\mathbf{x},\mathbf{y}}f_{\mathbf{y}}\label{eq:dots}\\
 & +\sum_{\mathbf{x},\mathbf{y},\mathbf{z},\mathbf{w}}f_{\mathbf{x}}^{\dagger}f_{\mathbf{y}}^{\dagger}V_{\text{B}}^{\mathbf{x},\mathbf{y},\mathbf{z},\mathbf{w}}f_{\mathbf{z}}f_{\mathbf{w}}+\cdots,\nonumber 
\end{align}
where $K_{\text{B}}^{\mathbf{x},\mathbf{y}}$ (for all $\mathbf{x},\mathbf{y}\in X$)
is a local bosonic operator with range $r_{0}$, and vanishes unless
$\left|\mathbf{x}-\mathbf{y}\right|\leq r_{0}$, and similarly for
$V_{\text{B}}^{\mathbf{x},\mathbf{y},\mathbf{z},\mathbf{w}}$, which
vanishes unless $\mathbf{x},\mathbf{y},\mathbf{z},\mathbf{w}$ are
contained in a disk or radius $r_{0}$. In Eq.\eqref{eq:dots} dots
represent additional pairing terms of the form $ff$, $f^{\dagger}f^{\dagger}$,
or $ffff$, $f^{\dagger}f^{\dagger}f^{\dagger}f^{\dagger}$, as well
as terms with a higher number of fermions, all of which are $r_{0}$-local
and preserve the fermion parity.

Since locality is defined in terms of anti-commuting Fermi operators,
a local stoquastic basis is not expected to exist, and accordingly,
the sign problem appears in any QMC method in which the Boltzmann
weights are given in terms of Hamiltonian matrix elements  in a local
basis \citep{troyer2005computational,li2019sign}. For this reason,
the methods used to perform QMC in the presence of fermions are distinct
from the ones used in their absence. These are collectively referred
to as DQMC \citep{PhysRevD.24.2278,Assaad,Santos_2003,li2019sign,berg2019monte},
and lead to the imaginary time path integral representation of the
partition function $Z=\text{Tr}\left(e^{-\beta H}\right)$,
\begin{align}
Z & =\int D\phi D\psi e^{-S_{\phi}-S_{\psi,\phi}}\label{eq:2}\\
 & =\int D\phi e^{-S_{\phi}}\text{Det}\left(D_{\phi}\right)\nonumber \\
 & =\int D\phi e^{-S_{\phi}}\text{Det}\left(I+U_{\phi}\right),\nonumber 
\end{align}
involving a bosonic field $\phi$ with an action $S_{\phi}$, and
a fermionic (grassmann valued) field $\psi$, with a quadratic action
$S_{\psi,\phi}=\sum_{\mathbf{x},\mathbf{x}'}\int\text{d}\tau\overline{\psi}_{\mathbf{x},\tau}\left[D_{\phi}\right]_{\mathbf{x},\mathbf{y}}\psi_{\mathbf{y},\tau}$
defined by the $\phi$-dependent single-fermion operator $D_{\phi}$.
In the third line of Eq.\eqref{eq:2} we assumed the Hamiltonian form
$D_{\phi}=\partial_{\tau}+h_{\phi\left(\tau\right)}$, and used a
standard identity for the determinant in terms of the single-fermion
imaginary-time evolution operator $U_{\phi}=\text{TO}e^{-\int_{0}^{\beta}h_{\phi\left(\tau\right)}\text{d}\tau}$
\citep{PhysRevD.24.2278}, where $\text{TO}$ denotes the time ordering.
The field $\phi$ ($\psi$) is defined on a continuous imaginary-time
circle $\tau\in\mathbb{R}/\beta\mathbb{Z}$, with periodic (anti-periodic)
boundary conditions, and on the spatial lattice $X$. The second
and third lines of Eq.\eqref{eq:2} define the Monte Carlo phase space
$\left\{ \phi\right\} $ and Boltzmann weight 
\begin{align}
p\left(\phi\right) & =e^{-S_{\phi}}\text{Det}\left(D_{\phi}\right)\label{eq:18-2}\\
 & =e^{-S_{\phi}}\text{Det}\left(I+U_{\phi}\right).\nonumber 
\end{align}

In applications, the DQMC representation \eqref{eq:2} may be obtained
from the Hamiltonian $H$ in a number of ways. If a Yukawa type model
is assumed as a starting point \citep{berg2019monte}, i.e $H_{I}=H_{\text{B}}+\sum_{\mathbf{x},\mathbf{y}}f_{\mathbf{x}}^{\dagger}K_{\text{B}}^{\mathbf{x},\mathbf{y}}f_{\mathbf{y}}$,
then the action $S_{\phi}$ is obtained from the Hamiltonian $H_{\text{B}}$,
and $h_{\phi\left(\tau\right)}=h_{0}+K_{\text{B}}$. Alternatively,
the representation \eqref{eq:2} may be obtained through a Hubbard-Stratonovich
decoupling and/or a  series expansion of fermionic self-interactions
\citep{PhysRevLett.82.4155,chandrasekharan2013fermion,wang2015split}.
Such is the case e.g when there are no bosons $\mathcal{H}=\mathcal{H}_{\text{F}}$,
and $H_{I}=\sum f_{\mathbf{x}}^{\dagger}f_{\mathbf{y}}^{\dagger}V^{\mathbf{x},\mathbf{y},\mathbf{z},\mathbf{w}}f_{\mathbf{z}}f_{\mathbf{w}}$. Note that for a given fermionic self-interaction, there are various possible DQMC representations, obtained e.g via a Hubbard-Stratonovich
decoupling in different channels. 

To take into account and generalize the above relations between $H$
and the corresponding DQMC representation, we will only assume (i)
that the effective single-fermion Hamiltonian $h_{\phi\left(\tau\right)}$
reduces to the free fermion matrix $h^{\left(0\right)}$ in the absence
of $\phi$, i.e $h_{\phi\left(\tau\right)=0}=h_{0}$, (ii) that the
boson field $\phi$ is itself an $r_{0}$-local object\footnote{Thus $\phi$ is a map from sets of lattice sites with diameter less
than $r_{0}$, such as links, plaquettes etc., to a fixed vector space
$\mathbb{C}^{k}$. Additionally, the $\phi$ integration in \eqref{eq:2}
runs over all such functions. As an example, restricting to constant
functions $\phi$ leads to non-local all to all interactions between
fermions.}, and (iii) that the $r_{0}$-locality of $h_{0}$ and $H_{I}$ implies
the $r$-locality of $S_{\phi}$ and $h_{\phi\left(\tau\right)}$,
where $r$ is some function of $r_{0}$, independent of system size.
The physical content of these assumptions is that the fields $\psi$
and operators $f$ correspond to the same physical fermion\footnote{Technically, via the fermionic coherent state construction of the
functional integral \citep{altland2010condensed}.}, and that the boson $\phi$ mediates \textit{all} fermionic interactions
$H_{I}$, and therefore corresponds to both the physical bosons in
$\mathcal{H}_{\text{B}}$ and to composite objects made of an even
number of fermions within a range $r_{0}$ (e.g a Cooper pair $\phi\sim ff$).

We can therefore write 
\begin{align}
S_{\phi} & =\sum_{\tau,\mathbf{x}}S_{\phi;\tau,\mathbf{x}},\label{eq:16}\\
h_{\phi\left(\tau\right)} & =\sum_{\mathbf{x}}h_{\phi\left(\tau\right);\mathbf{x}},\nonumber 
\end{align}
where each term $S_{\phi;\tau,x}$ depends only on the values of $\phi$
at points $\left(\mathbf{x}',\tau'\right)$ with $\left|\tau-\tau'\right|,\left|\mathbf{x}-\mathbf{x}'\right|\leq r$,
and similarly, each term $h_{\phi\left(\tau\right);\mathbf{x}}$ is
supported on a disk of radius $r$ around $\mathbf{x}$, and depends
on the values of $\phi\left(\tau\right)$ at points $\mathbf{x}$
within this disk.

Note that even though $H$ is Hermitian, we do not assume the same
for $h_{\phi\left(\tau\right)}$. Non-Hermitian $h_{\phi\left(\tau\right)}$s
naturally arise in Hubbard\textendash Stratonovich decouplings, see
e.g \citep{PhysRevB.71.155115,wang2015split}. Even when $h_{\phi\left(\tau\right)}$
is Hermitian for all $\phi$, its time-dependence implies that $U_{\phi}$
is non-Hermitian, and therefore $\text{Det}\left(I+U_{\phi}\right)$
in Eq.\eqref{eq:18-2} is generically complex valued \citep{PhysRevD.24.2278}.
This is the generic origin of the sign problem in DQMC. Section \ref{subsec:Local-and--homogeneous}
below describes the notion of \textit{fermionic design principles},
algebraic conditions on $U_{\phi}$ implying $\text{Det}\left(I+U_{\phi}\right)\geq0$,
and defines what it means for such design principles to be local and
homogenous. 

In the following analysis, we exclude the case of 'classically-interacting fermions',
where $\phi$ is time-independent and $h_{\phi}$ is Hermitian. In
this case the fermionic weight $\text{Det}\left(I+e^{-\beta h_{\phi}}\right)$
is trivially non-negative, and sign-free DQMC is always possible,
provided $S_{\phi}\in\mathbb{R}$. We view such models as 'exactly
solvable', on equal footing with free-fermion and commuting projector
models. Given a phase of matter, the possible existence of exactly
solvable models is independent of the possible existence of sign-free
models. Even when an exactly solvable model exists, QMC simulations
are of interest for generic questions, such as phase transitions due
to deformations of the model \citep{hofmann2019search}. In particular,
Ref.\citep{PhysRevLett.119.127204} utilized a classically-free description
of Kitaev's honeycomb model to obtain the thermal Hall conductance
and chiral central charge, which should be contrasted with the intrinsic
sign problem we find in the corresponding phase of matter, see Table \ref{tab:1} and Sec.\ref{sec:No-sign-free-DQMC}.

\subsubsection{Local and homogenous fermionic design principles\label{subsec:Local-and--homogeneous}}

The representation \eqref{eq:2} is sign-free if $p\left(\phi\right)=e^{-S_{\phi}}\text{Det}\left(I+U_{\phi}\right)\geq0$
for all $\phi$. A design principle then amounts to a set of polynomially
verifiable properties \footnote{That is, properties which can be verified in a polynomial-in-$\beta\left|X\right|$
time. As an example, given a local Hamiltonian, deciding whether there
exists a local basis in which it is stoquastic is NP-complete \citep{marvian2018computational,klassen2019hardness}.
In particular, one does not need to perform the exponential operation
of evaluating $p$ on every configuration $\phi$ to assure that $p\left(\phi\right)\geq0$.
Had this been possible, there would be no need for a Monte Carlo sampling
of the phase space $\left\{ \phi\right\} $.} of $S_{\phi}$ and $h_{\phi\left(\tau\right)}$ that guarantee that
the complex phase of $\text{Det}\left(I+U_{\phi}\right)$ is opposite
to that of $e^{-S_{\phi}}$. For the sake of presentation, we restrict
attention to the case where $S_{\phi}$ is manifestly real valued,
and $\text{Det}\left(I+U_{\phi}\right)\geq0$ due to an algebraic
condition on the operator $U_{\phi}$, which we write as $U_{\phi}\in\mathcal{C}_{U}$.
This is assumed to follow from an algebraic condition on $h_{\phi\left(\tau\right)}$,
written as $h_{\phi\left(\tau\right)}\in\mathcal{C}_{h}$, manifestly
satisfied for all $\phi\left(\tau\right)$. The set $\mathcal{C}_{h}$
is assumed to be closed under addition, while $\mathcal{C}_{U}$ is
closed under multiplication: $h_{1}+h_{2}\in\mathcal{C}_{h}$ for
all $h_{1},h_{2}\in\mathcal{C}_{h}$, and $U_{1}U_{2}\in\mathcal{C}_{U}$
for all $U_{1},U_{2}\in\mathcal{C}_{U}$. 

The simplest example, where $\mathcal{C}_{U}=\mathcal{C}_{h}$ is
the set of matrices obeying a fixed time reversal symmetry, is discussed
in Sec.\ref{subsec:Example:-time-reversal}. In Appendix \ref{subsec:Locality-of-known}
we review all other design principles known to us, demonstrate that
most of them are of the simplified form above, and generalize our
arguments to those that are not. Comparing with the bosonic Hamiltonians
treated in Sec.\ref{sec:No-stoquastic-Hamiltonians}, we note that
$\mathcal{C}_{h}$ is analogous to the set of stoquastic Hamiltonians
$H$ in a fixed basis, while $\mathcal{C}_{U}$ is analogous to the
resulting set of matrices $e^{-\beta H}$ with non-negative entries.

Design principles, as defined above (and in the literature), are purely
algebraic conditions, which carry no information about the underlying
geometry of space-time. However, as demonstrated in Sec.\ref{subsec:Example:-time-reversal},
in order to allow for local interactions, mediated by an $r_{0}$-local
boson $\phi$, a design principle must also be local in some sense.
We will adopt the following definitions, which are shown to be satisfied
by all physical applications of design principles that we are aware
of, in Sec.\ref{subsec:Example:-time-reversal} and Appendix \ref{subsec:Locality-of-known}.

\paragraph*{Definition (term-wise sign-free):}

We say that a DQMC representation is term-wise sign-free due to a
design principle $\mathcal{C}_{h}$, if each of the local terms $S_{\phi;\tau,\mathbf{x}},h_{\phi\left(\tau\right);\mathbf{x}}$
obey the design principle separately, rather than just they sums
$S_{\phi},h_{\phi\left(\tau\right)}$. Thus $S_{\phi;\tau,\mathbf{x}}$
is real valued, and $h_{\phi\left(\tau\right);\mathbf{x}}\in\mathcal{C}_{h}$,
for all $\tau,\mathbf{x}$.

\medskip{}

This is analogous to the requirement in Sec.\ref{subsec:Setup} that
$H'$ be term-wise stoquastic. Note that even when a DQMC representation
is term-wise sign-free, the resulting Boltzmann weights $p\left(\phi\right)$
are sign-free in a non-local manner: $\text{Det}\left(I+U_{\phi}\right)$
involves the values of $\phi$ at all space-time points, and splitting
the determinant into a product of local terms by the Leibniz formula
reintroduces signs, which capture the fermionic statistics. In this
respect, the ``classical'' Boltzmann weights $p\left(\phi\right)$
are always non-local in DQMC.

\paragraph*{Definition (on-site homogeneous design principle):\label{par:on-site-homogeneous}}

A design principle is said to be on-site homogenous if any permutation
of the lattice sites $\sigma\in S_{X}$ obeys it. That is, the operator
\begin{align}
 & O_{\left(\mathbf{x},\alpha\right),\left(\mathbf{x}',\alpha'\right)}^{\left(\sigma\right)}=\delta_{\mathbf{x},\sigma\left(\mathbf{x}'\right)}\delta_{\alpha,\alpha'},\label{eq:20-0}
\end{align}
viewed as a single-fermion imaginary-time evolution operator, obeys
the design principle: $O^{\left(\sigma\right)}\in\mathcal{C}_{U}$,
for all $\sigma\in S_{X}$.

\medskip{}

This amounts to the statement that the design principle treats all
lattice sites on equal footing, since it follows that $U_{\phi}\in\mathcal{C}_{U}$ if and only if $O^{\left(\sigma\right)}U_{\phi}O^{\left(\tilde{\sigma}\right)}\in\mathcal{C}_{U}$,
for all permutations $\sigma,\tilde{\sigma}$. It may be that a design
principle is on-site homogeneous only with respect to a sub-lattice $X'\subset X$.
In this case we simply treat $X'$ as the spatial lattice, and add
the finite set $X/X'$ to the $\mathsf{d}_{\text{F}}$ internal degrees
of freedom. Comparing with Sec.\ref{subsec:Setup}, on-site homogeneous
design principles are analogous to the set of Hamiltonians $H'$ which
are stoquastic in an on-site homogeneous basis - any qudit permutation
operator has non-negative entries in this basis, like the imaginary
time evolution $e^{-\beta H'}$.

\medskip{}

With these two notions of locality and homogeneity in design principles,
we now define the DQMC analog of locally stoquastic Hamiltonians (see
Sec.\ref{sec:No-stoquastic-Hamiltonians}). 

\medskip{}

\paragraph*{Definition (locally sign-free DQMC):}

Given a local fermion-boson Hamiltonian $H$, we say that $H$ allows
for a locally sign-free DQMC simulation, if there exists a local unitary
$U$, such that $H'=UHU^{\dagger}$ has a local DQMC representation
\eqref{eq:2}, which is term-wise sign-free due to an on-site homogeneous
design principle.

\medskip{}

Note that the DQMC representation \eqref{eq:2} is not of the Hamiltonian
but of the partition function, and clearly $Z'=\text{Tr}\left(e^{-\beta H'}\right)=\text{Tr}\left(e^{-\beta H}\right)=Z$.
What the above definition entails, is that it is $H'$, rather than
$H$, from which the DQMC data $S_{\phi},h_{\phi\left(\tau\right)}$
is obtained, as described in Sec.\ref{subsec:Local-determinantal-QMC}.
This data is then assumed to be term-wise sign-free due to an on-site
homogeneous design principle. The local unitary $U$ appearing in
the above definition is generally fermionic \citep{PhysRevB.91.125149}:
it can be written as a finite time evolution $U=\text{TO}e^{-i\int_{0}^{1}\tilde{H}\left(t\right)dt}$,
where $\tilde{H}$ is a local fermion-boson Hamiltonian, which is
either piecewise-constant or smooth as a function of $t$, c.f Sec.\ref{subsec:Setup}.

\medskip{}

\subsubsection{Example: time reversal design principle\label{subsec:Example:-time-reversal}}

To demonstrate the above definitions in a concrete setting, consider
the time-reversal design principle, defined by an anti-unitary operator
$\mathsf{T}$ acting on the single-fermion Hilbert space $\mathcal{H}_{1\text{F}}\cong\mathbb{C}^{\left|X\right|}\otimes\mathbb{C}^{\mathsf{d}_{\text{F}}}$,
such that $\mathtt{\mathsf{T}}^{2}=-I$. The set $\mathcal{C}_{h}$
contains all $\mathsf{T}$-invariant matrices, $\left[\mathsf{T},h_{\phi\left(\tau\right)}\right]=0$.
It follows that $\left[\mathsf{T},U_{\phi}\right]=0$, so that $\mathcal{C}_{U}=\mathcal{C}_{h}$
in this case, and this implies $\text{Det}\left(I+U_{\phi}\right)\geq0$
\citep{Hands_2000,PhysRevB.71.155115}. 

A sufficient condition on $\mathsf{T}$ that guarantees that the design
principle it defines is on-site homogenous is that it is of the form
$\mathsf{T}_{0}=I_{\left|X\right|}\otimes\mathsf{t}$, where $I_{\left|X\right|}$
is the identity matrix on $\mathbb{C}^{\left|X\right|}$, and $\mathsf{t}$
is an anti-unitary on $\mathbb{C}^{\mathsf{d}_{\text{F}}}$ that squares
to $-I_{\mathsf{d}_{F}}$. Equivalently, $\mathsf{T}$ is block diagonal,
with identical blocks $\mathsf{t}$ corresponding to the lattice sites
$\mathbf{x}\in X$. It is then clear that the permutation matrices
$O^{\left(\sigma\right)}$ defined in Eq.\eqref{eq:20-0} commute with
$\mathsf{T}$, so $O^{\left(\sigma\right)}\in\mathcal{C}_{U}$ for
all $\sigma\in S_{X}$. Note that the design principle $\mathsf{T}$
may correspond to a \textit{physical} time-reversal $\mathcal{T}$,
discussed in Sec.\ref{sec:Spontaneous-chirality}, only if it is on-site
homogenous, which is why we distinguish the two in our notation.

Additionally, if the operator $\mathsf{T}$ is $r_{\mathsf{T}}$-local
with some range $r_{\mathsf{T}}\geq0$, then any local $h_{\phi\left(\tau\right)}$
which is sign-free due to $\mathsf{T}$ can be made term-wise sign-free.
Indeed, if $\left[\mathsf{T},h_{\phi\left(\tau\right)}\right]=0$
then
\begin{align}
h_{\phi\left(\tau\right)} & =\frac{1}{2}\left(h_{\phi\left(\tau\right)}+\mathsf{T}h_{\phi\left(\tau\right)}\mathsf{T}^{-1}\right)\label{eq:21-0}\\
 & =\sum_{\mathbf{x}}\frac{1}{2}\left(h_{\phi\left(\tau\right);\mathbf{x}}+\mathsf{T}h_{\phi\left(\tau\right);\mathbf{x}}\mathsf{T}^{-1}\right)\nonumber \\
 & =\sum_{\mathbf{x}}\tilde{h}_{\phi\left(\tau\right);\mathbf{x}},\nonumber 
\end{align}
where $\tilde{h}{}_{\phi\left(\tau\right);\mathbf{x}}$ is now supported
on a disk of radius $r+2r_{\mathsf{T}}$ and commutes with $\mathsf{T}$,
for all $\mathbf{x}$. We see that the specific notion of $r_{\mathsf{T}}$-locality
coincides with the general notion of 'term-wise sign free'. In particular,
$\mathsf{T}=\mathsf{T}_{0}$ has a range $r_{\mathsf{T}}=0$, and
can therefore be applied term-wise. 

The above statements imply that if $\mathsf{T}=u\mathsf{T}_{0}u^{\dagger}$,
where $u$ is a single-fermion local unitary, and $H$ has a local
DQMC representation which is sign-free due to $\mathsf{T}$, then
$H$ allows for a locally sign-free DQMC simulation. Indeed, extending
$u$ to a many-body local unitary $U$, we see that $H'=UHU^{\dagger}$
admits a local DQMC representation where $\left[\mathsf{T}_{0},h_{\phi\left(\tau\right)}'\right]=0$.
Since $\mathsf{T}_{0}$ is on-site homogenous, and $h_{\phi\left(\tau\right)}'$
can be assumed term-wise sign-free (see Eq.\eqref{eq:21-0}), we have
the desired result. As demonstrated in Appendix \ref{subsec:Locality-of-known},
much of the above analysis carries over to other known design principles.

All realizations of $\,\mathsf{T}$ presented in Ref.\citep{PhysRevB.71.155115}
in the context of generalized Hubbard models, and in Ref.\citep{berg2019monte}
in the context of quantum critical metals, have the on-site homogeneous
form $\mathsf{T}_{0}$, and therefore correspond to locally sign-free
DQMC simulations.

We now consider a few specific time-reversal design principles $\mathsf{T}$.
The physical spin-1/2 time reversal $\mathsf{T}=\mathcal{T}^{\left(1/2\right)}$,
where $\mathcal{T}_{\left(\mathbf{x},\alpha\right),\left(\mathbf{x}',\alpha'\right)}^{\left(1/2\right)}=\delta_{\mathbf{x},\mathbf{x}'}\varepsilon_{\alpha\alpha'}\mathcal{K}$,
and $\alpha,\alpha'\in\left\{ \uparrow,\downarrow\right\} $ correspond
to up and down spin components,  is an on-site homogeneous design
principle, which accounts for the absence of signs in the attractive
Hubbard model \citep{PhysRevB.71.155115}. The composition $\mathsf{T}=\mathcal{M}\mathcal{T}^{\left(1/2\right)}$
of $\mathcal{T}^{\left(1/2\right)}$ with a modulo 2 translation,
$\mathcal{M}_{\left(\mathbf{x},\alpha\right),\left(\mathbf{x}',\alpha'\right)}=\delta_{\left(-1\right)^{x},\left(-1\right)^{x'+1}}\delta_{x_{e},x_{e}'}\delta_{y,y'}\delta_{\alpha,\alpha'}$,
where $x_{e}=2\left\lfloor x/2\right\rfloor $ is the even part of
$x$, is an on-site homogeneous design principle with respect to the
sub-lattice$X'=\left\{ \left(2x_{1},x_{2}\right):\;\mathbf{x}\in X\right\} $,
but not with respect to $X$. On the other hand, the composition $\mathsf{T}=\mathcal{P}^{\left(0\right)}\mathcal{T}^{\left(1/2\right)}$
of $\mathcal{T}^{\left(1/2\right)}$ with a spin-less reflection (or
parity) $\mathcal{P}_{\left(\mathbf{x},\alpha\right),\left(\mathbf{x}',\alpha'\right)}^{\left(0\right)}=\delta_{x,-x'}\delta_{y,y'}\delta_{\alpha\alpha'}$,
is not on-site homogeneous with respect to any sub-lattice. 

\begin{figure}[t]
\begin{centering}
\includegraphics[width=0.6\columnwidth]{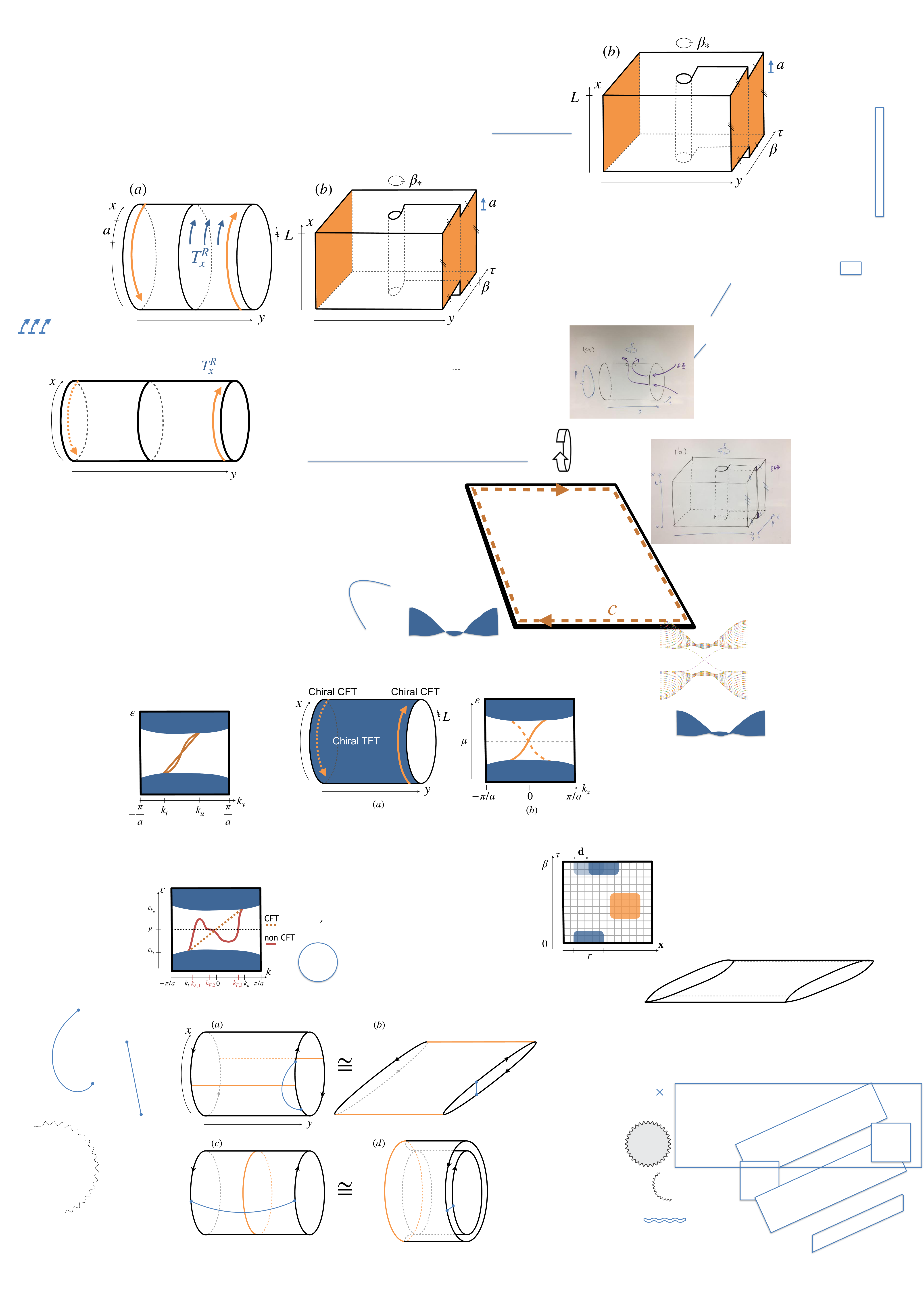}
\par\end{centering}
\caption{$\mathcal{P}\mathcal{T}$ symmetry as a 'non-local design principle'
for chiral topological matter. (a), (c): $\mathcal{P}\mathcal{T}$
symmetry, where $\mathcal{P}$ is a reflection (with respect to the
orange lines) and $\mathcal{P}\mathcal{T}$ is an on-site time-reversal,
is a natural symmetry in chiral topological phases. If $\left(\mathcal{P}\mathcal{T}\right)^{2}=-I$,
as is the case when $\mathcal{P}=\mathcal{P}^{\left(0\right)}$ is
spin-less and $\mathcal{T}=\mathcal{T}^{\left(1/2\right)}$ is spin-full,
it implies the non-negativity of fermionic determinants. Nevertheless,
as $\mathcal{P}\mathcal{T}$ is non-local, it only allows for QMC
simulations with $\mathcal{P}\mathcal{T}$ invariant bosonic fields,
which mediate non-local interactions (blue lines) between fermions.
Arrows indicate the chirality of boundary degrees of freedom. (b),
(d): Such non-local interactions effectively fold the system into
a non-chiral locally-interacting system supported on half the cylinder,
where $\mathcal{P}\mathcal{T}$ acts as an on-site time reversal.
In particular, the boundary degrees of freedom are now non-chiral.
Thus, $\mathcal{P}\mathcal{T}$ does not allow for sign-free QMC simulations
of chiral topological matter. More generally, fermionic design principles
must be local in order to allow for sign-free DQMC simulations of
local Hamiltonians. \label{fig:-symmetry-as}}
\end{figure}

The latter example is clearly non-local, and we use it to demonstrate
the necessity of locality in design principles. As discussed in Sec.\ref{sec:Spontaneous-chirality},
the breaking of $\mathcal{P}$ and $\mathcal{T}$ down to $\mathcal{P}\mathcal{T}$
actually defines the notion of chirality, and therefore $\mathcal{P}\mathcal{T}$
is a natural symmetry in chiral topological matter. Accordingly, the
design principle $\mathsf{T}=\mathcal{P}^{\left(0\right)}\mathcal{T}^{\left(1/2\right)}$
applies to a class of models for chiral topological phases, see Appendix
\ref{subsec:A-non-local-design}. This seems to allow, from the naive
algebraic perspective, for a sign-free DQMC simulation of certain
chiral topological phases. However, the weights $p\left(\phi\right)$
will only be non-negative for bosonic configurations $\phi$ which
are is invariant under $\mathsf{T}=\mathcal{P}^{\left(0\right)}\mathcal{T}^{\left(1/2\right)}$.
Restricting the $\phi$ integration in Eq.\eqref{eq:2} to such configurations
leads to non-local interactions between fermions $\psi$, coupling
the points $\left(x,y\right)$ and $\left(-x,y\right)$. These interactions
effectively fold the non-local chiral system into a local non-chiral
system on half of space, see Fig.\ref{fig:-symmetry-as}. Thus, $\mathsf{T}=\mathcal{P}^{\left(0\right)}\mathcal{T}^{\left(1/2\right)}$
does \textit{not} allow for sign-free DQMC simulations of chiral topological
matter.

\subsubsection{Sign-free geometric manipulations in DQMC\label{sec:Sign-free-geometric-manipulation}}

Let $Z$ be a partition function in a local DQMC form \eqref{eq:2},
on the discrete torus $X=\mathbb{Z}_{N_{x}}\times\mathbb{Z}_{N_{y}}$
and imaginary time circle $S_{\beta}^{1}=\mathbb{R}/\beta\mathbb{Z}$,
which is term-wise sign-free due to an on-site homogenous design principle.
In this section we show that it is possible to cut $X$ to the cylinder
$C$, and subsequently introduce a screw dislocation in the space-time
$C\times S_{\beta}^{1}$, which corresponds to the momentum polarization
\eqref{eq:12-3-1}, while maintaining the DQMC weights $p\left(\phi\right)$
non-negative.

\paragraph*{Introducing spatial boundaries}

Given a translation $T^{\mathbf{d}}$ ($\mathbf{d}\in X$), we can
cut the torus $X$ along a line $l$ parallel to $\mathbf{d}$, and
obtain a cylinder $C$ where $T^{\mathbf{d}}$ acts as a translation
within each boundary component, as in Sec.\ref{sec:No-stoquastic-Hamiltonians}.
Given the DQMC representation \eqref{eq:2} on $X$, the corresponding representation
on $C$ is obtained by eliminating all local terms $S_{\phi;\tau,\mathbf{x}},h_{\phi\left(\tau\right);\mathbf{x}}$
whose support overlaps $l$, as in Fig.\ref{fig:cutting}. This procedure
may render $S_{\phi},h_{\phi\left(\tau\right)}$ independent of certain
degrees of freedom $\phi\left(\mathbf{x},\tau\right),\psi\left(\mathbf{x},\tau\right)$,
with $\mathbf{x}$ within a range $r$ of $l$, in which case we simply
remove such degrees of freedom from the functional integral \eqref{eq:2}\footnote{For $r_{0}$-local $\phi$, which is defined on links, plaquettes,
etc., we also remove from the functional $\phi$ integration those
links, plaquettes, etc. which overlap $l$.}. Since $S_{\phi;\tau,\mathbf{x}},h_{\phi\left(\tau\right);\mathbf{x}}$
obey the design principle for every $\mathbf{x},\tau$, the resulting
$S_{\phi},h_{\phi\left(\tau\right)}$ still obey the design principle
and the weights $p\left(\phi\right)$ remain real and non-negative.

\paragraph*{Introducing a screw dislocation in space-time}

Let us now restrict attention to $\mathbf{d}=\left(1,0\right)$, and
make contact with the momentum polarization \eqref{eq:12-3-1}. Given
a partition function on the space-time $C\times S_{\beta}^{1}$, consider
twisting the boundary conditions in the time direction,
\begin{align}
 & \phi_{\tau+\beta,x,y}=\phi_{\tau,x-\lambda\Theta\left(y\right),y},\label{eq:5-1-0}\\
 & \psi_{\tau+\beta,x,y}=-\psi_{\tau,x-\lambda\Theta\left(y\right),y}.\nonumber 
\end{align}
Note that $\lambda\in\mathbb{Z}_{N_{x}}$, since $x\in\mathbb{Z}_{N_{x}}$.
In particular, the full twist $\lambda=N_{x}$ is equivalent to the
untwisted case $\lambda=0$, which is equivalent to the statement
that the modular parameter of the torus is defined mod 1 (see e.g
example 8.2 of \citep{nakahara2003geometry}). The case $\lambda=0$
gives the standard boundary conditions, where the partition function
is, in Hamiltonian terms, just $Z=\text{Tr}\left(e^{-\beta H}\right)$.
In this case $Z>0$ since $H$ is Hermitian, though its QMC representation
$Z=\sum_{\phi}p\left(\phi\right)$ will generically involve complex
valued weights $p$. The twisted case $\lambda=1$ includes the insertion
of the half-translation operator 
\begin{align}
\tilde{Z} & =\text{Tr}\left(T_{R}e^{-\beta H}\right),\label{eq:7-2}
\end{align}
which appears in the momentum polarization \eqref{eq:12-3-1}. Since
$T_{R}$ is unitary rather than hermitian, $\tilde{Z}$ itself will
generically be complex. However,

\paragraph*{Claim:}

If $Z$ has a local DQMC representation \eqref{eq:2}, which is
term-wise sign-free due to an on-site homogeneous design principle,
then $\tilde{Z}$ also has a sign-free QMC representation: $\tilde{Z}=\sum_{\phi}\tilde{p}\left(\phi\right)$,
with $\tilde{p}\left(\phi\right)\geq0$. In particular, $\tilde{Z}\geq0$.

Proof of the claim is provided below. It revolves around two physical
points: (i) For the boson $\phi$, we only use the fact that all boundary
conditions, and those in Eq.\eqref{eq:5-1-0} in particular, are locally
invisible. (ii) For the fermion $\psi$, the local invisibility of
boundary conditions does not suffice, and the important point is that
translations do not act on internal degrees of freedom, and therefore
correspond to permutations of the lattice sites. The same holds for
the half translation $T_{R}$. This distinguishes translations from
internal symmetries, as well as from all other spatial symmetries,
which involve point group elements, and generically act non-trivially
on internal degrees of freedom. For example, a $C_{4}$ rotation will
act non-trivially on spin-full fermions.

\paragraph*{Proof:}

We first consider the fermionic part of the Boltzmann weight, $\text{Det}\left(I+U_{\phi}\right)$.
The Hamiltonian $h_{\phi\left(\tau\right)}$ depends on the values
of $\phi$ at a single time slice $\tau$, and is therefore unaffected
by the twist in bosonic boundary conditions. It follows that $U_{\phi}$
is independent of the twist in bosonic boundary conditions. On the
other hand, the fermionic boundary conditions in \eqref{eq:5-1-0} correspond
to a change of the time evolution operator $U_{\phi}\mapsto T_{R}U_{\phi}$,
in analogy with \eqref{eq:7-2}. Since the design-principle $\mathcal{C}_{U}$ is assumed to be on-site homogeneous,
and $T_{R}=O^{\left(\sigma\right)}$ is a permutation operator, with
$\sigma:\left(x,y\right)\mapsto\left(x+\Theta\left(y\right),y\right)$,
we have $T_{R}U_{\phi}\in\mathcal{C}_{U}$, and $\text{Det}\left(I+T_{R}U_{\phi}\right)\geq0$.

\begin{figure}[t]
\begin{centering}
\includegraphics[width=0.3\columnwidth]{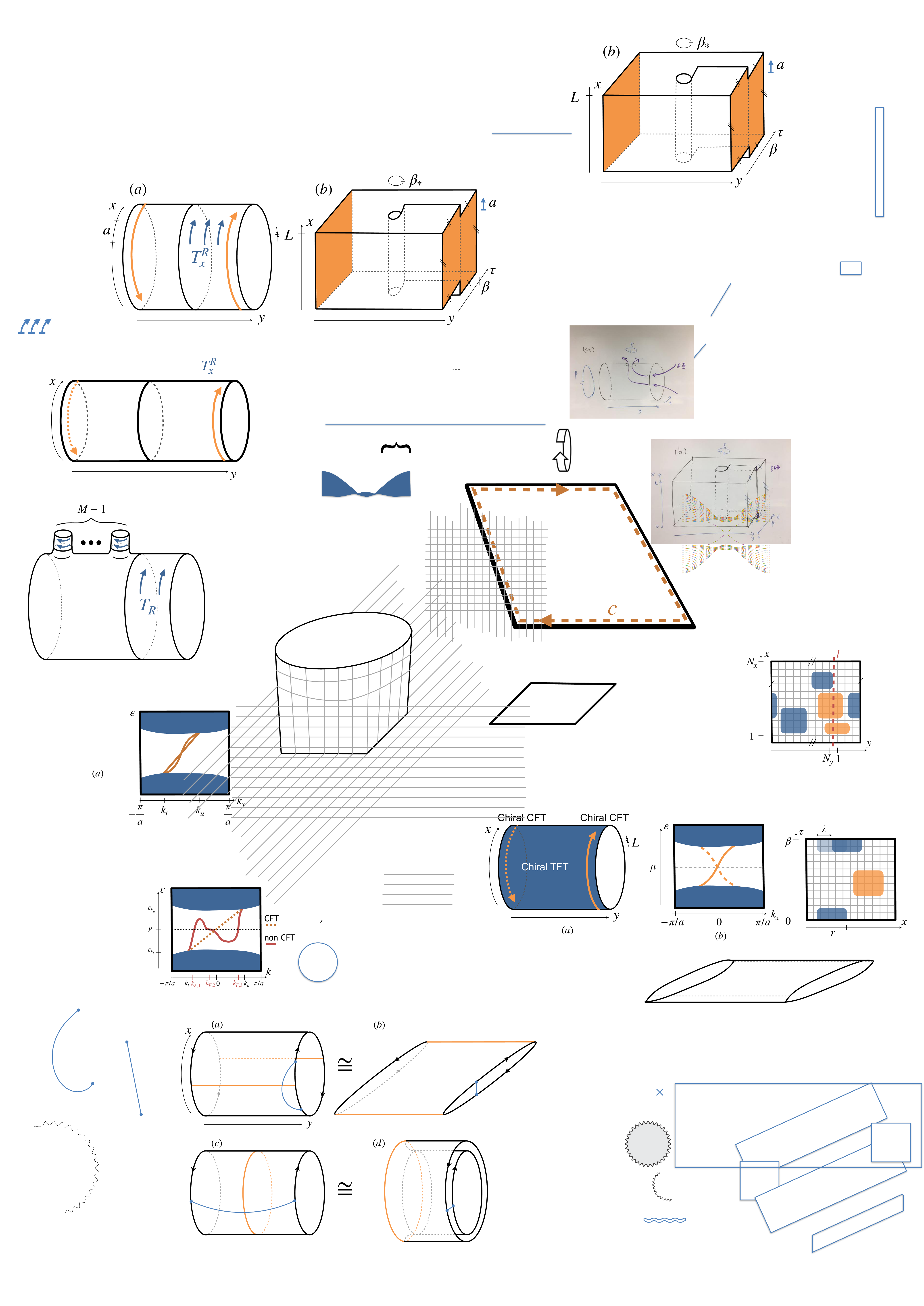}
\par\end{centering}
\caption{\label{fig:BoundaryConditions}Implementing the bosonic boundary conditions
\eqref{eq:5-1-0}. The lattice lies in the $x-\tau$ plane, at $y>0$
where the boundary conditions are non trivial. The orange area marks
the support, of diameter $r$, of a local term $S_{\phi;\tau,\mathbf{x}}$
which is unaffected by the boundary conditions. Blue areas correspond
to the support of a local term which is affected by the boundary conditions,
with pale blue indicating the un-twisted case $\lambda=0$.}
\end{figure}

Let us now consider the bosonic part of the Boltzmann weight $e^{-S_{\phi}}$,
where each of the local terms $S_{\phi;\tau,\mathbf{x}}$ is manifestly
real valued for all $\phi$. We  assume that the imaginary time
circle $S_{\beta}^{1}$ is discretized, such that the total number
of space-time points $\left(\tau,\mathbf{x}\right)=u\in U$ is finite.
Such a discretization is common in DQMC algorithms \citep{PhysRevD.24.2278,chandrasekharan2013fermion},
and the continuum case can be obtained by taking the appropriate limit.
The term $S_{\phi;\tau,\mathbf{x}}$ can then be written as a composition
$f\circ g_{V}$, where $f$ is a real valued function, and $g_{V}:\left(\phi_{u}\right)_{u\in U}\mapsto\left(\phi_{u}\right)_{u\in V}$
chooses the values of $\phi$ on which $S_{\phi;\tau,\mathbf{x}}$
depends, where $V\subset U$ is the support of $S_{\phi;\tau,\mathbf{x}}$.
The bosonic boundary conditions \eqref{eq:5-1-0} then amount to a modification
of the support $V\mapsto V_{\lambda}$, as depicted in Fig.\ref{fig:BoundaryConditions},
but not of the function $f$, which remains real valued. In particular,
for $\lambda=1$ we have $S_{\phi;\tau,\mathbf{x}}\mapsto\tilde{S}_{\phi;\tau,\mathbf{x}}=f\circ g_{V_{1}}$,
and $S_{\phi}\mapsto\tilde{S}_{\phi}=\sum_{\tau,\mathbf{x}}\tilde{S}_{\phi;\tau,\mathbf{x}}\in\mathbb{R}$.

 Combining the above conclusions for the bosonic and fermionic parts of $\tilde{p}\left(\phi\right)=e^{-\tilde{S}_{\phi}}\text{Det}\left(I+T_{R}U_{\phi}\right)$,
we find that $\tilde{p}\left(\phi\right)\geq0$ for all $\phi$.

\subsection{Excluding sign-free DQMC for chiral topological matter\label{sec:No-sign-free-DQMC}}

We are now ready to demonstrate the existence of an intrinsic sign
problem in chiral topological matter comprised of bosons \textit{and}
fermions, using the machinery of Sections \ref{sec:No-stoquastic-Hamiltonians}-\ref{sec:Determinantal-quantum-Monte}. 

Let $H$ be a gapped local fermion-boson Hamiltonian on the discrete
torus, which allows for a locally sign-free DQMC simulation. Unpacking
the definition, this means that $H'=UHU^{\dagger}$ has a local DQMC
representation which is term-wise sign-free due to an on-site homogeneous
design principle. As shown in Sec.\ref{sec:Sign-free-geometric-manipulation},
this implies that $\tilde{Z}':=\text{Tr}\left(T_{R}e^{-\beta H'}\right)$,
written on the cylinder, also has a local DQMC representation, obeying
a local and on-site design principle, and as a result, $\tilde{Z}'>0$.
Now, as shown in Sec.\ref{sec:No-stoquastic-Hamiltonians}, the positivity
of $\tilde{Z}'$ implies $\theta_{a}=e^{2\pi ic/24}$ for some anyon
$a$. We therefore have the fermionic version of Result \hyperref[Result 1]{1},
\begin{description}
\item [{Result$\;$1F\label{Result 1F}}] If a local fermion-boson Hamiltonian
$H$, which is in a chiral topological phase of matter, allows for
a locally sign-free DQMC simulation, then one of the corresponding
topological spins satisfies $\theta_{a}=e^{2\pi ic/24}$. Equivalently,
a chiral topological phase of matter where $e^{2\pi ic/24}$ is not
the topological spin of some anyon, i.e $e^{2\pi ic/24}\notin\left\{ \theta_{a}\right\} $,
admits no local fermion-boson Hamiltonians for which locally sign-free
DQMC simulation is possible.
\end{description}
As shown in Sec.\ref{sec:Spontaneous-chirality}, the positivity of
$\tilde{Z}'$ implies $\theta_{a}=e^{2\pi ic/24}$ for some anyon
$a$, even if chirality appears only spontaneously. We therefore obtain
the fermionic version of Result \hyperref[Result 2]{2},
\begin{description}
\item [{Result$\;$2F\label{Result 2F}}] If a local fermion-boson Hamiltonian
$H$, which is in a spontaneously-chiral topological phase of matter,
allows for a locally sign-free DQMC simulation, then one of the corresponding
topological spins satisfies $\theta_{a}=e^{2\pi ic/24}$. Equivalently,
a spontaneously-chiral topological phase of matter where $e^{2\pi ic/24}$
is not the topological spin of some anyon, i.e $e^{2\pi ic/24}\notin\left\{ \theta_{a}\right\} $,
admits no local fermion-boson Hamiltonians which allow for a locally
sign-free DQMC simulation.\textit{}
\end{description}

In stating these results, we do not restrict to fermionic phases,
because bosonic phases may admit a fermionic description, for which
DQMC is of interest. When a bosonic phase admits a fermionic description,
the bosonic field $\phi$ in Eq.\eqref{eq:2} will contain a $\mathbb{Z}_{2}$
gauge field that couples to the fermion parity $\left(-1\right)^{N_{f}}$
of $\psi$. An important series of examples is given by the non-abelian Kitaev
spin liquids, which admit a description in terms of gapped Majorana
fermions with an odd Chern number $\nu$, coupled to a $\mathbb{Z}_{2}$
gauge field \citep{kitaev2006anyons}. As described in Table \ref{tab:1}, the criterion $e^{2\pi ic/24}\notin\left\{ \theta_{a}\right\} $ applies to the Kitaev spin liquid, for all  $\nu\in2\mathbb{Z}-1$. Result \hyperref[Result 1]{1} then excludes
the possibility of locally stoquastic Hamiltonians for the microscopic
description in terms of spins, while Result \hyperref[Result 1F]{1F}
excludes the possibility of locally sign-free DQMC simulations in
the emergent fermionic description.

\subsection{Conjectures: beyond chiral matter\label{sec:Generalization-and-extension} }

In Sec.\ref{sec:No-stoquastic-Hamiltonians} and Appendices \ref{sec:Spontaneous-chirality}-\ref{sec:No-sign-free-DQMC} we established a criterion for the existence of intrinsic sign problems
in chiral topological matter: if $e^{2\pi ic/24}\notin\left\{ \theta_{a}\right\} $,
or equivalently $1\notin\text{Spec}\left(\mathbf{T}\right)$ (see
Result \hyperref[Result 1']{1'}), then an intrinsic sign problem exists.
Even if taken at face value, this criterion never applies to non-chiral bosonic topological phases, where
$c=0$, due to the vacuum topological spin $1\in\left\{ \theta_{a}\right\} $.
The same statement applies to all bosonic phases with $c\in24\mathbb{Z}$.
In this section we propose a refined criterion for intrinsic sign
problems in topological matter, which non-trivially applies to both
chiral \textit{and} non-chiral cases, and also unifies the results
of this section with those obtained by other means in a parallel work
\citep{PhysRevResearch.2.033515}.

Reference \citep{PhysRevLett.115.036802} proposed the 'universal
wave-function overlap' method for characterizing topological order
from any basis $\left\{ \ket i\right\} $ for the ground state subspace
of a local gapped Hamiltonian $H$ on the torus $X$. The method is
based on the conjecture 
\begin{align}
\bra i\mathbf{T}_{\text{m}}\ket j= & e^{-\alpha_{\mathbf{T}}A+o\left(A^{-1}\right)}\mathbf{T}_{ij},\label{eq:1-3}
\end{align}
where $A$ is the area of the torus, $\alpha_{\mathbf{T}}$ is a non-universal
complex number with non-negative real part, the microscopic Dehn-twist
operator $\mathbf{T}_{\text{m}}$ implements the Dehn twist $\left(x,y\right)\mapsto\left(x+y,y\right)$
on the Hilbert space, and $\mathbf{T}_{ij}$ are the entries of the
topological $\mathbf{T}$-matrix that characterizes the phase of $H$,
in the basis $\left\{ \ket i\right\} $. The same statement applies
to any element $\mathbf{M}$ of the mapping class group of the torus,
isomorphic to $SL\left(2,\mathbb{Z}\right)$, with $\mathbf{M}$ in
place of $\mathbf{T}$ in Eq.\eqref{eq:1-3}. The non-universal exponential
suppression of the overlap is expected because $\mathbf{M}_{\text{m}}$
will not generically map the ground-state subspace to itself, but
if $\mathbf{M}_{\text{m}}$ happens to be a symmetry of $H$, then
$\alpha_{\mathbf{M}}=0$ \citep{PhysRevB.85.235151,PhysRevLett.110.067208}.
Though we are not aware of a general analytic derivation of Eq.\eqref{eq:1-3},
it was verified analytically and numerically in a large number of
examples in Refs.\citep{PhysRevLett.115.036802,PhysRevB.91.125123,PhysRevB.91.075114,PhysRevB.90.205114,PhysRevB.95.235107},
for Hamiltonians in both chiral and non-chiral phases.  

Note the close analogy between Eq.\eqref{eq:1-3} and the momentum
polarization \eqref{eq:12-3-1}, where the microscopic Dehn-twist
$\mathbf{T}_{\text{m}}$ on the torus and the half translation $T_{R}$
on the cylinder play a similar role, and non-universal extensive contributions
are followed by sub-extensive universal data. To make this analogy
clearer, and make contact with the analysis of Sections \ref{sec:No-stoquastic-Hamiltonians}
and \ref{sec:No-sign-free-DQMC}, we consider the object $Z_{\mathbf{T}}=\text{Tr}\left(\mathbf{T}_{\text{m}}e^{-\beta H}\right)$,
which satisfies 
\begin{align}
Z_{\mathbf{T}}= & Ze^{-\alpha_{\mathbf{T}}A+o\left(A^{-1}\right)}\text{Tr}\left(\mathbf{T}\right),\label{eq:25-0}
\end{align}
and can be interpreted as either the (unnormalized) thermal expectation
value of $\mathbf{T}_{\text{m}}$, or the partition function on a
space-time twisted by $\mathbf{T}$, in analogy with Sec.\ref{subsec:Momentum-polarization}.
 Equation \eqref{eq:25-0} is valid for temperatures $\Delta E\ll1/\beta\ll E_{\text{g}}$,
much lower than the bulk gap $E_{\text{g}}$ and much higher than
any finite size splitting in the ground state-subspace, $\Delta E=o\left(A^{-1}\right)$.

Just like $T_{R}$, the operator $\mathbf{T}_{\text{m}}$ acts as
a permutation of the lattice sites. Therefore, following Section
\ref{sec:No-stoquastic-Hamiltonians} and Appendix \ref{sec:No-sign-free-DQMC},
if $H$ is either locally stoquastic, or admits a locally sign-free
DQMC simulation, then $\text{Tr}\left(\mathbf{T}\right)\ge0$. In
terms of $c$ and $\left\{ \theta_{a}\right\} $, this implies $e^{-2\pi ic/24}\sum_{a}\theta_{a}=\text{Tr}\left(\mathbf{T}\right)\geq0$,
where the sum runs over all topological spins. 

The last statement applies to both bosonic and fermionic Hamiltonians.
For bosonic Hamiltonians, it can be strengthened by means of the Frobenius-Perron
theorem. If $H'=UHU^{\dagger}$ is stoquastic in the on-site basis
$\ket s$, Hermitian, and has a degenerate ground state subspace,
then this subspace can be spanned by an orthonormal basis $\ket{i'}$
with positive entries in the on-site basis, $\braket s{i'}\geq0$,
see e.g Ref.\citep{PhysRevResearch.2.033515}. This implies that 
\begin{align}
0\leq\bra{i'}\mathbf{T}_{\text{m}}\ket{j'}= & e^{-\alpha_{\mathbf{T}}'A+o\left(A^{-1}\right)}\mathbf{T}_{i'j'},\label{eq:2-3-1}
\end{align}
where $\alpha_{\mathbf{T}}'$ is generally different from $\alpha_{\mathbf{T}}$,
but the matrix $\mathbf{T}_{i'j'}$ has the same spectrum as $\mathbf{T}_{ij}$
in Eq.\eqref{eq:1-3}. This is a stronger form of \eqref{eq:25},
which implies $\mathbf{T}_{i'j'}\geq0$. Since $\mathbf{T}_{i'j'}$
is also unitary, it is a permutation matrix, $\mathbf{T}_{i'j'}=\delta_{i',\sigma\left(j'\right)}$
for some $\sigma\in S_{N}$, where $N$ is the number of ground states.
In turn, this implies that the spectrum of $\mathbf{T}$ is a disjoint
union of complete sets of roots of unity,
\begin{align}
\left\{ \theta_{a}e^{-2\pi ic/24}\right\} _{a=1}^{N} & =\text{Spec}\left(\mathbf{T}\right)=\bigcup_{k=1}^{K}R_{n_{k}},\label{eq:27-1}
\end{align}
where $R_{n_{k}}$ is the set of $n_{k}$th roots of unity, $n_{k},K\in\mathbb{N}$,
and $\sum_{k=1}^{K}n_{k}=N$.  Therefore, 
\begin{description}
\item [{Conjecture$\;$1}] A bosonic topological phase of matter where
$\left\{ \theta_{a}e^{-2\pi ic/24}\right\} $ is not a disjoint union
of complete sets of roots of unity, admits no local Hamiltonians which
are locally stoquastic.
\end{description}
In particular, this implies an intrinsic sign problem whenever $1\notin\left\{ \theta_{a}e^{-2\pi ic/24}\right\} $,
thus generalizing Result \hyperref[Result 1]{1}. Moreover, the above
statement applies non-trivially to phases with $c\in24\mathbb{Z}$.
In particular, for non-chiral phases, where $c=0$, it reduces to
the result established in Ref.\citep{PhysRevResearch.2.033515}, thus generalizing it
as well. The simplest example for a non-chiral phase
with an intrinsic sign problem is the doubled semion phase, where
$\left\{ \theta_{a}\right\} =\left\{ 1,i,-i,1\right\} $ \citep{PhysRevB.71.045110}. 

Though we are currently unaware of an analog of the Frobenius-Perron
theorem that applies to DQMC, we expect that an analogous result can
be established for fermionic Hamiltonians.
\begin{description}
\item [{Conjecture$\;$1F}] A topological phase of matter where $\left\{ \theta_{a}e^{-2\pi ic/24}\right\} $
is not a complete set of roots of unity, admits no local fermion-boson
Hamiltonians for which locally sign-free DQMC simulation is possible.
\end{description}
The above conjectures suggest a substantial improvement over the criterion
$e^{2\pi ic/24}\notin\left\{ \theta_{a}\right\} $. To demonstrate
this, we go back to the 
$1/q$ Laughlin phases and $SU\left(2\right)_{k}$
Chern-Simons theories considered in Table \ref{tab:1}. We find a conjectured
intrinsic sign problem in \textit{all} of the first one-thousand bosonic
Laughlin phases ($q$ even), fermionic Laughlin phases ($q$ odd),
and $SU\left(2\right)_{k}$ Chern-Simons theories. In particular, we note that the prototypical $1/3$ Laughlin phase is not captured by the criterion $e^{2\pi ic/24}\notin\left\{ \theta_{a}\right\}$, but is conjectured to be intrinsically sign-problematic. 

\subsection{Discussion\label{subsec:Discussion}}

In this section we established the existence of intrinsic sign problems
in a broad class of chiral topological phases, namely those where
$e^{2\pi ic/24}$ does not happen to be the topological spin of an
anyon. Since these intrinsic sign problems persist even when chirality appears spontaneously, they are
rooted in the macroscopic and observable data $c$, $\left\{ \theta_{a}\right\} $,
rather than the microscopic absence (or presence) of time reversal
symmetry. Going beyond the simple setting of stoquastic Hamiltonians, we provided
the first treatment of intrinsic sign problems in fermionic systems.
In particular, we constructed a general framework which describes
all DQMC algorithms and fermionic design principles that we are aware
of, including the state of art design principles \citep{wang2015split,wei2016majorana,li2016majorana,wei2017semigroup}
which are only beginning to be used by practitioners. Owing to its
generality, it is likely that our framework will apply to additional
design principles which have not yet been discovered, insofar as they
are applied locally. We also presented conjectures that strengthen our results, and unify
them with those obtained in Refs.\citep{hastings2016quantum,PhysRevResearch.2.033515},
under a single criterion in terms of $c$ and $\left\{ \theta_{a}\right\} $.
These conjectures also imply intrinsic sign problems in many topological
phases not covered by existing results. 

Conceptually, our results  show that the sign problem
is not \textit{only} a statement of computational complexity: it is,
in fact, intimately connected with the physically observable properties
of quantum matter. Such a connection has long been heuristically appreciated
by QMC practitioners, and is placed on a firm and quantitative footing
by the discovery of intrinsic sign problems. Despite the progress made here, our understanding of intrinsic sign
problems is still in its infancy, and many open questions remain:

\paragraph*{Quantum computation and intrinsic sign problems}

Intrinsic sign-problems relate the physics of topological phases to
their computational complexity, in analogy with the classification
of topological phases which enable universal quantum computation \citep{Freedman:2002aa,nayak2008non}.
As we have seen, many phases of matter that are known to be universal
for quantum computation are also intrinsically sign-problematic, supporting
the paradigm of 'quantum advantage' or 'quantum supremacy' \citep{Arute:2019aa}.
Determining whether intrinsic sign-problems appear in \textit{all}
phases of matter which are universal for quantum computation is an
interesting open problem. Additionally, we identified intrinsic sign problems in many topological phases which are not universal for quantum computation. The intermediate complexity of such phases between classical and quantum computation is another interesting direction for future work. 

\paragraph*{Unconventional superconductivity and intrinsic sign problems}

As described in the introduction, a major motivation for the study of
intrinsic sign problems comes from long standing open problems in
fermionic many-body systems, the nature of high temperature superconductivity
 in particular. It is currently believed that many high temperature
superconductors, and the associated repulsive Hubbard models, are
 \textit{non-chiral} $d$-wave superconductors \citep{kantian2019understanding,berg2019monte},
in which we did not identify an intrinsic sign problem. The optimistic possibility that the sign problem can in fact
be cured in repulsive Hubbard models is therefore left open, though this has not yet been
accomplished in the relevant regime of parameters, away from half filling, despite intense research efforts 
\citep{PhysRevX.5.041041}. Nevertheless, the state of the art DMRG
results of Ref.\citep{kantian2019understanding} do not exclude the
possibility of a \textit{chiral} $d$-wave superconductor ($\ell=\pm 2$ in Table \ref{tab:1}). In this case we do find an intrinsic sign problem, which would account for the notorious sign problems observed in repulsive Hubbard models. More speculatively, it is possible that the mere proximity of repulsive Hubbard models to a chiral $d$-wave phase stands behind their notorious sign problems. The possible effect of an intrinsic sign problem in a given phase on the larger phase diagram was recently studied in Ref.\citep{zhang2020non}. There is also evidence
for chiral $d$-wave superconductivity in doped graphene and related
materials  \citep{PhysRevB.84.121410,Black_Schaffer_2014}, and our results therefore suggest the impossibility of sign-free QMC simulations of these. We believe that the study
of intrinsic sign problems in the context of unconventional superconductivity is a promising direction for future work. 

\paragraph*{Non-locality as a possible route to sign-free QMC }

The intrinsic sign problems identified in this thesis add to existing
evidence for the complexity of chiral topological phases - these do
not admit local commuting projector Hamiltonians \citep{PhysRevB.89.195130,potter2015protection,PhysRevB.98.165104,kapustin2019thermal},
nor do they admit local Hamiltonians with a PEPS state as an exact
ground state \citep{PhysRevLett.111.236805,PhysRevB.90.115133,PhysRevB.92.205307,PhysRevB.98.184409}. Nevertheless, relaxing the locality requirement does lead to positive
results for the simulation of chiral topological matter using commuting
projectors or PEPS. First, commuting projector Hamiltonians can be
obtained if the local bosonic or fermionic degrees of freedom are
replaced by anyonic (and therefore non-local) excitations of an underlying
chiral topological phase \citep{PhysRevB.97.245144}. Second, chiral
topological Hamiltonians can have a PEPS ground state if they include
interactions (or hopping amplitudes) that slowly decay as a power-law
with distance. One may therefore hope that sign-free QMC simulations of chiral topological
matter can also be performed if the locality requirements made in
Sec.\ref{sec:No-stoquastic-Hamiltonians} are similarly relaxed.
Do such 'weakly-local' sign-free models exist?

\paragraph*{Easing intrinsic sign problems}

In this section we proved the existence of an intrinsic sign problem
in chiral topological phases of matter, but we did not quantify the
\textit{severity} of this sign problem, which is an important concept
in both practical applications and theory of QMC. The severity of
a sign problem is quantified by the smallness of the average sign of the QMC weights $p$ with respect to the distribution $\left|p\right|$, i.e $\left\langle \text{sgn}\right\rangle :=\sum p/\sum\left|p\right|$.  Since $\left\langle \text{sgn}\right\rangle $ can be viewed as the
ratio of two partition functions, it  obeys the generic scaling 
$\left\langle \text{sgn}\right\rangle \sim e^{-\Delta\beta N}$, with
$\Delta\geq0$, as $\beta N\rightarrow\infty$ \citep{troyer2005computational,hangleiter2019easing}.
A sign problem exists when $\Delta>0$, in which case QMC simulations
require exponential computational resources, and this is what the
intrinsic sign problem we identified implies for 'most' chiral topological
phases of matter. From the point of view of computational complexity,
all that matters is whether $\Delta=0$ or $\Delta>0$, but for practical
applications the value of $\Delta$ is very important, see e.g \citep{PhysRevB.84.121410}.
One may hope for a possible refinement of our results that provides
a lower bound $\Delta_{0}>0$ for $\Delta$, but since we have studied
\textit{topological} phases of matter, we view this as unlikely. It
may therefore be possible to obtain fine-tuned models and QMC methods
that lead to a $\Delta$ small enough to be \textit{practically} useful.
More generally, it may be possible to search for such models and methods algorithmically,
thus \textit{easing} the intrinsic sign problem \citep{hangleiter2019easing,torlai2019wavefunction,PhysRevD.97.094510,alex2020complex}. We also note that the results presented in this thesis do not exclude approaches to the sign-problem based on a modified or constrained Monte Carlo sampling \citep{PhysRevLett.83.3116,fixed-node,constrained-path}, as well as machine-learning aided QMC \citep{ML+QMC}, and infinite-volume diagrammatic QMC \citep{PhysRevLett.119.045701}.

\paragraph*{Possible extensions }

The chiral central charge only appears modulo 24 in our results. Nevertheless,
for full value of $c$ is physically meaningful, as reviewed in the
introduction. Does an intrinsic sign problem exist in all phases with
$c\neq0$? The results of Ref.\citep{ringel2017quantized} strongly
suggest this. 

The arguments of Appendix \ref{sec:Generalization-and-extension} apply
equally well to any element of the modular group, rather than just
the topological $\mathbf{T}$-matrix, implying that the spectrum of
all elements decomposes into full sets of roots of unity. Does this
imply a tighter constraint on the TFT data than conjectured in Appendix \ref{sec:Generalization-and-extension}?
Moreover, the 'universal wave-function overlap' conjecture, on which
Appendix \ref{sec:Generalization-and-extension} relies, applies also to
space-time dimensions $D>2+1$, which suggests intrinsic sign problems
in these space-time dimensions, including the beloved $D=3+1$. 


Another promising direction involves symmetry protected or enriched topological phases.  In particular, the sign problem was cured in a number of bosonic SPT Hamiltonians \citep{PhysRevB.95.174418,PhysRevB.85.045114, PhysRevB.86.045106,gazit2016bosonic}, and all SPT ground states can, by definition, be made non-negative in a local basis. Nevertheless, a `symmetry protected' intrinsic sign problem, where \textit{symmetric} local bases are excluded,  was recently discovered \citep{ellison2020symmetryprotected}. Such  constraints  may  be more  useful  for designing sign-free models than the stronger intrinsic sign problems discussed in this thesis. 

\pagebreak{}

\section{Outlook \label{sec:Discussion-and-outlook}}

In this thesis we studied the geometric physics of chiral topological
phases, and related this physics to the complexity of simulating such
phases using quantum Monte Carlo algorithms. The analysis in Sec.\ref{sec:Main-section-1:}-\ref{sec:Main-section-2:}
was focused on chiral superfluids and superconductors, and revealed
an intricate interplay of symmetry breaking, topology, and geometry.
Though we stressed the Higgs mode as an emergent geometry in the relativistic
analysis of Sec.\ref{sec:Main-section-1:}, the fuller non-relativistic
treatment in Sec.\ref{sec:Main-section-2:} was carried out at energy
scales below the Higgs mass, where the emergent geometry simply follows
the background geometry (or strain). We believe that there is beautiful
physics to be revealed at higher energy scales, where the dynamics
of the Higgs mode will be described by a non-relativistic, parity
odd, and in part topological, quantum geometry similar to the Girvin-MacDonald-Platzman
(GMP) mode in fractional quantum Hall states \citep{girvin1986magneto,haldane2009hall,haldane2011geometrical,you2014theory,gromov2017bimetric}.
In particular, this physics should be important near a nematic phase
transition where the spin-2 Higgs mode becomes light, which can be
tuned by attractive Landau interactions \citep{hsiao2018universal}.
We also believe that, through `composite fermions' \citep{read2000paired,PhysRevX.5.031027,Son_2018},
this will lead to a new description of the GMP mode in paired quantum
Hall states, including the illusive particle-hole invariant Pfaffian,
and in analogy with the treatment of the GMP mode in Jain states close
to half filling \citep{PhysRevB.48.17368,PhysRevLett.117.216403,PhysRevB.97.195103,PhysRevB.97.195314}.

Regarding intrinsic sign problems, we believe that the results obtained
in Sec.\ref{sec:Main-section-3:}, as well as in Refs.\citep{hastings2016quantum,ringel2017quantized,PhysRevResearch.2.033515,ellison2020symmetryprotected},
represent the tip of an iceberg. Concrete extensions of these results
were proposed in Sec.\ref{subsec:Discussion}. Beyond these, are there
intrinsically sign-problematic phases which are not gapped, not topological,
or both? Does the physics of high temperature superconductivity, or
of dense nuclear matter, imply an intrinsic sign problem, thus accounting
for the persistent sign problems observed by QMC practitioners in
relevant models? Taking a broader perspective, intrinsic sign problems
form a bridge between the notions of phases of matter and computational
complexity. This should be contrasted with most results in quantum
complexity \citep{TCS-066,RevModPhys.90.015002}, which are established
for classes of Hamiltonians defined by microscopic conditions such
as locality, non-frustration, and stoquasticity, as well as energy
gap assumptions, or for specific canonical models, but usually not
for phases of matter. In fact, refining statements in quantum complexity
to the level of phases was the original motivation for the study of
intrinsic sign problems \citep{hastings2016quantum}. Additional statements
regarding the complexity of phases are given by the classification
of topological phases which enable universal quantum computation by
anyon braiding \citep{Freedman:2002aa,nayak2008non}, and the proof
that whenever the area law for entanglement entropy holds for a gapped
Hamiltonian, it holds in its entire phase \citep{PhysRevLett.111.170501}.
We find this theme to be promising for both the study of many-body
quantum systems, and their use as computational devices.

\newpage
\appendix
\addcontentsline{toc}{section}{Appendix} 
\part{Appendix} 
\parttoc 


\addtocontents{toc}{\protect\setcounter{tocdepth}{0}}


\pagebreak{}

\section{Boundary fermions and gravitational anomaly \label{sec:Boundary-fermions-and}}

It is well known that the $p$-wave SC has localized degrees of freedom
on curves in space where the Chern number $\nu$ jumps, due to boundaries,
or domain walls in $\Delta$ or $\mu$, which at low energies are
$D=1+1$ chiral Majorana spinors \cite{read2000paired}. In this section
we derive the action for the boundary spinor in the presence of a
space-time dependent order parameter, and describe its gravitational
anomaly and corresponding anomaly inflow. 

We start by deriving the boundary action in the geometric description
in Appendix \ref{subsec:Boundary-states-in}, then review the relevant
facts regarding the boundary gravitational anomaly within the gravitational
description in Appendix \ref{subsec:Boundary-gravitational-anomaly},
and finally translate the results back to the $p$-wave SC language
in Appendix \ref{subsec:Implication-for-the}. 

The form of the boundary action in both the geometric description
\eqref{eq:87} and in the $p$-wave SC language \eqref{eq:137} is
not surprising, and within the geometric description the gravitational
anomaly and anomaly inflow are well known. It is the implication of
gravitational anomaly and anomaly inflow for the $p$-wave SC, through
the emergent geometry described in sections \ref{sec:Emergent-Riemann-Cartan-geometry}
and \ref{sec:Symmetries,-currents,-and}, which is the result of this
section.

\subsection{Boundary fermions in a product geometry\label{subsec:Boundary-states-in}}

We take the space time manifold to be $\mathbb{R}\times\mathbb{R}^{2}$,
and assume that the vielbein has a product form with respect to the
spatial coordinate $y$, 
\begin{align}
  e^{A}=e_{\;\alpha}^{A}\left(x^{\alpha}\right)\text{d}x^{\alpha},\;e^{y}=o\text{d}y,\label{eq:81}
\end{align}
 where $\alpha,\beta,\dots\in\left\{ t,x\right\} $ and $A,B,\dots\in\left\{ 0,1\right\} $
(unlike the notation of section \ref{sec:Emergent-Riemann-Cartan-geometry}
where $A,B,\dots\in\left\{ 1,2\right\} $). To account for the orientation
$o=\text{sgn}\left(\text{det}e_{\;\mu}^{a}\right)$ of the vielbein
explicitly, we assumed $e_{\;\alpha}^{A}$ has a positive orientation,
and wrote $e^{y}=o\text{d}y$. To be concrete we take $o=1$ for now.
It follows that the metric also has the product form $\text{d}s^{2}=g_{\alpha\beta}\left(x^{\alpha}\right)\text{d}x^{\alpha}\text{d}x^{\beta}-\text{d}y^{2}$
where $g_{\alpha\beta}=e_{\;\alpha}^{A}\eta_{AB}e_{\beta}^{B}$. The
form of the vielbein implies that the LC spin connection only has
the nonzero components $\tilde{\omega}_{AB\alpha}$, which only depend
on $t,x$. We also assume that the spin connection only has nonzero
components $\omega_{AB\alpha}$ and depends only on $t,x$. Under
these assumptions $c=C_{abc}\varepsilon^{abc}=0$, and therefore torsion
simply drops out from the action, as can be seen from the form \eqref{43-1}.
This is a result of the low dimensionality of the problem. We further
assume that the mass has the form of a flat domain wall in the $y$
direction. By this we mean $m=m\left(y\right)$ with boundary conditions
$m\rightarrow\pm m_{0}$ as $y\rightarrow\pm\infty$, and $m_{0}\neq0$,
which corresponds to an interface between two distinct phases. To
be concrete we take $m_{0}>0$ for now. $S_{\text{RC}}$ then takes
the form 
\begin{align}
  S_{\text{RC}}=\frac{1}{2}\int\mbox{d}^{3}x\left|e\right|\overline{\chi}\left[ie_{A}^{\;\alpha}\gamma^{A}\tilde{D}_{\alpha}+i\gamma^{2}\partial_{y}-m\left(y\right)\right]\chi.
\end{align}
This separable form implies the decomposition described in \cite{chandrasekharan1994anomaly,fosco1999dirac},
which we now apply to the present situation. Defining $a=\partial_{y}-m\left(y\right),\;a^{\dagger}=-\partial_{y}-m\left(y\right),\;P_{\pm}=\frac{1}{2}\left(1\pm i\gamma^{2}\right)$,
the action takes the form 
\begin{align}
  S_{\text{RC}}=\frac{1}{2}\int\mbox{d}^{3}x\left|e\right|\overline{\chi}\left[ie_{A}^{\;\alpha}\gamma^{A}\tilde{D}_{\alpha}+aP_{+}+a^{\dagger}P_{-}\right]\chi.\label{eq:73}
\end{align}
The operators $h_{+}=a^{\dagger}a$ and $h_{-}=aa^{\dagger}$ are
hermitian and non negative. The positive parts of their spectrum coincide.
We denote the positive eigenvalues by $\lambda^{2}>0$, including
both the discrete and continuous parts of the spectrum, with the corresponding
eigenfunctions $\phi_{\lambda,\pm}$ satisfying $h_{\pm}\phi_{\lambda,\pm}=\lambda^{2}\phi_{\lambda,\pm}$.
These eigenfunctions of $h_{\pm}$ are related by $\phi_{\lambda,+}=\frac{1}{\lambda}a^{\dagger}\phi_{\lambda,-},\;\phi_{\lambda,-}=\frac{1}{\lambda}a\phi_{\lambda,+}$,
where the sign chosen for $\lambda$ is arbitrary, and for concreteness
we take $\lambda>0$. Each set of eigenfunctions can be assumed to
be orthonormal $\int_{-\infty}^{\infty}\text{d}y\phi_{\lambda,\pm}^{*}\phi_{\lambda',\pm}=\delta_{\lambda\lambda'}$.
Apart from the positive part of the spectrum, there can also be a
unique eigenfunction with eigenvalue zero, a zero mode, for $h_{+}$
or $h_{-}$ but not both. The only candidates are $\phi_{0,\pm}\left(y\right)\propto e^{\pm\int_{0}^{y}m\left(s\right)\text{d}s}$,
and a zero mode exists when one of these functions is normalizable.
With our choice of boundary conditions for $m$, only $\phi_{0,-}$
is normalizable. In terms of these eigenfunctions, the natural orthogonal
 decomposition of the spinor $\chi$ is 
\begin{align}
  \chi\left(x,y,t\right)=&P_{+}\chi\left(x,y,t\right)+P_{-}\chi\left(x,y,t\right)\nonumber\\
  =&\begin{subarray}{c}
\sum\end{subarray}_{\lambda>0}\left[\chi_{\lambda,+}\left(x,t\right)\phi_{\lambda,+}\left(y\right)+\chi_{\lambda,-}\left(x,t\right)\phi_{\lambda,-}\left(y\right)\right]\nonumber\\
&+\chi_{0,-}\left(x,t\right)\phi_{0,-}\left(y\right),\label{eq:83}
\end{align}
where $\chi_{\lambda,\pm}$ are spinors of definite chirality, $P_{\pm}\chi_{\lambda,\pm}=\chi_{\lambda,\pm}$.
Inserting this decomposition into \eqref{eq:73} we obtain 
\begin{align}
 S_{\text{RC}}=\frac{1}{2}&\int\mbox{d}^{2}x\left|e\right|\overline{\chi_{0,-}}ie_{A}^{\;\alpha}\gamma^{A}\tilde{D}_{\alpha}\chi_{0,-}\label{eq:84}\\
 &+\sum_{\lambda>0}\frac{1}{2}\int\mbox{d}^{2}x\left|e\right|\overline{\chi_{\lambda}}\left[ie_{A}^{\;\alpha}\gamma^{A}\tilde{D}_{\alpha}+\lambda\right]\chi_{\lambda},\nonumber
\end{align}
where $\chi_{\lambda}=\chi_{\lambda,-}+\chi_{\lambda,+}$. Thus the
action splits into an infinite sum of actions for independent $D=1+1$
spinors, coupled to RC geometry, which in the $D=1+1$ case is the
same as the coupling to Riemannian geometry. The spinor corresponding
to the zero mode is chiral, massless, and exponentially localized
on the domain wall as can be seen from the expression $\phi_{0,-}\left(y\right)\propto e^{-\int_{0}^{y}m\left(s\right)\text{d}s}$.
It represents the robust boundary state that exists between two distinct
topological phases. The chiral boundary spinor exhibits a gravitational
anomaly, which we describe in the following.

All other spinors are non chiral and massive with masses $\lambda\neq0$.
It is useful to think of the eigenvalue problems $h_{\pm}\phi_{\lambda,\pm}=\left(-\partial_{y}^{2}+m^{2}\left(y\right)\pm m'\left(y\right)\right)\phi_{\lambda,\pm}$
as one dimensional time independent Schrodinger problems to understand
the eigenvalues $\lambda$ and eigenfunctions $\phi_{\lambda,\pm}$
\cite{chandrasekharan1994anomaly,fosco1999dirac}.  Almost all of
the massive spinors correspond to delocalized bulk degrees of freedom,
with the functions $\phi_{\lambda,\pm}\left(y\right)$ corresponding
to ``scattering states'' of the ``Hamiltonians'' $h_{\pm}$. Additionally,
there can be a finite number of ``bound states'' $\phi_{\lambda,\pm}$,
in which case $\chi_{\lambda}$ corresponds to an additional non-chiral
boundary state, which is not robust, and can always be removed by
making the domain wall narrower, or the bulk masses $\pm m_{0}$ smaller. 

Since the action splits into a sum of $D=1+1$ fermionic actions and
the decomposition \eqref{eq:83} is orthogonal, the effective action
also splits into a sum 
\begin{align}
W_{\text{RC}}\left[e,\omega\right]=W_{\text{R}}^{-}\left[e\right]+\sum_{\lambda>0}W_{\text{R}}\left[e;\lambda\right],
\end{align}
where $W_{\text{R}}\left[e;\lambda\right]$ is the effective action
obtained by integrating over a $D=1+1$ Majorana spinor with mass
$\lambda\neq0$ coupled to Riemannian geometry, and $W_{\text{R}}^{\pm}\left[e\right]$
is the effective action obtained by integrating over a $D=1+1$ massless
chiral Majorana spinor coupled to Riemannian geometry, with chirality
$\pm$. Above we assumed $m\left(\pm\infty\right)=\pm m_{0}$ with
$m_{0}>0$ and $o=1$. Generalizing slightly,  the net chirality of
the boundary spinors is given by 
\begin{align}
  C=\frac{o}{2}\text{sgn}\left(m\left(\infty\right)\right)-\frac{o}{2}\text{sgn}\left(m\left(-\infty\right)\right).\label{eq:101}
\end{align}

The action $S_{\text{R}}^{\pm}=\frac{1}{2}\int\mbox{d}^{2}x\left|e\right|\overline{\chi_{0,\pm}}ie_{A}^{\;\alpha}\gamma^{A}\tilde{D}_{\alpha}\chi_{0,\pm}$
for a single chiral Majorana spinor coupled to Riemannian geometry
can be simplified by using a \textit{Majorana representation} for
the Clifford algebra, as described in appendix
\ref{subsec:Charge-conjugation-(Appendix)}. In the Majorana representation
$\chi_{0,\pm}=\xi v_{\pm}$ where $\xi$ is a single-component \textit{real}
Grassmann field and $v_{\pm}$ are the normalized eigenvectors of
$i\gamma^{2}$, $\left(i\gamma^{2}\right)v_{\pm}=\pm v_{\pm}$. The
action $S_{\text{R}}^{\pm}$ then reduces to 
\begin{align}
 S_{\text{R}}^{\pm}=\frac{i}{2}\int\mbox{d}^{2}x\left|e\right|\xi e_{\mp}^{\;\alpha}\partial_{\alpha}\xi,\label{eq:87}
\end{align}
where $e_{\mp}^{\;\alpha}=e_{0}^{\;\alpha}\mp e_{1}^{\;\alpha}$.

\subsection{Boundary gravitational anomaly and anomaly inflow\label{subsec:Boundary-gravitational-anomaly}}

The chiral boundary spinor does not couple to the spin connection
$\omega$, and therefore does not distinguish the RC background from
a Riemannian background described by the vielbein. This can be seen
by examining the $1+1$ dimensional version of the conservation laws
described in section \ref{subsec:Currents,-symmetries,-and} for the
energy-momentum tensor $\mathsf{j}_{\;A}^{\alpha}=\frac{1}{\left|e\right|}\frac{\delta S_{\text{R}}^{\pm}}{\delta e_{\;\alpha}^{A}}$
and spin current $\mathsf{j}^{AB\alpha}=\frac{1}{\left|e\right|}\frac{\delta S_{\text{R}}^{\pm}}{\delta\omega_{AB\alpha}}$
of the boundary spinor. As in section \ref{subsec:Currents,-symmetries,-and},
these follow from the $Diff$ and Lorentz gauge symmetries of the
``classical'' action $S_{\text{R}}^{\pm}$. Since the boundary
fermion does not couple to $\omega$, its spin current vanishes, $\mathsf{j}^{AB\alpha}=0$.
Therefore \eqref{eq:57} takes the form 
\begin{align}
 \mathsf{j}^{[AB]}=0,
\end{align}
expressing the symmetry of the boundary energy-momentum tensor, as
in Riemannian geometry. The energy-momentum conservation law \eqref{eq:71}
then takes the form $\nabla_{\alpha}\mathsf{j}_{\;\beta}^{\alpha}-\mathsf{j}_{\;\beta}^{\alpha}T_{\gamma\alpha}^{\gamma}=T_{\beta\alpha}^{A}\mathsf{j}_{\;A}^{\alpha}$,
which reduces to 
\begin{align}
  \tilde{\nabla}_{\alpha}\mathsf{j}^{\alpha\beta}=C_{AB}^{\;\;\;\;\beta}\mathsf{j}^{[AB]}=0,
\end{align}
where $\tilde{\nabla}$ is the LC covariant derivative. This is the
energy-momentum conservation law in a background Riemannian geometry.
The energy-momentum tensor is given explicitly by 
\begin{align}
 \mathsf{j}_{\;\beta}^{\alpha}=-\frac{i}{2}e_{\mp}^{\;\alpha}\xi\partial_{\beta}\xi,
\end{align}
up to a term that vanishes on the equation of motion for $\xi$, $ie_{\mp}^{\;\alpha}\partial_{\alpha}\xi+\frac{i}{2}\xi\left|e\right|^{-1}\partial_{\alpha}\left(\left|e\right|e_{\mp}^{\;\alpha}\right)=0$,
which can also be written in a manifestly covariant form. One can
verify that $\mathsf{j}_{\;\beta}^{\alpha}$ is conserved, symmetric,
and traceless on the equation of motion. 

Chiral Majorana fermions in $D=1+1$ coupled to Riemannian geometry
exhibit a gravitational anomaly, which implies that while the ``classical''
action $S_{\text{R}}^{\pm}$ is invariant under both $Diff$ and Lorentz
gauge transformations, the corresponding effective action $W_{\text{R}}^{\pm}$
is not \footnote{$W_{\text{R}}^{\pm}$ is an example for the nonlocal boundary functional
$F$ discussed in section \ref{subsec:Gauge-symmetry-of}.}. A physical manifestation of this phenomena is that the ``classical''
conservation law $\tilde{\nabla}_{\alpha}\mathsf{j}_{\;\beta}^{\alpha}=0$
is violated quantum mechanically, $\tilde{\nabla}_{\alpha}\left\langle \mathsf{j}_{\;\beta}^{\alpha}\right\rangle \neq0$.
The anomaly can be calculated by various techniques \cite{bertlmann2000anomalies,bastianelli2006path},
the simplest of which is the calculation of a single Feynman graph,
as was originally done in \cite{alvarez1984gravitational} for the
two dimensional Weyl spinor, and is reviewed in \cite{bastianelli2006path}
part 5.1.2 for the case of a Majorana-Weyl spinor relevant here. The gravitational anomaly\footnote{There are a few ambiguities in describing what the gravitational anomaly
is from an intrinsic boundary point of view. First, there is the issue
of covariant versus consistent anomalies which also exists in gauge
anomalies \cite{bertlmann2000anomalies}. See also \cite{stone2012gravitational}
and part 2 of \cite{jensen2013thermodynamics} for a short review.
Then, for the consistent gravitational anomaly, there is the issue
of Lorentz anomalies versus Einstein ($Diff$) anomalies where one
can obtain an effective boundary action that is invariant under local
Lorentz transformations but not under $Diff$, or vice versa \cite{bertlmann2000anomalies}.
It is also useful to discuss linear combinations of the Einstein and
Lorentz anomalies, related to the symmetry of the effective action
under the Lorentz-covariant $Diff$ action \eqref{eq:62}, see part
6.3 of \cite{bastianelli2006path}. All of these ambiguities are resolved
when calculating the boundary energy-momentum tensor within the anomaly
inflow mechanism: the bulk gCS term contributes to the boundary energy-momentum
tensor, assuring it is symmetric and covariant, so that the physically
relevant gravitational anomaly is the covariant Einstein anomaly \cite{stone2012gravitational},
which is what we refer to here as ``the gravitational anomaly''.} is given by \cite{bertlmann2000anomalies}
\begin{align}
 \tilde{\nabla}_{\alpha}\left\langle \mathsf{j}^{\alpha\beta}\right\rangle =\frac{\nu/2}{96\pi}\frac{1}{\left|e\right|}\varepsilon^{y\beta\alpha}\partial_{\alpha}\tilde{\mathcal{R}}.\label{eq:88}
\end{align}

The physical interpretation of the anomaly, within the gravitational
theory, is obtained by identifying the right hand side with the energy-momentum
inflow from the bulk, \eqref{eq:111}. Then \eqref{eq:88} can be
written as
\begin{align}
  \tilde{\nabla}_{\alpha}\left\langle \mathsf{j}^{\alpha\beta}\right\rangle =\left\langle \mathsf{J}^{y\beta}\right\rangle ,
\end{align}
which, together with the bulk conservation equation \eqref{eq:71},
is just the statement of energy-momentum conservation for a system
with a boundary. This is the anomaly inflow mechanism, recasting what
appears to be energy-momentum non-conservation in a $D=1+1$ system,
as energy-momentum conservation in a $D=2+1$ system with a boundary. 

\subsection{Implication for the $p$-wave SC\label{subsec:Implication-for-the}}

Let us now apply the above to the $p$-wave SC with a flat domain
wall in the chemical potential, $\mu\left(y\right)$, which physically
represents a fixed chemical potential and an additional $y$-dependent
electric potential. To obtain an emergent geometry which is a product
geometry, we take the order parameter to be of the form $\Delta=\left(\Delta^{x},\Delta^{y}\right)=\Delta_{0}e^{i\theta\left(t,x\right)}\left(1+f\left(t,x\right),\pm i\right)$
\footnote{Assuming that $\mu$ depends on $y$ but $\Delta$ is independent
of $y$ may not be self consistent. Nevertheless, it is a simple ansatz
that allows for a description of the boundary fermion and its anomaly,
which is fixed within a topological phase \cite{read2000paired}.} with $\Delta_{0}>0$ and small $f$. We also assume that $A_{y}=0$
and $A_{t},A_{x}$ are functions of $t,x$. This corresponds to a
perturbation of the $p_{x}\pm ip_{y}$ configuration. Note that assuming
$A_{y}=0$ involves a partial $U\left(1\right)$ gauge fixing, leaving
only $y$ independent gauge transformations $\alpha\left(t,x\right)$.
These are the $U\left(1\right)$ gauge transformations that will be
considered in this subsection. After further $U\left(1\right)$ gauge
fixing such that $\theta\mapsto0$, using a gauge transformation $\alpha\left(t,x\right)=-\theta\left(t,x\right)/2$
\footnote{Here we are explicitly assuming that there are no vortices, such that
$\alpha=-\theta/2$ is a gauge transformation.}, the inverse vielbein will be of the form 
\begin{eqnarray}
 &  & e_{a}^{\;\mu}=\left(\begin{array}{ccc}
1 & 0 & 0\\
0 & 1+f\left(t,x\right) & 0\\
0 & 0 & \pm1
\end{array}\right),\label{eq:88-1}
\end{eqnarray}
so the vielbein is of the product form \eqref{eq:81} with $o=\pm1$.
The corresponding inverse metric is given by 
\begin{eqnarray}
 &  & g^{\mu\nu}=\left(\begin{array}{ccc}
1 & 0 & 0\\
0 & -\left(1+f\left(t,x\right)\right)^{2} & 0\\
0 & 0 & -1
\end{array}\right).
\end{eqnarray}
We will also need the Ricci scalar for this metric, 
\begin{eqnarray}
 &  & \tilde{\mathcal{R}}=\frac{2\left(\left(1+f\right)\partial_{t}^{2}f-2\left(\partial_{t}f\right)^{2}\right)}{\left(1+f\right)^{2}}.
\end{eqnarray}
Recalling that $\mu$ determines the bulk masses $m_{n}$, we can
use the formula $\nu=\frac{1}{2}\sum_{n=1}^{4}o_{n}\text{sgn}\left(m_{n}\right)$
for the Chern number in terms of the low energy data, and \eqref{eq:101},
to express the net chirality of the boundary spinors as $C=\sum_{n=1}^{4}C_{n}=\Delta\nu$,
where $C_{n}=\frac{o_{n}}{2}\text{sgn}\left(m_{n}\left(y=\infty\right)\right)-\frac{o_{n}}{2}\text{sgn}\left(m\left(y=-\infty\right)\right)$
and $\Delta\nu=\nu\left(y=\infty\right)-\nu\left(y=-\infty\right)$.
This relation between the boundary net chirality $C$ and the Chern
number difference $\Delta\nu$ is the well known bulk-boundary correspondence.
It can be derived from index theorems as described in \cite{volovik2009universe},
but in the following we will place it on a more physical footing by
describing it as a consequence of energy-momentum conservation. 
Let us now rewrite the action \eqref{eq:87} in terms of the $p$-wave
SC quantities and in physical units (without setting the emergent
speed of light $\Delta_{0}$ to 1, but with $\hbar=1$), 
\begin{align}
 S_{\text{e}}^{\pm}=\frac{i}{2}\int\mbox{d}t\text{d}x\tilde{\xi}\left(\partial_{t}\mp\left|\Delta^{x}\left(t,x\right)\right|\partial_{x}\right)\tilde{\xi}.\label{eq:137}
\end{align}
Here $\left|\Delta^{x}\right|=\Delta_{0}\left(1+f\right)$ and $\tilde{\xi}=\left|e\right|^{1/2}\xi$
is a chiral Majorana spinor \textit{density} from the geometric point
of view, but a chiral Majorana spinor from the physical flat space
point of view. As an operator $\tilde{\xi}$ satisfies $\left\{ \tilde{\xi}\left(x_{1}\right),\tilde{\xi}\left(x_{2}\right)\right\} =\delta\left(x_{1}-x_{2}\right)$.
We see that $\left|\Delta^{x}\right|$ acts as a space-time dependent
velocity for the boundary fermions, which reduces to a constant $\Delta_{0}$
in the $p_{x}\pm ip_{y}$ configuration. Note that both fields $\left|\Delta^{x}\right|,\tilde{\xi}$
are uncharged under $U\left(1\right)$. This is clear for $\left|\Delta^{x}\right|$,
and to see this explicitly for $\tilde{\xi}$ we relate it to the
original spin-less fermion $\psi$ and the (phase of the) order parameter
$\Delta$, 
\begin{eqnarray}
 &  & \psi\left(t,x,y\right)\propto\tilde{\xi}\left(t,x\right)e^{i\theta\left(t,x\right)/2}\phi_{0,\pm}\left(y\right)+\cdots
\end{eqnarray}
where $\phi_{0,\pm}\left(y\right)\propto e^{\pm\int_{0}^{y}m\left(s\right)\text{d}s}$
was defined in Appendix \ref{subsec:Boundary-states-in} and the dots
represent the massive bulk modes and additional non robust massive
boundary modes. From this expression it is clear that $\tilde{\xi}$
is uncharged even though $\psi$ is.

Let us now consider the energy-momentum conservation law for the boundary.
The expression $\tilde{\nabla}_{\alpha}\left\langle \mathsf{j}_{\;\beta}^{\alpha}\right\rangle =0$
involves the covariant derivative, and is therefore inappropriate
from the $p$-wave SC point of view, where space-time is flat and
$e$ is just the order parameter and has no geometric role. We already
described how to interpret covariant energy-momentum conservation
laws from the flat space-time point of view in section \ref{subsec:Energy-momentum},
where we studied the bulk conservation laws. Here we simply repeat
the procedure. We first relate the energy-momentum tensor $\mathsf{j}_{\;\beta}^{\alpha}$
to the canonical boundary (or edge) energy-momentum tensor $t_{\text{e}\;\beta}^{\alpha}$,
and write it in terms of $\tilde{\xi}$
\begin{align}
  t_{\text{e}\;\beta}^{\alpha}=-\left|e\right|\mathsf{j}_{\;\beta}^{\alpha}=&\frac{i}{2}e_{\mp}^{\;\alpha}\tilde{\xi}\partial_{\beta}\tilde{\xi}\nonumber\\
  =&\begin{cases}
\frac{i}{2}\tilde{\xi}\partial_{\beta}\tilde{\xi} & \alpha=t\\
\mp\frac{i}{2}\left|\Delta^{x}\left(t,x\right)\right|\tilde{\xi}\partial_{\beta}\tilde{\xi} & \alpha=x
\end{cases}.
\end{align}
This is the correct notion of energy and momentum from the physical
flat space-time point of view. Note that the relation between $t_{\text{e}\;\beta}^{\alpha}$
and $\mathsf{j}_{\;\beta}^{\alpha}$ is the same as for the bulk quantities
\eqref{eq:56}, and that since $\tilde{\xi}$ is uncharged the canonical
energy-momentum tensor $t_{\text{e}\;\beta}^{\alpha}$ is automatically
$U\left(1\right)$-covariant. We then write the conservation law $\tilde{\nabla}_{\alpha}\left\langle \mathsf{j}_{\;\beta}^{\alpha}\right\rangle =0$
in terms of $t_{\text{e}\;\beta}^{\alpha}$ and using partial derivatives
as $\partial_{\alpha}t_{\text{e}\;\beta}^{\alpha}+\frac{i}{2}\tilde{\xi}\partial_{\alpha}\tilde{\xi}\partial_{\beta}e_{\mp}^{\;\alpha}=0$,
or more explicitly, 
\begin{eqnarray}
 &  & \partial_{\alpha}t_{\text{e}\;\beta}^{\alpha}\mp\frac{i}{2}\tilde{\xi}\partial_{x}\tilde{\xi}\partial_{\beta}\left|\Delta^{x}\right|=0.
\end{eqnarray}
This is just a special case of the usual conservation law \eqref{eq:18}
for the canonical energy-momentum tensor. As usual, it describes the
space-time dependence of the background field $\left|\Delta^{x}\right|$
as a source of energy-momentum for the boundary fermion $\tilde{\xi}$.
This is the ``classical'' analysis of energy momentum-conservation
for the boundary fermion. Quantum mechanically, this equation acquires
a correction due to the anomaly and the presence of the bulk. Translating
the anomaly equation \eqref{eq:88} to the flat space-time point of
view, we obtain 
\begin{align}
 \partial_{\alpha}\left\langle t_{\text{e}\;\beta}^{\alpha}\right\rangle \mp\frac{i}{2}\left\langle \tilde{\xi}\partial_{x}\tilde{\xi}\right\rangle \partial_{\beta}\left|\Delta^{x}\right|=-\frac{\nu}{192\pi}g_{\beta\gamma}\varepsilon^{y\gamma\alpha}\partial_{\alpha}\tilde{\mathcal{R}}.
\end{align}

As in the gravitational point of view, the right hand side is actually
the inflow of energy-momentum from the bulk \eqref{eq:104}, 
\begin{align}
 \partial_{\alpha}\left\langle t_{\text{e}\;\beta}^{\alpha}\right\rangle \mp\frac{i}{2}\left\langle \tilde{\xi}\partial_{x}\tilde{\xi}\right\rangle \partial_{\beta}\left|\Delta^{x}\right|=\left\langle t_{\text{cov }\beta}^{y}\right\rangle. \label{eq:89-1-1}
\end{align}

This equation expresses the conservation of energy ($\beta=t$) and
$x$-momentum ($\beta=x$) on the domain wall. Along with the bulk
conservation equation \eqref{32}, $\partial_{\mu}\left\langle t_{\text{cov}\;\nu}^{\mu}\right\rangle =\frac{1}{2}\left\langle \psi^{\dagger}\partial_{j}\psi^{\dagger}\right\rangle D_{\nu}\Delta^{j}+h.c+F_{\nu\mu}\left\langle J^{\mu}\right\rangle $
\footnote{We note that the domain wall acts as a source for $y$-momentum, which
is included in the term $F_{\nu\mu}\left\langle J^{\mu}\right\rangle $
since $\mu\left(y\right)$ is part of the electric potential $A_{t}$.}, it expresses the sense in which energy-momentum is conserved in
a $p$-wave SC in the presence of a boundary, or domain wall. 

We thus obtain the equation $\Delta\nu=C$, usually referred to as
bulk boundary correspondence, as a direct consequence of bulk+boundary
energy-momentum conservation in the presence of a space-time dependent
order parameter. 


\section{Emergent Riemann-Cartan geometry: further details}

\subsection{\label{subsec:Equivalent-forms-of}Equivalent forms of $S_{\text{RC}}$
and equality to $S_{\text{rSF}}$ }

It is useful to write the action $S_{\text{RC}}$ in a few equivalent
forms \cite{bertlmann2000anomalies,hughes2013torsional}. To pass
between these equivalent forms one only needs the identity 
\begin{eqnarray}
 &  & \partial_{\nu}\left(\left|e\right|e_{a}^{\;\nu}\right)=\left|e\right|\tilde{\omega}_{\;ab}^{b}\label{40}
\end{eqnarray}
 relating $e$ to the LC spin connection, and the following identity,
which holds for any spin connection $\omega$ but relies on the property
$\gamma^{a}\gamma^{b}\gamma^{c}=i\varepsilon^{abc}$ of $\gamma$
matrices in 2+1 dimensions, 
\begin{align}
  ie_{a}^{\;\mu}\gamma^{a}\omega_{\mu}=&\frac{1}{4}ie_{a}^{\;\mu}\omega_{bc\mu}\left\{ \gamma^{a},\Sigma^{bc}\right\} +\frac{1}{4}ie_{a}^{\;\mu}\omega_{bc\mu}\left[\gamma^{a},\Sigma^{bc}\right]\nonumber\\
  =&-\frac{1}{4}\omega_{abc}\varepsilon^{abc}+\frac{1}{2}i\omega_{\;ab}^{b}\gamma^{a}.\label{eq:135}
\end{align}
The most explicit form of the action is
\begin{align}
 S_{\text{RC}}=\frac{1}{2}\int\mbox{d}^{2+1}x\left|e\right|\overline{\chi}\left[\frac{1}{2}ie_{a}^{\;\mu}\gamma^{a}\overleftrightarrow{\partial_{\mu}}-\frac{1}{4}\omega_{abc}\varepsilon^{abc}-m\right]\chi,\label{eq:40}
\end{align}
where the derivatives act only on the spinors. Here we see that in
2+1 dimensions the spin connection only enters through the scalar
$\omega_{abc}\varepsilon^{abc}$ as a correction to the mass. It also
makes it rather simple to see why $S_{\text{RC}}$ is equal to $S_{\text{rSF}}$
from \eqref{eq:14-0}, 
\begin{align}
  &S_{\text{RC}}  =\frac{1}{2}\int\mbox{d}^{2+1}x\left|e\right|\overline{\chi}\left[\frac{1}{2}ie_{a}^{\;\mu}\gamma^{a}\overleftrightarrow{\partial_{\mu}}-\frac{1}{4}\omega_{abc}\varepsilon^{abc}-m\right]\chi\nonumber\\
   & =\frac{1}{2}\int\mbox{d}^{2+1}x\Psi^{\dagger}\gamma^{0}\left[\frac{1}{2}ie_{a}^{\;\mu}\gamma^{a}\overleftrightarrow{\partial_{\mu}}-\frac{1}{4}\omega_{abc}\varepsilon^{abc}-m\right]\Psi\nonumber \\
   & =\frac{1}{2}\int\mbox{d}^{2+1}x\Psi^{\dagger}\gamma^{0}\left[\frac{i}{2}\gamma^{0}\overleftrightarrow{\partial_{t}}+\frac{1}{2}ie_{A}^{\;j}\gamma^{A}\overleftrightarrow{\partial_{j}}+A_{t}-m\right]\Psi\nonumber \\
   & =\frac{1}{2}\int\mbox{d}^{2+1}x\Psi^{\dagger}\begin{pmatrix}\frac{i}{2}\overleftrightarrow{\partial_{t}}+A_{t}-m & \frac{1}{2}\Delta^{j}\overleftrightarrow{\partial_{j}}\\
-\frac{1}{2}\Delta^{j*}\overleftrightarrow{\partial_{j}} & \frac{i}{2}\overleftrightarrow{\partial_{t}}-A_{t}+m
\end{pmatrix}\Psi\nonumber \\
 & =\frac{1}{2}\int\mbox{d}^{2+1}x\Psi^{\dagger}\begin{pmatrix}i\partial_{t}-m+A_{t} & \frac{1}{2}\left\{ \Delta^{j},\partial_{j}\right\} \\
-\frac{1}{2}\left\{ \Delta^{j*},\partial_{j}\right\}  & i\partial_{t}+m-A_{t}
\end{pmatrix}\Psi\nonumber\\&=S_{\text{rSF}},\label{eq:43}
\end{align}
where we have used the dictionary \eqref{17}, and also integrated by parts. In going from the third to the fourth line we
have reinstated the emergent speed of light $c_{\text{light}}=\frac{\Delta_{0}}{\hbar}$,
but kept $\hbar=1$. This completes that proof of the equality $S_{\text{rSF}}=S_{\text{RC}}$,
which was stated and explained in section \ref{sec:Emergent-Riemann-Cartan-geometry}. 

Before we move on, an important comment is in order. Since $A_{j}$ does not appear in $S_{\text{rSF}}$, it is clear that for the above equality of
 actions only the identification $\omega_{t}=-2A_{t}\Sigma^{12}$ is required, rather than the full $\omega_{\mu}=-2A_{\mu}\Sigma^{12}$ of \eqref{eq:17}. Accordingly,  $\omega_{j}$ does not appear in $S_{\text{RC}}$ when $\omega, e$ are both spatial ($\omega_{0A\mu}=0, e_{0}^{\;\mu}=\delta_{t}^{\mu}$), because then 
\begin{align}
\omega_{abc}\varepsilon^{abc}=2e_{0}^{\mu}\omega_{12\mu}=2\omega_{12t}.\label{eq:400}
\end{align}
Thus, for the equality of actions $S_{\text{rSF}}=S_{\text{RC}}$  it is not required that $\omega_{j}=-2A_{j}\Sigma^{12}$. Nevertheless, we are actually identifying two QFTs as equal, and there is more to a QFT than its classical action. One must also compare symmetries, observables, and path integral measures (the latter is discussed in appendix \ref{subsec:Equality-of-path}). The mapping of symmetries 
and observables is the subject of section \ref{sec:Symmetries,-currents,-and}, and only holds if the full identification $\omega_{\mu}=-2A_{\mu}\Sigma^{12}$ is made:

In section \ref{spin}, we identify the physical $U\left(1\right)$ symmetry group with the $Spin\left(2\right)$ subgroup of $Spin\left(1,2\right)$ in Riemann-Cartan geometry. For this reason $A_{\mu}$, which is $U\left(1\right)$ connection, really maps to a $Spin\left(2\right)$ connection in the geometric point of view, even if certain components of it do not appear in the action $S_{\text{rSF}}$. 

In section \ref{currents} we discuss the mapping of observables. In particular, even though $A_{j}$ disappears from the action in the relativistic limit, it does not disappear from the energy-momentum tensor (see \eqref{eq:56}, \eqref{eq:49}, where the derivative $D_{\mu}$ contains $A_{\mu}$). Moreover, as explained below \eqref{eq:71-1}, even though the order parameter $\Delta$ corresponds to the spatial vielbein in \eqref{17}, in order to obtain the expectation value of full energy-momentum tensor we must take derivatives of the effective action with respect to all components of the vielbein, not only the spatial ones obtained from $\Delta$. This corresponds to adding to $S_{\text{rSF}}$ a fictitious background field $e_{0}^{\;\mu}$ which is set to zero after the expectation value is computed. In the presence of $e_{0}^{\;\mu}$ the potential $A_{t}$ generalizes to $e_{0}^{\;\mu}A_{\mu}$, and so $A_{j}$ does appear in $S_{\text{rSF}}$. Accordingly, with a general $e_{0}^{\;\mu}$ we see from \eqref{eq:400} that $\omega_{12j}$ appears in $S_{\text{RC}}$. The equality $S_{\text{RC}}=S_{\text{rSF}}$ in the presence of $e_{0}^{\;\mu}$ is then obtained only if $\omega_{j}=-2A_{j}\Sigma^{12}$. 

To close this discussion, we note that the identification of $\Delta^{j}$ as a spatial vielbein and $A_{\mu}$ as a $Spin\left(2\right)$ connection actually holds beyond the relativistic limit, though this was not discussed in Sec.\ref{sec:Main-section-1:}. Beyond the relativistic limit  $A_{j}$ will appear in both the action and observables, and identifying the full $\omega_{12\mu}$ with $A_{\mu}$ will be crucial also at the level of the fermionic action.  

Going back to equivalent forms of $S_{\text{RC}}$,  if we wish to isolate the effect of torsion, we can also write 
\begin{align}
  S_{\text{RC}} =\frac{1}{2}\int\mbox{d}^{2+1}x\left|e\right|\overline{\chi}\left[\frac{1}{2}\right.ie_{a}^{\;\mu}\gamma^{a}\overleftrightarrow{\partial_{\mu}}-\frac{1}{4}\tilde{\omega}_{abc}\varepsilon^{abc}
  \left.-\frac{1}{4}C_{abc}\varepsilon^{abc}-m\right]\chi, 
\end{align}
or 
\begin{align}
    S_{\text{RC}} =\frac{1}{2}\int\mbox{d}^{2+1}x\left|e\right|\overline{\chi}\left[\frac{1}{2}\right.ie_{a}^{\;\mu}\left(\gamma^{a}\tilde{D}_{\mu}-\overleftarrow{\tilde{D}_{\mu}}\gamma^{a}\right)
    \left.-\left(m+\frac{1}{4}c\right)\right]\chi,
\end{align}
where we see that in 2+1 dimensions torsion enters only trough the
scalar $c=C_{abc}\varepsilon^{abc}$ as a correction to the mass.
One can also integrate by parts in order
to obtain a form from which it is simple to derive the equation of
motion, 
\begin{align}
  S_{\text{RC}} & =\frac{1}{2}\int\mbox{d}^{2+1}x\left|e\right|\overline{\chi}\left[ie_{a}^{\;\mu}\gamma^{a}\tilde{D}_{\mu}-\frac{1}{4}C_{abc}\varepsilon^{abc}-m\right]\chi\nonumber\\
   & =\frac{1}{2}\int\mbox{d}^{2+1}x\left|e\right|\overline{\chi}\left[ie_{a}^{\;\mu}\gamma^{a}D_{\mu}-\frac{1}{2}iC_{\;ab}^{b}\gamma^{a}-m\right]\chi.\label{43-1}
\end{align}
The form in the first equation is special to 2+1 dimensions, but
the form in the second equation holds in any dimension.

\subsection{Dirac and BdG equations\label{subsec:Dirac-and-BdG}}

Since the $p$-wave SF action is equal to $S_{\text{RC}}$ in the
relativistic limit, and the fermions $\chi$ and $\Psi$ are related
simply, the equation of motion for $\chi$, which is the Dirac equation
in RC background, maps to the equation of motion for $\Psi$, which
is the BdG equation (in the relativistic limit). 

The equation of motion for the Majorana spinor $\chi$ needs to be
derived carefully, because $\chi$ is Grassmann valued and $\overline{\chi}=\chi^{T}\gamma^{0}$ cannot
be treated as independent of $\chi$. Nevertheless, if the operator between $\chi^{T}$ and $\chi$
is particle-hole symmetric, the equations of motion are the same as those of a Dirac
spinor, which are easy to read from \eqref{43-1}, 
\begin{align}
 0=\left[ie_{a}^{\;\mu}\gamma^{a}D_{\mu}-\frac{1}{2}iC_{\;ab}^{b}\gamma^{a}-m\right]\chi.
\end{align}
This is the Dirac equation in RC background. When inserting $\chi=\left|e\right|^{-1/2}\Psi$
and using the identity 
\begin{align}
  \partial_{\mu}\left|e\right|=\left|e\right|\Gamma_{\mu\rho}^{\rho}=\left|e\right|\tilde{\Gamma}_{\mu\rho}^{\rho},\label{47}
\end{align}
we obtain 
\begin{align}
 0=\left[i\gamma^{\mu}\left(D_{\mu}-\frac{1}{2}\Gamma_{\mu\rho}^{\rho}\right)-\frac{1}{2}iC_{\;ab}^{b}\gamma^{a}-m\right]\Psi.
\end{align}
The expression in brackets is the appropriate covariant derivative
for a spinor density of weight 1/2 \cite{ortin2004gravity}, which
is what $\Psi=\left|e\right|^{1/2}\chi$ is from the geometric point
if view. Simplifying this equation using \eqref{40} and \eqref{47},
we arrive at 
\begin{align}
 0=\left[\frac{1}{2}i\gamma^{a}\left\{ e_{a}^{\;\mu},\partial_{\mu}\right\} -\frac{1}{4}\omega_{abc}\varepsilon^{abc}-m\right]\Psi.
\end{align}
By using the dictionary \eqref{17} and multiplying by $\gamma^{0}=\sigma^{z}$
 this reduces to 
\begin{align}
 0=\begin{pmatrix}i\partial_{t}+A_{t}-m & \frac{1}{2}\left\{ \Delta^{j},\partial_{j}\right\} \\
-\frac{1}{2}\left\{ \Delta^{j*},\partial_{j}\right\}  & i\partial_{t}-A_{t}+m
\end{pmatrix}\Psi,
\end{align}
which is the BdG equation in the relativistic limit.  Thus the BdG equation in the relativistic limit is not quite the Dirac equation,
because $\Psi$ is a spinor density, though it is the Dirac equation for the spinor $\chi$.

\subsection{Equality of path integrals \label{subsec:Equality-of-path}}

In appendix \ref{subsec:Equivalent-forms-of} we showed that the action
for the $p$-wave SF in the relativistic limit, is equal to the action
for a Majorana fermion coupled to RC geometry. To conclude that the
corresponding fermionic path integrals are equal, we also need to
verify that the path integral measure for the $p$-wave SF is equal
to that of the Majorana fermion in RC background. For the $p$-wave
SF \eqref{eq:10}, the path integral measure is written formally as
$\text{D}\psi^{\dagger}\text{D}\psi=\prod_{x}\text{d}\psi^{\dagger}\left(x\right)\text{d}\psi\left(x\right)$
where $x$ runs over all points in space time. In the BdG formalism
we work with the Nambu (or Majorana) spinor $\Psi=\left(\psi,\psi^{\dagger}\right)^{T}$,
in terms of which the measure takes the form $\text{D}\psi^{\dagger}\text{D}\psi=\text{D}\Psi$.
As described in section \ref{sec:Emergent-Riemann-Cartan-geometry}
and appendix \ref{subsec:Dirac-and-BdG}, from the geometric point
of view $\Psi$ is a Majorana spinor density of weight 1/2, and $\chi=\left|e\right|^{-1/2}\Psi$
is a Majorana spinor. In terms of the spinor $\chi$, the measure
takes the form $\text{D}\Psi=\text{D}\left(\left|e\right|^{1/2}\chi\right)$,
which is the correct measure for a matter field in curved background
\cite{hawking1977zeta,fujikawa1980comment,abanov2014electromagnetic}.
With this measure, the path integral over the Majorana spinor $\chi$
formally computes functional pfaffians as in flat space, $e^{iW_{M}\left[A\right]}=\int\text{D}\left(\left|e\right|^{1/2}\chi\right)e^{\frac{i}{2}\int\text{d}^{d}x\left|e\right|\chi^{T}A\chi}=\text{Pf}\left(iA\right)=\sqrt{\text{Det}iA}$,
where $A$ is an antisymmetric hermitian operator with respect to
the inner product $\left\langle f,g\right\rangle =\int\text{d}^{d}x\left|e\right|f^{\dagger}Ag$,
and the determinant $\text{Det}$ is defined by the product of eigenvalues.
For a Dirac spinor $\chi$ the fermionic path integral formally computes
functional determinants, $e^{iW_{D}\left[D\right]}=\int\text{D}\left(\left|e\right|^{1/2}\chi^{\dagger}\right)\text{D}\left(\left|e\right|^{1/2}\chi\right)e^{i\int\text{d}^{d}x\left|e\right|\chi^{\dagger}D\chi}=\text{Det}\left(iD\right)$,
where $D$ is hermitian. In particular, the effective action for a
Majorana spinor is half that of a Dirac spinor with the same operator,
$W_{M}\left[A\right]=\frac{1}{2}W_{D}\left[A\right]$.

\subsection{Explicit formulas for certain geometric quantities\label{subsec:Explicit-formulas-for}}

Using  $\tilde{\omega}_{\;b\mu}^{a}=e_{\;\alpha}^{a}\left(\partial_{\mu}e_{b}^{\;\alpha}+\tilde{\Gamma}_{\;\beta\mu}^{\alpha}e_{b}^{\;\beta}\right)$
we can calculate the LC spin connection for a vielbein of the form
\begin{align}
 e_{a}^{\;\mu}=\frac{1}{\Delta_{0}}\left(\begin{array}{ccc}
\Delta_{0} & 0 & 0\\
0 & \mbox{Re}(\Delta^{x}) & \mbox{Re}(\Delta^{y})\\
0 & \mbox{Im}(\Delta^{x}) & \text{Im}(\Delta^{y})
\end{array}\right)=\left(\begin{array}{cc}
1\\
 & e_{A}^{\;j}
\end{array}\right)
\end{align}
that occurs in the $p$-wave SC,

\begin{align}
  \tilde{\omega}_{A0t}=&0,\\
  \tilde{\omega}_{A0j}=&e_{A}^{\;i}\frac{1}{2}\partial_{t}g_{ij},\nonumber \\
  \tilde{\omega}_{12t}=&\frac{1}{2}\varepsilon^{AB}e_{Ai}\partial_{t}e_{B}^{\;i}=-\frac{1}{2}\frac{1}{\text{det}\left(e\right)}\varepsilon^{ij}e_{Ai}\partial_{t}e_{\;j}^{A},\nonumber \\
 \tilde{\omega}_{12j}=&\frac{1}{2}\left(\varepsilon^{AB}e_{Ai}\partial_{j}e_{B}^{\;i}-\frac{1}{\text{det}\left(e\right)}\varepsilon^{kl}\partial_{k}g_{lj}\right)=-\frac{1}{2}\frac{1}{\text{det}\left(e\right)}\varepsilon^{kl}\left(e_{Ak}\partial_{j}e_{\;l}^{A}+\partial_{k}g_{lj}\right).\nonumber 
\end{align}
In terms of the parameterization $\Delta=e^{i\theta}\left(\left|\Delta^{x}\right|,e^{i\phi}\left|\Delta^{y}\right|\right)$,
as in section \ref{subsec:The-order-parameter}, the $SO\left(2\right)$
part can be written as

\begin{eqnarray}
 &  & \tilde{\omega}_{12t}=o\left[\frac{1}{2}\cot\left|\phi\right|\partial_{t}\log\frac{\left|\Delta^{y}\right|}{\left|\Delta^{x}\right|}-\frac{1}{2}\partial_{t}\left|\phi\right|\right]-\partial_{t}\theta, \label{eq:146-1}\\
 &  & \tilde{\omega}_{12x}=o\left[\frac{\left|\Delta^{y}\right|}{\left|\Delta^{x}\right|}\frac{\cot\left|\phi\right|}{\sin\left|\phi\right|}\partial_{y}\left|\phi\right|+\left(\frac{1}{\sin^{2}\left|\phi\right|}-1\right)\partial_{x}\left|\phi\right|+\cot\left|\phi\right|\partial_{x}\log\left|\Delta^{y}\right|+\frac{1}{\sin\left|\phi\right|}\frac{\left|\Delta^{y}\right|}{\left|\Delta^{x}\right|}\partial_{y}\log\left|\Delta^{x}\right|\right]-\partial_{x}\theta,\nonumber \\
 &  & \tilde{\omega}_{12y}=o\left[-\frac{\left|\Delta^{x}\right|}{\left|\Delta^{y}\right|}\frac{\cot\left|\phi\right|}{\sin\left|\phi\right|}\partial_{x}\left|\phi\right|-\left(\frac{1}{\sin^{2}\left|\phi\right|}\right)\partial_{y}\left|\phi\right|-\cot\left|\phi\right|\partial_{y}\log\left|\Delta^{x}\right|-\frac{1}{\sin\left|\phi\right|}\frac{\left|\Delta^{x}\right|}{\left|\Delta^{y}\right|}\partial_{x}\log\left|\Delta^{y}\right|\right]-\partial_{y}\theta,\nonumber 
\end{eqnarray}

where $o=\text{sgn}\phi$ is the orientation. Note that the terms
in square brackets only depend on the metric degrees of freedom $\left|\phi\right|,\left|\Delta^{x}\right|,\left|\Delta^{y}\right|$,
and that this reduces to $\tilde{\omega}_{12\mu}=-\partial_{\mu}\theta$
in the $p_{x}\pm ip_{y}$ configuration. We can then obtain explicit
formulas for the contorsion using $\omega_{ab\mu}=-2A_{\mu}\left(\delta_{a}^{1}\delta_{b}^{2}-\delta_{b}^{1}\delta_{a}^{2}\right)$
and $C_{ab\mu}=\omega_{ab\mu}-\tilde{\omega}_{ab\mu}$, 
\begin{eqnarray}
 &  & C_{12\mu}=-2A_{\mu}-\tilde{\omega}_{12\mu}, \label{eq:147-1}\\
 &  & C_{A0j}=-e_{A}^{\;i}\frac{1}{2}\partial_{t}g_{ij}.\nonumber 
\end{eqnarray}

We also consider the quantity $c=\varepsilon^{abc}C_{abc}$ which
appears in certain forms of the action $S_{\text{RC}}$ \eqref{eq:43-1},
and of the effective action \eqref{eq:76}. Evaluated in terms of
$\Delta$ and $A$ we find 
\begin{align}
  \frac{1}{2}c=C_{12t}
  =\partial_{t}\theta-2A_{t}-o\left[\frac{1}{2}\cot\left|\phi\right|\partial_{t}\log\frac{\left|\Delta^{y}\right|}{\left|\Delta^{x}\right|}-\frac{1}{2}\partial_{t}\left|\phi\right|\right],
\end{align}
which reduces to $\frac{1}{2}c=D_{t}\theta=\partial_{t}\theta-2A_{t}$
in the $p_{x}\pm ip_{y}$ configuration.

\subsection{\label{subsec:Discrete-symmetries}Discrete symmetries}

\subsubsection{Charge conjugation and particle-hole \label{subsec:Charge-conjugation-(Appendix)}}

Our conventions for gamma matrices and spinors follow appendix B
of \cite{ortin2004gravity}. In three dimensions, if the matrices
$\gamma^{a}$ define a representation of the Clifford algebra then
$-\left(\gamma^{a}\right)^{T}$ define an equivalent representation.
The matrix $\mathcal{C}$ relating the two representations by $-\left(\gamma^{a}\right)^{T}=\mathcal{C}_{\;b}^{a}\gamma^{b}=\mathcal{C}\gamma^{a}\mathcal{C}^{-1}$
is called charge conjugation. In our representation $\gamma^{0}=\sigma^{z},\;\gamma^{1}=-i\sigma^{x},\;\gamma^{2}=i\sigma^{y}$,
one finds that $\mathcal{C}=\sigma^{y}$ up to a phase and $\mathcal{C}_{\;b}^{a}=\text{diag}\left[-1,-1,1\right]$,
so we see that $\mathcal{C}$ is unitary and $\mathcal{C}^{2}=1$.
Likewise, the matrices $\left(\gamma^{a}\right)^{\dagger}$ also define
an equivalent representation, and are therefore related by $\left(\gamma^{a}\right)^{\dagger}=\mathcal{D}_{\;b}^{a}\gamma^{b}=\mathcal{D}\gamma^{a}\mathcal{D}^{-1}$
where $\mathcal{D}$ is the Dirac conjugation. In any unitary representation
$\mathcal{D}=i\gamma^{0}$ up to a phase and $\mathcal{D}_{\;b}^{a}=\text{diag}\left[1,-1,-1\right]$.
Using $\mathcal{D}$ we define the conjugate spinor $i\overline{\Psi}=\Psi^{\dagger}\mathcal{D}$.
We also note that $-\left(\gamma^{a}\right)^{*}=\mathcal{B}_{\;b}^{a}\gamma^{b}=\mathcal{B}\gamma^{a}\mathcal{B}^{-1}$
with $\mathcal{B}=\mathcal{D}\mathcal{C}$, which will also show up
in our discussion of time reversal. In our representation, $\mathcal{B}=\sigma^{x}$
and $\mathcal{B}_{\;b}^{a}=\text{diag}\left[-1,1-1\right]$. 

A spinor $\Psi$ is called a Majorana spinor if it satisfies the reality
condition $i\overline{\Psi}=\Psi^{T}\mathcal{C}$, which can also
be written as $\Psi^{\dagger}\mathcal{B}=\Psi^{T}$. In our representation
this condition reads $\Psi^{\dagger}=\Psi^{T}\sigma^{x}$, which is
the reality condition satisfied by the Nambu spinor $\Psi=\left(\psi,\psi^{\dagger}\right)^{T}$.
We see that the Nambu spinor is a Majorana spinor. The reality condition
can also be written as $\Psi=P\Psi$ where $P=\sigma^{x}K$ and $K$
is the complex conjugation. $P$ is usually referred to as a particle-hole
symmetry \cite{ryu2010topological}, and it is anti-unitary and $P^{2}=1$.
Eventually, the particle-hole symmetry of the $p$-wave SC maps to
the charge conjugation symmetry of the relativistic Majorana fermion,
with the differences between the two being a matter of convention. 

For any Hamiltonian $H=\frac{1}{2}\int\text{d}^{2}x\Psi^{\dagger}\left(x\right)H_{\text{BdG}}\Psi\left(x\right)$,
the BdG Hamiltonian $H_{\text{BdG}}$ can be assumed to satisfy a
reality condition, $\left\{ H_{\text{BdG}},P\right\} =0$. An example
is given by \eqref{eq:14-0}. To make a similar statement for actions,
where $\psi,\psi^{\dagger}$ are Grassmann valued, we need to clarify
how the conjugation $K$ acts on the Grassmann algebra generated $\psi,\psi^{\dagger}$.
This is defined by $K\psi=\psi^{\dagger},\;K\psi^{\dagger}=\psi$,
anti-linearity, and a reversal of the ordering of Grassmann numbers.
For example, $K\left(\psi\psi^{\dagger}\right)=K\psi^{\dagger}K\psi=\psi\psi^{\dagger}$,
$K\left(i\psi\right)=-i\psi^{\dagger}$. It is under this complex
conjugation that a fermionic action, such as \eqref{eq:10}, is ``real'',
$K\left(S_{\text{SF}}\left[\psi,\psi^{\dagger},\Delta,A\right]\right)=S_{\text{SF}}\left[\psi,\psi^{\dagger},\Delta,A\right]$,
and it is due to this reality of $S_{\text{SF}}$ that we expect to
obtain a real effective action after integrating out the fermions
\cite{wetterich2011spinors}. Then, for any action $S=\frac{1}{2}\int\text{d}^{2+1}x\Psi^{\dagger}\left(x\right)S_{\text{BdG}}\Psi\left(x\right)$,
the operator $S_{\text{BdG}}$ can then be assumed to satisfy $\left\{ S_{\text{BdG}},P\right\} =0$,
and an example is given by the Dirac operator in \eqref{43-1}.

When working with Majorana fermions it is useful to use gamma matrices
$\gamma^{a}$ that form a \textit{Majorana representation} \cite{ortin2004gravity},
which means that $\gamma^{a}$ are all imaginary . In a Majorana
representation $\Psi^{\dagger}\mathcal{B}=\Psi^{T}$ simplifies to
$\Psi^{\dagger}=\Psi^{T}$, so a Majorana spinor in a Majorana representation
has real components.  To obtain a Majorana representation from our
representation we change basis in the space of spinors using the unitary
matrix $U=\frac{1}{\sqrt{2}}\begin{pmatrix}1 & 1\\
-i & i
\end{pmatrix}$. Then $\gamma^{a}\mapsto\tilde{\gamma}^{a}=U\gamma^{a}U^{\dagger}$
and $\Psi\mapsto\tilde{\Psi}=U\Psi$. Explicitly, $\tilde{\gamma}^{0}=-\sigma^{y},\;\tilde{\gamma}^{1}=-i\sigma^{z},\;\tilde{\gamma}^{2}=-i\sigma^{x}$,
and the Nambu spinor $\Psi=\left(\psi,\psi^{\dagger}\right)^{T}$
maps to $\tilde{\Psi}=\begin{pmatrix}\tilde{\Psi}_{1}\\
\tilde{\Psi}_{2}
\end{pmatrix}=\frac{1}{\sqrt{2}}\begin{pmatrix}\psi+\psi^{\dagger}\\
\frac{1}{i}\left(\psi-\psi^{\dagger}\right)
\end{pmatrix}$, where $\tilde{\Psi}_{1},\tilde{\Psi}_{2}$ are both real as Grassmann
valued fields. As operators $\tilde{\Psi}_{1},\tilde{\Psi}_{2}$ are
hermitian and $\left\{ \tilde{\Psi}_{i},\tilde{\Psi}_{j}\right\} =\delta_{ij}$,
so they are \textit{Majorana operators} in the sense of \cite{kitaev2009periodic}.
In the Majorana representation $H_{\text{BdG}}$ is imaginary and
antisymmetric, and so is $S_{\text{BdG}}$.

\subsubsection{Spatial reflection and time reversal in the $p$-wave superfluid\label{subsec:Spatial-reflections-and}}

In section \ref{subsec:Symmetries,-currents,-and} we discussed the
sense in which energy, momentum, and angular momentum are conserved
in a $p$-wave SF, which followed from the symmetry of the $p$-wave
SF action under space-time translations and spatial rotations. There
are also discrete (or large) space-time transformations which are
of interest. Spatial reflections reverse the orientation of space,
and are generated by a single arbitrary reflection, which we take
to be $R:y\mapsto-y$, followed by the spatial rotations and translations
described previously. $R$ acts naturally on the fields $\psi,\Delta,A$:
\begin{align}
  \psi\left(y\right)\mapsto&\psi\left(-y\right),\label{eq:35}\\
  \left(\Delta^{x},\Delta^{y}\right)\left(y\right)\mapsto&\left(\Delta^{x},-\Delta^{y}\right)\left(-y\right),\nonumber \\
  \left(A_{t},A_{x},A_{y}\right)\left(y\right)\mapsto&\left(A_{t},A_{x},-A_{y}\right)\left(-y\right),\nonumber 
\end{align}
where we suppressed the dependence on the coordinated $t,x$ which do not transform. One can verify that $R$ is a symmetry of the $p$-wave SF action
\eqref{eq:10}. The best way to understand these transformations is
to identify the fields as space-time tensors: $\psi$ is a scalar,
$\Delta^{j}\partial_{j}$ is a vector field, and $A_{\mu}\text{d}x^{\mu}$
is a differential 1-form. The above transformation laws are then a
special case of how space-time transformations act on space-time tensors,
by the pullback/push forward. 

Time reversal transformations reverse the orientation of time, and
are generated by a single arbitrary time reversal, which we take to
be $T:t\mapsto-t$, followed by the time translations described previously.
The action of $T$ on the fields includes the transformation laws
analogous to \eqref{eq:35}, but additionally involves a complex conjugation,
as follows from the Schrodinger equation in the Fock space $i\partial_{t}\ket{\Omega\left(t\right)}=H\left(t;A,\Delta\right)\ket{\Omega\left(t\right)}$.
In our case $H\left(t;A,\Delta\right)$ is the $p$-wave SF Hamiltonian
\eqref{eq:9}, in a notation that stresses the time dependence through
the background fields. On the Fock space the complex conjugation is
the usual complex conjugation of coefficients in the position basis,
defined by $K\psi\left(x,y\right)K^{-1}=\psi\left(x,y\right),\;K\psi^{\dagger}\left(x,y\right)K^{-1}=\psi^{\dagger}\left(x,y\right),\;K\ket 0=\ket 0$
and anti-linearity. Acting with it on the $p$-wave SF Hamiltonian
\eqref{eq:9} we find that the action of $T$ on the background fields
$\Delta,A$ is 
\begin{align}
  \left(\Delta^{x},\Delta^{y}\right)\left(t\right)\mapsto&\Delta^{T}\left(t\right)=\left(\Delta^{x},\Delta^{y}\right)^{*}\left(-t\right),\label{eq:29-0}\\
 \left(A_{t},A_{x},A_{y}\right)\left(t\right)\mapsto& A^{T}\left(t\right)=-\left(-A_{t},A_{x},A_{y}\right)\left(-t\right),\nonumber 
\end{align}
where we suppressed the dependence on the coordinates $x,y$ which do not transform. If $\ket{\Omega\left(t\right)}$ satisfies the Schrodinger equation
with Hamiltonian $H\left(t;A,\Delta\right)$ and initial condition
$\ket{\Omega}$ then $K\ket{\Omega\left(-t\right)}$ satisfies the
Schrodinger equation with time reversed Hamiltonian $KH\left(-t;A,\Delta\right)K^{-1}=H\left(t;A^{T},\Delta^{T}\right)$
and time reversed initial state $K\ket{\Omega}$.  As a result one
obtains the following relation between expectation values of operators,
\begin{align}
 \bra{\Omega}O_{A,\Delta}\left(-t\right)\ket{\Omega}=\bra{K\Omega}\left(KOK\right)_{A^{T},\Delta^{T}}\left(t\right)\ket{K\Omega}.\label{eq:30-1-0}
\end{align}
Here $O$ is a Schrodinger operator considered as an operator at
time $t=0$, and $O_{A,\Delta}\left(t\right)$ is its time evolution
using $H\left(t;A,\Delta\right)$. $\ket{K\Omega}=K\ket{\Omega}$
is the time reversed state, and $KOK$ is the time reversed Schrodinger
operator.   

 To describe how time reversal acts on the action, we need to use
the complex conjugation $K$ on the Grassmann algebra, described in
\ref{subsec:Charge-conjugation-(Appendix)}. We then define the action
of time reversal on the Grassmann fields $\psi,\psi^{\dagger}$ as
the analog of \eqref{eq:35}, but with an additional conjugation by
$K$,
\begin{align}
  \psi\left(t,x,y\right)\mapsto&\psi^{T}\left(t,x,y\right)=\psi^{\dagger}\left(-t,x,y\right),\label{eq:30-0}\\
  \psi^{\dagger}\left(t,x,y\right)\mapsto&\left(\psi^{\dagger}\right)^{T}\left(t,x,y\right)=\psi\left(-t,x,y\right).\nonumber 
\end{align}
Using the transformations \eqref{eq:29-0},\eqref{eq:30-0} and the ``reality''
of the action \eqref{eq:10} one finds
\begin{align}
  S_{\text{SF}}\left[\psi^{T},\left(\psi^{\dagger}\right)^{T},\Delta^{T},A^{T}\right]=&-K\left(S_{\text{SF}}\left[\psi,\psi^{\dagger},\Delta,A\right]\right)\nonumber\\
  =&-S_{\text{SF}}\left[\psi,\psi^{\dagger},\Delta,A\right],
\end{align}
so that up to a sign, time reversal is a symmetry of the action. 
It was shown in \cite{wetterich2011spinors} that, at least formally,
this sign does not effect the value of the fermionic functional integral,
and can therefore be ignored. Then time reversal symmetry defined
by \eqref{eq:29-0}, \eqref{eq:30-0} can be regraded as a symmetry of
the action in the usual sense, and one can use this fact to derive
\eqref{eq:30-1-0} using functional integrals.  

\subsubsection{Spatial reflection and time reversal in the geometric description\label{subsec:relativisitc Spatial-reflection-and}}

In this section we map and slightly generalize $R,T$, as defined
in appendix \ref{subsec:Spatial-reflections-and}, to the geometric
description of the $p$-wave SC in terms of a Majorana spinor in RC
space, given in section \ref{sec:Emergent-Riemann-Cartan-geometry}.
We will see that there is a difference between the standard notion
of $R,T$ for a spinor in $2+1$ dimensions \cite{witten2015fermion}
and the notion of $R,T$ for the $p$-wave SC, described in appendix
\ref{subsec:Spatial-reflections-and}. The reason is that our mapping
of the $p$-wave SC to a Majorana spinor maps charge to spin, and
charge is $R,T$-even, while spin is $R,T$-odd. This is a general
property of the BdG formalism. The main point is that the physical
$R,T$, coming from the $p$-wave SC, leave the mass $m$ invariant
and flip the orientation $o$, as opposed to the standard $R,T$ for
a spinor in $2+1$ dimensions, which map $m\mapsto-m$ and leave $o$
invariant. Thus, the contribution $\frac{1}{2}o\cdot\text{sgn}\left(m\right)$
of a single Majorana spinor to the Chern number is $R,T$-odd under
both notions of $R,T$, but for different reasons. 

First, by spatial reflection we mean an element of the Diffeomorphism
group that reverses the orientation of space but not of time, and
not to an internal Lorentz transformation. Since the composition
of any spatial reflection with $Diff_{0}$ is again a spatial reflection,
it suffices to consider a single spatial reflection $R$. Since spatial
reflections are just diffeomorphisms, their action on the fields is
already defined by \eqref{eq:51}, which is just the pullback 
\begin{eqnarray}
 &  & \chi\mapsto R^{*}\chi,\;e^{a}\mapsto R^{*}e^{a},\;\omega\mapsto R^{*}\omega,\label{eq:67}
\end{eqnarray}
and is a symmetry of the action $S_{\text{RC}}$. If space-time is
$\mathbb{R}_{t}\times\mathbb{R}^{2}$ it suffices to consider $R:\;y\mapsto-y$,
as was done in appendix \ref{subsec:Spatial-reflections-and}. Then
\eqref{eq:67} takes the explicit form 
\begin{align}
  \chi\left(y\right)\mapsto&\chi\left(-y\right),\label{eq:51-1}\\
  \left(e_{\;t}^{a},e_{\;x}^{a},e_{\;y}^{a}\right)\left(y\right)\mapsto&\left(e_{\;t}^{a},e_{\;x}^{a},-e_{\;y}^{a}\right)\left(-y\right),\nonumber \\
 \left(\omega_{t},\omega_{x},\omega_{y}\right)\left(y\right)\mapsto&\left(\omega_{t},\omega_{x},-\omega_{y}\right)\left(-y\right),\nonumber 
\end{align}
which maps to the transformation laws \eqref{eq:35} of the $p$-wave
SF. The orientation of space-time $o=\text{sgn}\left(\text{det}e\right)$
is odd under spatial reflections, like the orientation of space.
Note that even the flat vielbein $e_{\;\mu}^{a}=\delta_{\mu}^{a}$
transforms under $R$, which corresponds to the mapping of a $p_{x}+ip_{y}$
order parameter to a $p_{x}-ip_{y}$ order parameter by $R$.

A time reversal is any diffeomorphism that reverses the orientation
of time but not of space. It suffices to consider a single representative,
and since we work with space-times of the form $\mathbb{R}_{t}\times M_{2}$
we may take $\tau:t\mapsto-t$. Apart from the pullback by $\tau$
analogous to \eqref{eq:51-1}, $T$ also includes additional ``external''
transformations of the fields, which all trace back to the complex
conjugation included in the time reversal operation in quantum mechanics,
as in appendix \ref{subsec:Spatial-reflections-and}. As reviewed
in appendix \ref{subsec:Charge-conjugation-(Appendix)}, a complex
conjugation of the gamma matrices is implemented by $-\left(\gamma^{a}\right)^{*}=\mathcal{B}\gamma^{a}\mathcal{B}^{-1}=\mathcal{B}_{\;b}^{a}\gamma^{b}$
where $\mathcal{B}=\sigma^{x}$, $\mathcal{B}_{\;b}^{a}=\text{diag}\left[-1,1-1\right]$
in our representation. We then define the action of $T$ on the fields
by 
\begin{align}
  \chi\mapsto K\left(\tau^{*}\chi\right),\;e^{a}\mapsto\mathcal{B}_{\;b}^{a}\tau^{*}e^{b},\;\omega\mapsto\mathcal{B}\left(\tau^{*}\omega\right)\mathcal{B}^{-1},
\end{align}
where $\tau^{*}$ is the pullback by $\tau$, and $K$ is the complex
conjugation of the Grassmann algebra defined in appendix \ref{subsec:Charge-conjugation-(Appendix)}.
One can check that this is a symmetry of the action $S_{\text{RC}}$
up to an irrelevant sign already explained in appendix \ref{subsec:Spatial-reflections-and},
and that this action of $T$ reduces to the transformation laws \eqref{eq:29-0}
and \eqref{eq:30-0} of the $p$-wave SF fields.

The standard time reversal for spinors in 2+1 dimensions is given
by $T_{\text{s}}=i\sigma^{y}K$, where the phase $i$ is a matter
of convention. It is anti-unitary and $T_{\text{s}}^{2}=-1$. This
is related to $T$ through the charge conjugation matrix defined in
\ref{subsec:Charge-conjugation-(Appendix)}, 
\begin{eqnarray}
 &  & T_{\text{s}}=i\mathcal{C}T\text{ or }T=-i\mathcal{C}T_{\text{s}}.
\end{eqnarray}
This relates the time reversal $T$ that is natural in the $p$-wave SC,
to the standard time reversal $T_{\text{s}}$ and standard charge
conjugation $\mathcal{C}$. 

\textcolor{red}{}

\subsection{Global structures and obstructions\label{subsec:Global-structures-and}}

We already described the emergent geometry in a $p$-wave SC  locally
in section \ref{sec:Emergent-Riemann-Cartan-geometry}. Here we complete
the description by considering global aspects. We use some elements
from the theory of fiber bundles and characteristic classes, which
are reviewed in \cite{friedrich2000dirac,nakahara2003geometry} for
example. 

We work with space-time manifolds of the form $M_{3}=\mathbb{R}_{t}\times M_{2}$,
which represent the world volume of the $p$-wave SF. $M_{2}$ is
the sample, the two dimensional spatial surface occupied by the $p$-wave
SF, and $\mathbb{R}_{t}$ is the real line parameterizing time. Because
the order parameter is locally a vector $\Delta^{j}$ with $U\left(1\right)$
charge 2, at any time $t\in\mathbb{R}_{t}$ it is globally a map between
vector bundles $\Delta:T^{*}M_{2}\rightarrow E^{2}$, that acts by
$v_{j}\mapsto\Delta^{j}v_{j}$ \footnote{Equivalently, $\Delta$ is globally a section of $TM_{2}\otimes E^{2}$.}.
Here $T^{*}M_{2}$ is the co-tangent bundle of the sample $M_{2}$
and $E^{2}$ is the square of the electromagnetic $U\left(1\right)$
vector bundle. $E$ has fibers $\mathbb{C}$ and $U\left(1\right)$-valued
transition functions, and its topology is labeled by the monopole
number (first Chern number)  $\Phi=\frac{1}{2\pi}\int_{M_{2}}F\in\mathbb{Z}$
if $M_{2}$ has no boundary, and it is otherwise trivial. $E^{2}$
is obtained from $E$ by replacing every transition function by its
square, and therefore the topology of $E^{2}$ is labeled by $2\Phi\in2\mathbb{Z}$.
If $M_{2}$ has no boundary, the topology of the tangent bundle $TM_{2}$
(and that of $T^{*}M_{2}$) is labeled by the Euler characteristic
$\chi=2\left(1-g\right)\in2\mathbb{Z}$ where $g$ is the genus of
$M_{2}$. 

As a map $\Delta:T^{*}M_{2}\rightarrow E^{2}$, if $\Delta$ is non
singular in the sense of section \ref{subsec:The-order-parameter}
($\text{det}e\neq0$), it defines three geometric structures on $M_{2}$:
a metric, which is $g^{ij}$, an orientation, $o=\text{sgn}\left(\text{det}e\right)$,
and a spin structure, which follows from the fact that $\Delta$ has
charge 2.

To see this, we can think of $E^{2}$ as an $SO\left(2\right)$ vector
bundle, with fibers $\mathbb{R}^{2}$ and $SO\left(2\right)$ valued
transition functions. The map $\Delta$ then gives a reduction of
the structure group of $T^{*}M_{2}$ from $GL\left(2\right)$ to $SO\left(2\right)$,
thus defining a metric and an orientation. Since the transition functions
of $E^{2}$ are obtained by squaring the transition functions of $E$,
it is natural to think of $E$ as a $Spin\left(2\right)$ vector bundle\footnote{Both $Spin\left(2\right)$ and $SO\left(2\right)$ are isomorphic
as Lie groups to $U\left(1\right)$, but are related by the double
cover $Spin\left(2\right)\ni e^{i\alpha}\mapsto e^{2i\alpha}\in SO\left(2\right)$.}. $E^{2}$ therefore naturally carries a spin structure \cite{nakahara2003geometry},
and the mapping $\Delta:T^{*}M_{2}\rightarrow E^{2}$ then endows
$M_{2}$ with a spin structure. 

 The different possible spin structures correspond to an assignment
of signs $\pm1$ to non contractible loops in $M_{2}$, or more precisely
to elements of $H^{1}\left(M_{2},\mathbb{Z}_{2}\right)$. Generally,
this identification of spin structures with $H^{1}\left(M_{2},\mathbb{Z}_{2}\right)$
is not canonical, which means that there is no natural way to declare
one of the spin structures as ``trivial''.

In the simple case where $TM_{2}$ is trivial as in the case of the
torus $M_{2}=\mathbb{R}^{2}/\mathbb{Z}^{2}$, spin structures correspond
canonically to elements of $H^{1}\left(M_{2},\mathbb{Z}_{2}\right)$,
which in turn correspond to a choice of periodic or anti-periodic
boundary conditions for spinors around the non contractible loops.
The boundary condition for the BdG spinor $\Psi=\left(\psi,\psi^{\dagger}\right)^{T}$
follows from that of the microscopic spin-less fermion $\psi$, for
which it is natural to take fully periodic boundary conditions, which
is the ``trivial'' spin structure. Other boundary conditions have
been discussed in \cite{read2000paired,read2009non}.

A non singular $\Delta$ is not always possible. First, it requires
that $M_{2}$ be orientable. If $M_{2}$ is not orientable $\Delta$
would have singularities $sing\left(\Delta\right)$ such that $M_{2}-sing\left(\Delta\right)$
is orientable. $p$-wave SF on non orientable surfaces were considered
in \cite{quelle2016edge}. The other obstruction is a mismatch in
the topology of $E^{2}$ and $TM_{2}$, and is given by $2\Phi+o\chi$,
or $\Phi-\left(g-1\right)o$ \cite{read2000paired}. If the topological
invariant $2\Phi+o\chi$ does not vanish then $\Delta$ must have
singularities. A simple way to obtain this condition is to assume
$\tilde{\omega}_{12\mu}=\omega_{12\mu}=-2A_{\mu}$, which implies
$\frac{1}{2}o\sqrt{g}\tilde{\mathcal{R}}\text{d}^{2}x=\tilde{R}_{12}=\text{d}\tilde{\omega}_{12}=-2\text{d}A=-2F$,
and use the Gauss-Bonet formula $\chi=2\left(1-g\right)=\frac{1}{4\pi}\int_{M_{2}}\tilde{\mathcal{R}}\sqrt{g}\mbox{d}^{2}x$
for the Euler characteristic. The simplest example is $M_{2}=S^{2}$
the sphere, where there must be a monopole $\Phi=o=\pm1$ for a non
singular order parameter with orientation $o$. Possible singularities
of the order parameter on the sphere without a monopole have been
studied in \cite{moroz2016chiral}. There are no obstructions to the
existence of a metric and (in the two dimensional case) of a spin
structure. 

A simple way to handle singularities of $\Delta$ is to exclude them
by working with $M_{2}-sing\left(\Delta\right)$ instead of $M_{2}$.
Then $\Delta$ defines on $M_{2}-sing\left(\Delta\right)$ and orientation,
metric, and spin structure.

The emergent geometry of space-time follows from that of space due
to the simple product structure $M_{3}=\mathbb{R}_{t}\times M_{2}$.
Thus the order parameter corresponds to the (inverse) space-time vielbein
\eqref{17}, which is globally a map $T^{*}M_{3}\rightarrow E^{2},\;v_{\mu}\mapsto e_{a}^{\;\mu}v_{\mu}$
where $E^{2}$ is now viewed as an $SO\left(1,2\right)$ vector bundle.
In other words, $e$ is globally a Solder form.

\section{Quantization of coefficients for a sum of gravitational Chern-Simons terms\label{subsec:quntization-of-coefficients}}

As stated in section \ref{subsec:quantization}, gauge invariance of 
\begin{align}
K=\alpha_{1}\int_{M_{3}}Q_{3}\left(\tilde{\omega}_{\left(1\right)}\right)+\alpha_{2}\int_{M_{3}}Q_{3}\left(\tilde{\omega}_{\left(2\right)}\right)
\end{align}
for all closed $M_{3}$ implies $\alpha_{1}+\alpha_{2}\in\frac{1}{192\pi}\mathbb{Z}$. Here we sketch the derivation, following \cite{witten2007three} (section 2.1 and the discussion leading to equation (2.27)). First, only the gauge invariance of $e^{iK}$ is required, because $K$ is a contribution to the effective action, obtained by taking the logarithm of the fermionic path integral, which is a gauge invariant object. Second, the gCS term on a general $M_{3}$ is only locally given by $\alpha\int Q_{3}\left(\tilde{\omega}\right)$,
not globally. It is convenient to globally define gCS on a given
$M_{3}$ as $\alpha\int_{M_{4}}\text{tr}\left(\tilde{R}^{2}\right)$,
where $M_{4}$ is some four manifold with $M_{3}$ as a boundary,
$\partial M_{4}=M_{3}$. This is based on the fact that locally on
$M_{4}$ we have $\text{d}Q_{3}\left(\tilde{\omega}\right)=\text{tr}\left(\tilde{R}^{2}\right)$.
With this definition, we have 
\begin{align}
e^{iK_{M_{4}}}=e^{i\left[\alpha_{1}\int_{M_{4}}\text{tr}\left(\tilde{R}_{\left(1\right)}^{2}\right)+\alpha_{2}\int_{M_{4}}\text{tr}\left(\tilde{R}_{\left(2\right)}^{2}\right)\right]},
\end{align}
which is clearly gauge invariant, but we must ensure that it is also
independent of the arbitrary choice of $M_{4}$. In fact, changing
$M_{4}$ corresponds precisely to performing a large gauge transformation
on $M_{3}$, see \cite{deser1998definition} for a more direct approach.
For $M_{4}\neq M_{4}'$ such that $\partial M_{4}=M_{3}=\partial M_{4}'$,
we have 
\begin{align}
e^{iK_{M_{4}}}/e^{iK_{M_{4}'}}=e^{i\left[\alpha_{1}\int_{X_{4}}\text{tr}\left(\tilde{R}_{\left(1\right)}^{2}\right)+\alpha_{2}\int_{X_{4}}\text{tr}\left(\tilde{R}_{\left(2\right)}^{2}\right)\right]},
\end{align}
where $X_{4}$ is a closed manifold obtained by gluing $M_{4},M_{4}'$
along their shared boundary, after reversing the orientation on $M_{4}'$.
Since we start with a spin manifold $M_{3}$, we assume that $M_{4},M_{4}'$
are also spin manifolds, and therefore so is $X_{4}$. On the closed spin manifold
$X_{4}$, the Atiah-Singer index theorem implies 
\begin{align}
 \int_{X_{4}}\text{tr}\left(\tilde{R}_{\left(1\right)}^{2}\right)=\int_{X_{4}}\text{tr}\left(\tilde{R}_{\left(2\right)}^{2}\right)\in2\pi\times192\pi\mathbb{Z}.
\end{align}
In particular, one can choose $M_{4}'$ such that the integer on the right hand side is 1, in which case 
\begin{align}
 e^{iK_{M_{4}}}/e^{iK_{M_{4}'}}=e^{2\pi i\left(\alpha_{1}+\alpha_{2}\right)192\pi}.
\end{align}
An $M_{4}$-independent $e^{iK_{M_{4}}}=e^{iK}$ then requires $\alpha_{1}+\alpha_{2}\in\frac{1}{192\pi}\mathbb{Z}$.

\section{Calculation of gravitational pseudo Chern-Simons currents\label{subsec:Calculation-of-certain}}

Here we derive the contributions \eqref{eq:92-1} to the bulk currents,
which come from the gpCS term $-\beta_{1}\int_{M_{3}}\tilde{\mathcal{R}}e^{a}De_{a}$
in the effective action. We write
\begin{align}
  \delta\int_{M_{3}}\tilde{\mathcal{R}}e^{a}De_{a}=\int_{M_{3}}\left(e^{a}De_{a}\right)\delta\tilde{\mathcal{R}}+\int_{M_{3}}\tilde{\mathcal{R}}\delta\left(e^{a}De_{a}\right).\label{eq:171}
\end{align}
It's convenient to calculate the first contribution in terms of scalars
using $e^{a}De_{a}=-oc\left|e\right|\text{d}^{3}x$. We need the formula
$\delta\tilde{\mathcal{R}}=-\delta g_{\mu\nu}\tilde{\mathcal{R}}^{\mu\nu}+\left(\nabla_{\mu}\nabla_{\nu}-g_{\mu\nu}\nabla^{2}\right)\delta g_{\mu\nu}$
relating the curvature variation to the metric variation, and $\delta g_{\mu\nu}=2e_{a(\nu}\delta e_{\;\mu)}^{a}$
relating the metric variation to the vielbein variation. We find 
\begin{align}
 \int_{M_{3}}\left(e^{a}De_{a}\right)\delta\tilde{\mathcal{R}}
 =-2o\int_{M_{3}}\left|e\right|\left[\left(\tilde{\nabla}^{\mu}\tilde{\nabla}^{\nu}-g^{\mu\nu}\tilde{\nabla}^{2}\right)c-\tilde{\mathcal{R}}^{\mu\nu}c\right]e_{a\nu}\delta e_{\;\mu}^{a}.
\end{align}
The second contribution in \eqref{eq:171} is simpler to calculate
in terms of differential forms \cite{hughes2013torsional}, 
\begin{align}
  \delta\int\tilde{\mathcal{R}}&e_{a}De^{a}\label{eq:394-1-1}\\
 =&\int_{M}\tilde{\mathcal{R}}\left(\delta e^{a}T_{a}+e^{a}\mbox{d}\delta e_{a}+e^{a}\delta\omega_{ab}e^{b}+e^{a}\omega_{ab}\delta e^{b}\right)\nonumber \\
 =&\int_{M}\left(\tilde{\mathcal{R}}2\delta e_{a}T^{a}-\tilde{\mathcal{R}}\delta\omega_{ab}e^{a}e^{b}-\delta e_{a}e^{a}\mbox{d}\tilde{\mathcal{R}}\right)+\int_{\partial M}\delta e_{a}\tilde{\mathcal{R}}e^{a},\nonumber 
\end{align}
which implies 
\begin{align}
 *&\mathsf{J}^{a}=-\beta_{1}\left(2\tilde{\mathcal{R}}T^{a}-e^{a}\mbox{d}\tilde{\mathcal{R}}\right),\;*\mathsf{J}^{ab}=-\beta_{1}\left(-\tilde{\mathcal{R}}e^{a}e^{b}\right),\nonumber\\
*&\mathsf{j}^{a}=-\beta_{1}\tilde{\mathcal{R}}e^{a},\;*\mathsf{j}^{ab}=0. 
\end{align}
Here we kept track of boundary terms and calculated the contributions
to boundary currents $\mathsf{j}^{a}=\mathsf{j}_{\mu}^{\;a}\mbox{d}x^{\mu},\;\mathsf{j}^{ab}=\mathsf{j}_{\;\;\mu}^{ab}\text{d}x^{\mu}$, which are relevant for our discussion in section \ref{subsec:Additional-contributions}.
Collecting all of the bulk contributions one finds \eqref{eq:92-1}.

In section \ref{subsec:Additional-contributions} we wrote down \eqref{eq:92-1}
for a product geometry with respect to the coordinate $z$, and assumed
torsion vanishes. Here we generalize to non-zero torsion. With non
zero torsion, \eqref{eq:94-1} generalizes to 
\begin{align}
  \left\langle \mathsf{J}^{\alpha z}\right\rangle _{\text{gpCS}}=&-\beta_{1}\frac{1}{\left|e\right|}\varepsilon^{z\alpha\beta}\partial_{\beta}\tilde{\mathcal{R}},\label{eq:11-2-2}\\
 \left\langle \mathsf{J}^{z\alpha}\right\rangle _{\text{gpCS}}=&\beta_{1}\left[\frac{1}{\left|e\right|}\varepsilon^{z\alpha\beta}\partial_{\beta}\tilde{\mathcal{R}}+\frac{1}{\left|e\right|}\varepsilon^{z\beta\gamma}C_{\beta\gamma}^{\;\;\alpha}\tilde{\mathcal{R}}\right].\nonumber 
\end{align}
For $z=t$, which describes a time independent situation, we find
\begin{align}
  \left\langle J_{E}^{i}\right\rangle _{\text{gpCS}}=&\beta_{1}\varepsilon^{ij}\partial_{j}\tilde{\mathcal{R}},\label{eq:11-2-3}\\
 \left\langle P_{i}\right\rangle _{\text{gpCS}}=&-\beta_{1}\left[g_{ik}\varepsilon^{kj}\partial_{j}\tilde{\mathcal{R}}+2o\left|e\right|\tilde{\mathcal{R}}C_{12i}\right],\nonumber 
\end{align}
which generalizes \eqref{eq:92-2}. Explicit expressions for the contorsion
$C_{12i}$ are given in appendix \ref{subsec:Explicit-formulas-for}.
Equation \eqref{eq:11-2-3} is compatible with the operator equation
\eqref{eq:49}, and the density response \eqref{eq:98}.

In the case $z=y$, the inflow \eqref{eq:104-1} generalizes to
\begin{align}
  \left\langle t_{\text{cov}\;\alpha}^{y}\right\rangle _{\text{gpCS}}&=-\left|e\right|\left\langle \mathsf{J}_{\;\alpha}^{y}\right\rangle _{\text{gpCS}}\\
  &=-\beta_{1}\left[g_{\alpha\beta}\varepsilon^{\beta\gamma y}\partial_{\gamma}\tilde{\mathcal{R}}+2o\left|e\right|C_{01\alpha}\tilde{\mathcal{R}}\right].\nonumber
\end{align}
For the order parameter $\Delta=\Delta_{0}e^{i\theta\left(t,x\right)}\left(1+f\left(t,x\right),\pm i\right)$
that we consider in this case, we find using appendix \ref{subsec:Explicit-formulas-for}
that $C_{01t}=0,\;C_{01x}=e_{1}^{\;i}\frac{1}{2}\partial_{t}g_{ij}$.
The boundary current \eqref{eq:98-3} is unchanged, but the bulk+boundary
conservation equation \eqref{eq:99} is generalized to 
\begin{align}
  \tilde{\nabla}_{\alpha}\left\langle \mathsf{j}_{\;\beta}^{\alpha}\right\rangle _{\text{gpCS}}-C_{ab\beta}\left\langle \mathsf{j}^{[ab]}\right\rangle _{\text{gpCS}}=\left\langle \mathsf{J}^{y\beta}\right\rangle _{\text{gpCS}},
\end{align}
so that bulk+boundary conservation still holds for the current from
gpCS, in the presence of torsion.

\section{Perturbative calculation of the relativistic effective action \label{subsec:Perturbative-calculation-of}}

Here we present a perturbative calculation of the effective action
for the RC background fields $e,\omega$ induced by a Majorana spinor
in 2+1 dimensions. A perturbative calculation requires three types
of input: free propagators, interaction vertices, and a renormalization
scheme to handle UV divergences. In our case the propagator and vertices
are standard in the context of the coupling of relativistic fermions
to gravity, but the renormalization scheme will not be standard in
this context.

The standard renormalization schemes used in the literature are aimed
at preserving Lorentz symmetry, obtaining properly quantized coefficients
for CS terms, and obtaining finite results that do not depend on a
regulator \cite{hughes2013torsional,parrikar2014torsion}.
This is usually done as follows. First, one introduces a Lorentz invariant
regulator, such as a frequency and wave-vector cutoff $\Lambda_{\text{rel}}$,
then one introduces Pauli-Villars regulators, and tunes their masses
such that the limit $\Lambda_{\text{rel}}\rightarrow\infty$ produces
finite results and properly quantized CS coefficients. 

In contrast, we take the lattice model \eqref{eq:2-1} as a microscopic
description of the $p$-wave SC, and the relativistic continuum limit
as an approximation of it. As we obtained naturally in sections \ref{subsec:Band-structure-and}
and \ref{subsec:Coupling-the-Lattice}, this means that there are
four Majorana spinors, with different orientations and masses, and
a wave-vector cutoff $\Lambda_{UV}\sim1/a$, but no frequency cutoff,
as dictated by the lattice model. Note that these multiple Majorana
spinors are not Pauli-Villars regulators, simply because they are
all fermions. None of them has the ``wrong statistics''. The cutoff
$\Lambda_{UV}$ is a physical parameter of the model and we do not
wish to take it to infinity. Thus wave-vector integrals cannot diverge.
In contrast, since time is continuous, there is no physical frequency
cutoff, and divergences in frequency integrals do appear. These divergences
are unphysical, and can be viewed as a byproduct of the construction
of the path integral by time discretization. These divergences need
to be renormalized in the usual sense, and we do this by minimal subtraction.


To set up the perturbative calculation, we write the action $S_{\text{RC}}$
in terms of the spinor densities $\Psi=\left|e\right|^{1/2}\chi$,
and using the explicit form \eqref{eq:40}, 
\begin{align}
  S_{\text{RC}}  =\frac{1}{2}\int\mbox{d}^{3}x\overline{\Psi}&\left[\frac{1}{2}ie_{a}^{\;\mu}\gamma^{a}\overleftrightarrow{\partial_{\mu}}-\frac{1}{4}\omega_{abc}\varepsilon^{abc}-m\right]\Psi&\label{eq:40-1} \\
   =\frac{1}{2}\int\mbox{d}^{3}x\overline{\Psi}&\left[ie_{a}^{\;\mu}\gamma^{a}\partial_{\mu}+\frac{i}{2}\left(\partial_{\mu}e_{a}^{\;\mu}\right)\gamma^{a}\right.
 \left.-\frac{1}{4}\omega_{abc}\varepsilon^{abc}-m\right]\Psi.\nonumber
\end{align}
Assuming for now that the vielbein has a positive orientation, we
insert $e_{a}^{\;\mu}=\delta_{a}^{\mu}+h_{a}^{\;\mu}$ with small
$h$, and split the action into an inverse  propagator $G^{-1}$ and vertices $V$,
\begin{align}
  S_{\text{RC}}=&\frac{1}{2}\int\mbox{d}^{3}x\Psi^{\dagger}\gamma^{0}\left[G^{-1}+V\right]\Psi,\\
  G^{-1}=&i\delta_{a}^{\;\mu}\gamma^{a}\partial_{\mu}-m,\;V=V_{1}+V_{2},\nonumber \\
  V_{1}=&i\gamma^{a}h_{a}^{\;\mu}\partial_{\mu}+\frac{i}{2}\gamma^{a}\left(\partial_{\mu}h_{a}^{\;\mu}\right),\;V_{2}=-\frac{1}{4}\omega_{abc}\varepsilon^{abc}.\nonumber 
\end{align}
The vertex $V_{1}$ is first order in the perturbation $h$. The
vertex $V_{2}$ is given explicitly by 
\begin{align}
  V_{2}=-\frac{1}{4}\omega_{ab\mu}e_{c}^{\;\mu}\varepsilon^{abc}=-\frac{1}{4}\omega_{ab\mu}\delta_{c}^{\mu}\varepsilon^{abc}-\frac{1}{4}\omega_{ab\mu}h_{c}^{\;\mu}\varepsilon^{abc},
\end{align}
and therefore contains a term of order $\omega$ and a term quadratic
in the perturbations, of order $h\omega$. Terms in vertices which
are nonlinear in perturbations are sometimes called contact terms,
and the above contribution to $V_{2}$ is the only contact term in
our scheme. Note that there is no vertex related to the volume element
$\left|e\right|$, because the fundamental fermionic degree of freedom
is the spinor density $\Psi$, see appendix \ref{subsec:Equality-of-path}.
In expressions written in terms of $h_{a}^{\;\mu}$ we use $\eta_{\mu\nu}$
to raise and lower coordinate indices and $\delta_{a}^{\mu}$ to map
internal indices to coordinate indices, so in practice there is no
difference between these indices in such expressions. 

The perturbative expansion of the effective action is given by 
\begin{align}
  2W_{\text{RC}}  =&-2i\log\mbox{Pf}\left(i\gamma^{0}\left(G^{-1}+V\right)\right)\\
  =&-i\mbox{Tr}\left(\log i\gamma^{0}G^{-1}\right)\nonumber\\
  &-i\mbox{Tr}\left(GV\right)+\frac{i}{2}\mbox{Tr}\left(GV\right)^{2}+O\left(V^{3}\right),\nonumber 
\end{align}
which, apart from the first term, is a sum over Feynman diagrams with
a fermion loop and any number of vertices $V$. We will be interested
in $W_{\text{RC}}$ to second order in the perturbations $h$ and
$\omega$ and up to third order in derivatives. Terms of first oder
in $h,\omega$ correspond to properties of the unperturbed ground
state, or vacuum, while terms of second order correspond to linear
response coefficients. The first term is independent of $h,\omega$
and corresponds to the ground state energy of the unperturbed system.
This information can also be obtained from the term linear in $h$,
and we therefore ignore $\mbox{Tr}\log i\gamma^{0}G^{-1}$ in the
following. Expanding the vertices, 
\begin{align}
  2W_{\text{RC}}=&-i\mbox{Tr}\left(GV_{1}\right)-i\mbox{Tr}\left(GV_{2}\right) +\frac{i}{2}\mbox{Tr}\left(GV_{1}\right)^{2}\label{eq:80}\\
 &+\frac{i}{2}\mbox{Tr}\left(GV_{2}\right)^{2}+i\mbox{Tr}\left(GV_{1}GV_{2}\right)+O\left(V^{3}\right).\nonumber 
\end{align}
These functional traces can now be written as integrals over Fourier
components and traces over spinor indices, 

\begin{align}
 \mbox{Tr}\left(GV_{1}\right)=&-h_{a}^{\;\mu}\left(p=0\right)\int_{q}q_{\mu}\mbox{tr}\left(\gamma^{a}G_{q}\right),\\
  \mbox{Tr}\left(GV_{2}\right)=&\omega\left(p=0\right)\int_{q}\mbox{tr}\left(G_{q}\right),\nonumber \\
  \mbox{Tr}\left(GV_{1}\right)^{2}=&\int_{p}h_{a}^{\;\mu}\left(p\right)h_{b}^{\;\nu}\left(-p\right)\int_{q}\left(q+\frac{1}{2}p\right)_{\mu}\left(q+\frac{1}{2}p\right)_{\nu}\text{tr}\left(\gamma^{a}G_{q}\gamma^{b}G_{p+q}\right),\nonumber \\
  \mbox{Tr}\left(GV_{2}\right)^{2}=&\int_{p}\omega\left(p\right)\omega\left(-p\right)\int_{q}\text{tr}\left(G_{q}G_{p+q}\right),\nonumber\\
  \mbox{Tr}\left(GV_{1}GV_{2}\right)=&-\int_{p}h_{a}^{\;\mu}\left(p\right)\omega\left(-p\right)\int_{q}\left(q+\frac{1}{2}p\right)_{\mu}\text{tr}\left(\gamma^{a}G_{q}G_{p+q}\right),\nonumber 
\end{align}

where $\omega=-\frac{1}{4}\omega_{ab\mu}e_{c}^{\;\mu}\varepsilon^{abc}$,
and $\int_{p}=\int\frac{\text{d}^{3}p}{\left(2\pi\right)^{3}}$. Our
conventions for the Fourier transform of a function $f$ is $f\left(x\right)=\int_{q}e^{iq_{\mu}x^{\mu}}f\left(q\right)$.
The Fourier transform of the Greens function is then $G_{q}=-\frac{1}{ \cancel{q}+m}=-\frac{\cancel{q}-m}{q^{2}-m^{2}}$, where $\cancel{q}=q_a\gamma^a$.
The spinor traces are evaluated using the usual identities for gamma
matrices in 2+1 dimensions, 
\begin{align}
  \text{tr}&\left(\gamma^{a}\right)=0,\;\text{tr}\left(\gamma^{a}\gamma^{b}\right)=2\eta^{ab},\;\text{tr}\left(\gamma^{a}\gamma^{b}\gamma^{c}\right)=\pm2i\varepsilon^{abc},\nonumber\\
  \text{tr}&\left(\gamma^{a}\gamma^{b}\gamma^{c}\gamma^{d}\right)=2\left(\eta^{ab}\eta^{cd}-\eta^{ac}\eta^{bd}+\eta^{ad}\eta^{bc}\right).\label{eq:177} 
\end{align}
The sign $\pm$ distinguishes the two inequivalent representations
of gamma matrices in 2+1 dimensions, and with our chosen representation,
$\text{tr}\left(\gamma^{a}\gamma^{b}\gamma^{c}\right)=2i\varepsilon^{abc}$.
Using these identities yields for the single vertex diagrams 
\begin{eqnarray}
 &  & \mbox{Tr}\left(GV_{1}\right)=2\eta^{ab}h_{a}^{\;\mu}\left(p=0\right)\int_{q}\frac{q_{\mu}q_{b}}{q^{2}-m^{2}},\label{eq:83-1}\\
 &  & \mbox{Tr}\left(GV_{2}\right)=2m\omega\left(p=0\right)\int_{q}\frac{1}{q^{2}-m^{2}}.\nonumber 
\end{eqnarray}
The expressions for the diagrams with two vertices are more complicated,
so let us start by analyzing the single vertex diagrams. This will
suffice to demonstrate our renormalization scheme and compare it to
direct calculations within the lattice model and to renormalizations
which are more natural in the context of relativistic QFT.

\subsection{Single vertex diagrams }

From \eqref{eq:80} and \eqref{eq:83-1} it follows that 
\begin{align}
 W_{\text{RC}}=\Lambda_{\;\mu}^{a}\int\text{d}^{3}xh_{a}^{\;\mu}+s\int\text{d}^{3}x\omega+O\left(V^{2}\right),
\end{align}
where
\begin{align}
  \Lambda_{\;\mu}^{a}=-i\eta^{ab}\int\frac{\text{d}^{3}q}{\left(2\pi\right)^{3}}\frac{q_{\mu}q_{b}}{q^{2}-m^{2}},\;s=-i\int\frac{\text{d}^{3}q}{\left(2\pi\right)^{3}}\frac{m}{q^{2}-m^{2}},
\end{align}
can now be recognized as the energy-momentum tensor and spin density
of the unperturbed ground state, 
\begin{align}
    \left\langle \mathsf{J}_{\;\mu}^{a}\right\rangle =&-\Lambda_{\;\mu}^{a},\\
\left\langle \mathsf{J}^{ab\mu}\right\rangle =&-\frac{1}{4}\frac{1}{2}\left\langle \overline{\chi}\chi\right\rangle \delta_{c}^{\mu}\varepsilon^{abc}=-\frac{1}{4}s\delta_{c}^{\mu}\varepsilon^{abc}.\nonumber
\end{align}
 Preforming a Wick rotation $q_{0}\mapsto iq_{0}$, 
\begin{align}
  \Lambda_{\;\mu}^{a}=\delta^{ab}\int\frac{\text{d}^{3}q}{\left(2\pi\right)^{3}}\frac{q_{\mu}q_{b}}{\left|q\right|^{2}+m^{2}},\;s=-\int\frac{\text{d}^{3}q}{\left(2\pi\right)^{3}}\frac{m}{\left|q\right|^{2}+m^{2}},\label{eq:181}
\end{align}
where $\left|\cdot\right|$ is the euclidian norm. We start by calculating
$s$ in our lattice motivated renormalization scheme. In this scheme
the integral reads 
\begin{align}
  s=-\int_{\left|\boldsymbol{q}\right|<\Lambda_{UV}}\frac{\text{d}^{2}\boldsymbol{q}}{\left(2\pi\right)^{2}}\int_{-\infty}^{\infty}\frac{\text{d}q_{0}}{2\pi}\frac{m}{q_{0}^{2}+\left|\boldsymbol{q}\right|^{2}+m^{2}},
\end{align}
where $\Lambda_{UV}$ is a physical cutoff related to the lattice
spacing by $\Lambda_{UV}\sim a^{-1}$. The $q_{0}$ integral converges,
and does not require renormalization. It yields the result within
the lattice motivated scheme, 
\begin{align}
 s=-\frac{1}{2}\int_{\left|\boldsymbol{q}\right|<\Lambda_{UV}}\frac{\text{d}^{2}\boldsymbol{q}}{\left(2\pi\right)^{2}}\frac{m}{\sqrt{\left|\boldsymbol{q}\right|^{2}+m^{2}}},\label{eq:183}
\end{align}
and adding the operator ordering correction gives the ground state
charge density 
\begin{align}
  \rho=\left\langle J^{t}\right\rangle =-\frac{1}{2}\int_{\left|\boldsymbol{q}\right|<\Lambda_{UV}}\frac{\text{d}^{2}\boldsymbol{q}}{\left(2\pi\right)^{2}}\left(1-\frac{m}{\sqrt{\left|\boldsymbol{q}\right|^{2}+m^{2}}}\right).
\end{align}
After summing over low energy Majorana spinors and restoring units, this coincides
with the relativistic limit of the exact ground state charge density of the
lattice model \cite{read2000paired}, 
\begin{align}
  \rho=-\frac{1}{2}\int_{BZ}\frac{\mbox{d}^{2}\boldsymbol{q}}{\left(2\pi\right)^{2}}\left(1-\frac{h_{\boldsymbol{q}}}{\sqrt{\left|\Delta_{\boldsymbol{q}}\right|^{2}+h_{\boldsymbol{q}}^{2}}}\right),
\end{align}
where $h_{\boldsymbol{q}},\Delta_{\boldsymbol{q}}$ were defined
in section \ref{subsec:Band-structure-and}. 

For comparison we calculate the $s$ integral in a standard renormalization
scheme of relativistic QFT. In this approach the integral does not
converge. We introduce a frequency and wave-vector cutoff $\Lambda_{\text{rel}}$,
and restrict the integration to $\left|q\right|<\Lambda_{\text{rel}}$.
This yields 
\begin{align}
  s=-\int_{\left|q\right|<\Lambda_{\text{rel}}}\frac{\text{d}^{3}q}{\left(2\pi\right)^{3}}\frac{m}{\left|q\right|^{2}+m^{2}}
  =-\frac{\Lambda_{\text{rel}}m}{2\pi^{2}}+\frac{m^{2}\text{sgn}m}{4\pi}+O\left(\frac{m}{\Lambda_{\text{rel}}}\right).
\end{align}
A simple way to proceed is to preform minimal subtraction, which means
we remove the diverging piece, and take $\Lambda_{\text{rel}}/m\rightarrow\infty$.
This gives the fully relativistic result
\begin{eqnarray}
 s=\frac{m^{2}\text{sgn}m}{4\pi}.
\end{eqnarray}
Comparing with \eqref{eq:72} we find $\zeta_{H}=s=\frac{m^{2}\text{sgn}m}{4\pi}$
for a positive orientation which is essentially the torsional Hall
viscosity of \cite{hughes2013torsional}\footnote{It is not exactly the same result because we did not use the same
relativistic renormalization scheme.}. The relativistic result can also be obtained by expanding the lattice
result \eqref{eq:183} in $\Lambda_{UV}$ and keeping the $O\left(1\right)$
piece. This is a general feature, the $O\left(1\right)$ piece of
any coefficient in the effective action is always relativistic. 

Let us now turn to the calculation of the ground state energy-momentum
tensor $\Lambda_{\;\mu}^{a}$. With a relativistic regulator $\Lambda_{\;\mu}^{a}$ is $O\left(3\right)$ invariant and must therefore be proportional to the identity,
\begin{align}
\Lambda_{\;\mu}^{a}=\delta^{ab}\int_{\left|q\right|<\Lambda_{\text{rel}}}\frac{\text{d}^{3}q}{\left(2\pi\right)^{3}}\frac{q_{\mu}q_{b}}{\left|q\right|^{2}+m^{2}}=\delta_{\mu}^{a}\frac{\Lambda}{2\kappa_{N}},
\end{align}
with the cosmological constant 
\begin{align}
  \frac{\Lambda}{2\kappa_{N}}=\frac{1}{3}\int_{\left|q\right|<\Lambda_{\text{rel}}}\frac{\text{d}^{3}q}{\left(2\pi\right)^{3}}\frac{\left|q\right|^{2}}{\left|q\right|^{2}+m^{2}}
  =\frac{1}{3}\left[\frac{\Lambda_{\text{rel}}^{3}}{6\pi^{2}}-\frac{\Lambda_{\text{rel}}m^{2}}{2\pi^{2}}+\frac{\left|m\right|^{3}}{4\pi}+O\left(\frac{m}{\Lambda_{\text{rel}}}\right)\right].
\end{align}
Keeping the $O\left(1\right)$ piece we find the relativistic expression
\begin{align}
   \Lambda_{\;\mu}^{a}=\delta_{\mu}^{a}\frac{\Lambda}{2\kappa_{N}}=\delta_{\mu}^{a}\frac{\left|m\right|^{3}}{6\pi},
\end{align}
which again, is essentially the result of \cite{hughes2013torsional}.
With the lattice motivated renormalization scheme, 
\begin{align}
  \Lambda_{\;\mu}^{a}=\delta^{ab}\int_{\left|\boldsymbol{q}\right|<\Lambda_{UV}}\frac{\text{d}^{2}\boldsymbol{q}}{\left(2\pi\right)^{2}}\int_{-\infty}^{\infty}\frac{\text{d}q_{0}}{2\pi}\frac{q_{\mu}q_{b}}{q_{0}^{2}+\left|\boldsymbol{q}\right|^{2}+m^{2}}.
\end{align}
Here the $q_{0}$ integral for $\Lambda_{\;t}^{0}$ does not converge,
and needs to be regularized. We do this by introducing a frequency
cutoff $\Lambda_{0}$, 
\begin{align}
  \Lambda_{\;\mu}^{a}=\delta^{ab}\int_{\left|\boldsymbol{q}\right|<\Lambda_{UV}}\frac{\text{d}^{2}\boldsymbol{q}}{\left(2\pi\right)^{2}}\int_{-\Lambda_{0}}^{\Lambda_{0}}\frac{\text{d}q_{0}}{2\pi}\frac{q_{\mu}q_{b}}{q_{0}^{2}+\left|\boldsymbol{q}\right|^{2}+m^{2}}.
\end{align}
Unlike $\Lambda_{UV}$ which is a physical parameter of the model,
$\Lambda_{0}$ is a fictitious cutoff which we take to infinity at
the end of the calculation. The $q_{0}$ divergence can be interpreted
as an artifact of time discretization \cite{shankar1994renormalization}. At this point
the domain of integration is not a ball in Euclidian Fourier space
but a cylinder, so it is not invariant under $O\left(3\right)$, only
under $O\left(2\right)$\footnote{More accurately, the domain of integration for each lattice fermion
is not the disk $\left\{ \left|\boldsymbol{q}\right|<\Lambda_{UV}\right\} $
but the square $\left[-\Lambda_{UV}/2,\Lambda_{UV}/2\right]^{2}$
with $\Lambda_{UV}=\pi/a$, which is a quarter of the Brillouin
zone $BZ$, see section \ref{subsec:Coupling-the-Lattice}. The symmetry
group of this domain is not $O\left(2\right)$ but the point group
symmetry of the lattice $D_{4}\subset O\left(2\right)$. This subtlety
has no effect on the following. } and the reflection $q\mapsto-q$. This implies that the tensor $\Lambda_{\;\mu}^{a}$
takes the form 
\begin{align}
  \Lambda_{\;t}^{0}=&\int_{\left|\boldsymbol{q}\right|<\Lambda_{UV}}\frac{\text{d}^{2}\boldsymbol{q}}{\left(2\pi\right)^{2}}\int_{-\Lambda_{0}}^{\Lambda_{0}}\frac{\text{d}q_{0}}{2\pi}\frac{q_{0}^{2}}{q_{0}^{2}+\left|\boldsymbol{q}\right|^{2}+m^{2}},\\
  \Lambda_{\;j}^{A}=&\frac{1}{2}\delta_{j}^{A}\int_{\left|\boldsymbol{q}\right|<\Lambda_{UV}}\frac{\text{d}^{2}\boldsymbol{q}}{\left(2\pi\right)^{2}}\int_{-\Lambda_{0}}^{\Lambda_{0}}\frac{\text{d}q_{0}}{2\pi}\frac{\left|\boldsymbol{q}\right|^{2}}{q_{0}^{2}+\left|\boldsymbol{q}\right|^{2}+m^{2}},\nonumber
\end{align}
with all other components vanishing. The $q_{0}$ integral for
the energy density $\Lambda_{\;t}^{0}$ gives 
\begin{align}
  \Lambda_{\;t}^{0}=\frac{1}{2}\int_{\left|\boldsymbol{q}\right|<\Lambda_{UV}}\frac{\mbox{d}^{2}\boldsymbol{q}}{\left(2\pi\right)^{2}}\left[\frac{2\Lambda_{0}}{\pi}-\sqrt{\left|\boldsymbol{q}\right|^{2}+m^{2}}+O\left(\frac{m}{\Lambda_{0}}\right)\right].
\end{align}
Keeping the $O\left(1\right)$ piece we find, within the lattice motivated scheme, 
\begin{eqnarray}
  \Lambda_{\;t}^{0}=-\frac{1}{2}\int_{\left|\boldsymbol{q}\right|<\Lambda_{UV}}\frac{\mbox{d}^{2}\boldsymbol{q}}{\left(2\pi\right)^{2}}\sqrt{\left|\boldsymbol{q}\right|^{2}+m^{2}},\label{eq:196}
\end{eqnarray}
which is the familiar expression for the ground state energy of a
single Majorana fermion, which is half the energy of a filled Dirac
sea. The $q_{0}$ integral for $\Lambda_{\;j}^{A}$ converges and
gives the pressure 
\begin{align}
  \Lambda_{\;j}^{A}=\frac{1}{2}\delta_{j}^{A}\frac{1}{2}\int_{\left|\boldsymbol{q}\right|<\Lambda_{UV}}\frac{\mbox{d}^{2}\boldsymbol{q}}{\left(2\pi\right)^{2}}\frac{\left|\boldsymbol{q}\right|^{2}}{\sqrt{\left|\boldsymbol{q}\right|^{2}+m^{2}}}.
\end{align}
We see that the ground state energy density and pressure are no longer
equal. In other words the ground state energy-momentum tensor is not
Lorentz invariant, due to the lattice renormalization
scheme. It may be surprising that the expression \eqref{eq:196} for
the energy density is part of an energy momentum tensor which is not
Lorentz invariant. This has been discussed in the literature in the
context of the cosmological constant problem \cite{ossola2003considerations,koksma2011cosmological,visser2016lorentz}. 

Let us now compare the above with the lattice model. For the energy
density we need to add the operator ordering correction, 

\begin{align}
  \varepsilon=\left\langle t_{\text{cov}\;t}^{t}\right\rangle =\frac{1}{2}\int_{\left|\boldsymbol{q}\right|<\Lambda_{UV}}\frac{\mbox{d}^{2}\boldsymbol{q}}{\left(2\pi\right)^{2}}\left(m-\sqrt{\left|\boldsymbol{q}\right|^{2}+m^{2}}\right).
\end{align}
Restoring units and summing over Dirac points, we recognize the above
as the relativistic approximation of the ground state energy density
of the lattice model \cite{volovik2009universe}, 
\begin{align}
  \varepsilon=\frac{1}{2}\int_{BZ}\frac{\text{d}^{2}\boldsymbol{q}}{\left(2\pi\right)^{2}}\left(h_{\boldsymbol{q}}-E_{\boldsymbol{q}}\right).
\end{align}
The above calculations of simple ground state properties serve as
consistency checks. We have seen explicitly that these quantities
are UV sensitive. With the lattice motivated renormalization scheme
the effective action produces physical quantities that approximate
those of the lattice model, which are distinct from those obtained
with a relativistic scheme. In the following we will focus on UV insensitive
terms. In doing so we will also ignore operator ordering corrections,
because these always contain $\delta^{2}\left(0\right)\sim\int_{\left|\boldsymbol{q}\right|<\Lambda_{UV}}\frac{\text{d}^{2}\boldsymbol{q}}{\left(2\pi\right)^{2}}\sim\Lambda_{UV}^{2}$
and are therefore UV sensitive.

\subsection{Two vertex diagrams}

Let us now turn to the calculation of the more interesting second
order terms, which correspond to linear responses. After preforming
the traces over gamma matrices one finds

\begin{align}
  \mbox{Tr}\left(GV_{1}\right)^{2}=&  -2im\varepsilon^{abc}\int_{p}h_{a}^{\;\mu}\left(p\right)h_{b}^{\;\nu}\left(-p\right)p_{c}\int_{q}\frac{\left(q+\frac{1}{2}p\right)_{\mu}\left(q+\frac{1}{2}p\right)_{\nu}}{\left(q^{2}-m^{2}\right)\left(\left(p+q\right)^{2}-m^{2}\right)}\label{eq:200}\\
  &+2m^{2}\eta^{ab}\int_{p}h_{a}^{\;\mu}\left(p\right)h_{b}^{\;\nu}\left(-p\right)\int_{q}\frac{\left(q+\frac{1}{2}p\right)_{\mu}\left(q+\frac{1}{2}p\right)_{\nu}}{\left(q^{2}-m^{2}\right)\left(\left(p+q\right)^{2}-m^{2}\right)}\nonumber \\
  &+2\left(\eta^{ac}\eta^{bd}-\eta^{ab}\eta^{cd}+\eta^{ad}\eta^{cb}\right)\int_{p}h_{a}^{\;\mu}\left(p\right)h_{b}^{\;\nu}\left(-p\right)\int_{q}\frac{\left(q+\frac{1}{2}p\right)_{\mu}\left(q+\frac{1}{2}p\right)_{\nu}q_{c}\left(p+q\right)_{d}}{\left(q^{2}-m^{2}\right)\left(\left(p+q\right)^{2}-m^{2}\right)},\nonumber\\ 
 \mbox{Tr}\left(GV_{2}\right)^{2}=&\int_{p}\omega\left(p\right)\omega\left(-p\right)\int_{q}\frac{2\eta^{ab}q_{a}\left(p+q\right)_{b}+2m^{2}}{\left(q^{2}-m^{2}\right)\left(\left(p+q\right)^{2}-m^{2}\right)},\\
 \mbox{Tr}\left(GV_{1}GV_{2}\right)= & -2i\varepsilon^{abc}\int_{p}h_{a}^{\;\mu}\left(p\right)\omega\left(-p\right)p_{c}\int_{q}\frac{\left(q+\frac{1}{2}p\right)_{\mu}q_{b}}{\left(q^{2}-m^{2}\right)\left(\left(p+q\right)^{2}-m^{2}\right)}\label{eq:202}\\
 &+2m\eta^{ab}\int_{p}h_{a}^{\;\mu}\left(p\right)\omega\left(-p\right)\int_{q}\frac{\left(q+\frac{1}{2}p\right)_{\mu}\left(2q+p\right)_{b}}{\left(q^{2}-m^{2}\right)\left(\left(p+q\right)^{2}-m^{2}\right)}.\nonumber 
\end{align}

One is then left with the calculation of the integrals over the loop
momenta $q$ in the above equations. The first step in doing so is
Wick rotating to euclidian signature by changing $q_{0}\mapsto iq_{0},\;p_{0}\mapsto ip_{0}$
in the $q$ integrals. 

At this point one can use Feynman parameters to simplify the form
of the integrands, but since we are only interested in the effective
action to low orders in derivatives of the background fields, we find
it simpler to expand the integrands in powers of $p/m$. 

We start with the first integral in \eqref{eq:200}, which contains
the gCS term. Expanding the integrand in $p/m$ we find

\begin{align}
  \frac{\left(q+\frac{1}{2}p\right)_{\mu}\left(q+\frac{1}{2}p\right)_{\nu}}{\left(q^{2}+m^{2}\right)\left(\left(p+q\right)^{2}+m^{2}\right)}=&
 \frac{q_{\mu}q_{\nu}}{\left(m^{2}+q^{2}\right)^{2}} 
+\left[\frac{p_{(\mu}q_{\nu)}}{\left(m^{2}+q^{2}\right)^{2}}-2\frac{q_{\mu}q_{\nu}p\cdot q}{\left(m^{2}+q^{2}\right)^{3}}\right] \label{eq:H33}\\
  &+\left[\frac{p_{\mu}p_{\nu}}{4\left(m^{2}+q^{2}\right)^{2}}-\frac{p^{2}q_{\mu}q_{\nu}}{\left(m^{2}+q^{2}\right)^{3}}-\frac{2p_{(\mu}q_{\nu)}p\cdot q}{\left(m^{2}+q^{2}\right)^{3}}+\frac{4q_{\mu}q_{\nu}\left(p\cdot q\right)^{2}}{\left(m^{2}+q^{2}\right)^{4}}\right]
 +O\left(p^{3}\right)\nonumber 
\end{align}

where terms are grouped according to their order in $p/m$.The $q$ integral over
the $O\left(1\right)$ terms diverges, and therefore produces
a UV sensitive term in the effective action. With a relativistic renormalization
we find 
\begin{align}
  2W_{\text{RC}}=\frac{m^{2}\text{sgn}\left(m\right)}{4\pi}\int\text{d}^{3}xh_{a}^{\;\mu}\eta_{\mu\nu}\varepsilon^{abc}\partial_{b}h_{c}^{\;\nu}+\cdots
\end{align}
Comparing with \eqref{eq:72} and using $e_{a}De^{a}=\varepsilon^{abc}h_{a\nu}\partial_{b}h_{c}^{\;\nu}\text{d}^{3}x+\cdots$
we find again the torsional Hall viscosity $\zeta_{H}=\frac{m^{2}\text{sgn}\left(m\right)}{4\pi}$,
for positive orientation. With the lattice renormalization the $\eta_{\mu\nu}$
in the above is replaced by a non Lorentz invariant tensor, but in Sec.\ref{sec:Main-section-1:} we are only interested in UV insensitive responses, so we will not discuss it here further (see Appendix \ref{zeroth-order}). 

The $q$ integral over the $O\left(p/m\right)$ terms vanishes because
it is odd under the reflection $q\mapsto-q$.

The $O\left(p^{2}/m^{2}\right)$ contributions are most interesting for
us. The $q$ integral over these converges, and therefore produces
UV insensitive terms in the effective action. Instead of calculating
the integral with the finite physical $\Lambda_{UV}$, we can calculate
it with $\Lambda_{UV}/m=\infty$ at the expense of producing small $O\left(m/\Lambda_{UV}\right)$
corrections. Then the calculation
reduces to a standard calculation within relativistic QFT which has
appeared a few times in the literature with slightly different conventions  \cite{goni1986massless,van1986topological,vuorio1986parity,vuorio1986parityErr}, and which is done below for completeness. See also \cite{kurkov2018gravitational} for a recent heat kernel calculation and review of the literature, and \cite{bonetti2013one} for similar computations in 4+1 dimensions. 
With $\Lambda_{UV}/m=\infty$ the integral is Lorentz invariant and
this implies the standard reductions to radial functions such as $q_{\mu}q_{\nu}\mapsto\frac{1}{3}\eta_{\mu\nu}q^{2}$.
The $O\left(p^{2}/m^{2}\right)$ contributions in \eqref{eq:H33} then reduce to
\begin{align}
  \frac{1}{4}&\frac{p_{\mu}p_{\nu}}{\left(m^{2}+q^{2}\right)^{2}}-\frac{1}{3}\frac{p^{2}\eta_{\mu\nu}q^{2}}{\left(m^{2}+q^{2}\right)^{3}}-\frac{2}{3}\frac{p_{\mu}p_{\nu}q^{2}}{\left(m^{2}+q^{2}\right)^{3}}\nonumber\\
  &+\frac{4}{15}\frac{p^{2}\eta_{\mu\nu}+2p_{\mu}p_{\nu}}{\left(m^{2}+q^{2}\right)^{4}}q^{4},
\end{align}
and preforming the $q$ integral yields 
\begin{align}
  \frac{1}{96\pi\left|m\right|}\left(p_{\mu}p_{\nu}-p^{2}\eta_{\mu\nu}\right).
\end{align}
This corresponds to the following term in the effective action
\begin{align}
  2W_{\text{RC}}=&\frac{\text{sgn}\left(m\right)}{2}\frac{1}{96\pi}2\int\text{d}^{3}xh_{a}^{\;\mu}\varepsilon^{abc}\partial_{c}\left(\partial_{\mu}\partial_{\nu}-\partial^{2}\eta_{\mu\nu}\right)h_{b}^{\;\nu}\nonumber\\
  &+\cdots
\end{align}
To identify this term it is easiest to fix a Lorentz gauge where
$h_{\left[\mu\nu\right]}=0$. In terms of the $p$-wave SC this corresponds
to $U\left(1\right)$ gauge fixing the phase $\theta$ of the order
parameter to $0$, along with an additional boost which is only a symmetry 
in the relativistic limit. Then $h$ corresponds also to the first
order metric perturbation, $g_{\mu\nu}=\eta_{\mu\nu}-2h_{(\mu\nu)}=\eta_{\mu\nu}-2h_{\mu\nu}$,
and the above corresponds to the expansion of the gCS term 
\begin{eqnarray}
 &  & 2W_{\text{RC}}=\frac{\text{sgn}\left(m\right)}{2}\frac{1}{96\pi}\int Q_{3}\left(\tilde{\Gamma}\right)+\cdots\label{eq:H38}
\end{eqnarray}
to second order in $h$. In preforming such expansions we found the
Mathematica package xAct very useful \cite{brizuela2009xpert,xAct}.
Equation \eqref{eq:H38} corresponds to $\kappa_{H}=\frac{1}{48\pi}\frac{\text{sgn}\left(m\right)}{2}$. We note that within the perturbative calculation there is no difference between $\int Q_{3}\left(\tilde{\Gamma}\right)$
and $\int Q_{3}\left(\tilde{\omega}\right)$, see \eqref{eq:70}.

The above result is valid for a vielbein $e_{a}^{\;\mu}=\delta_{a}^{\mu}+h_{a}^{\;\mu}$
which has a positive orientation. A vielbein with a negative orientation
can be written as $e_{a}^{\;\mu}=L_{\;a}^{b}\left(\delta_{b}^{\mu}+h_{b}^{\;\mu}\right)$
where $L$ is a Lorentz transformation with $\text{det}L=-1$. We
can deal with such vielbeins by absorbing $L$ into the gamma matrices,
$\gamma^{a}\mapsto L_{\;a}^{b}\gamma^{a}$. The only effect that this
change has on the traces \eqref{eq:177} is changing $\text{tr}\left(\gamma^{a}\gamma^{b}\gamma^{c}\right)=2i\varepsilon^{abc}$
to $\text{tr}\left(\gamma^{a}\gamma^{b}\gamma^{c}\right)=-2i\varepsilon^{abc}$.
The metric is independent of the orientation and so is $\tilde{\Gamma}$,
so the result valid for both orientations is 
\begin{eqnarray}
 &  & 2W_{\text{RC}}=\frac{\text{sgn}\left(m\right)o}{2}\frac{1}{96\pi}\int Q_{3}\left(\tilde{\Gamma}\right)+\cdots
\end{eqnarray}
where $o=\text{sgn}\left(\text{det}e\right)$ is the orientation.
The second and third lines in \eqref{eq:200} correspond, with a relativistic
regulator, to $O\left(h^{2}\right)$ contributions to the cosmological
constant and E-H term which are UV sensitive. 

One can compute the other traces in the same manner. The only additional
UV insensitive contribution comes from the second integral in \eqref{eq:202}.
It is given by 
\begin{align}
  2W_{\text{RC}}=&\frac{\text{sgn}\left(m\right)}{2}\frac{1}{96\pi}\int\text{d}^{3}x4\omega\left(\partial_{\mu}\partial_{\nu}-\partial^{2}\eta_{\mu\nu}\right)2h^{\mu\nu}\nonumber\\
  &+\cdots
\end{align}
This corresponds to the expansion of the gpCS term to second
order in the vertices,
\begin{align}
  2W_{\text{RC}}  =\frac{\text{sgn}\left(m\right)}{2}\frac{1}{96\pi}\int\text{d}^{3}x\left|e\right|\tilde{\mathcal{R}}c+\cdots
\end{align}
where we have used \eqref{eq:76}, the expansion of the curvature
$\tilde{\mathcal{R}}=-2\left(\partial_{a}\partial_{b}-\partial^{2}\eta_{ab}\right)h^{ab}+O\left(h^{2}\right)$,
the definition $c=\varepsilon^{abc}\left(\omega_{abc}-\tilde{\omega}_{abc}\right)$,
and the expansion $\tilde{\omega}_{abc}\varepsilon^{abc}=-\varepsilon^{abc}\partial_{a}h_{bc}+O\left(h^{2}\right)$
of the LC spin connection. Note that in the Lorentz gauge $h_{\left[ab\right]}=0$,
$\tilde{\omega}_{abc}\varepsilon^{abc}$ vanishes to first order.
This completes the calculation of the UV insensitive terms in the
effective action which we have studied in Sec.\ref{sec:Main-section-1:}.

\section{Non-relativistic geometric quantities and their perturbative expansion\label{sec:Geometric quantities}}

We write $E_{A}^{\;i}=\delta_{A}^{\;i}+H_{A}^{\;i}$ for the inverse vielbein, and expand the relevant geometric quantities in $H$. For the inverse metric $G^{ij}=E_{A}^{\;i}\delta^{AB}E_{B}^{\;j}$ and volume element $\sqrt{G}=\left|E\right|=\left|\text{det}\left(E_{A}^{\;i}\right)\right|$ we find  \begin{align} G^{ij}= & \delta^{ij}+2H^{(ij)}+H_{A}^{\;i}H^{Aj}\label{eq:48-0}\\ = & \delta^{ij}+\delta G^{ij},\nonumber \\ \sqrt{G}= & 1-H_{A}^{\;A}+\frac{1}{2}H_{A}^{\;A}H_{B}^{\;B}+\frac{1}{2}H_{A}^{\;B}H_{B}^{\;A}+O\left(H^{3}\right),\nonumber \\ \log\sqrt{G}= & -H_{A}^{\;A}+\frac{1}{2}H_{A}^{\;B}H_{B}^{\;A}+O\left(H^{3}\right),\nonumber  \end{align} where, in expanded expressions, all index manipulations are trivial, and in particular, there is no difference between coordinate indices $i,j$ and $SO\left(2\right)_{L}$ indices $A,B$. Note that the strain used in Sec.\ref{sec:Main-section-2:} is given by $u_{ij}=\left(G_{ij}-\delta_{ij}\right)/2=-H_{(ij)}+O\left(H^{2}\right)$. We use the notation $\varepsilon^{\mu\nu\rho}$ for the totally anti-symmetric (pseudo) tensor, normalized such that $\varepsilon^{xyt}=1/\sqrt{G}$, as well as $\varepsilon^{ij}=\varepsilon^{ijt}$. 
The non-relativistic spin connection used in Sec.\ref{sec:Main-section-2:} is the $SO\left(2\right)_{L}$ connection \begin{align} \omega_{t}= & \frac{1}{2}\varepsilon^{AB}E_{Ai}\partial_{t}E_{B}^{\;i}\label{eq:49-0}\\ = & -\frac{1}{2}\partial_{t}\left(\varepsilon^{AB}H_{AB}\right)-\frac{1}{2}\varepsilon^{AB}H_{iA}\partial_{t}H_{B}^{\;i}+O\left(H^{3}\right),\nonumber \\ \omega_{j}= & \frac{1}{2}\left(\varepsilon^{AB}E_{Ai}\partial_{j}E_{B}^{\;i}-\frac{1}{E}\varepsilon^{kl}\partial_{k}G_{lj}\right)\nonumber \\ = & -\frac{1}{2}\partial_{j}\left(\varepsilon^{AB}H_{AB}\right)-\partial_{\bot}^{l}H_{(lj)}-\frac{1}{2}\varepsilon^{AB}H_{iA}\partial_{j}H_{B}^{\;i}+O\left(H^{3}\right),\nonumber  \end{align} where $\partial_{\bot}^{l}=\varepsilon^{lk}\partial_{k}$, which is obtained naturally within Newton-Cartan geometry \cite{moroz2015effective,gromov2016boundary}. This connection is torsion-full, but has a vanishing ``reduced torsion'' \cite{bradlyn2015low}. In Sec.\ref{sec:Main-section-2:}, a term $B/2m$ was implicitly added to $\omega_{t}$, but here we will add it explicitly when writing expressions for $S_{\text{eff}}$ and $S_{\text{ind}}$. Such a term appears in the presence of an additional background field $E_{0}^{\;i}$ which couples to momentum density $P_{i}$ \cite{bradlyn2015low,moroz2015effective}, and can be identified with $G^{ij}A_{j}/m$ in a Galilean invariant system, where $P_{i}=mG_{ij}J^{j}$. The Ricci scalar is given by \begin{align} R= & 2\varepsilon^{ij}\partial_{i}\omega_{j}\\ = & 2\partial_{\bot}^{i}\partial_{\bot}^{j}H_{ij}+O\left(H^{2}\right)\nonumber \\ = & -2\left(\partial^{i}\partial^{j}-\partial^{2}\delta^{ij}\right)H_{ij}+O\left(H^{2}\right).\nonumber  \end{align}

\section{Odd viscosity at non-zero wave-vector: generalities}

\subsection{Definition and $T$ symmetry\label{subsec: B.1}}

We define the viscosity tensor as the linear response of stress to
strain rate 
\begin{align}
 & T^{ij}\left(t,\mathbf{x}\right)=\int\text{d}t\text{d}^{2}\mathbf{x}'\eta^{ij,kl}\left(t,\mathbf{x},t',\mathbf{x}'\right)\partial_{t'}H_{kl}\left(t',\mathbf{x}'\right),
\end{align}
where 
\begin{align}
 & \eta^{[ij],kl}=0=\eta^{ij,[kl]}.\label{eq:p1}
\end{align}
In a translationally invariant system we can pass to Fourier components
$T^{ij}\left(\omega,\mathbf{q}\right)=i\omega\eta^{ij,kl}\left(\omega,\mathbf{q}\right)H_{kl}\left(\omega,\mathbf{q}\right)$.
By definition, $\eta^{ij,kl}\left(t,\mathbf{x},t',\mathbf{x}'\right)$
is real, and therefore 
\begin{align}
\eta^{ij,kl}\left(\omega,\mathbf{q}\right)= & \eta^{ij,kl}\left(-\omega,-\mathbf{q}\right)^{*}.\label{eq:p5}
\end{align}
Under time reversal $T$, 
\begin{align}
 & \eta^{ij,kl}\left(\omega,\mathbf{q}\right)\mapsto\eta_{T}^{ij,kl}\left(\omega,\mathbf{q}\right)=\eta^{kl,ij}\left(\omega,-\mathbf{q}\right).
\end{align}
The even and odd viscosities are then defined by $\eta_{\text{e},\text{o}}=\left(\eta\pm\eta_{T}\right)/2$,
and satisfy $\left(\eta_{\text{e},\text{o}}\right)_{T}=\pm\eta_{\text{e},\text{o}}$.
More explicitly, 
\begin{align}
 & \eta_{\text{e}}^{ij,kl}\left(\omega,\mathbf{q}\right)=+\eta_{\text{e}}^{kl,ij}\left(\omega,-\mathbf{q}\right),\\
 & \eta_{\text{o}}^{ij,kl}\left(\omega,\mathbf{q}\right)=-\eta_{\text{o}}^{kl,ij}\left(\omega,-\mathbf{q}\right).\nonumber 
\end{align}
We will see below that in isotropic (or $SO\left(2\right)$ invariant)
systems $\eta$ is even in $\mathbf{q}$, so that 
\begin{align}
 & \eta_{\text{e}}^{ij,kl}\left(\omega,\mathbf{q}\right)=+\eta_{\text{e}}^{kl,ij}\left(\omega,\mathbf{q}\right),\label{eq:p3}\\
 & \eta_{\text{o}}^{ij,kl}\left(\omega,\mathbf{q}\right)=-\eta_{\text{o}}^{kl,ij}\left(\omega,\mathbf{q}\right),\label{eq:p2}
\end{align}
which is identical to the definition of $\eta_{\text{e},\text{o}}$
at $\mathbf{q}=0$ \cite{PhysRevLett.75.697,Avron1998,PhysRevB.86.245309,hoyos2014hall,PhysRevE.89.043019}. 

\subsection{$SO\left(2\right)$ and $P$ symmetries}

Complex tensors satisfying \eqref{eq:p1} and \eqref{eq:p2}, in $2$
spatial dimensions, form a vector space $V\cong\mathbb{C}^{3}$ which
can be spanned by \cite{Avron1998} 
\begin{align}
 & \sigma^{ab}=2\sigma^{[a}\otimes\sigma^{b]},\;\;a,b=0,x,z,\label{eq:29-1}
\end{align}
where $\sigma^{x},\sigma^{z}$ are the symmetric Pauli matrices, and
$\sigma^{0}$ is the identity matrix. Thus every odd viscosity tensor
can be written as 
\begin{align}
\eta_{\text{o}}\left(\omega,\mathbf{q}\right) & =\eta_{xz}\left(\omega,\mathbf{q}\right)\sigma^{xz}+\eta_{x0}\left(\omega,\mathbf{q}\right)\sigma^{x0}+\eta_{z0}\left(\omega,\mathbf{q}\right)\sigma^{z0},\label{eq:30-2}
\end{align}
with complex coefficients $\eta_{ab}\left(\omega,\mathbf{q}\right)$.
Under a rotation $R=e^{i\alpha\left(i\sigma^{y}\right)}\in SO\left(2\right)$
the metric perturbation and stress tensor transform as
\begin{align}
 & H_{ij}\left(\omega,\mathbf{q}\right)\mapsto R_{\;i}^{k}R_{\;j}^{l}H_{kl}\left(\omega,R^{-1}\cdot\mathbf{q}\right),\\
 & T^{ij}\left(\omega,\mathbf{q}\right)\mapsto R_{\;k}^{i}R_{\;k}^{j}T^{kl}\left(\omega,R^{-1}\cdot\mathbf{q}\right),\nonumber 
\end{align}
where $\left(R\cdot\mathbf{q}\right)^{i}=R_{\;j}^{i}q^{j}$. The same
transformation rules apply for $R\in O\left(2\right)$, which defines
the parity transformation $P$, in flat space. It follows that 
\begin{align}
 & \eta^{ij,kl}\left(\omega,\mathbf{q}\right)\mapsto R_{\;i'}^{i}R_{\;j'}^{j}R_{\;k'}^{k}R_{\;l'}^{l}\eta^{i'j',k'l'}\left(\omega,R^{-1}\cdot\mathbf{q}\right)\label{eq:32-1}
\end{align}
under $O\left(2\right)$, which is compatible with \eqref{eq:p1},
and the decomposition $\eta=\text{\ensuremath{\eta}}_{\text{o}}+\eta_{\text{e}}$.
In particular, equation \eqref{eq:32-1} shows that the viscosity
tensor is $P$-even, or more accurately, a tensor under $P$ rather
than a pseudo-tensor. In an $SO\left(2\right)$-invariant system,
the viscosity tensor will also be $SO\left(2\right)$-invariant
\begin{align}
 & \eta^{ij,kl}\left(\omega,\mathbf{q}\right)=R_{\;i'}^{i}R_{\;j'}^{j}R_{\;k'}^{k}R_{\;l'}^{l}\eta^{i'j',k'l'}\left(\omega,R^{-1}\cdot\mathbf{q}\right),\;R\in SO\left(2\right).\label{eq:33-1}
\end{align}
Note that this holds even when $SO\left(2\right)$ symmetry is \textit{spontaneously}
broken, as in $\ell$-wave SFs. At $\mathbf{q}=\mathbf{0}$, there
is a unique tensor satisfying \eqref{eq:33-1}, namely 
\begin{align}
\left(\sigma^{xz}\right)^{ij,kl} & =-\frac{1}{2}\left(\varepsilon^{ik}\delta^{jl}+\varepsilon^{jk}\delta^{il}+\varepsilon^{il}\delta^{jk}+\varepsilon^{jl}\delta^{ik}\right),\label{34}
\end{align}
leaving a single odd viscosity coefficient $\eta_{xz}\left(\omega\right)=\eta_{\text{o}}^{\left(1\right)}\left(\omega\right)$
\cite{PhysRevLett.75.697,Avron1998,PhysRevB.86.245309,hoyos2014hall,PhysRevE.89.043019}.

A non-zero $\mathbf{q}$, however, along with the  tensors $\delta^{ij}$
and $\varepsilon^{ij}$, can be used to construct additional $SO\left(2\right)$-invariant
odd viscosity tensors, beyond $\sigma^{xz}$. From the data $\mathbf{q},\delta^{ij},\varepsilon^{ij}$,
three linearly independent, symmetric, rank-2 tensors can be constructed,
which we take to be 
\begin{align}
\left(\tau^{0}\right)^{ij}= & q^{2}\delta^{ij},\label{eq:35-1}\\
\left(\tau^{x}\right)^{ij}= & -2q_{\bot}^{(i}q^{j)}/q^{2},\nonumber \\
\left(\tau^{z}\right)^{ij}= & 2q^{i}q^{j}/q^{2}-\delta^{ij},\nonumber 
\end{align}
where $q_{\bot}^{i}=\varepsilon^{ij}q_{j}$. The notation above is
due to the relation
\begin{align}
\begin{pmatrix}\tau^{x}\\
\tau^{z}
\end{pmatrix} & =\begin{pmatrix}\cos2\theta & -\sin2\theta\\
\sin2\theta & \cos2\theta
\end{pmatrix}\begin{pmatrix}\sigma^{x}\\
\sigma^{z}
\end{pmatrix}\label{eq:35-0}\\
 & =\frac{1}{q^{2}}\begin{pmatrix}q_{x}^{2}-q_{y}^{2} & -2q_{x}q_{y}\\
2q_{x}q_{y} & q_{x}^{2}-q_{y}^{2}
\end{pmatrix}\begin{pmatrix}\sigma^{x}\\
\sigma^{z}
\end{pmatrix},\nonumber 
\end{align}
where $\theta=\text{arg}\left(\mathbf{q}\right)$, so that $\tau^{x},\tau^{z}$
are a rotated version of $\sigma^{x},\sigma^{z}$. Moreover, all three
$\tau$s are $SO\left(2\right)$-invariant, $\tau^{ij}\left(\mathbf{q}\right)=R_{\;i'}^{i}R_{\;j'}^{j}\tau^{i'j'}\left(R^{-1}\cdot\mathbf{q}\right)$,
and can therefore be used to construct three $SO\left(2\right)$-invariant
odd viscosity tensors 
\begin{align}
 & \tau^{ab}=2\tau^{[a}\otimes\tau^{b]},\;a,b=0,x,z,
\end{align}
which form a basis for $V$. Any odd viscosity tensor (at $\mathbf{q}\neq\mathbf{0}$)
can then be written as 
\begin{align}
\eta_{\text{o}}\left(\omega,\mathbf{q}\right) & =\eta_{\text{o}}^{\left(1\right)}\left(\omega,\mathbf{q}\right)\tau^{xz}+\eta_{\text{o}}^{\left(2\right)}\left(\omega,\mathbf{q}\right)\tau^{0x}+\eta_{\text{o}}^{\left(3\right)}\left(\omega,\mathbf{q}\right)\tau^{0z}.
\end{align}
Furthermore, for an $SO\left(2\right)$-invariant $\eta_{\text{o}}$,
the coefficients $\eta_{\text{o}}^{\left(1\right)},\eta_{\text{o}}^{\left(2\right)},\eta_{\text{o}}^{\left(3\right)}$
depend on $\mathbf{q}$ through its norm, owing to the $SO\left(2\right)$-invariance
of $\tau^{ab}$. We therefore arrive at the general form of an $SO\left(2\right)$-invariant
odd viscosity tensor, 
\begin{align}
\eta_{\text{o}}\left(\omega,\mathbf{q}\right)= & \eta_{\text{o}}^{\left(1\right)}\left(\omega,q^{2}\right)\tau^{xz}+\eta_{\text{o}}^{\left(2\right)}\left(\omega,q^{2}\right)\tau^{0x}+\eta_{\text{o}}^{\left(3\right)}\left(\omega,q^{2}\right)\tau^{0z}.\label{eq:39-2}
\end{align}
In particular, we see that $\eta_{\text{o}}$ is even in $\mathbf{q}$
(and the same applies also to the even viscosity $\eta_{\text{e}}$).
To determine the small $\omega,\mathbf{q}$ behavior of the coefficients
we change to the $\mathbf{q}$-independent basis of $\sigma$s, 
\begin{align}
\eta_{\text{o}}\left(\omega,\mathbf{q}\right)= & \eta_{\text{o}}^{\left(1\right)}\left(\omega,q^{2}\right)\sigma^{xz}\label{eq:39-1}\\
 & +\left[\eta_{\text{o}}^{\left(2\right)}\left(\omega,q^{2}\right)\left(q_{x}^{2}-q_{y}^{2}\right)+\eta_{\text{o}}^{\left(3\right)}\left(\omega,q^{2}\right)\left(2q_{x}q_{y}\right)\right]\sigma^{0x}\nonumber \\
 & +\left[\eta_{\text{o}}^{\left(2\right)}\left(\omega,q^{2}\right)\left(-2q_{x}q_{y}\right)+\eta_{\text{o}}^{\left(3\right)}\left(\omega,q^{2}\right)\left(q_{x}^{2}-q_{y}^{2}\right)\right]\sigma^{0z}.\nonumber 
\end{align}
In gapped systems (such as QH states) $\eta_{\text{o}}$ will be regular
around $\omega=0=q$, and so will the coefficients $\eta_{\text{o}}^{\left(1\right)},\eta_{\text{o}}^{\left(2\right)},\eta_{\text{o}}^{\left(3\right)}$.
In gapless systems (such as $\ell$-wave SFs) there will be a singularity
at $\omega=0=q$, but the limit $q\rightarrow0$ at $\omega\neq0$
will be regular. In both cases, the limit $q\rightarrow0$ at $\omega\neq0$
of \eqref{eq:39-1} reduces to the known result $\eta_{\text{o}}\left(\omega,\mathbf{0}\right)=\eta_{\text{o}}^{\left(1\right)}\left(\omega,0\right)\sigma^{xz}$
\cite{PhysRevLett.75.697,Avron1998,PhysRevB.86.245309,hoyos2014hall,PhysRevE.89.043019}. 

\subsection{$PT$ symmetry}

The combination $PT$ of parity and time-reversal is a symmetry in
any system in which $T$ is broken (perhaps spontaneously) due to
some kind of angular momentum, as in QH states, $\ell$-wave SFs,
and active chiral fluids \cite{banerjee2017odd}. Here we consider
the implications of $PT$ symmetry on \eqref{eq:39-1}. 

From the definition \eqref{eq:35-1} it is clear that $\tau^{0}$
and $\tau^{z}$ are $P$-even, while $\tau^{x}$ is $P$-odd. Therefore,
$\tau^{xz},\tau^{0x}$ are $P$-odd while $\tau^{0z}$ is $P$-even
(and all three are $T$-even). Since $\eta_{\text{o}}$ is $T$-odd
and $P$-even, and using \eqref{eq:39-2}, it follows that $\eta_{\text{o}}^{\left(1\right)}$
and $\eta_{\text{o}}^{\left(2\right)}$ are $P,T$-odd, while $\eta_{\text{o}}^{\left(3\right)}$
is $T$-odd but $P$-even. In particular, $\eta_{\text{o}}^{\left(3\right)}$
is $PT$-odd, and must vanish in $PT$-symmetric systems. The odd
viscosity tensor in $SO\left(2\right)$ and $PT$ symmetric systems
is therefore given by 
\begin{align}
\eta_{\text{o}}\left(\omega,\mathbf{q}\right)= & \eta_{\text{o}}^{\left(1\right)}\left(\omega,q^{2}\right)\sigma^{xz}\label{eq:39-1-1}\\
 & +\eta_{\text{o}}^{\left(2\right)}\left(\omega,q^{2}\right)\left[\left(q_{x}^{2}-q_{y}^{2}\right)\sigma^{0x}-2q_{x}q_{y}\sigma^{0z}\right].\nonumber 
\end{align}
This form is confirmed by previous results for QH states \cite{abanov2014electromagnetic},
and by the results presented in Sec.\ref{sec:Main-section-2:} for $\ell$-wave SFs.
The same form is obtained at $\mathbf{q}=\mathbf{0}$, but in the
presence of vector, or pseudo-vector, anisotropy $\mathbf{b}$, in
which case we find 
\begin{align}
\eta_{\text{o}}\left(\omega\right)= & \eta_{\text{o}}^{\left(1\right)}\left(\omega\right)\sigma^{xz}\label{eq:39-1-1-1}\\
 & +\eta_{\text{o}}^{\left(2\right)}\left(\omega\right)\left[\left(b_{x}^{2}-b_{y}^{2}\right)\sigma^{0x}-2b_{x}b_{y}\sigma^{0z}\right],\nonumber 
\end{align}
which explains the tensor structure found in \cite{PhysRevB.99.045141,PhysRevB.99.035427}.

\subsection{Frequency dependence and reality conditions \label{subsec: B.4}}

In closed and clean systems, like the $\ell$-wave SFs studied in Sec.\ref{sec:Main-section-2:}, the viscosity can be obtained from an induced action
\begin{align}
S_{\text{ind}} & \supset\frac{1}{2}\int\text{d}t\text{d}t'\text{d}^{2}\mathbf{x}\text{d}^{2}\mathbf{x}'H_{ij}\left(t,\mathbf{x}\right)\eta^{ij,kl}\left(t-t',\mathbf{x}-\mathbf{x}'\right)\partial_{t'}H_{kl}\left(t',\mathbf{x}'\right)\label{eq:38}\\
 & =\frac{1}{2}\int\frac{\text{d}\omega}{2\pi}\frac{\text{d}^{2}\mathbf{q}}{\left(2\pi\right)^{2}}H_{ij}\left(-\omega,-\mathbf{q}\right)i\omega\eta^{ij,kl}\left(\omega,\mathbf{q}\right)H_{kl}\left(\omega,\mathbf{q}\right).\nonumber 
\end{align}
As a result, $\eta$ satisfies the additional property, 
\begin{align}
\eta^{ij,kl}\left(\omega,\mathbf{q}\right)= & -\eta^{kl,ij}\left(-\omega,-\mathbf{q}\right),\label{eq:p4}
\end{align}
which, along with \eqref{eq:p2}-\eqref{eq:p3} and the fact that
$\eta$ is even in $\mathbf{q}$, implies that $\eta_{\text{o}}$
($\eta_{\text{e}}$) is even (odd) in $\omega$,
\begin{align}
\eta_{\text{e}}^{ij,kl}\left(\omega,\mathbf{q}\right)= & -\eta_{\text{e}}^{ij,kl}\left(-\omega,\mathbf{q}\right),\label{eq:44}\\
\eta_{\text{o}}^{ij,kl}\left(\omega,\mathbf{q}\right)= & +\eta_{\text{o}}^{ij,kl}\left(-\omega,\mathbf{q}\right).\nonumber 
\end{align}
 This result, along with \eqref{eq:p5} and the fact that $\eta$
is even in $\mathbf{q}$, implies that $\eta_{\text{o}}$ ($\eta_{\text{e}}$)
is real (imaginary),
\begin{align}
\eta_{\text{e}}^{ij,kl}\left(\omega,\mathbf{q}\right)\in & i\mathbb{R},\label{eq:45}\\
\eta_{\text{o}}^{ij,kl}\left(\omega,\mathbf{q}\right)\in & \mathbb{R}.\nonumber 
\end{align}
These general properties are satisfied by the odd viscosity tensor
 computed in Sec.\ref{sec:Main-section-2:}. These are also compatible with the examples
worked out in \cite{PhysRevB.86.245309}, as well with viscosity-conductivity
relations that hold in Galilean invariant systems (in conjugation
with known properties of the conductivity) \cite{hoyos2012hall,PhysRevB.86.245309,hoyos2014effective}. 

We note that some care is required when interpreting \eqref{eq:44}-\eqref{eq:45}
around singularities of $\eta$. For example, the first equation in
\eqref{eq:44} \textit{naively} implies that $\eta_{\text{e}}\left(0,\mathbf{q}\right)=0$,
which in particular implies that the bulk and shear viscosities $\eta_{\text{e}}\left(0,\mathbf{0}\right)=\zeta\sigma^{0}\otimes\sigma^{0}+\eta^{\text{s}}\left(\sigma^{x}\otimes\sigma^{z}+\sigma^{z}\otimes\sigma^{x}\right)$
 vanish in the closed, clean, case. This however, is not quite correct,
due to a possible singularity of $\eta_{\text{e}}$ at $\omega=0$,
as well as the usual infinitesimal imaginary part of $\omega$ required
to obtain the retarded response. For example, for free fermions, reference
\cite{PhysRevB.86.245309} finds $\eta^{\text{s}}\left(\omega,\mathbf{0}\right)\sim\frac{i}{\omega+i\epsilon}=\pi\delta\left(\omega\right)+i\text{PV}\frac{1}{\omega}$
(where $\text{PV}$ is the principle value), which has an infinite
real part at $\omega=0$, in analogy with the Drude behavior of the
conductivity.  

\subsection{Odd viscosity from Gaussian integration: a technical result\label{subsec:Odd-viscosity-from}}

We now restrict attention to $\ell$-wave SFs. The effective Lagrangian,
perturbatively expanded to second order, and in the absence of the
$U\left(1\right)$ background, takes the form 
\begin{align}
\mathcal{L}_{\text{eff}}= & \frac{1}{2}\theta\mathcal{G}^{-1}\theta+\mathcal{V}\theta+\mathcal{C},\label{eq:r1}
\end{align}
where the Green's function $\mathcal{G}$ is independent of $H$,
the vertex $\mathcal{V}$ is linear in $H$, and the contact term
$\mathcal{C}$ is quadratic in $H$. Performing Gaussian integration
over $\theta$ yields the induced Lagrangian 
\begin{align}
\text{\ensuremath{\mathcal{L}}}_{\text{ind}}= & -\frac{1}{2}\mathcal{V}\mathcal{G}\mathcal{V}+\mathcal{C},\label{eq:44-1}
\end{align}
and comparing with \eqref{eq:38} one can read off $\eta_{\text{o}}$.
In Appendix \ref{subsec:Effective-action-and} we write explicit expressions
for a Galilean invariant $\mathcal{L}_{\text{eff}}$, which we then
expand to obtain explicit expressions for $\mathcal{G}^{-1},\mathcal{V},\mathcal{C}$.
Appendix \ref{subsec:Obtaining--from} then describes the resulting
$\mathcal{L}_{\text{ind}}$. Here we take a complementary approach
and obtain the general form of $\eta_{\text{o}}$ from \eqref{eq:44-1},
using the formalism developed above, based only on $SO\left(2\right)$
and $PT$ symmetries. 

The motivation for the analysis in this section is the following.
The power counting described in Sec.\ref{sec: effective field theory}  is designed such that
the $O\left(p^{n}\right)$ Lagrangian $\mathcal{L}_{n}\subset\mathcal{L}_{\text{eff}}$
produces $O\left(p^{n}\right)$ contributions to $\mathcal{L}_{\text{ind}}$.
Therefore, naively, one expects the $O\left(q^{2}\right)$ odd viscosity
to depend on $\mathcal{L}_{0},\mathcal{L}_{2},$ and $\mathcal{L}_{3}$
(since $\mathcal{L}_{1}=0$). Using the notation $\eta_{\text{o}}=\eta_{\mathcal{V}}+\eta_{\mathcal{C}}$
for the parts of $\eta_{\text{o}}$ due to $-\mathcal{V}\mathcal{G}\mathcal{V}/2$
and $\mathcal{C}$, respectively, the result of this section is that
$\eta_{\mathcal{V}}$, to $O\left(q^{2}\right)$, is actually independent
of $\mathcal{L}_{3}$. 

We now describe the details. For $\eta_{\mathcal{C}}$, we cannot
do better than the general discussion thus far - it is given by \eqref{eq:39-1},
with $\eta_{\text{o}}^{\left(3\right)}=0$, and both $\eta_{\text{o}}^{\left(1\right)},\eta_{\text{o}}^{\left(2\right)}$
are real and regular at $\omega=0=q$, since $\mathcal{L}_{\text{eff}}$,
and $\mathcal{C}$ in particular, are obtained by integrating out
gapped degrees of freedom (the Higgs modes and the fermion $\psi$).
For $\eta_{\mathcal{V}}$, however, we can do better. We first write
more explicitly 
\begin{align}
\theta\mathcal{G}^{-1}\theta & =\frac{1}{2}\theta\left(-\omega,-\mathbf{q}\right)\mathcal{G}^{-1}\left(\omega,\mathbf{q}\right)\theta\left(\omega,\mathbf{q}\right),\label{eq:r2}\\
\mathcal{V}\theta & =\theta\left(-\omega,-\mathbf{q}\right)V^{ij}\left(\omega,\mathbf{q}\right)H_{ij}\left(\omega,\mathbf{q}\right).\nonumber 
\end{align}
Based on $SO\left(2\right)$ and $PT$ symmetries, the objects $\mathcal{G}^{-1},V^{ij}$
take the forms 
\begin{align}
\mathcal{G}^{-1}\left(\omega,\mathbf{q}\right)= & D\left(\omega^{2},q^{2}\right),\label{eq:r3}\\
V^{ij}\left(\omega,\mathbf{q}\right)= & i\omega a\left(\omega^{2},q^{2}\right)\left(\rho^{0}\right)^{ij}+i\omega b\left(\omega^{2},q^{2}\right)\left(\rho^{z}\right)^{ij}+s_{\theta}c\left(\omega^{2},q^{2}\right)\left(\rho^{x}\right)^{ij},\nonumber 
\end{align}
where 
\begin{align}
\left(\rho^{0}\right)^{ij}= & \delta^{ij},\label{eq:35-1-1-1-1}\\
\left(\rho^{x}\right)^{ij}= & q_{\bot}^{(i}q^{j)},\nonumber \\
\left(\rho^{z}\right)^{ij}= & q^{i}q^{j},\nonumber 
\end{align}
are, in this context, more convenient than the $\tau$s \eqref{eq:35-1},
and $a,b,c,D$ are general functions of their arguments which are
$P,T$-even, real, and regular at $\omega=0=q$, as follows from the
same properties of $\mathcal{L}_{\text{eff}}$. In particular, we
will use the following expansions 
\begin{align}
a\left(0,q^{2}\right) & =a_{0}+a_{1}q^{2}+O\left(q^{4}\right),\label{eq:r4}\\
b\left(0,q^{2}\right) & =b_{0}+O\left(q^{2}\right),\nonumber \\
c\left(0,q^{2}\right) & =c_{0}+c_{1}q^{2}+O\left(q^{4}\right),\nonumber \\
D\left(0,q^{2}\right) & =D_{1}q^{2}+D_{2}q^{4}+O\left(q^{6}\right),\nonumber 
\end{align}
where $D_{0}=0$ because $\theta$ enters $\mathcal{L}_{\text{eff}}$
only through its derivatives. The odd viscosity $\eta_{\mathcal{V}}$
is then given by 
\begin{align}
\eta_{\mathcal{V}}\left(\omega,\mathbf{q}\right)= & -\frac{1}{2i\omega}\frac{V\left(-\omega,-\mathbf{q}\right)\otimes V\left(\omega,\mathbf{q}\right)-V\left(\omega,\mathbf{q}\right)\otimes V\left(-\omega,-\mathbf{q}\right)}{D\left(\omega,\mathbf{q}\right)}\\
= & \frac{2s_{\theta}c\left(\omega^{2},q^{2}\right)}{D\left(\omega^{2},q^{2}\right)}\left[a\left(\omega^{2},q^{2}\right)\rho^{0x}+b\left(\omega^{2},q^{2}\right)\rho^{zx}\right],\nonumber 
\end{align}
which is of the form \eqref{eq:39-1}, with $\eta_{\text{o}}^{\left(3\right)}=0$
and 
\begin{align}
\eta_{\mathcal{V}}^{\left(1\right)}\left(\omega,q^{2}\right)= & -\frac{s_{\theta}c\left(\omega^{2},q^{2}\right)}{2D\left(\omega^{2},q^{2}\right)}b\left(\omega^{2},q^{2}\right)q^{4},\\
\eta_{\text{\ensuremath{\mathcal{V}}}}^{\left(2\right)}\left(\omega,q^{2}\right)= & \frac{s_{\theta}c\left(\omega^{2},q^{2}\right)}{D\left(\omega^{2},q^{2}\right)}\left[a\left(\omega^{2},q^{2}\right)+b\left(\omega^{2},q^{2}\right)q^{2}\right].\nonumber 
\end{align}
Setting $\omega=0$ and expanding in $q$, we find
\begin{align}
\eta_{\text{\ensuremath{\mathcal{V}}}}^{\left(1\right)}\left(0,q^{2}\right)= & -\frac{s_{\theta}c_{0}b_{0}}{2D_{1}}q^{2}+O\left(q^{4}\right),\label{eq:22-2}\\
\eta_{\text{\ensuremath{\mathcal{V}}}}^{\left(2\right)}\left(0,q^{2}\right)= & \frac{s_{\theta}}{D_{1}}\left[a_{0}c_{0}q^{-2}+\left(a_{0}c_{1}+a_{1}c_{0}+b_{0}c_{0}-a_{0}c_{0}\frac{D_{2}}{D_{1}}\right)\right]+O\left(q^{2}\right).\nonumber 
\end{align}
Having identified the coefficients $a_{0},a_{1},b_{0},c_{0},c_{1},D_{1},D_{2}$
that determine $\eta_{\mathcal{V}}$ to $O\left(q^{2}\right)$, we
now determine the order in the derivative expansion  of $\mathcal{L}_{\text{eff}}$
in which these enter. Explicitly, the above coefficients are defined
by 
\begin{align}
\mathcal{L}_{\text{eff}}\supset & \frac{1}{2}\theta\left(-\omega,-\mathbf{q}\right)\left(D_{1}q^{2}+D_{2}q^{4}\right)\theta\left(\omega,\mathbf{q}\right)\label{eq:130}\\
 & +\theta\left(-\omega,-\mathbf{q}\right)\left[i\omega\left(a_{0}+a_{1}q^{2}\right)\delta^{ij}+i\omega b_{0}q^{i}q^{j}+s_{\theta}\left(c_{0}+c_{1}q^{2}\right)q^{i}q_{\bot}^{j}\right]H_{ij}\left(\omega,\mathbf{q}\right).\nonumber 
\end{align}
We see that $c_{1}$ enters $\mathcal{L}_{\text{eff}}$ at $O\left(p^{3}\right)$,
while all other coefficients enter at a lower order, and come from
$\mathcal{L}_{0},\mathcal{L}_{2}$. In particular, $\eta_{\text{\ensuremath{\mathcal{V}}}}^{\left(1\right)}$
in \eqref{eq:22-2} is independent of $\mathcal{L}_{3}$. Even though
$c_{1}$ is the coefficient of an $O\left(p^{3}\right)$ term, it
is actually due to $\mathcal{L}_{2}$. Using \eqref{eq:49-0} we identify
$c_{0}\theta q^{2}q^{i}q_{\bot}^{j}H_{ij}=-s_{\theta}c_{1}\partial^{i}\theta\partial^{2}\omega_{i}$,
which must be a part of 
\begin{align}
 & \frac{c_{1}}{2}\left(\partial_{i}\theta-A_{i}-s_{\theta}\omega_{i}\right)\partial^{2}\left(\partial_{i}\theta-A_{i}-s_{\theta}\omega_{i}\right).
\end{align}
This is an $O\left(p^{2}\right)$ term, and in fact comes from $\mathcal{L}_{2}^{\left(2\right)}\subset\mathcal{L}_{2}$,
see \eqref{eq:54-0}. Thus, both $\eta_{\text{\ensuremath{\mathcal{V}}}}^{\left(1\right)},\eta_{\text{\ensuremath{\mathcal{V}}}}^{\left(2\right)}$
in \eqref{eq:22-2} are completely independent of $\mathcal{L}_{3}$.

\section{Effective action and its perturbative expansion \label{subsec:Effective-action-and}}

\subsection{Zeroth order\label{zeroth-order}}

It is useful to write the zeroth order scalar $X$ as 
\begin{align}
X= & \left(\partial_{t}\theta-\mathcal{A}_{t}-\frac{s_{\theta}}{2m}B\right)-\frac{1}{2m}G^{ij}\left(\partial_{i}\theta-\mathcal{A}_{i}\right)\left(\partial_{j}\theta-\mathcal{A}_{j}\right),\label{eq:69}
\end{align}
where
\begin{align}
\mathcal{A}_{\mu} & =A_{\mu}+s_{\theta}\omega_{\mu}.\label{eq:70-0}
\end{align}
We will also use $\mathcal{B}=B+\frac{s_{\theta}}{2}R,\;\mathcal{E}_{i}=E_{i}+s_{\theta}E_{\omega,i}$
for the magnetic and electric fields obtained from $\mathcal{A}_{\mu}$,
where $E_{\omega,i}=\partial_{t}\omega_{i}-\partial_{i}\omega_{t}$.
Expanding $\mathcal{L}_{0}=P\left(X\right)$  to second order in
the fields, one finds (up to total derivatives)
\begin{align}
\sqrt{G}\mathcal{L}_{0}= & \frac{1}{2}\frac{n_{0}}{m}\theta\left[\partial^{2}-c_{s}^{-2}\partial_{t}^{2}\right]\theta\label{eq:51-0}\\
 & +\left[-\frac{n_{0}}{m}\left(\partial_{i}\mathcal{A}^{i}-c_{s}^{-2}\partial_{t}\left(\mathcal{A}_{t}+\frac{s_{\theta}}{2m}B\right)\right)-n_{0}\partial_{t}\sqrt{G}\right]\theta\nonumber \\
 & +\left[-n_{0}\sqrt{G}\mathcal{A}_{t}-\frac{1}{2}\frac{n_{0}}{m}\left(\mathcal{A}^{2}-c_{s}^{-2}\left(\mathcal{A}_{t}+\frac{s_{\theta}}{2m}B\right)^{2}\right)+P_{0}\sqrt{G}\right]\nonumber \\
= & \frac{1}{2}\theta\mathcal{G}^{-1}\theta+\mathcal{V}\theta+\mathcal{C},\nonumber 
\end{align}
where $\partial^{2}=\partial^{i}\partial_{i},\;\mathcal{A}^{2}=\mathcal{A}_{i}\mathcal{A}^{i}$,
and we defined the inverse Green's function $\mathcal{G}^{-1}$, vertex
$\mathcal{V}$, and contact terms $\mathcal{C}$, respectively. These
are used in section \ref{subsec:Obtaining--from} below to obtain
$S_{\text{ind}}$. 

In \eqref{eq:51-0}, the geometric objects $\sqrt{G}$ and $\omega_{\mu}$
should be interpreted as expanded to the required order according
to \eqref{eq:48-0}-\eqref{eq:49-0}. In particular, the term $-n_{0}\sqrt{G}\mathcal{A}_{t}$
includes $-s_{\theta}n_{0}\sqrt{G}\omega_{t}$, which produces the
leading contribution to $\eta_{\text{o}}^{\left(1\right)}$. To see
this, we expand 

\begin{align}
\sqrt{G}\omega_{t}= & -\frac{1}{2}\partial_{t}\left(\varepsilon^{AB}H_{AB}\right)+\frac{1}{2}\partial_{t}\left(\varepsilon^{AB}H_{AB}\right)H_{i}^{\;i}-\frac{1}{2}\varepsilon^{AB}H_{iA}\partial_{t}H_{B}^{\;i}+O\left(H^{3}\right)\label{eq:50}\\
= & -\frac{1}{2}\partial_{t}\left(\varepsilon^{AB}H_{AB}\right)-\frac{1}{2}\varepsilon^{AB}H_{Ai}\partial_{t}H_{B}^{\;i}+O\left(H^{3}\right),\nonumber 
\end{align}
which is identical to the expansion \eqref{eq:48-0} of $\omega_{t}$,
apart from $H_{iA}\leftrightarrow H_{Ai}$. Ignoring total derivatives,
this reduces to 
\begin{align}
\sqrt{G}\mathcal{L}_{0} & \supset-s_{\theta}n_{0}\sqrt{G}\omega_{t}\label{eq:43-0}\\
 & =-\frac{1}{2}s_{\theta}n_{0}\left[\partial_{t}\left(\varepsilon^{AB}H_{AB}\right)H_{i}^{\;i}-\varepsilon^{AB}\delta^{ij}H_{(Ai)}\partial_{t}H_{(Bj)}\right]+O\left(H^{3}\right)\nonumber \\
 & =\frac{1}{2}s_{\theta}n_{0}\varepsilon^{AB}H_{Ai}\partial_{t}H_{B}^{\;i}+O\left(H^{3}\right).\nonumber 
\end{align}
Comparing with \eqref{34} and \eqref{eq:38}, the second term in
the second line corresponds to $\eta_{\text{o}}^{\left(1\right)}=-s_{\theta}n_{0}/2$.
The first term in the second line depends on the anti-symmetric part
of $H$, and shows that the full expression \eqref{eq:43-0} actually
corresponds to a \textit{torsional} Hall (or odd) viscosity \cite{hughes2011torsional,hughes2013torsional}
$\zeta_{H}=-s_{\theta}n_{0}$, which can be read off from the third
line. The appearance of the torsional Hall viscosity at the level
of $S_{\text{eff}}$ (but not at the level of $S_{\text{ind}}$, see
section \ref{subsec:Obtaining--from}) can be understood from the
mapping of \cite{PhysRevB.98.064503} of the $p$-wave SF to a Majorana
spinor in Riemann-Cartan space-time. 

\subsection{Second order \label{subsec: Second order effective action}}

The full expression for $\mathcal{L}_{2}$ is given by $\mathcal{L}_{2}=\sum_{i=1}^{6}\mathcal{L}_{2}^{\left(i\right)}$
, where \cite{son2006general}
\begin{align}
\mathcal{L}_{2}^{\left(1\right)}= & F_{1}\left(X\right)R,\label{eq:49-2}\\
\mathcal{L}_{2}^{\left(2\right)}= & F_{2}\left(X\right)\left(mK_{\;i}^{i}-\nabla^{2}\theta\right)^{2},\nonumber \\
\mathcal{L}_{2}^{\left(3\right)}= & F_{3}\left(X\right)\left\{ -m^{2}\left(G^{ij}\partial_{t}K_{ij}-K^{ij}K_{ij}\right)-m\nabla_{i}E^{i}+\frac{1}{4}F^{ij}F_{ij}\right.\nonumber \\
 & \left.+2m\left[\partial_{i}K_{\;j}^{j}-\nabla^{j}\left(K_{ji}+\frac{1}{2m}F_{ji}\right)\right]\nabla^{i}\theta+R_{ij}\nabla_{i}\theta\nabla_{j}\theta\right\} ,\nonumber \\
\mathcal{L}_{2}^{\left(4\right)}= & F_{4}\left(X\right)G^{ij}\partial_{i}X\partial_{j}X,\nonumber \\
\mathcal{L}_{2}^{\left(5\right)}= & F_{5}\left(X\right)\left[\left(\partial_{t}-\frac{1}{m}\nabla^{i}\theta\partial_{i}\right)X\right]^{2},\nonumber \\
\mathcal{L}_{2}^{\left(6\right)}= & F_{6}\left(X\right)\left(mK_{\;i}^{i}-\nabla^{2}\theta\right)\left[\left(\partial_{t}-\frac{1}{m}\nabla^{i}\theta\partial_{i}\right)X\right].\nonumber 
\end{align}
 The terms $\mathcal{L}_{2}^{\left(5\right)}$ and $\mathcal{L}_{2}^{\left(6\right)}$
were not written explicitly in \cite{son2006general} because, on
shell (on the equation of motion for $\theta$), they are proportional
to $\mathcal{L}_{2}^{\left(4\right)}$ up to $O\left(p^{4}\right)$
corrections, and can therefore be eliminated by a redefinition of
$F_{4}$. However, for the purpose of comparing the general $S_{\text{eff}}$
with the microscopic expression \eqref{eq:39}, it is convenient to
work off shell and keep all terms explicit. 

Specializing to 2+1 dimensions and expanding to second order in fields,
one finds
\begin{align}
\sqrt{G}\mathcal{L}_{2}^{\left(1\right)}= & F_{1}'\left(\mu\right)R\left(\partial_{t}\theta-\mathcal{A}_{t}-\frac{s_{\theta}}{2m}B\right).\label{eq:54-0}\\
\sqrt{G}\mathcal{L}_{2}^{\left(2\right)}= & F_{2}\left(\mu\right)\left[-m^{2}H_{i}^{\;i}\partial_{t}^{2}H_{j}^{\;j}+2m\partial_{t}H_{k}^{\;k}\partial^{j}\left(\partial_{j}\theta-\mathcal{A}_{j}\right)-\left(\partial_{i}\theta-\mathcal{A}_{i}\right)\partial^{i}\partial^{j}\left(\partial_{i}\theta-\mathcal{A}_{i}\right)\right],\nonumber \\
\sqrt{G}\mathcal{L}_{2}^{\left(3\right)}= & F_{3}\left(\mu\right)\left(m^{2}H^{(ij)}\partial_{t}^{2}H_{\left(ij\right)}+\frac{1}{2}B^{2}-2m\varepsilon^{ij}\omega_{i}\partial_{t}\left(\partial_{j}\theta-\mathcal{A}_{j}\right)-B\mathcal{B}\right)\nonumber \\
 & +F_{3}'\left(\mu\right)\left(\partial_{t}\theta-\mathcal{A}_{t}-\frac{s_{\theta}}{2m}B\right)\left(m^{2}\partial_{t}^{2}H_{i}^{\;i}-m\partial_{i}E^{i}\right),\nonumber \\
\sqrt{G}\mathcal{L}_{2}^{\left(4\right)}= & -F_{4}\left(\mu\right)\left(\partial_{t}\theta-\mathcal{A}_{t}-\frac{s_{\theta}}{2m}B\right)\partial^{2}\left(\partial_{t}\theta-\mathcal{A}_{t}-\frac{s_{\theta}}{2m}B\right),\nonumber \\
\sqrt{G}\mathcal{L}_{2}^{\left(5\right)}= & -F_{5}\left(\mu\right)\left(\partial_{t}\theta-\mathcal{A}_{t}-\frac{s_{\theta}}{2m}B\right)\partial_{t}^{2}\left(\partial_{t}\theta-\mathcal{A}_{t}-\frac{s_{\theta}}{2m}B\right),\nonumber \\
\sqrt{G}\mathcal{L}_{2}^{\left(6\right)}= & -F_{6}\left(\mu\right)\left[m\partial_{t}H_{i}^{\;i}+\partial^{j}\left(\partial_{j}\theta-\mathcal{A}_{j}\right)\right]\partial_{t}\left(\partial_{t}\theta-\mathcal{A}_{t}-\frac{s_{\theta}}{2m}B\right),\nonumber 
\end{align}
from which one can easily extract the second order corrections to
$\mathcal{G}^{-1},\mathcal{V},\mathcal{C}$, of \eqref{eq:51-0}. Note
that $\mathcal{L}_{2}^{\left(3\right)}$ includes a term $\propto\varepsilon^{ij}\omega_{i}\partial_{t}\mathcal{A}_{j}=\varepsilon^{ij}\omega_{i}\partial_{t}\left(A_{j}+s_{\theta}\omega_{j}\right)$.
Comparing with \eqref{eq:54-2} below, it is clear that distinguishing
$\mathcal{L}_{2}^{\left(3\right)}$ from $\mathcal{L}_{\text{gCS}}$
is non-trivial. This is in fact the same problem of extracting the
central charge from the Hall viscosity addressed in Sec.\ref{sec: induced action and linear response},
but at the level of $S_{\text{eff}}$ (where $\theta$ is viewed as
a background field) rather than $S_{\text{ind}}$ (where $\theta$
has been integrated out). Accordingly, the central charge can be computed
by applying Eq.\eqref{eq:16-2-1}  to the response functions
obtained from $S_{\text{eff}}$. Additionally, relying on LGS, one
can extract $F_{3}$ as the coefficient of $H^{(ij)}\partial_{t}^{2}H_{\left(ij\right)}$.
Both approaches produce the same central charge \eqref{eq:c2} in
the perturbative computation of Sec.\ref{subsec:Benchmarking-the-effective} and Appendix \ref{subsec:Perturbative-expansion}.

\subsection{Gravitational Chern-Simons term}

The gCS Lagrangian  is given explicitly by
\begin{align}
\mathcal{L}_{\text{gCS}} & =-\frac{c}{48\pi}\left[\left(\omega_{t}+\frac{B}{2m}\right)R-\varepsilon^{ij}\omega_{i}\partial_{t}\omega_{j}\right]\label{eq:54-2}\\
 & =-\frac{c}{48\pi}\left[\omega\text{d}\omega+\frac{1}{2m}BR\right].\nonumber 
\end{align}
Its expansion to second order in fields, using \eqref{eq:48-0}-\eqref{eq:49-0},
is
\begin{align}
\sqrt{G}\mathcal{L}_{\text{gCS}} & =-\frac{c}{48\pi}\left[\varepsilon^{AB}H_{\left(Ai\right)}\partial_{\bot}^{i}\partial_{\bot}^{j}\partial_{t}H_{\left(Bj\right)}-\frac{1}{m}A_{i}\partial_{\bot}^{i}\partial_{\bot}^{j}\partial_{\bot}^{k}H_{(jk)}\right].
\end{align}
As opposed to $\sqrt{G}\omega_{t}$, the gCS term $\mathcal{L}_{\text{gCS}}$
is (locally) $SO\left(2\right)_{L}$ gauge invariant, and accordingly
depends only on the metric, or, within the perturbative expansion,
on the symmetric part $H_{(ij)}$. From this expansion one can read
off the gCS contributions to the odd viscosity $\eta_{\text{o}}$
, and to the odd, mixed, static susceptibility $\chi_{TJ,\text{o}}$,
described in Sec.\ref{sec: induced action and linear response}.

\subsection{Additional terms at third order \label{subsec:Additional terms at third order}}

To obtain reliable results at $O\left(p^{3}\right)$ we, in principle,
need the full Lagrangian $\mathcal{L}_{3}$, which includes, but is
not equal to, $\mathcal{L}_{\text{gCS}}$. Nevertheless, we argue
that $\mathcal{L}_{3}-\mathcal{L}_{\text{gCS}}$ does not contribute
to the quantity of interest in this section - $\eta_{\text{o}}$ to
$O\left(q^{2}\right)$. We already demonstrated in Appendix \ref{subsec:Odd-viscosity-from}
that the vertex part of the odd viscosity $\eta_{\mathcal{V}}$ is
independent of $\mathcal{L}_{3}$, and it remains to show that the
contact term part $\eta_{\mathcal{C}}$ is independent of $\mathcal{L}_{3}-\mathcal{L}_{\text{gCS}}$.
We do not have a general proof, but we address this issue in two ways: 
\begin{enumerate}
\item Within the microscopic model \eqref{eq:3-1-1}, the perturbative computation
of Appendix  \ref{subsec:Perturbative-expansion} provides an explicit
expression for $\eta_{\mathcal{C}}$, which is completely saturated
by the effective action presented thus far. Thus $\eta_{\mathcal{C}}$
is independent of $\mathcal{L}_{3}-\mathcal{L}_{\text{gCS}}$ in the
particular realization \eqref{eq:3-1-1}. 
\item The term $\mathcal{L}_{3}$ is $P,T$-odd, and therefore vanishes
in an $s$-wave SF. On the other hand, it suffices to consider the
$g$s-wave SF where $s_{\theta}=0$ (but $\ell\neq0$), since for
$s_{\theta}\neq0$ the spin connection included in $\nabla_{\mu}\theta$
will only produce $O\left(p^{4}\right)$ corrections. By contracting
Galilean vectors, we were able to construct four $P,T$-odd terms
in $\mathcal{L}_{3}-\mathcal{L}_{\text{gCS}}$ for the $g$s-wave
SF, 
\begin{align}
\mathcal{L}_{3}-\mathcal{L}_{\text{gCS}}\supset & \ell\left[C_{1}\left(X\right)\tilde{E}_{i}E_{\omega}^{i}+C_{2}\left(X\right)\varepsilon^{ij}\tilde{E}_{i}E_{\omega,j}+C_{3}\left(X\right)\partial_{i}XE_{\omega}^{j}+C_{4}\left(X\right)\varepsilon^{ij}\partial_{i}XE_{\omega,j}\right].
\end{align}
where $\tilde{E}_{i}$ is the electric field of the improved $U\left(1\right)$
connection $\tilde{A}_{t}=A_{t}+\frac{1}{2m}\nabla^{i}\theta\nabla_{i}\theta,\;\tilde{A}_{i}=\partial_{i}\theta-s_{\theta}\omega_{i}$
\cite{hoyos2014effective}. Perturbatively expanding these, we do
not find any $O\left(q^{2}\right)$ contributions to $\eta_{\mathcal{C}}$
(or to $\eta_{\text{\ensuremath{\mathcal{V}}}}$, in accordance with
Appendix \ref{subsec:Odd-viscosity-from}).
\end{enumerate}

\section{Induced action \label{subsec:Obtaining--from}}

The arguments presented in Sec.\ref{sec:Main-section-2:} suffice to establish the
quantization of $\tilde{\eta}_{\text{o}}$ and $\tilde{\chi}_{TJ,\text{o}}$
directly from $S_{\text{eff}}$ - an explicit expression for $S_{\text{ind}}$
is not required. Nevertheless, it is instructive to compute certain
contributions in $S_{\text{ind}}$ to demonstrate these results 
explicitly, and also to reproduce simpler properties of $\ell$-wave
SFs. Here we will compute the contribution of $\mathcal{L}_{0}+\mathcal{L}_{2}^{\left(1\right)}\subset\mathcal{L}_{\text{eff}}$
to the induced Lagrangian $\mathcal{L}_{\text{ind}}$, and, along
the way, demonstrate explicitly the relation between $\text{var}s=0$
QH states and CSFs alluded to in Sec.\ref{sec:discussion}.

The starting point is the induced action due to $\mathcal{L}_{0}=P\left(X\right)$,
obtained from \eqref{eq:51-0}. It is given by 
\begin{align}
\mathcal{L}_{\text{ind}}= & -\frac{1}{2}\mathcal{V}\mathcal{G}\mathcal{V}+\mathcal{C}\label{eq:75}\\
= & P_{0}\sqrt{G}-n_{0}\mathcal{A}_{t}\nonumber \\
 & +\frac{1}{2}\frac{n_{0}}{m}\frac{\mathcal{B}^{2}-c_{s}^{-2}\mathcal{E}^{2}+\frac{s_{\theta}c_{s}^{-2}}{m}\mathcal{E}^{i}\partial_{i}B-\frac{s_{\theta}^{2}c_{s}^{-2}}{4m^{2}}\left(\partial B\right)^{2}}{\partial^{2}-c_{s}^{-2}\partial_{t}^{2}}\nonumber \\
 & -n_{0}\frac{m\left(\partial_{t}\sqrt{G}\right)^{2}/2+\left(\mathcal{E}^{i}-\frac{s_{\theta}}{2m}\partial_{i}B\right)\partial_{i}\sqrt{G}}{\partial^{2}-c_{s}^{-2}\partial_{t}^{2}}.\nonumber 
\end{align}
This expression contains, rather compactly, the entire linear response
of the $\ell$-wave SF to $O\left(p\right)$ in the derivative expansion,
as well as certain $O\left(p^{2}\right)$ contributions \cite{hoyos2014effective},
and should be interpreted as expanded to second order using \eqref{eq:48-0}-\eqref{eq:49-0}.
In using \eqref{eq:49-0}, one can set $H_{[AB]}=0$, since $S_{\text{ind}}$
is $SO\left(2\right)_{L}$ invariant and the anti-symmetric part $H_{[AB]}$
corresponds to the $SO\left(2\right)_{L}$ phase of the vielbein $E_{A}^{\;i}$.
Technically, $H_{[AB]}$ always appears in the combination $\partial_{\mu}\left(\theta+s_{\theta}\varepsilon^{AB}H_{AB}/2\right)\subset\nabla_{\mu}\theta$,
so that integrating out $\theta$ eliminates $H_{[AB]}$. 

Note that, diagrammatically, equation \eqref{eq:75} corresponds to
linear response at tree-level. Higher orders in $\theta$ will generate
diagrams with $\theta$ running in loops, which can be shown to produce
$O\left(p^{3}\right)$ corrections above the leading order to any
observable \cite{son2006general}, and are therefore irrelevant for
the purpose of $q^{2}$ corrections to $\eta_{\text{o}}$. 

The $O\left(p^{0}\right)$ part of \eqref{eq:75} is obtained by setting
$s_{\theta}=0$, as in an $s$-wave SF, 

\begin{align}
\mathcal{L}_{\text{ind},0}= & P_{0}\sqrt{G}-n_{0}A_{t}+\frac{1}{2}\frac{n_{0}}{m}\frac{B^{2}-c_{s}^{-2}E^{2}}{\partial^{2}-c_{s}^{-2}\partial_{t}^{2}}\\
 & -n_{0}\frac{m\left(\partial_{t}\sqrt{G}\right)^{2}/2+E^{i}\partial_{i}\sqrt{G}}{\partial^{2}-c_{s}^{-2}\partial_{t}^{2}}\nonumber 
\end{align}
The first line contains the ground state pressure and density $P_{0},n_{0}$,
as well as the London diamagnetic function $\rho_{\text{e}}=\frac{n_{0}}{m}\frac{1}{q^{2}-c_{s}^{-2}\omega^{2}}$
and the ideal Drude longitudinal conductivity $\sigma_{\text{e}}=-\frac{n_{0}}{m}\frac{i\omega c_{s}^{-2}}{q^{2}-c_{s}^{-2}\omega^{2}}$
of the SF \cite{hoyos2014effective}. The second line contains the
mixed response and mixed static susceptibility 
\begin{align}
\kappa_{\text{e}}^{ij,k} & =-n_{0}\delta^{ij}\frac{iq^{k}}{q^{2}-c_{s}^{-2}\omega^{2}},\\
\chi_{TJ,\text{e}}^{ij,t} & =n_{0}\delta^{ij}\frac{q^{2}}{q^{2}-c_{s}^{-2}\omega^{2}},\nonumber 
\end{align}
defined in Sec.\ref{sec: induced action and linear response}, as well as the inverse compressibility
$K^{-1}=-n_{0}m\frac{\omega^{2}}{q^{2}-c_{s}^{-2}\omega^{2}}$ (which
agrees with the thermodynamic expression $K^{-1}=n_{0}^{2}\frac{\partial\mu}{\partial n_{0}}=n_{0}mc_{s}^{2}$
at $q=0$). In particular, the $\ell$-wave SF is indeed a superfluid
- the even viscosity $\eta_{\text{e}}$ vanishes to zeroth order in
derivatives (see \cite{PhysRevB.86.245309} for a subtlety in separating
$K^{-1}$ from $\eta_{\text{e}}$). 

The $O\left(p\right)$ part of the \eqref{eq:75} is $P,T$-odd and
vanishes when $s_{\theta}=0$. It is given by 
\begin{align}
\mathcal{L}_{\text{ind},1}= & -s_{\theta}n_{0}\omega_{t}\label{eq:76-1-1}\\
 & +\frac{1}{2}\frac{s_{\theta}n_{0}}{m^{2}c_{s}^{2}}\frac{E^{i}\partial_{i}B}{\partial^{2}-c_{s}^{-2}\partial_{t}^{2}}\nonumber \\
 & -s_{\theta}n_{0}\frac{\left(E_{\omega}^{i}-\frac{1}{2m}\partial_{i}B\right)\partial_{i}\sqrt{G}}{\partial^{2}-c_{s}^{-2}\partial_{t}^{2}}.\nonumber 
\end{align}
The first and third lines produce the following odd viscosity \cite{hoyos2014effective},
\begin{align}
\eta_{\text{o}}^{\left(1\right)}= & -\frac{1}{2}s_{\theta}n_{0},\label{eq:14-1-0}\\
\eta_{\text{o}}^{\left(2\right)}= & \frac{1}{2}s_{\theta}n_{0}\frac{1}{q^{2}-c_{s}^{-2}\omega^{2}},\nonumber 
\end{align}
and setting $\omega=0$ one obtains the leading terms in equation
\eqref{eq:14-1}.  By using the identity (up to a total derivative)
\begin{align}
 & E^{i}\partial_{i}B=\frac{1}{2}\varepsilon^{\mu\nu\rho}A_{\mu}\partial_{\nu}\partial^{2}A_{\rho},\label{eq:81-0}
\end{align}
 the second line of \eqref{eq:76-1-1} can be written as a non-local
CS term 

\begin{align}
\mathcal{L}_{\text{ind}}\supset & \frac{1}{2}\sigma_{\text{o}}\left(\omega,q\right)\varepsilon^{\mu\nu\rho}A_{\mu}ip_{\nu}A_{\rho}\label{eq:81-1-0}
\end{align}
with the odd (or Hall) conductivity $\sigma_{\text{o}}\left(\omega,q\right)=\sigma_{\text{o}}^{0}q^{2}/\left(q^{2}-c_{s}^{-2}\omega^{2}\right)$,
$\sigma_{\text{o}}^{0}=s_{\theta}n_{0}/2m^{2}c_{s}^{2}$ \cite{volovik1988quantized,goryo1998abelian,goryo1999observation,furusaki2001spontaneous,stone2004edge,roy2008collective,lutchyn2008gauge,hoyos2014effective},
with $\sigma_{\text{o}}\left(0,q\right)=\sigma_{\text{o}}^{0}$ unquantized,
and $\sigma_{\text{o}}\left(\omega,0\right)=0$, in accordance with
the boundary $U\left(1\right)_{N}$-neutrality \cite{PhysRevB.98.064503}.

To demonstrate explicitly that $c$ cannot be extracted from the odd
viscosity alone, it suffices to add the $O\left(p^{2}\right)$ term
$\mathcal{L}_{2}^{\left(1\right)}=F_{1}\left(X\right)R\subset\mathcal{L}_{2}$.
The situation is particularly simple for the special case $F_{1}\left(X\right)=-s_{\theta}^{2}P'\left(X\right)/4m$.
Then
\begin{align}
P\left(X-\frac{s_{\theta}^{2}}{4m}R\right) & =P\left(X\right)-\frac{s_{\theta}^{2}}{4m}P'\left(X\right)R+O\left(p^{4}\right)\\
 & =P\left(X\right)+F_{1}\left(X\right)R+O\left(p^{4}\right),\nonumber 
\end{align}
which shows that $F_{1}\left(X\right)R$ can be absorbed into $P\left(X\right)$
by a modification of $X$. The scalar $X-\frac{s_{\theta}^{2}}{4m}R$
is useful because, unlike $X$, it depends on $A_{\mu}$ and $\omega_{\mu}$
\textit{only} through the combination $\mathcal{A}_{\mu}=A_{\mu}+s_{\theta}\omega_{\mu}$.
This is evident in \eqref{eq:69}, where $B$ rather than $\mathcal{B}=B+\frac{s}{2}R$
appears. It is then clear that, to $O\left(p^{3}\right)$, adding
$\mathcal{L}_{2}^{\left(1\right)}=F_{1}\left(X\right)R=-\frac{s_{\theta}^{2}}{4m}P'\left(X\right)R$
to $\mathcal{L}_{0}=P\left(X\right)$ amounts to changing $B$ to
$\mathcal{B}$ in the induced Lagrangian \eqref{eq:75}, 
\begin{align}
\mathcal{L}_{\text{ind}}= & P_{0}\sqrt{G}-n_{0}\mathcal{A}_{t}\\
 & +\frac{1}{2}\frac{n_{0}}{m}\frac{\mathcal{B}^{2}-c_{s}^{-2}\mathcal{E}^{2}+\frac{s_{\theta}c_{s}^{-2}}{m}\mathcal{E}^{i}\partial_{i}\mathcal{B}-\frac{s_{\theta}^{2}c_{s}^{-2}}{4m^{2}}\left(\partial\mathcal{B}\right)^{2}}{\partial^{2}-c_{s}^{-2}\partial_{t}^{2}}\nonumber \\
 & -n_{0}\frac{m\left(\partial_{t}\sqrt{G}\right)^{2}/2+\left(\mathcal{E}^{i}-\frac{s_{\theta}}{2m}\partial_{i}\mathcal{B}\right)\partial_{i}\sqrt{G}}{\partial^{2}-c_{s}^{-2}\partial_{t}^{2}}.\nonumber 
\end{align}
The only contribution to $\eta_{\text{o}}$, beyond \eqref{eq:14-1-0},
comes from the term proportional to $\mathcal{E}^{i}\partial_{i}\mathcal{B}$.
By using the identity \eqref{eq:81-0} for $\mathcal{A}_{\mu}$, this
term can be written as the sum of non-local CS, WZ1, and WZ2 terms,
which generalizes \eqref{eq:81-1-0} to 
\begin{align}
\mathcal{L}_{\text{ind}}\supset & \frac{1}{2}\sigma_{\text{o}}\left(\omega,q\right)\varepsilon^{\mu\nu\rho}\left(A_{\mu}+s_{\theta}\omega_{\mu}\right)ip_{\nu}\left(A_{\rho}+s_{\theta}\omega_{\rho}\right).\label{eq:16-1-1}
\end{align}
Most importantly, this includes a non-local version of WZ2, which
is indistinguishable from $\mathcal{L}_{\text{gCS}}$ at $\omega=0$,
where $\sigma_{\text{o}}\left(0,q\right)=\sigma_{\text{o}}^{0}$ is
a constant. Noting that $F_{1}'=-s_{\theta}^{2}P''/4m=-\left(s_{\theta}/2\right)\sigma_{\text{o}}^{0}$,
and comparing to \eqref{eq:54-2}, it follows that $c$ and $F_{1}'$
will enter the $\omega=0$ odd viscosity only through the combination
$c+48\pi s_{\theta}F_{1}'$. In more detail, the odd viscosity tensor
due to $\mathcal{L}_{0}+\mathcal{L}_{2}^{\left(1\right)}+\mathcal{L}_{\text{gCS}}$,
is given by 

\begin{align}
\eta_{H}^{\left(1\right)}\left(\omega,q^{2}\right)= & -\frac{1}{2}s_{\theta}n_{0}-\left(\frac{c}{24}\frac{1}{4\pi}+\frac{s_{\theta}}{2}F_{1}'\frac{q^{2}}{q^{2}-c_{s}^{-2}\omega^{2}}\right)q^{2}+O\left(q^{4}\right),\label{eq:86}\\
\eta_{H}^{\left(2\right)}\left(\omega,q^{2}\right)= & \frac{1}{2}s_{\theta}n_{0}\frac{1}{q^{2}-c_{s}^{-2}\omega^{2}}+\left(\frac{c}{24}\frac{1}{4\pi}+\frac{s_{\theta}}{2}F_{1}'\frac{q^{2}}{q^{2}-c_{s}^{-2}\omega^{2}}\right)+O\left(q^{2}\right),\nonumber 
\end{align}
which, at $\omega=0$, is a special case of equation (12) of the main
text. 

Equation \eqref{eq:86} remains valid away from the special point
$F_{1}=-s_{\theta}^{2}P'/4m$, even though \eqref{eq:16-1-1} does
not. Examining the perturbatively expanded $\mathcal{L}_{0}$ \eqref{eq:51-0}
and $\mathcal{L}_{2}^{\left(1\right)}$ \eqref{eq:54-0}, we see that
a general $F_{1}$ amounts to replacing $B$ in \eqref{eq:75} with
$B+\alpha\frac{s_{\theta}}{2}R$, where $\alpha=-\frac{4mF_{1}'}{s_{\theta}^{2}P''}\neq1$
generically (as well as in the microscopic model \eqref{eq:c1}-\eqref{eq:c2}).
The general induced Lagrangian due to $\mathcal{L}_{0}+\mathcal{L}_{2}^{\left(1\right)}$,
valid to $O\left(p^{3}\right)$, is then given by
\begin{align}
\mathcal{L}_{\text{ind}}= & P_{0}\sqrt{G}-n_{0}\mathcal{A}_{t}\label{eq:87-0}\\
 & +\frac{1}{2}\frac{n_{0}}{m}\frac{\mathcal{B}^{2}-c_{s}^{-2}\mathcal{E}^{2}+\frac{s_{\theta}c_{s}^{-2}}{m}\mathcal{E}^{i}\partial_{i}\left(B+\alpha\frac{s_{\theta}}{2}R\right)-\frac{s_{\theta}^{2}c_{s}^{-2}}{4m^{2}}\left(\partial B\right)^{2}}{\partial^{2}-c_{s}^{-2}\partial_{t}^{2}}\nonumber \\
 & -n_{0}\frac{m\left(\partial_{t}\sqrt{G}\right)^{2}/2+\left[\mathcal{E}^{i}-\frac{s_{\theta}}{2m}\partial_{i}\left(B+\alpha\frac{s_{\theta}}{2}R\right)\right]\partial_{i}\sqrt{G}}{\partial^{2}-c_{s}^{-2}\partial_{t}^{2}},\nonumber 
\end{align}
and, along with the $\mathcal{L}_{\text{gCS}}$ , produces the odd
viscosity \eqref{eq:86}. This expression does not depend on $A_{\mu},\omega_{\mu}$
only through $\mathcal{A}_{\mu}$, but the terms contributing to \eqref{eq:86}
still vanish  $s_{\theta}=0$, which is why the \textit{improved}
odd viscosity due to \eqref{eq:87-0} vanishes. In addition to $\mathcal{L}_{2}^{\left(1\right)}$,
the second order terms $\mathcal{L}_{2}^{\left(2\right)},\mathcal{L}_{2}^{\left(3\right)}$
\eqref{eq:49-2} also produce $q^{2}$ corrections to the odd viscosity,
but not to the improved odd viscosity.

Though equation \eqref{eq:16-1-1} describes only a part of $\mathcal{L}_{\text{ind}}$,
and is non-generic, it does reveal the analogy between $\ell$-wave
SFs and $\text{var}s=0$ QH states, described in Sec.\ref{sec:discussion}, in a very simple setting. Indeed, comparing \eqref{eq:16-1-1}
with Eq.\eqref{eq:11}, we see that $\ell$-wave SFs
are analogous to $\text{var}s=0$ QH states, with $\overline{s}=s_{\theta}=\ell/2$,
but with a non-local, non-quantized, Hall conductivity, in place of
the filling factor $\nu/2\pi$. Additionally, both QH states and $\ell$-wave
SFs have the same gCS term \eqref{eq:54-2}, with $c$ the boundary
chiral central charge.

\section{Detailed analysis of the microscopic model Eq.\eqref{eq:3-1-1} \label{sec:microscopic model}}


\subsection{Symmetry \label{subsec:Symmetry}}
The microscopic action $S_{\text{m}}$ of Eq.\eqref{eq:3-1-1} is invariant under $U\left(1\right)_{N}$ gauge transformations, \begin{align}  & \psi\mapsto e^{-i\alpha}\psi,\;\Delta^{j}\mapsto e^{-2i\alpha}\Delta^{j},\;A_{\mu}\mapsto A_{\mu}+\partial_{\mu}\alpha, \end{align} which implies the current conservation $\partial_{\mu}(\sqrt{G}J^{\mu})=0$, where $\sqrt{G}J^{\mu}=-\delta S/\delta A_{\mu}$. It is also clear that $S_{\text{m}}$ is invariant under \textit{time-independent} spatial diffeomorphisms, generated by $\delta x^{i}=\xi^{i}\left(\mathbf{x}\right)$, if $\psi$ transforms as a function, $A_{\mu}$ as a 1-form, $\Delta^{j}$ as a vector, and $G_{ij}$ as a rank-2 tensor. As described in section \ref{sec: building blocks}, due to its Galilean symmetry in flat space, $S_{\text{m}}$ is also invariant under time-dependent spatial diffeomorphisms $\delta x^{i}=\xi^{i}\left(\mathbf{x},t\right)$, provided one modifies the transformation rule of $A_{i}$ to Eq.\eqref{eq:4-3-1-1}.
\subsection{Effective action and fermionic Green's function }
Starting with the microscopic action \eqref{eq:3-1-1}, the effective action for the order parameter $\Delta$ in the $A,G$ background is obtained by integrating out the (generically) gapped fermion $\psi$, \begin{align}  & e^{iS_{\text{eff,m}}\left[\Delta;A,G\right]}=\int\text{D}\left(G^{1/4}\psi\right)\text{D}\left(G^{1/4}\psi^{\dagger}\right)e^{iS_{\text{m}}\left[\psi;\Delta,A,G\right]},\label{eq:20} \end{align} where $G^{1/4}=\left(\text{det}G_{ij}\right)^{1/4}$ is the square root of the volume element $\sqrt{G}$. The form of the measure is fixed by the fact that the fundamental fermionic degree of freedom is the fermion-density $\tilde{\psi}=G^{1/4}\psi$, which satisfies the usual canonical commutation relation $\left\{ \tilde{\psi}^{\dagger}\left(\mathbf{x}\right),\tilde{\psi}\left(\mathbf{y}\right)\right\} =\delta^{\left(2\right)}\left(\mathbf{x}-\mathbf{y}\right)$ as an operator \cite{hawking1977zeta,fujikawa1980comment,abanov2014electromagnetic,PhysRevB.98.064503}. This is to be contrasted with $\left\{ \psi^{\dagger}\left(\mathbf{x}\right),\psi\left(\mathbf{y}\right)\right\} =\delta^{\left(2\right)}\left(\mathbf{x}-\mathbf{y}\right)/\sqrt{G\left(\mathbf{x}\right)}$ which ties the fermion to the background metric. 
In terms of $\tilde{\psi}$ the action \eqref{eq:3-1-1} takes the form \begin{align} S_{\text{m}} & =\int\text{d}^{2}x\text{d}t\left[\tilde{\psi}^{\dagger}\frac{i}{2}\overleftrightarrow{\nabla}_{t}\tilde{\psi}-\frac{1}{2m}G^{ij}\nabla_{i}\tilde{\psi}^{\dagger}\nabla_{j}\tilde{\psi}+\left(\frac{1}{2}\Delta^{i}\tilde{\psi}^{\dagger}\nabla_{i}\tilde{\psi}^{\dagger}+h.c\right)-\mathcal{U}\right],\label{eq:78} \end{align} where $\nabla_{\mu}=\partial_{\mu}+iA_{\mu}-\frac{1}{4}\partial_{\mu}\log G$ is the covariant derivative for densities, and $\mathcal{U}=\frac{1}{2\lambda}\sqrt{G}G_{ij}\Delta^{i*}\Delta^{j}$.   Passing to the BdG form of the fermionic part of the action, in terms of the Nambu spinor-density $\tilde{\Psi}^{\dagger}=\left(\tilde{\psi}^{\dagger},\tilde{\psi}\right)$ (which is a Majorana spinor-density \cite{PhysRevB.98.064503}), one finds  \begin{align} S_{\text{m}}=\int\text{d}^{2}x\text{d}t & \left\{ \frac{1}{2}\tilde{\Psi}^{\dagger}\gamma^{0}\left[i\gamma^{0}\partial_{t}-A_{t}+\frac{1}{2m}\nabla_{i}G^{ij}\nabla_{j}\right.\right.\label{eq:22}   +\left.\left.\frac{i}{2}\gamma^{\tilde{A}}\left(e_{\tilde{A}}^{\;i}\partial_{i}+\partial_{i}e_{\tilde{A}}^{\;i}\right)\right]\tilde{\Psi}-\mathcal{U}\right\} \nonumber \\ =\int\text{d}^{2}x\text{d}t & \left\{ \frac{1}{2}\tilde{\Psi}^{\dagger}\gamma^{0}\mathcal{G}^{-1}\tilde{\Psi}-\mathcal{U}\right\} ,\nonumber  \end{align} where derivatives act on all fields to the right; $\tilde{A}=1,2$ is an index for $U\left(1\right)_{N}$, viewed as a copy of $SO\left(2\right)$; the gamma matrices are $\gamma^{0}=\sigma^{z},\;\gamma^{1}=-i\sigma^{x},\;\gamma^{2}=i\sigma^{y}$, satisfying $\left\{ \gamma^{\mu},\gamma^{\nu}\right\} =2\eta^{\mu\nu}$ with $\eta^{\mu\nu}=\text{diag}\left[1,-1,-1\right]$, and $\text{tr}\left(\gamma^{0}\gamma^{1}\gamma^{2}\right)=2i$; and  \begin{align}  & e_{\tilde{A}}^{\;i}=\begin{pmatrix}\text{Re}\Delta^{x} & \text{Re}\Delta^{y}\\ \text{Im}\Delta^{x} & \text{Im}\Delta^{y} \end{pmatrix} \end{align} is the \textit{emergent} vielbein \cite{volovik1990gravitational,PhysRevB.98.064503}, to be distinguished from the background vielbein $E_{A}^{\;i}$ (with an $SO\left(2\right)_{L}$ index $A=1,2$) that was introduced in Sec.\ref{sec: building blocks} and that will be used momentarily. We also defined the inverse Green's function $\mathcal{G}^{-1}$. The effective action \eqref{eq:20} is then given by the logarithm of the Pfaffian  \begin{align} S_{\text{eff,m}}= & -i\log\text{Pf}\left(i\gamma^{0}\mathcal{G}^{-1}\right)-\int\text{d}^{2}x\text{d}t\mathcal{U}\label{eq:24-1}\\ = & -\frac{i}{2}\log\text{Det}\left(i\gamma^{0}\mathcal{G}^{-1}\right)-\int\text{d}^{2}x\text{d}t\mathcal{U}.\nonumber  \end{align}
\subsection{Fermionic ground state topology }
For a given $\Delta^{j}$, the fermion $\psi$ is gapped, unless the chemical potential $\mu$ or chirality $\ell=\text{sgn}\left(\text{Im}\left(\Delta^{x*}\Delta^{y}\right)\right)$  are tuned to 0, and forms a fermionic topological phase characterized by the bulk Chern number.  Assuming $A_{\mu}=0$ and space-time independent $\Delta^{i},G^{ij}$, it is given by \cite{volovik2009universe} \begin{align} C= & \frac{1}{24\pi^{2}}\mbox{tr}\int\mbox{d}^{3}q\varepsilon^{\alpha\beta\gamma}\left(\mathcal{G}\partial_{\alpha}\mathcal{G}^{-1}\right)\left(\mathcal{G}\partial_{\beta}\mathcal{G}^{-1}\right)\left(\mathcal{G}\partial_{\gamma}\mathcal{G}^{-1}\right)\in\mathbb{Z},\label{eq:28-1} \end{align} and determines the boundary chiral central charge $c=C/2$ \cite{read2000paired,kitaev2006anyons,volovik2009universe,ryu2010topological}. Here the fermionic Green's function $\mathcal{G}$ is Fourier transformed to Euclidian 3-momentum $q=\left(iq_{0},\mathbf{q}\right)$ (see \eqref{eq:184-1}). For the particular model \eqref{eq:3-1-1} one finds  \begin{align}  & c=-\left(\ell/4\right)\left(\text{sgn}\left(\mu\right)+\text{sgn}\left(m\right)\right)\in\left\{ 0,\pm1/2\right\} ,\label{eq:29} \end{align} see \cite{read2000paired,volovik2009universe,PhysRevB.98.064503} for similar expressions. Note that the central charge is well defined for both $m>0$ and $m<0$, even though the single particle dispersion is not bounded from below in the latter, and many physical quantities naively diverge (we will see below that certain physical quantities diverge also with $m>0$). A negative mass can occur as an effective mass in lattice models, in which case the lattice spacing provides a natural cutoff (which must be smooth in momentum space for \eqref{eq:28-1} to hold). In any case, a negative mass makes it possible to obtain both fundamental central charges $c=\pm1/2$, for fixed $\ell$, within the model \eqref{eq:3-1-1}. All possible $c\in\left(1/2\right)\mathbb{Z}$ can then be obtained by stacking layers of the model \eqref{eq:3-1-1} with the same $\ell$ but different $m,\mu$. Thus the model \eqref{eq:3-1-1} suffices to generate a representative for all topological phases of the $p$-wave CSF. For concreteness, below we will work only with $m>0$, in which case $c$ is given by Eq.\eqref{eq:12-1-2}. 
\subsection{Symmetry breaking and bosonic ground state in the presence of a background metric\label{subsec:Symmetry-breaking-and}}
For time independent fields $A,G,\Delta$ the effective action reduces to  \begin{align} S_{\text{eff,m}}\left[\Delta;G\right] =&-\int\text{d}^{2}x\text{d}t\varepsilon_{0}\left[\Delta;G\right], \end{align} where $\varepsilon_{0}$ is the ground-state energy-density as a function of the fields. In flat space $G_{ij}=\delta_{ij}$, with $A_{t}=-\mu$ and $A_{i}=0$, and assuming $\Delta$ is constant, it is given by \cite{volovik2009universe,PhysRevB.98.064503}  \begin{align} \varepsilon_{0}= & \frac{1}{2}\int\frac{\text{d}^{2}\mathbf{q}}{\left(2\pi\right)^{2}}\left(\xi_{\mathbf{q}}-\sqrt{\xi_{\mathbf{q}}^{2}+g^{ij}q_{i}q_{j}}\right)+\frac{1}{2\lambda}\delta_{ij}g^{ij},\label{eq:27} \end{align} where  \begin{align}  & \xi_{\mathbf{q}}=\left|\mathbf{q}\right|^{2}/2m-\mu\label{eq:33} \end{align} is the single particle dispersion, and $g^{ij}=\Delta^{(i}\Delta^{j)*}=\delta^{\tilde{A}\tilde{B}}e_{\tilde{A}}^{\;i}e_{\tilde{B}}^{\;j}$ is the \textit{emergent metric} - a dynamical metric to be distinguished from the background metric $G^{ij}$. The ground state configuration of $g^{ij}$ is determined by minimizing $\varepsilon_{0}$, while the overall phase $\theta$ of the order parameter and the chirality $\ell$, of which $g^{ij}$ is independent, are left undetermined. Thus $g^{ij}$ corresponds to a massive Higgs field, while $\theta$ is a Goldstone field. The energy-density \eqref{eq:27} is UV divergent, and requires regularization. We do this in the simplest manner, by introducing a momentum cutoff $q^{2}<\Lambda^{2}$. Since the divergence disappears for $g^{ij}=0$ (assuming $m>0$), this can be thought of as a small, but non-vanishing, range $1/\Lambda$ for the interaction mediated by $\Delta$. With a finite $\Lambda$, the energy-density is well defined and has a unique global minimum at $g^{ij}=\Delta_{0}^{2}\delta^{ij}$, with $\Delta_{0}$ determined by the self-consistent equation  \begin{align}  & \frac{1}{4}\int^{\Lambda}\frac{\text{d}^{2}\boldsymbol{q}}{\left(2\pi\right)^{2}}\frac{\left|\boldsymbol{q}\right|^{2}}{\sqrt{\xi_{\boldsymbol{q}}^{2}+\Delta_{0}^{2}\left|\boldsymbol{q}\right|^{2}}}=\frac{1}{\lambda}.\label{eq:28} \end{align} For $\mu>0$ the non-interacting system has a Fermi surface, and a solution exists for all $\lambda>0$, which is the statement of the BCS instability. For $\mu<0$, the non-interacting system is gapped, and a solution exists if the interaction is large enough compared with the gap, $\lambda\Lambda^{-4}\gtrsim\left|\mu\right|$.
Consider now the case of a general constant metric $G_{ij}$, and let us introduce a constant vielbein $E$ such that $G_{ij}=E_{\;i}^{A}\delta_{AB}E_{\;j}^{B}$. The inverse transpose $E^{-T}=\left(E^{-1}\right)^{T}$ is given in coordinates by $E_{A}^{\;i}$. We also introduce the internal order parameter $\Delta^{A}=E_{\;i}^{A}\Delta^{i}$. The action \eqref{eq:78} then reduces to  \begin{align} S_{\text{m}} & =\int\text{d}^{2}x\text{d}t\left[\tilde{\psi}^{\dagger}i\partial_{t}\tilde{\psi}-\frac{\delta^{AB}}{2m}E_{A}^{\;i}\partial_{i}\tilde{\psi}^{\dagger}E_{B}^{\;j}\partial_{j}\tilde{\psi}+\left(\frac{1}{2}\Delta^{A}E_{A}^{\;i}\tilde{\psi}^{\dagger}\partial_{i}\tilde{\psi}^{\dagger}+h.c\right)-\sqrt{G}\frac{1}{2\lambda}\delta_{AB}\Delta^{A*}\Delta^{B}\right].\label{eq:78-1} \end{align} This is identical to the flat space case, with $\partial_{i}$ replaced by $E_{A}^{\;i}\partial_{i}$. We also need to change the UV cutoff to $\delta^{AB}E_{A}^{\;i}q_{i}E_{B}^{\;j}q_{j}=G^{ij}q_{i}q_{j}<\Lambda^{2}$. This in natural since we interpret $\Lambda^{2}$ as a range of the interaction mediated by $\Delta$, which should be defined in terms of the geodesic distance rather than the Euclidian distance. It follows that the flat space result \eqref{eq:27} is modified to  \begin{align} \varepsilon_{0} & =\frac{1}{2}\int_{\left|E^{-T}\mathbf{q}\right|^{2}<\Lambda^{2}}\frac{\text{d}^{2}\mathbf{q}}{\left(2\pi\right)^{2}}\left(\xi_{E^{-T}\mathbf{q}}-\sqrt{\xi_{E^{-T}\mathbf{q}}^{2}+g^{AB}E_{A}^{\;i}E_{B}^{\;j}q_{i}q_{j}}\right)+\sqrt{G}\frac{1}{2\lambda}\delta_{AB}g^{AB}\label{eq:30}\\  & =\sqrt{G}\left[\frac{1}{2}\int_{q^{2}<\Lambda^{2}}\frac{\text{d}^{2}\mathbf{k}}{\left(2\pi\right)^{2}}\left(\xi_{\mathbf{k}}-\sqrt{\xi_{\mathbf{k}}^{2}+g^{AB}k_{A}k_{B}}\right)+\frac{1}{2\lambda}\delta_{AB}g^{AB}\right],\nonumber  \end{align} where $\mathbf{k}=E^{-T}\mathbf{q}$, or $k_{A}=E_{A}^{\;i}q_{i}$, and $g^{AB}=\Delta^{(A}\Delta^{B)*}=\delta^{\tilde{A}\tilde{B}}e_{\tilde{A}}^{\;A}e_{\tilde{B}}^{\;B}$ is the \textit{internal} emergent metric. This is identical to the $G_{ij}=\delta_{ij}$ result \eqref{eq:27}, apart from the volume element $\sqrt{G}$, and the fact that it is the internal metric $g^{AB}$ that appears, rather than $g^{ij}$. It is then clear that minimizing \eqref{eq:30} with respect to $g^{AB}$ gives  \begin{align}  & g^{AB}=\Delta_{0}^{2}\delta^{AB}\text{, or }g^{ij}=\Delta_{0}^{2}G^{ij}, \end{align} with the same $\Delta_{0}$ of \eqref{eq:28}, which is $G$ independent. Thus, the emergent metric is proportional to the background metric in the ground state.  This solution corresponds to emergent vielbeins $e_{\tilde{A}}^{\;A}\in O\left(2\right)$, or order parameters $\Delta^{A}=\Delta_{0}e^{2i\theta}\left(1,\pm i\right)$, which is the $p_{x}\pm ip_{y}$ configuration, and implies the SSB pattern  \begin{align}  & \left(\mathbb{Z}_{2,T}\ltimes U\left(1\right)_{N}\right)\times\left(\mathbb{Z}_{2,P}\ltimes SO\left(2\right)_{L}\right)\rightarrow\begin{cases} \mathbb{Z}_{2,PT}\ltimes U\left(1\right)_{L-\frac{\ell}{2}N} & \ell\in2\mathbb{Z}+1\\ \mathbb{Z}_{2,PT}\ltimes U\left(1\right)_{L-\frac{\ell}{2}N}\times\mathbb{Z}_{2,\left(-1\right)^{N}} & \ell\in2\mathbb{Z} \end{cases},\label{eq:4-2-1-1} \end{align} described less formally in Eq.\eqref{eq:2-1-1}. Note that fermion parity $\mathbb{Z}_{2,\left(-1\right)^{N}}$ is the $\mathbb{Z}_{2}$ subgroup of $U\left(1\right)_{L-\frac{\ell}{2}N}$ for odd $\ell$. For $\Delta^{j}$, we find the ground state configuration Eq.\eqref{eq:12-3}. 
As described in Sec.\ref{subsec:Benchmarking-the-effective}, we will ignore the massive Higgs fluctuations, and obtain $S_{\text{eff}}\left[\theta;A,G\right]$ by plugging the ground state configuration \eqref{eq:12-3} into the functional Pfaffian \eqref{eq:24-1}.
\subsection{Perturbative expansion\label{subsec:Perturbative-expansion}}
We now write $E_{A}^{\;i}=\delta_{A}^{i}+H_{A}^{\;i}$ and $e_{\tilde{A}}^{\;A}=\Delta_{0}\delta_{\tilde{A}}^{A}$ (which corresponds to $\Delta^{A}=\Delta_{0}\left(1,i\right)^{A}$) and expand \eqref{eq:22} to second order in $H,A$. Due to $SO\left(2\right)_{L}$ gauge symmetry, the anti-symmetric part of $H$ can be interpreted as the Goldstone field, $\theta=\left(s_{\theta}/2\right)\varepsilon_{AB}H^{AB}$. The $p_{x}-ip_{y}$ configuration $\Delta^{A}=\Delta_{0}\left(1,-i\right)^{A}$ can be incorporated by changing the sign of one of the gamma matrices $\gamma^{\tilde{A}}$. The expansion in $H,A$ produces a splitting of the propagator into an unperturbed propagator and vertices, $\mathcal{G}^{-1}=\mathcal{G}_{0}^{-1}+\mathcal{V}$, where $\mathcal{V}$ further splits as $\mathcal{V}=\mathcal{V}_{1}+\mathcal{V}_{2}$, where $\mathcal{V}_{1}$ ($\mathcal{V}_{2}$) is first (second) order in the fields. The terms in $\mathcal{V}_{2}$ are often referred to as contact terms. Using \eqref{eq:48-0} we find the explicit form of $\mathcal{G}_{0}^{-1},\mathcal{V}_{1},\mathcal{V}_{2}$ in Fourier components, \begin{align} \mathcal{G}_{0}^{-1}\left(q\right)= & -\gamma^{0}q_{0}-\Delta_{0}\gamma^{j}q_{j}-\xi_{\mathbf{q}},\label{eq:184-1}\\ \mathcal{V}_{1}\left(q,p\right)= & -A_{t,p}-\Delta_{0}\gamma^{A}\left(H_{A}^{\;i}\right)_{p}q_{i}\nonumber \\  & -\frac{1}{m}\left[q_{i}q_{j}-\frac{1}{4}\left(p_{i}p_{j}-\delta_{ij}p^{2}\right)\right]H_{p}^{ij}+\gamma^{0}\frac{1}{m}A_{p}^{j}q_{j},\nonumber \\ \mathcal{V}_{2}\left(q,0\right)= & -\frac{1}{2m}\left(H_{A}^{\;i}H^{Aj}\right)_{p=0}q_{i}q_{j}-\frac{1}{8m}\left(\partial^{j}H_{A}^{\;A}\partial_{j}H_{B}^{\;B}\right)_{p=0}\nonumber \\  & -\gamma^{0}\frac{2}{m}\left(A_{i}H^{(ij)}\right)_{p=0}q_{j}-\frac{1}{2m}\left(A^{j}A_{j}\right)_{p=0}.\nonumber  \end{align} Here $\left(\cdots\right)_{p}$ denotes the $p$ Fourier component of the field $\left(\cdots\right)$, and we set $p=0$ in $\mathcal{V}_{2}$ since only this component will be relevant. The unperturbed Greens's function is given explicitly by  \begin{align}  & \mathcal{G}_{0}\left(q\right)=-\frac{q_{0}\gamma^{0}+\Delta_{0}q_{i}\gamma^{i}-\xi_{\mathbf{q}}}{q_{0}^{2}-q_{i}q^{i}-\xi{}_{\mathbf{q}}^{2}}. \end{align} The perturbative expansion of $S_{\text{eff}}$ is obtained from \eqref{eq:24-1} by using $\text{log}\left[\text{Det}\left(\cdot\right)\right]=\text{Tr}\left[\log\left(\cdot\right)\right]$, and expanding the logarithm in $\mathcal{V}$, \begin{eqnarray}  & S_{\text{eff,m}} & =-i\text{Tr}\left\{ \log\left[i\gamma^{0}\left(\mathcal{G}_{0}^{-1}+\mathcal{V}\right)\right]\right\} \label{eq:25}\\  &  & =-\frac{i}{2}\mbox{Tr}\left(\log i\gamma^{0}\mathcal{G}_{0}^{-1}\right)-\frac{i}{2}\mbox{Tr}\left(\mathcal{G}_{0}\mathcal{V}\right)+\frac{i}{4}\mbox{Tr}\left(\mathcal{G}_{0}\mathcal{V}\right)^{2}+O\left(\mathcal{V}^{3}\right)\nonumber \\  &  & =-\frac{i}{2}\text{Tr}\left(\mathcal{G}_{0}\mathcal{V}_{1}\right)-\frac{i}{2}\text{Tr}\left(\mathcal{G}_{0}\mathcal{V}_{2}\right)+\frac{i}{4}\text{Tr}\left(\mathcal{G}_{0}\mathcal{V}_{1}\mathcal{G}_{0}\mathcal{V}_{1}\right)+\cdots,\nonumber  \end{eqnarray} where in the last line we kept explicit only terms at first and second order in $H,A$ (the term of zeroth order was described in the previous section). Writing the functional traces as integrals over Fourier components and traces over spinor indices, we then find  \begin{align} S_{\text{eff,m}}= & -\frac{i}{2}\text{tr}\int_{q}\mathcal{V}_{1}\left(q,0\right)\mathcal{G}_{0}\left(q\right)-\frac{i}{2}\text{tr}\int_{q}\mathcal{V}_{2}\left(q,0\right)\mathcal{G}_{0}\left(q\right)\label{eq:39}\\  & +\frac{i}{4}\text{tr}\int_{p,q}\mathcal{G}_{0}\left(q-\frac{1}{2}p\right)\mathcal{V}_{1}\left(q,-p\right)\mathcal{G}_{0}\left(q+\frac{1}{2}p\right)\mathcal{V}_{1}\left(q,p\right)+\cdots,\nonumber  \end{align} where $\int_{q}=\int\frac{\text{d}^{2}q\text{d}q_{0}}{\left(2\pi\right)^{3}}$. We are interested in $S_{\text{eff}}$ to third order in derivatives, which amounts to expanding the above expression to $O\left(p^{3}\right)$, and evaluating the resulting traces and integrals. These computations were performed systematically using Mathematica, and can be found in the supplemental material of Ref.\citep{PhysRevB.100.104512}. 
The result, focusing on terms relevant for $\eta_{\text{o}},\tilde{\eta}_{\text{o}}$ to $O\left(q^{2}\right)$, is compatible with the general effective action of Sec.\ref{sec: effective field theory}, as confirmed by comparing \eqref{eq:39} to the perturbatively expanded $S_{\text{eff}}$.  This comparison provides explicit expressions for all of the coefficients that appear in $S_{\text{eff}}$, as we now describe. The ground state pressure $P\left(\mu\right)$ diverges logarithmically, and is given by \begin{align} P & =\frac{1}{2}\int^{\Lambda}\frac{\text{d}^{2}q}{\left(2\pi\right)^{2}}\left[\frac{q^{2}}{2m}-\frac{\frac{1}{2}\Delta_{0}^{2}q^{2}+\frac{q^{2}}{2m}\left(\frac{q^{2}}{2m}-\mu\right)}{\sqrt{\Delta_{0}^{2}q^{2}+\left(\frac{q^{2}}{2m}-\mu\right)^{2}}}\right]\\  & =-\frac{m^{3}\Delta_{0}^{4}}{4\pi}\left(1-2\frac{\mu}{m\Delta_{0}^{2}}\right)\log\Lambda+O\left(\Lambda^{0}\right).\nonumber  \end{align} Directly computing the ground state density $n_{0}$ and leading odd viscosity $\eta_{\text{o}}^{\left(1\right)}$ one finds  \begin{align} n_{0} & =\frac{1}{2}\int\frac{\text{d}^{2}q}{\left(2\pi\right)^{2}}\left[1-\frac{\left(\frac{q^{2}}{2m}-\mu\right)}{\sqrt{\Delta_{0}^{2}q^{2}+\left(\frac{q^{2}}{2m}-\mu\right)^{2}}}\right]\\  & =\frac{m^{2}\Delta_{0}^{2}}{2\pi}\log\Lambda+O\left(\Lambda^{0}\right),\nonumber \\ \eta_{\text{o}}^{\left(1\right)} & =-\frac{\ell}{16}\int\frac{\text{d}^{2}q}{\left(2\pi\right)^{2}}\frac{\Delta_{0}^{2}q^{2}\left(\frac{q^{2}}{2m}+\mu\right)}{\left[\left(\frac{q^{2}}{2m}-\mu\right)^{2}+q^{2}\Delta_{0}^{2}\right]^{3/2}}\\  & =-\frac{\ell m^{2}\text{\ensuremath{\Delta}}_{0}^{2}}{8\pi}\log\Lambda+O\left(\Lambda^{0}\right),\nonumber  \end{align} so the relations $n_{0}=P'\left(\mu\right)$, and $\eta_{\text{o}}^{\left(1\right)}=-\left(\ell/4\right)n_{0}$, described in Sec.\ref{sec:Main-section-2:}, are maintained to leading order in the cutoff. 
As explained in Appendix \ref{subsec:Symmetry-breaking-and}, the cutoff $\Lambda$ corresponds to a non-vanishing interaction range, which softens the contact interaction in the model \eqref{eq:3-1-1}. With a space-independent metric, a smooth cutoff can easily be implemented by replacing  \begin{align}  & \Delta^{A}E_{A}^{\;j}\tilde{\psi}_{-\mathbf{q}}^{\dagger}iq_{j}\tilde{\psi}_{\mathbf{q}}^{\dagger}\mapsto\Delta^{A}E_{A}^{\;j}\tilde{\psi}_{-\mathbf{q}}^{\dagger}\left(iq_{j}e^{-q_{k}q_{l}G^{kl}/\Lambda^{2}}\right)\tilde{\psi}_{\mathbf{q}}^{\dagger}, \end{align} for example, in the Fourier transformed Eq.\eqref{eq:78-1}, and should lead to the \textit{exact }relations $n_{0}=P'\left(\mu\right)$, $\eta_{\text{o}}^{\left(1\right)}=-\left(\ell/4\right)n_{0}$. However, a computation of the $q^{2}$ correction to $\eta_{\text{o}}$ requires a space-dependent metric, where a non-vanishing interaction range involves the geodesic distance and complicates the vertex $\mathcal{V}$ in \eqref{eq:184-1} considerably. Moreover, all other coefficients in $S_{\text{eff}}$ converge, and we can therefore work with the simple contact interaction, $\Lambda=\infty$. 
For the second derivative $P''$, we find  \begin{align} \frac{P''}{m}= & \frac{1}{2\pi}\begin{cases} 1\\ \frac{1}{1+2\kappa} \end{cases},\label{eq:c1} \end{align} where $\kappa=\left|\mu\right|/m\Delta_{0}^{2}>0$, and the cases refer to $\mu>0$ and $\mu<0$. This coefficient  determines the  odd (or Hall) conductivity $\sigma_{\text{o}}^{0}=\left(\ell/2\right)P''/2m$ and has been computed previously in the literature \cite{volovik1988quantized,goryo1998abelian,goryo1999observation,furusaki2001spontaneous,stone2004edge,roy2008collective,lutchyn2008gauge,hoyos2014effective,ariad2015effective}. Note that $P''$ is continuous at $\mu=0$, while $P'''$ is not, in accordance with the third order phase transition found in an exact solution of the model \eqref{eq:3-1-1} in the absence of background fields \cite{PhysRevB.82.224510}.

The coefficients $P'',F_{1}',F_{2},F_{3}$ were presented in Sec.\ref{subsec:Benchmarking-the-effective}. The remaining coefficients $F_{4},F_{5},F_{6}$, are irrelevant for the quantities discussed in Sec.\ref{sec:Main-section-2:},  and are presented here for completeness, \begin{align} F_{4}= & \frac{1}{24\pi\mu}\begin{cases} \frac{\kappa-2}{2}\\ \frac{1}{1+2\kappa} \end{cases},\;F_{5}=\frac{1}{24\pi\mu\Delta_{0}^{2}}\begin{cases} 1\\ -\frac{1}{\left(1+2\kappa\right)^{2}} \end{cases},\;F_{6}=-\frac{\kappa}{24\pi\mu}\begin{cases} \frac{1}{2}\\ \frac{1}{\left(1+2\kappa\right)^{2}} \end{cases}.\label{eq:c3} \end{align}  
As stated in Sec.\ref{subsec:Benchmarking-the-effective},  there is a sense in which the relativistic limit  $\kappa\rightarrow0$, or $m\rightarrow\infty$, reproduces the  effective action of a massive Majorana spinor in Riemann-Cartan space-time (Sec.\ref{sec:Bulk-response}).  Taking the relativistic limit of the dimensionful coefficients \eqref{eq:c3}, one finds $F_{6}=0$, while $F_{4}=-\Delta_{0}^{2}F_{5}\neq0$ describe a relativistic term which is second order in torsion, and was not written explicitly in Sec.\ref{sec:Bulk-response}.

\section{Further details regarding Eq.\eqref{eq:12-3-1}\label{subsec:Further-details-regarding}}

This appendix involves basic facts in CFT, which can be found in e.g
\citep{ginsparg1988applied,di1996conformal}.

\subsection{Definition of $h_{0}$ and ambiguities in its value \label{subsec:Definition-of-}}

A chiral topological phase of matter has a finite-dimensional ground
state subspace on the spatial torus. A basis $\left\{ \ket a\right\} _{a=1}^{N}$
for the torus ground state subspace exists, such that each state $\ket a$
corresponds to a conformal family in the boundary CFT \citep{PhysRevB.85.235151,PhysRevB.88.195412,PhysRevLett.110.236801},
constructed over a primary with right/left moving conformal weights
$h_{a}^{\left(l\right)},h_{a}^{\left(r\right)}\geq0$. The corresponding
chiral and total conformal weights are then given by $h_{a}=h_{a}^{\left(l\right)}-h_{a}^{\left(r\right)}$
and $h_{a}^{+}=h_{a}^{\left(l\right)}+h_{a}^{\left(r\right)}$, respectively.
The chiral and total central charges of the CFT are similarly defined
in terms of the left/right moving central charges, $c=c^{\left(l\right)}-c^{\left(r\right)}$
and $c^{+}=c^{\left(l\right)}+c^{\left(r\right)}$. 

When the torus is cut to a cylinder with finite circumference $L$,
the ground state degeneracy is lifted, generically leaving a unique
ground state. The lowest energy eigenstates on the cylinder can also
be labeled as $\left\{ \ket a\right\} _{a=1}^{N}$. Each $\ket a$
corresponds to a non-universal choice of state in the conformal family
labeled by $h_{a}^{\left(l\right)},h_{a}^{\left(r\right)}$ , which
need not be the primary, as demonstrated explicitly in Appendix \ref{subsec:Beyond-the-assumption}
below. 

If the boundary is described by an idealized CFT, all $\ket a$s correspond
to primaries, and the corresponding energies are given by $E_{a}=\left(4\pi v/L\right)\left(h_{a}^{+}-c^{+}/24\right)$,
relative to the ground state energy on the torus, where $v$ is the
velocity of the CFT and $L$ is the circumference of the cylinder.
These expressions receive exponentially small corrections of $O\left(Le^{-R/\xi}\right)$
and $O\left(Re^{-L/\xi}\right)$, where $\xi$ is the bulk correlation
length and $R$ is the length of the cylinder \citep{PhysRevLett.110.067208}.
The cylinder ground state then corresponds to the CFT ground state,
the primary with minimal $h_{a}^{+}$. 

More generally, each state $\ket a$ corresponds to either a primary
or a descendent, and has conformal weights $h_{a}^{\left(l\right)}+n_{a}^{\left(l\right)},h_{a}^{\left(r\right)}+n_{a}^{\left(r\right)}$,
where $n_{a}^{\left(l\right)},n_{a}^{\left(r\right)}\in\mathbb{N}_{0}$.
The corresponding energies $E_{a}$ differ from the idealized $\left(4\pi v/L\right)\left(h_{a}^{+}-c^{+}/24\right)$,
and the choice of conformal family $a_{0}$ with minimal $E_{a_{0}}$
is non-universal. In terms of $n_{a}=n_{a}^{\left(l\right)}-n_{a}^{\left(r\right)}$,
we then define $h_{0}:=h_{a_{0}}+n_{a_{0}}$, the chiral conformal
weight associated with the cylinder ground state $\ket{a_{0}}$. The
value of $h_{0}$ therefore carries two ambiguities: a choice of conformal
family $a_{0}\in\left\{ a\right\} $, and the choice of a state in
the conformal family, $n_{a_{0}}\in\mathbb{N}_{0}$. As described
in Sec.\ref{subsec:Boundary-finite-size}, the only universal statement
is $\theta_{0}=e^{2\pi ih_{0}}\in\left\{ \theta_{a}\right\} $, where
$\theta_{a}=e^{2\pi ih_{a}}$ are the topological spins of bulk anyons. 

The result of Ref.\citep{PhysRevB.88.195412} for the momentum polarization
is given terms of the low lying cylinder eigenstates $\ket a$,
\begin{align}
 & \bra aT_{R}\ket a=\exp\left[\alpha N_{x}+\frac{2\pi i}{N_{x}}\left(h_{a}-\frac{c}{24}\right)+o\left(N_{x}^{-1}\right)\right],\label{eq:17}
\end{align}
where the lattice spacing is set to 1, $N_{x}=L$.  It follows that
the thermal expectation value $\tilde{Z}/Z=\text{Tr}\left(T_{R}e^{-\beta H}\right)/Z$
is equal to $\exp\left[\alpha N_{x}+\frac{2\pi i}{N_{x}}\left(h_{0}-\frac{c}{24}\right)+o\left(N_{x}^{-1}\right)\right]$,
if the temperature $\beta^{-1}$ is much lower than the boundary energy
differences $\sim N_{x}^{-1}$, namely $\beta^{-1}=o\left(N_{x}^{-1}\right)$,
as described in Sec.\ref{sec:Signs-from-geometric}.

\subsection{The value of $h_{0}$ in fermionic phases of matter\label{subsec:The-value-of-h0}}

Fermionic topological phases are microscopically comprised of fermions
(and possibly bosons), and have the fermion parity $\left(-1\right)^{N_{f}}$
as a global symmetry \citep{Kapustin:2015aa,freed2016reflection,PhysRevB.95.235140,aasen2019fermion}.
It is therefore useful to probe such phases with a background $\mathbb{Z}_{2}$
gauge field corresponding to $\left(-1\right)^{N_{f}}$, or a spin
structure.  For our purposes, this amounts to considering both periodic
and anti-periodic boundary conditions around non-contractible cycles
in space-time. 

In Sec.\ref{sec:Main-section-3:} we were only interested in locally sign-free QMC
representations of thermal partition functions, and sign-free geometric
manipulations that can be performed to these. We therefore restricted
attention to thermal boundary conditions in the imaginary time direction
(see Sec.\ref{subsec:Local-determinantal-QMC}), and to periodic boundary
conditions around the spatial cylinder. These boundary conditions
cannot generically be modified without introducing signs into the
QMC weights.

Here we provide a fuller picture by considering the behavior of $h_{0}$
with both periodic and anti-periodic boundary conditions, in the closed
$x$ direction of the spatial cylinder. Since $h_{0}$ is a ground
state property, the time direction is open and does not play a role. 

For a fermionic chiral topological phase, the boundary CFT is also
fermionic. The primary conformal weights $\left\{ h_{a}\right\} $
then depend on the choice of boundary conditions (in the $x$ direction),
and as a result, so will the set of topological spins $\left\{ \theta_{a}\right\} $
in which $\theta_{0}=e^{2\pi ih_{0}}$ is valued. In particular, the
vacuum spin $\theta_{I}=0$ will not be included in $\left\{ \theta_{a}\right\} $
for periodic boundary conditions, while for anti-periodic boundary
conditions, both the vacuum $\theta_{I}=1$ \textit{and} the spin
$\theta_{\psi}=-1$ of the microscopic fermion will appear \citep{ginsparg1988applied,PhysRevLett.110.236801}.
Note that $\theta_{\psi}$ does not correspond to an emergent fermion,
as in e.g the toric code \citep{KITAEV20032}, and therefore does
not imply an additional ground state on the torus. 

As an example, consider the series of Laughlin phases at filling $1/q$
, with $q\in\mathbb{N}$, all of which have the chiral central charge
$c=1$. These correspond to $U\left(1\right)_{q}$ Chern-Simons theories.
First, for $q\in2\mathbb{N}$ the phase is bosonic, and we consider
only periodic boundary conditions. The primary conformal weights are
given by $h_{a}=a^{2}/2q$ \citep{PhysRevB.89.125303,hu2020microscopic},
with $a\in\mathbb{N}_{0}$. The topological spins $\theta_{a}=e^{2\pi ih_{a}}$
depend only on $a\mod q$, and the $q$ spins $\left\{ \theta_{a}\right\} _{a=0}^{q-1}$
correspond to the $q$ degenerate ground states that appear on the
torus. In particular, the vacuum spin $\theta_{I}=1$ is obtained
for $a=0$. 

For $q\in2\mathbb{N}-1$ the phase is fermionic, and we consider both
periodic and anti-periodic boundary conditions. For periodic boundary
conditions the weights are given by $h_{a}=\left(a+1/2\right)^{2}/2q$
\citep{hu2020microscopic}. As in the bosonic case, $\theta_{a}=e^{2\pi ih_{a}}$
depend only on $a\mod q$, with $\left\{ \theta_{a}\right\} _{a=0}^{q-1}$
corresponding to the $q$ degenerate ground states on the torus. Unlike
the bosonic case, the vacuum spin $\theta_{I}=1$ is not included
in $\left\{ \theta_{a}\right\} _{a=0}^{q-1}$. For anti-periodic boundary
conditions, the weights are given by $h_{a}=a^{2}/2q$ as in the bosonic
case \citep{PhysRevB.89.125303}. The set $\left\{ \theta_{a}\right\} _{a=0}^{q-1}$
again corresponds to the $q$ torus ground states, but now $\theta_{\psi}=\theta_{a=q}=-1$
is an additional topological spin that corresponds to the physical
Fermion $\psi$ \citep{bonderson2007non}.

The simplest fermionic Laughlin phase is given by $q=1$, and corresponds
to an integer quantum Hall state, or a Chern insulator \citep{PhysRevLett.61.2015,qi2008topological},
which is studied in detail in Appendix \ref{subsec:Beyond-the-assumption}
below. The Chern insulator has a unique ground state on the torus,
and accordingly, there is a unique topological spin $\theta_{\sigma}=e^{2\pi i\left(1/8\right)}$
for periodic boundary conditions on the cylinder, and two topological
spins $\theta_{I}=1,\theta_{\psi}=-1$ for anti-periodic boundary
conditions. Here $\psi$ corresponds the physical fermions from which
the Chern insulator is comprised. The object carrying the spin $\theta_{\sigma}$
is the complex analog of the celebrated Majorana zero mode supported
on vortices in the bulk of a chiral $p$-wave superconductor \citep{read2000paired,kitaev2006anyons}.

\textcolor{red}{}

\section{Momentum polarization with non CFT boundaries\label{subsec:Beyond-the-assumption}}

\textcolor{red}{}

As reviewed in Sec.\ref{sec:Signs-from-geometric}, the existing analytic
derivation of Eq.\eqref{eq:12-3-1} relies on the CFT description
of the physical boundaries of the cylinder, and of the line $y=0$
where $T_{R}$ is discontinuous \citep{PhysRevB.88.195412}. In this
appendix we perform an analytic and numerical study that shows that,
at least for free fermions, the relevant CFT expressions and the resulting
Eq.\eqref{eq:12-3-1}, hold even if the boundary is not described
by an idealized CFT. We will however, find a number of subtleties
which have not been demonstrated in the literature, as already described
below Eq.\eqref{eq:2-2} and in Appendix \ref{subsec:Further-details-regarding}. 

\subsection{CFT finite-size correction in non CFT boundaries\label{subsec:CFT-finite-size-correction}}

The main ingredient in the analytic derivation of Eq.\eqref{eq:12-3-1}
is the expression \eqref{eq:2-2} for the finite size correction to
the momentum density in CFT \citep{PhysRevB.88.195412}. In this appendix
we show that, at least in the non-interacting case, Equation \eqref{eq:2-2}
remains valid, with $\theta_{0}=e^{2\pi ih_{0}}\in\left\{ \theta_{a}\right\} $,
even when the boundary cannot be described by a CFT. 

We will consider a Chern insulator, such as the prototypical Haldane
model \citep{PhysRevLett.61.2015}. When the boundary degrees of freedom
can be described by a CFT, they correspond to the Weyl fermion CFT,
where $c=\pm1$ and the primary conformal weights are $h_{\sigma}=\pm1/8$
($h_{I}=0,h_{\psi}=\pm1/2$) for periodic (anti-periodic) boundary
conditions, as described in Appendix \ref{subsec:The-value-of-h0}.
The sign corresponds to the two possible chiralities. More generally,
on a lattice with spacing $1$, the boundary supports a complex fermion
with an energy dispersion $\varepsilon_{k}$, where $k=k_{x}$ takes
values in the Brillouin zone $\mathbb{R}/2\pi\mathbb{Z}$ for an infinite
circumference $L=\infty$, or its discretization $\left(2\pi/L\right)\mathbb{Z}_{L}$
($\left(2\pi/L\right)\left(\mathbb{Z}_{L}+1/2\right)$), for $L<\infty$
and periodic (anti-periodic) boundary conditions. The only requirement
on $\varepsilon_{k}$ is that it be \textit{chiral}, in the sense
that it connects the two separated bulk energy bands. If the intersections
$k_{l}$ and $k_{u}$ with the lower and upper bulk bands, respectively,
satisfy $k_{l}<k_{u}$ ($k_{l}>k_{u}$), we say that the boundary
is right (left) moving, or has a positive (negative) chirality, see
Fig.\ref{fig:Schematic-band-structure,}.
\begin{figure}[!th]
\begin{centering}
\includegraphics[width=0.35\columnwidth]{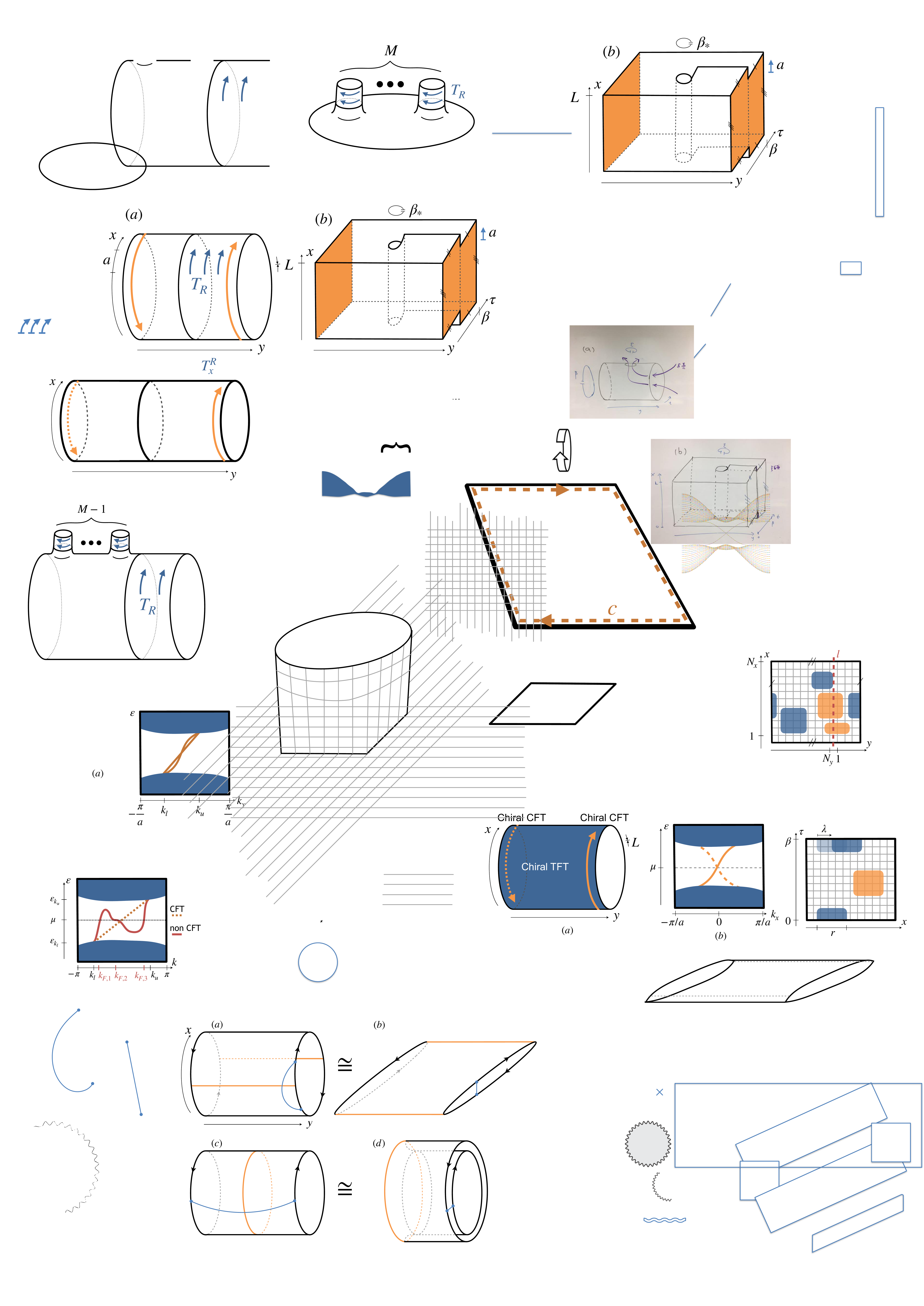}
\par\end{centering}
\caption{Schematic band structure, energy $\varepsilon$ as a function of momentum
$k=k_{x}$ in the periodic $x$ direction, of a Chern insulator on
the cylinder. The figure shows the bulk energy bands (blue) and the
chiral boundary dispersion, with two dispersion branches, on a single
boundary component (orange and red curves). The opposite chirality
branches on the second boundary component are not drawn. The momenta
$k_{l}$ and $k_{u}$ correspond to the intersections of the boundary
dispersion with the lower and upper bulk bands, respectively. Since
$k_{u}>k_{l}$, both dispersion branches have a positive chirality.
The orange line indicates the idealized linear dispersion with a single
Fermi momentum $k_{F}=0$, which corresponds to the Weyl fermion CFT.
The solid red curve corresponds to a more general chiral branch, with
three Fermi momenta $k_{F,1},k_{F,2},k_{F,3}$, where the dispersion
around $k_{F,2}$ takes a (non-generic) non-linear form. With periodic
boundary conditions around the cylinder, both dispersion branches
produce the same $L^{-2}$ correction to the momentum density in Eq.\eqref{eq:18-0},
with a positive chirality $+$, up to a mod 1 ambiguity: $1/12\protect\mapsto1/12+n,\;n\in\mathbb{N}$.\label{fig:Schematic-band-structure,}}
\end{figure}

 More generally, the dispersion will contain several dispersion branches
$\left\{ \varepsilon_{j,k}\right\} _{j=1}^{J}$, but since the momentum
density is additive in $j$ we restrict attention to a single branch.
Without loss of generality, we also fix the chemical potential $\mu=0$,
in which case the Fermi momentum $k_{F}$ satisfies $\varepsilon_{k_{F}}=0$.
The value of $k_{F}$ plays an important role is the subsequent analysis. 

The simplest dispersion that satisfies the above requirements is the
linear one $\varepsilon_{k}=v\left(k-k_{F}\right)$. For $k_{F}=0$
this corresponds to the Weyl fermion CFT. The presence of $k_{F}\neq0$
corresponds to the addition of a chemical potential $vk_{F}$, which
breaks the conformal symmetry. The generic form is $\varepsilon_{k}=v\left(k-k_{F}\right)+O\left(k-k_{F}\right)^{2}$.
A non-generic dispersion can take the form $\varepsilon_{k}=v_{3}\left(k-k_{F}\right)^{3}+O\left(k-k_{F}\right)^{4}$,
and there may be several Fermi momenta if the dispersion is non monotonic,
see Fig.\ref{fig:Schematic-band-structure,}. 

In all cases the many-body ground state momentum is given by summing
the momenta of all filled single Fermion states $p\left(L\right)=\frac{1}{L}\sum_{\varepsilon{}_{k}<0}k$,
where the sum runs over $k\in\left(2\pi/L\right)\mathbb{Z}_{L}$ such
that $\varepsilon_{k}$ is negative and in the bulk energy gap. In
order to obtain $p$ as a continuous function of $L$, we treat the
bulk energy gap as a smooth cutoff $p\left(L\right)=\frac{1}{L}\sum_{\varepsilon_{k}<0}kC\left(\varepsilon_{k}\right)$,
where the function $C\left(\varepsilon\right)$ goes to $1$ ($0$)
fast enough as $\varepsilon$ goes to $0$ ($\varepsilon_{k_{l}}$
or $\varepsilon_{k_{u}}$)\footnote{It suffices that $C'\left(\varepsilon\right)$ vanish at $\varepsilon=0,\varepsilon_{k_{l}},\varepsilon_{k_{u}}$.}.
The cutoff $C$ represents the smooth delocalization of boundary eigenstates
as their energy nears the bulk energy bands. 

To obtain the $L$ dependence of $p\left(L\right)$, we will use the
Euler-Maclaurin formula
\begin{align}
\sum_{n=n_{1}}^{n_{2}}f\left(n\right) & =\int_{n_{1}}^{n_{2}}f\left(x\right)dx+\frac{f\left(n_{2}\right)+f\left(n_{1}\right)}{2}\label{eq:19}\\
+ & \frac{1}{6}\frac{f'\left(n_{2}\right)-f'\left(n_{1}\right)}{2!}-\frac{1}{30}\frac{f'''\left(n_{2}\right)-f'''\left(n_{1}\right)}{4!}+\mathsf{R},\nonumber 
\end{align}
where the remainder satisfies $\left|\mathsf{R}\right|\leq\frac{2\zeta\left(5\right)}{\left(2\pi\right)^{5}}\int_{n_{1}}^{n_{2}}\left|f^{\left(5\right)}\left(x\right)\right|dx$.
We begin by considering periodic boundary conditions, where we set
$f\left(n\right)=\left(2\pi n/L^{2}\right)C\left(\varepsilon_{2\pi n/L}\right)$.
Assuming a single, vanishing, Fermi momentum $k_{F}=0$, we set $\left(n_{1},n_{2}\right)=\left(-\infty,0\right)$
for positive chirality, and $\left(n_{1},n_{2}\right)=\left(0,\infty\right)$
for negative chirality. Equation \eqref{eq:19} then gives 

\begin{align}
p\left(L\right)= & p\left(\infty\right)\pm\frac{2\pi}{L^{2}}\frac{1}{12}+O\left(\frac{1}{L^{4}}\right),\;\text{as }L\rightarrow\infty,\label{eq:18-0}
\end{align}
where $p\left(\infty\right)=\int_{\varepsilon_{k}<0}kC\left(\varepsilon_{k}\right)dk/2\pi$
and $\pm=\text{sgn}\left(k_{u}-k_{l}\right)$ is the chirality. The
$1/L^{2}$ correction in \eqref{eq:18-0} comes from $f'\left(0\right)=2\pi/L^{2}$
in \eqref{eq:19}. We see that the leading finite size correction
$h_{0}-c/24$ is unchanged from its CFT value $h_{\sigma}-c/24=\pm1/12$,
even when a CFT description does not apply. 

The case of a single non-zero Fermi momentum $k_{F}\neq0$ is more
interesting, as it demonstrates that the integer part of $h_{0}$
can change as a function of $L$ and $k_{F}$. The direct derivation
of the end result from the Euler-Maclaurin formula is surprisingly
lengthy, so we omit it and present a more direct route to the end
result. To be concrete, assume a positive chirality and $k_{F}>0$.
The Euler-Maclaurin formula leads to cutoff independent results, so
we can restrict attention to cutoff functions $C\left(\varepsilon_{k}\right)$
which are identically 1 for $0<k<k_{F}$. Since these can serve as
cutoff functions for the case $k_{F}=0$ as well, we can deduce the
$k_{F}\neq0$ momentum density $p\left(L,k_{F}\right)$ from the $k_{F}=0$
momentum density $p\left(L\right)$, 
\begin{align}
p\left(L,k_{F}\right) & =\frac{1}{L}\sum_{k<k_{F}}kC\left(\varepsilon_{k}\right)\\
 & =\frac{1}{L}\sum_{k<0}kC\left(\varepsilon_{k}\right)+\frac{1}{L}\sum_{0<k<k_{F}}k\nonumber \\
 & =p\left(L\right)+\frac{2\pi}{L^{2}}\sum_{l=1}^{n}l,\nonumber 
\end{align}
where $n=\left\lfloor k_{F}L/2\pi\right\rfloor $. Using Eq.\eqref{eq:18-0},
we then have 
\begin{align}
p\left(L,k_{F}\right)= & p\left(\infty\right)+\frac{2\pi}{L^{2}}\left[\frac{1}{12}+\sum_{l=1}^{n}l\right]+O\left(\frac{1}{L^{4}}\right),\label{eq:32-0}
\end{align}
where $p\left(\infty\right)$ is the momentum density at $L=\infty$
and $k_{F}=0$. We see that the value of $h_{0}-c/24$ is only equal
to the idealized CFT result $h_{\sigma}-c/24=1/12$ modulo 1, while
the integer part jumps periodically as a function of $k_{F}$ at fixed
number of sites $L$, or as the number of sites $L$ at fixed $k_{F}$.
Treating $k_{F}$ as fixed and valued in $(-\pi,\pi]$, the period
in $L$ is given by $q=\left|2\pi/k_{F}\right|\geq2$, which need
not be an integer. As described in Appendix \ref{subsec:Further-details-regarding},
the mod 1 ambiguity is attributed to $h_{0}$ rather than $c$, which
corresponds to the topological spin $\theta_{0}=\theta_{\sigma}=e^{2\pi i\left(1/8\right)}$. 

The interpretation of Eq.\eqref{eq:32-0} is straight forward. As the
number of sites $L$ increases, the single particle momenta $\left(2\pi/L\right)\mathbb{Z}_{L}$
become denser in the Brillouin zone $\mathbb{R}/2\pi\mathbb{Z}$.
The $n$th jump in $h_{0}$ correspond to the motion of a single particle
state with momentum $2\pi n/L$ through $k_{F}$ and into the Fermi
sea, adding a momentum density $2\pi n/L^{2}$ to the ground state. 

\begin{figure}[!th]
\begin{centering}
\includegraphics[width=0.6\columnwidth]{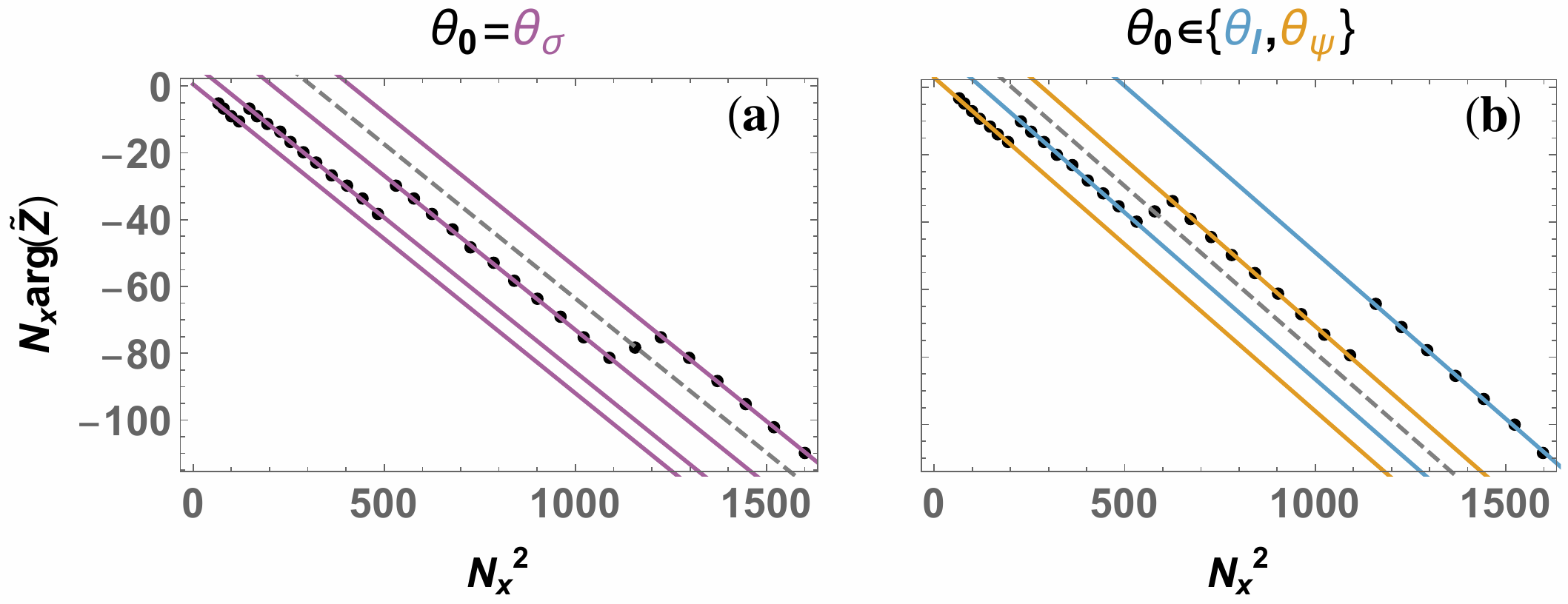}
\par\end{centering}
\caption{Numerical results for the momentum polarization Eq.\eqref{eq:12-3-1},
in a Chern insulator with $k_{F}\protect\neq0$. Black dots mark numerically
obtained values of $N_{x}\arg\tilde{Z}$ as a function of $N_{x}^{2}$.
(a) Periodic boundary conditions. Purple lines indicate linear fits,
with approximately the same slope and intercepts $2\pi\left(1/12+\sum_{l=1}^{n}l\right)$
with the $y$ axis, with $n=0,1,2,3$, in accordance with Eq.\eqref{eq:32-0}.
This allows for the extraction of the topological spin $\theta_{\sigma}=e^{2\pi i\left(1/8\right)}$.
To illustrate the possibility of accidental degeneracies, we choose
$k_{F}=3/34$, where a degeneracy occurs for $N_{x}=34$, and the
average value of $N_{x}\arg\tilde{Z}$ between the two ground states
is obtained. A grey dotted line indicates the average of the two neighboring
purple lines. (b) Anti-periodic boundary conditions. Colored lines
indicate linear fits, with approximately the same slope, and intercepts
$2\pi\left[-1/24+\sum_{l=1}^{n}\left(l-1/2\right)\right]$, with $n=1,2,3,4$,
in accordance with Eq.\eqref{eq:35-2}. The value $n=0$ is not obtained
as it occurs only for small circumferences $N_{x}<6$ where $N_{x}\arg\tilde{Z}$
is not computed. Orange lines correspond to the fermion spin $\theta_{\psi}=-1$,
while blue lines correspond to the vacuum spin $\theta_{I}=0$. To
illustrate the possibility of accidental degeneracies, we choose $k_{F}=\left(5/2\right)/24$,
where an accidental degeneracy occurs for $N_{x}=24$. \label{fig:Numerical-results-for}}
\end{figure}

Figure \ref{fig:Numerical-results-for} presents the results of numerical
computations of the momentum polarization Eq.\eqref{eq:12-3-1} in
a Chern insulator with $k_{F}\neq0$. Details of the model and computations
can be found in the supplemental material of Ref.\citep{PhysRevResearch.2.043032}. In particular,
Fig.\ref{fig:Numerical-results-for}(a) verifies Eq.\eqref{eq:32-0}.

A subtle point, not mentioned above, is that when $k_{F}L/2\pi\in\mathbb{Z}$,
which happens only when $k_{F}/2\pi=a/b$ is rational and $L\in b\mathbb{N}$,
a single particle state with momentum exactly $k_{F}$  exists, leading
to an accidental degeneracy on the cylinder, between two many-body
ground states with momentum densities given by Eq.\eqref{eq:32-0} with
$n$ and $n+1$. The momentum polarization \eqref{eq:12-3-1} then
gives the average momentum density in the two ground-states, as visualized
by the grey dotted line in Fig.\ref{fig:Numerical-results-for}(a).
For such system sizes the value $\theta_{0}=-\theta_{\sigma}$ may
be obtained rather than the generic $\theta_{0}=\theta_{\sigma}$. 

The same analysis can be performed for anti-periodic boundary conditions,
where we sum over single particle momenta $k\in\frac{2\pi}{L}\left(\mathbb{Z}_{L}+\frac{1}{2}\right)$.
Equation \eqref{eq:32-0} is then modified to 
\begin{align}
p\left(L,k_{F}\right)= & p\left(\infty\right)\label{eq:35-2}\\
 & +\frac{2\pi}{L^{2}}\left[-\frac{1}{24}+\sum_{l=1}^{n}\left(l-\frac{1}{2}\right)\right]+O\left(\frac{1}{L^{4}}\right),\nonumber 
\end{align}
where $n=\left\lfloor \frac{k_{F}L}{2\pi}-\frac{1}{2}\right\rfloor $.
As a function of $L$, jumps in $h_{0}-c/24$ occur with the same
period $q=\left|2\pi/k_{F}\right|\geq2$, but are shifted by $q/2$.
Moreover, $h_{0}-c/24$ now attains two values modulo 1, namely $h_{I}-c/24=-1/24$
and $h_{\psi}-c/24=1/2-1/24$. 

Equation \eqref{eq:35-2} therefore demonstrates explicitly the statements
made in Appendix \ref{subsec:Definition-of-}. For $k_{F}=0$, the
cylinder ground state of the Chern insulator corresponds to the idealized
Weyl fermion CFT. A single value $h_{0}=0$ is attained, which is
the conformal weight $h_{I}$ of the CFT vacuum. A non-vanishing $k_{F}$
corresponds to the addition of a chemical potential to the CFT, which
changes the energies of the CFT states, favoring a CFT exited state
over the CFT vacuum. The cylinder ground state of the Chern insulator
may then correspond to any CFT state in the conformal family of either
the vacuum $I$ or fermion $\psi$, which need not be primary. From
the bulk TFT perspective, we see that $\theta_{0}=e^{2\pi ih_{0}}$
may be equal to either of the topological spins $\theta_{I}=1,\theta_{\psi}=-1$
as a function of $L$. 

As in the case of periodic boundary conditions, accidental degeneracies
on the cylinder occur when $k_{F}L/2\pi\in\mathbb{Z}+1/2$, changing
the value of $h_{0}$ attained from the momentum polarization to its
average over the degenerate states. For such system sizes, the value
$\theta_{0}=\pm\sqrt{\theta_{I}\theta_{\psi}}$ is obtained than
the generic $\theta_{0}\in\left\{ \theta_{I},\theta_{\psi}\right\} $. 

Equation \eqref{eq:35-2} is verified numerically in Fig.\ref{fig:Numerical-results-for}(b),
which demonstrates that the value of $\theta_{0}=e^{2\pi ih_{0}}$,
obtained from the momentum polarization \eqref{eq:12-3-1}, takes
different values in the set $\left\{ \theta_{a}\right\} $ of topological
spins as a function of system size $L$, apart from accidental degeneracies. 

\subsection{No finite-size correction at finite temperature}

The line $y=0$ where $T_{R}$ jumps can be interpreted as an additional
boundary component at the 'entanglement temperature' $\beta_{*}^{-1}$.
Reference \citep{PhysRevB.88.195412} used the modular transformations
of CFT partition functions to demonstrate that when $\beta_{*}\ll L/v$,
this boundary component does not contribute to the $1/L$ correction
to $\log\tilde{Z}$. Here, we note that the same result holds for
free fermions with a general dispersion $\varepsilon_{k}$. The contribution
of the additional boundary component to $\log\tilde{Z}$ is given
by $\log\tilde{Z}{}_{*}\left(L\right)=Lf_{*}\left(L\right)$, with
the free energy density 
\begin{align}
f_{*}\left(L\right) & =\frac{1}{L}\sum_{k}\log\left(1+e^{iak}e^{-\beta_{*}\varepsilon_{k}}\right)C\left(\varepsilon_{k}\right).
\end{align}
Using Eq.\eqref{eq:19} one finds $f_{*}\left(L\right)=f_{*}\left(\infty\right)+O\left(L^{-4}\right)$
for both periodic and anti-periodic boundary conditions, which implies
$\log\tilde{Z}{}_{*}=Lf_{*}\left(\infty\right)+O\left(L^{-3}\right)$,
with no $1/L$ contribution. The complex number $f_{*}\left(\infty\right)=\int\log\left(1+e^{iak}e^{-\beta_{*}\varepsilon_{k}}\right)C\left(\varepsilon_{k}\right)\text{d}k/2\pi$
contributes to the non-universal $\alpha$ in Eq.\eqref{eq:12-3-1}.

\section{Cutting the torus along an arbitrary vector\label{subsec:Cutting-the-torus-2}}

For $\mathbf{d}=\left(d_{x},0\right)$ we restrict to $N_{x}=n_{x}d_{x},\;n\in\mathbb{N}$,
viewing $d_{x}$ as an enlarged lattice spacing, and treating $n$
as a reduced number of sites along the circumference in place of $N_{x}$.
The same logic applies to $\mathbf{d}=\left(0,d_{y}\right)$. For
$\mathbf{d}=\left(d_{x},d_{y}\right)$ with both $d_{x},d_{y}\neq0$,
we restrict to system sizes $\left(N_{x},N_{y}\right)=n\mathbf{d}$,
such that the line $l=\text{span}_{\mathbb{R}}\mathbf{d}$ is a diagonal
of the rectangle $nN_{x}\times nN_{y}$ and corresponds to a circle
on the torus $\left(\mathbb{R}/N_{x}\mathbb{Z}\right)\times\left(\mathbb{R}/N_{y}\mathbb{Z}\right)$.
Cutting $X$ along this line produces a cylinder $C$ of circumference
$L=n\left|\mathbf{d}\right|$. We then view $\left|\mathbf{d}\right|$
as a lattice spacing and $n$ as a the number of sites along the circumference.
Note that the distance between the boundary components of the resulting
cylinder is $R=nd_{x}d_{x}/\left|\mathbf{d}\right|$, and the thermodynamic
limit is indeed obtained as $n\rightarrow\infty$. With these identifications
the momentum polarization \eqref{eq:12-3-1} remains unchanged, apart
from a modification of the non-universal $\alpha$ to $\alpha\left|\mathbf{d}\right|^{2}$.\label{fn:For--we}

\section{Dealing with accidental degeneracies on the cylinder \label{subsec:Dealing-with-accidental}}

As demonstrated in Appendix \ref{subsec:Beyond-the-assumption}, for
certain system sizes $N_{x}$ accidental degeneracies occur on the
cylinder, and the function $\theta_{0}\left(N_{x}\right)=e^{2\pi ih_{0}\left(N_{x}\right)}$
obtained from Eq.\eqref{eq:12-3-1} may take values outside the set
$\left\{ \theta_{a}\right\} $, namely $\theta_{0}=\pm\sqrt{\theta_{a}\theta_{b}}$
for a two-fold degeneracy. In this appendix we complete the derivation
of Result \hyperref[Result 1]{1} and \hyperref[Result 1F]{1F} by considering
the possibility of such degeneracies. 

First, even in the presence of degeneracies, $\theta_{0}$ is valued
in a finite set. Therefore, Eq.\eqref{eq:6-1} still implies that
$\epsilon'=n/m$ is rational, and that $\theta_{0}\left(N_{x}\right)e^{-2\pi ic/24}=e^{-2\pi i\epsilon'N_{x}^{2}}$
periodically covers a subset $S\ni1$ of $m$th roots of unity, for
all large enough $N_{x}$. We denote by $\mathcal{N}\subset\mathbb{N}$
the set of circumferences $N_{x}$ for which a degeneracy appears
and $\theta_{0}\left(N_{x}\right)\notin\left\{ \theta_{a}\right\} $.
 If circumferences $N_{x}\in m\mathbb{N}$, where $e^{2\pi i\epsilon'N_{x}^{2}}=1$,
are not all contained in $\mathcal{N}$, then $1=\theta_{0}\left(N_{x}\right)e^{-2\pi ic/24}\in\left\{ \theta_{a}e^{-2\pi ic/24}\right\} $,
as stated in Results \hyperref[Result 1]{1} and \hyperref[Result 1F]{1F}. 

We are left with the complementary case, where degeneracies occur
for all $N_{x}\in m\mathbb{N}$, i.e $m\mathbb{N}\subset\mathcal{N}$.
Note that this case is highly fine tuned, as it ties together the
non-universal $\epsilon'=n/m$ and the set $\mathcal{N}$ of $N_{x}$s
where accidental degeneracies appear. In order to deal with this case,
we make use of the Frobenius-Perron theorem to resolve the degenerate
ground-state subspace, without introducing signs. The analysis applies
only to the bosonic setting of Result \hyperref[Result 1]{1}, and is
similar to that made in Sec.\ref{sec:Generalization-and-extension}.
We now have a Hamiltonian $H'$ on the cylinder, which has an exactly
degenerate ground-state subspace for all $N_{x}\in m\mathbb{N}$,
and has non-positive matrix elements in the on-site homogenous basis
$\ket s$. The Frobenius-Perron theorem implies that an orthonormal
basis $\ket i$ with non-negative entries may be chosen for the ground
state subspace, $\braket si\geq0$ for all $s,i$. It follows that
the matrix elements of $T_{R}$ in the basis $\ket i$ are non-negative,
$M_{ij}:=\bra iT_{R}\ket j\geq0$. Taking the $N_{x}$th matrix power
of $M$ we have $\left(M^{N_{x}}\right)_{ij}\geq0$. Equation \eqref{eq:12-3-1}
implies that the eigenvalues of $M^{N_{x}}$ are of the form $e^{-\delta'N_{x}^{2}}e^{-2\pi i\epsilon'N_{x}^{2}+o\left(1\right)}\theta_{a}e^{-2\pi ic/24}$,
 so we can write $\left(M^{N_{x}}\right)_{ij}=e^{-\delta'N_{x}^{2}}e^{-2\pi i\epsilon'N_{x}^{2}+o\left(1\right)}T_{ij}$,
where $T_{ij}$ has eigenvalues $\left\{ \lambda\right\} \subset\left\{ \theta_{a}e^{-2\pi ic/24}\right\} $.
In particular, $T_{ij}$ is unitary. Since $e^{2\pi i\epsilon'N_{x}^{2}}=1$
for all $N_{x}\in m\mathbb{N}$, we see that $T_{ij}$ also has non-negative
entires, and is therefore a permutation matrix, containing $1$ in
its spectrum (see Sec.\ref{sec:Generalization-and-extension}). It
follows that $1\in\left\{ \lambda\right\} \subset\left\{ \theta_{a}e^{-2\pi ic/24}\right\} $,
asserting Result \hyperref[Result 1]{1}. 

We are currently unaware of an analog of the Frobenius-Perron theorem
in the context of DQMC, that may be used to resolve the degenerate
ground-state subspace without introducing signs. Instead, we will
make a physical assumption under which Result \hyperref[Result 1F]{1F}
holds. Namely, we will assume that the fine tuned constraint $\epsilon'=n/m$\textit{
and} $m\mathbb{N}\subset\mathcal{N}$ may be lifted by a sign-free
perturbation. This includes (i) perturbations to the effective single-fermion
Hamiltonian $h_{\phi\left(\tau\right)}$ that do not violate the algebraic
condition $h_{\phi\left(\tau\right)}\in\mathcal{C}_{h}$, (ii) perturbations
to the bosonic action $S_{\phi}$ that maintain its reality, and (iii)
changes of the vector $\mathbf{d}$ along which the torus is cut to
a cylinder, as described in Appendix \ref{subsec:Cutting-the-torus-2},
which will generically change the details of the boundary spectrum,
including the non-universal number $\epsilon'$ and the set $\mathcal{N}$
of $N_{x}$s where accidental degeneracies appear. A robustness of
$\epsilon'$ and the set $\mathcal{N}$, both non-universal, under
all three of the above deformations certainly goes beyond the low
energy description of a chiral TFT in the bulk and a chiral CFT on
the boundary. We also adopt this assumption in the fermionic spontaneously-chiral
setting of Result \hyperref[Result 2F]{2F}. 

A stoquastic variant of the above assumption may also be adopted to
establish the bosonic spontaneously-chiral Result \hyperref[Result 2]{2},
but a stronger statement can in fact be made, by again making use
of the Frobenius-Perron theorem to resolve the accidentally degenerate
ground states. The Frobenius-Perron theorem does not immediately complete
the derivation of Result \hyperref[Result 2]{2}, since the former is
a ground state statement, while the latter made use of the finite
temperature $\Delta E\ll\beta^{-1}\ll N_{x}^{-1}$, where $\Delta E$
is the exponentially small finite-size splitting between low lying
symmetry breaking eigenstates. This difficulty does not arise in the
'classical symmetry breaking' scenario, where $\Delta E=0$. In the
generic case $\Delta E\neq0$, we can make progress under the assumption
that the $\mathcal{T},\mathcal{P}$-even state $W\left[\ket ++\ket -\right]$
has lower energy than the $\mathcal{T},\mathcal{P}$-odd state $W\left[\ket +-\ket -\right]$,
rather than the opposite possibility. The derivation of Result \hyperref[Result 2]{2}
in Sec.\ref{subsec:Momentum-polarization-for} can then be repeated
at zero temperature. In particular, Eq.\eqref{eq:10-00} and its analysis
are unchanged. 

\section{A 'non-local design principle' for chiral topological matter\label{subsec:A-non-local-design}}

As stated in Sec.\ref{subsec:Example:-time-reversal}, the composition
$\mathsf{T}=\mathcal{P}^{\left(0\right)}\mathcal{T}^{\left(1/2\right)}$
of the spin-less reflection with the spin-1/2 time-reversal, naturally
provides a design-principle for a class of models for chiral topological
matter. Here we describe such $\mathsf{T}$-invariant models for chiral
topological superconductors.

The simplest model is comprised of two copies, labeled by $\sigma=\uparrow,\downarrow$,
of a spin-less $p+ip$ superconductor,

\begin{align}
H & =\sum_{\mathbf{x},\mathbf{x}',\sigma,\sigma'}\left[\psi_{\sigma,\mathbf{x}}^{\dagger}h_{\mathbf{x},\mathbf{x}',\sigma,\sigma'}\psi_{\sigma,\mathbf{x}'}+\psi_{\sigma,\mathbf{x}}^{\dagger}\Delta_{\mathbf{x}\mathbf{x}',\sigma,\sigma'}\psi_{\sigma,\mathbf{x}'}^{\dagger}+h.c\right].
\end{align}
Here 
\begin{align}
\Delta= & \begin{pmatrix}\Delta_{0}\left(d^{x}+id^{y}\right) & 0\\
0 & \Delta_{0}\left(d^{x}+id^{y}\right)
\end{pmatrix},
\end{align}
where $d_{\mathbf{x}\mathbf{x}'}^{x}$ ($d_{\mathbf{x}\mathbf{x}'}^{y}$)
is the anti-symmetric $x$ ($y$) difference operator, 
\begin{align}
d_{\mathbf{x}\mathbf{x}'}^{x} & =\left(\delta_{x,x'+1}-\delta_{x+1,x}\right)\delta_{y,y'}/2,\\
d_{\mathbf{x}\mathbf{x}'}^{y} & =\delta_{x,x'}\left(\delta_{y,y'+1}-\delta_{y+1,y'}\right)/2,\nonumber 
\end{align}
and $\Delta_{0}\in\mathbb{R}-\left\{ 0\right\} $. Additionally,
\begin{align}
h= & \begin{pmatrix}t & 0\\
0 & t
\end{pmatrix},
\end{align}
and the hopping $t$ is real and reflection symmetric, e.g 
\begin{align}
t_{\mathbf{x},\mathbf{x}'} & =\frac{1}{2}t_{0}\left(\delta_{x,x'+1}+\delta_{x+1,x'}\right)\delta_{y,y'}+\left(x\leftrightarrow y\right)-\mu,
\end{align}
with $t_0>0,\;\mu\in\mathbb{R}$. It is well known that the chemical potential
$\mu$ can be used to tune the model between gapped SPT phases with
$c=0,-1,1,$ for $\left|\mu\right|>2t_{0},-2t_{0}<\mu<0,0<\mu<2t_{0}$, respectively,
see e.g \citep{PhysRevB.98.064503}. Additionally, the Hamiltonian
is invariant under the combination of the unitary spin-less reflection
$\mathcal{P}^{\left(0\right)}:\psi_{\sigma,\left(x,y\right)}\mapsto\psi_{\sigma,\left(x,-y\right)}$
and the anti-unitary spin-full time-reversal $\mathcal{T}^{\left(1/2\right)}:\psi_{\uparrow,\mathbf{x}}\mapsto\psi_{\downarrow,\mathbf{x}},\;\psi_{\downarrow,\mathbf{x}}\mapsto-\psi_{\uparrow,\mathbf{x}}$.

The model can be written in the BdG form 
\begin{align}
H & =\sum_{\mathbf{x},\mathbf{x}'}\Psi_{\mathbf{x}}^{\dagger}h_{\text{BdG}}^{\mathbf{x},\mathbf{x}'}\Psi_{\mathbf{x}'},
\end{align}
where $\Psi_{\mathbf{x}}^{T}=\left(\psi_{\uparrow\mathbf{x}},\psi_{\downarrow\mathbf{x}},\psi_{\uparrow\mathbf{x}}^{\dagger},\psi_{\downarrow\mathbf{x}}^{\dagger}\right)$
is the Nambu spinor (a Majorana spinor), and 
\begin{align}
h_{\text{BdG}} & =\begin{pmatrix}h & \Delta\\
-\Delta^{*} & -h^{*}
\end{pmatrix}.\label{eq:48}
\end{align}
The ``single-fermion'' space on which $h_{\text{BdG}}$ acts
is $\mathcal{H}_{1\text{F}}=\mathcal{H}_{X}\otimes\mathcal{H}_{\text{spin}}\otimes\mathcal{H}_{\text{Nambu}}\cong\mathbb{C}^{\left|X\right|}\otimes\mathbb{C}^{2}\otimes\mathbb{C}^{2}$.
The spin-less reflection acts on $\mathcal{H}_{1\text{F}}$ as $\mathcal{P}^{\left(0\right)}=\mathcal{P}_{X}^{\left(0\right)}\otimes I_{2}\otimes I_{2}$,
where $\mathcal{P}_{X}^{\left(0\right)}=\delta_{x,x'}\delta_{y,-y'}$.
The spin-full time-reversal acts by $\mathcal{T}^{\left(1/2\right)}=I_{\left|X\right|}\otimes iY\otimes I_{2}\mathcal{K}$,
where $Y$ is the Pauli matrix and $\mathcal{K}$ is the complex conjugation.
The operator $\mathsf{T}=\mathcal{P}^{\left(0\right)}\mathcal{T}^{\left(1/2\right)}$
satisfies $\mathsf{T}^{2}=-I$ and $\left[\mathsf{T},h_{\text{BdG}}\right]=0$,
and is therefore a time-reversal design principle which applies to
$h_{\text{BdG}}$, implying $\text{det}\left(\partial_{\tau}+h_{\text{BdG}}\right)\ge0$.
 Since $h_{\text{BdG}}$ acts on the Majorana spinor $\Psi$, the
relevant quantity is actually the Pfaffian $\text{Pf}\left(\partial_{\tau}+h_{\text{BdG}}\right)=\sqrt{\text{det}\left(\partial_{\tau}+h_{\text{BdG}}\right)}\geq0$,
where the principle branch of the square root is chosen.

The Hamiltonian $h_{\text{BdG}}$ can be considerably generalized
while maintaining $\left[\mathsf{T},h_{\text{BdG}}\right]=0$, by
taking 
\begin{align}
h= & \begin{pmatrix}t & r\\
-r^{*} & t^{*}
\end{pmatrix},\\
\Delta= & \begin{pmatrix}e^{i\alpha}\left(\left|\Delta_{x}\right|d^{x}+i\left|\Delta_{y}\right|d^{y}\right)^{\left|\ell\right|} & e^{i\tilde{\alpha}}\left(\left|\tilde{\Delta}_{x}\right|d^{x}+i\left|\tilde{\Delta}_{y}\right|d^{y}\right)^{\left|\tilde{\ell}\right|}\\
-e^{-i\tilde{\alpha}}\left(\left|\tilde{\Delta}_{x}\right|d^{x}+i\left|\tilde{\Delta}_{y}\right|d^{y}\right)^{\left|\tilde{\ell}\right|} & e^{-i\alpha}\left(\left|\Delta_{x}\right|d^{x}+i\left|\Delta_{y}\right|d^{y}\right)^{\left|\ell\right|}
\end{pmatrix},\nonumber 
\end{align}
 where $t_{\mathbf{x},\mathbf{x}'},r_{\mathbf{x},\mathbf{x}'}$
are general matrices, and $\ell\in2\mathbb{Z}+1$ ($\tilde{\ell}\in2\mathbb{Z}$)
is the angular momentum channel of the triplet (singlet) pairing.
This can be further generalized to a sum over all angular momentum
channels $\sum_{\ell\in2\mathbb{Z}+1}e^{i\alpha_{\ell}}\left(\left|\Delta_{\ell,x}\right|d^{x}+i\left|\Delta_{\ell,y}\right|d^{y}\right)^{\left|\ell\right|}$
and similarly for $\tilde{\ell}$. The model is Hermitian for $t=t^{\dagger}$,
$r=-r^{T}$, but this is not required to avoid the sign problem.

In order to obtain an interacting model, the parameters $\phi=\left\{ t,r,\alpha_{\ell},\tilde{\alpha}_{\tilde{\ell}},\left|\Delta_{\ell,x}\right|,\left|\Delta_{\ell,y}\right|,\left|\tilde{\Delta}_{\tilde{\ell},x}\right|,\left|\tilde{\Delta}_{\tilde{\ell},y}\right|\right\} $
can now be promoted to space-time dependent bosonic fields, with any
action $S_{\phi}\in\mathbb{R}$. The model will be sign-free as long
as $h_{\text{BdG}}$ remains $\mathsf{T}$-invariant for all configurations
$\phi$, which requires that only reflection-even configurations $\phi\left(\tau,x,y\right)=\phi\left(\tau,x,-y\right)$
are summed over. As discussed in Sec.\ref{subsec:Example:-time-reversal},
this implies non-local interactions, which effectively fold the chiral
system into a non-chiral system of half of space.

\section{Locality and homogeneity of known design principles \label{subsec:Locality-of-known}}

In this appendix we review all fermionic design principles known to
us, clarify their common features, and describe the conditions under
which they are \textit{on-site homogeneous}, imply a \textit{term-wise
sign-free} DQMC representation, and allow a \textit{locally sign-free
DQMC} simulation, as defined in Sec.\ref{subsec:Local-and--homogeneous}.
The design principles are stated as algebraic conditions satisfied
by the effective single-fermion Hamiltonian $h_{\phi}=h_{\phi\left(\tau\right)}$
and the corresponding imaginary-time evolution $U_{\phi}=\text{TO}e^{-\int_{0}^{\beta}h_{\phi\left(\tau\right)}\text{d}\tau}$,
or in terms of the operator $D_{\phi}=\partial_{\tau}+h_{\phi}$,
see Sec.\ref{subsec:Local-determinantal-QMC}.

\paragraph*{Contraction semi-groups and Majorana time reversals}

The time-reversal design principle covered in Sec.\ref{subsec:Example:-time-reversal}
is a special case of a broad class of design principles that were
recently discovered and unified \citep{li2016majorana,wei2016majorana,wei2017semigroup}.
These are stated in terms of Majorana fermions, where $\psi$ is real
and $\overline{\psi}=\psi^{T}$, in which case $h_{\phi}$ is anti-symmetric
and the determinants in \eqref{eq:2} are replaced by their square
roots. Reference \citep{wei2017semigroup} shows that if 
\begin{align}
\mathsf{J}_{1}h_{\phi}-h_{\phi}^{*}\mathsf{J}_{1} & =0,\label{eq:J1}\\
i\left(\mathsf{J}_{2}h_{\phi}-h_{\phi}^{*}\mathsf{J}_{2}\right) & \geq0,\label{eq:J2}
\end{align}
where the matrices $\mathsf{J}_{1},\mathsf{J}_{2}$ are real and orthogonal,
and obey $\mathsf{J}_{1}^{T}=\pm\mathsf{J}_{1}$, $\mathsf{J}_{2}^{T}=-\mathsf{J}_{2}$,
$\left\{ \mathsf{J}_{1},\mathsf{J}_{2}\right\} =0$, then $\text{Det}\left(I+U_{\phi}\right)\geq0$.
The equality \eqref{eq:J1} corresponds to an anti-unitary symmetry
$\mathsf{T}_{1}=\mathsf{J}_{1}\mathcal{K},\;\mathsf{T}_{1}^{2}=\pm I$,
where $\mathcal{K}$ is the complex conjugation. If the inequality
\eqref{eq:J2} is replaced by an equality, it corresponds to an additional
anti-unitary symmetry, $\mathsf{T}_{2}=\mathsf{J}_{2}\mathcal{K}$,
$\mathsf{T}_{2}^{2}=-I$. The case $\mathsf{T}_{1}^{2}=-I$ then reduces
to the standard time-reversal $\mathsf{T}$ described in Sec.\ref{subsec:Example:-time-reversal},
while $\mathsf{T}_{1}^{2}=I$ corresponds to the 'Majorana class'
of Ref.\citep{li2016majorana}. More generally, the inequality \eqref{eq:J2}
states that the left hand side is a positive semi-definite matrix,
and implies that $h_{\phi}$ is a generator of the contraction semi-group
defined by the Hermitian metric $\eta_{2}=i\mathsf{J}_{2},\;\eta_{2}^{2}=I$,
$\left[\mathsf{T}_{1},\eta_{2}\right]=0$. Explicitly, Eq.\eqref{eq:J1}-\eqref{eq:J2}
can be written as  
\begin{align}
\left[\mathsf{T}_{1},h_{\phi}\right]=0,\;\;\eta_{2}h_{\phi}+h_{\phi}^{\dagger}\eta_{2} & \geq0,\label{eq:eta2h}
\end{align}
and imply 
\begin{align}
\left[\mathsf{T}_{1},U_{\phi}\right]=0,\;\;\eta_{2}-U_{\phi}^{\dagger}\eta_{2}U_{\phi} & \geq0.\label{eq:eta2}
\end{align}

In the language of Sec.\ref{subsec:Local-and--homogeneous}, for fixed
$\mathsf{T}_{1},\eta_{2}$, the set $\mathcal{C}_{h}$ contains all
matrices $h_{\phi}$ satisfying \eqref{eq:eta2h}. It is clear that
this set is additive: $h_{1}+h_{2}\in\mathcal{C}_{h}$ for all $h_{1},h_{2}\in\mathcal{C}_{h}$.
The set $\mathcal{C}_{U}$ contains all matrices $U_{\phi}$ satisfying
Eq.\eqref{eq:eta2}, and is multiplicative: $U_{1}U_{2}\in\mathcal{C}_{U}$
for all $U_{1},U_{2}\in\mathcal{C}_{U}$. 

A sufficient condition on $\mathsf{T}_{1},\eta_{2}$ that guarantees
that the design principle they define is on-site homogenous is that
they are of the form $\mathsf{T}_{1}=I_{\left|X\right|}\otimes\mathsf{t}_{1},\eta_{2}=I_{\left|X\right|}\otimes e_{2}$,
written in terms of the decomposition $\mathcal{H}_{1\text{F}}\cong\mathbb{C}^{\left|X\right|}\otimes\mathbb{C}^{\mathsf{d}_{\text{F}}}$
of the single-fermion space.  The permutation matrices $O^{\left(\sigma\right)}$
defined in Eq.\eqref{eq:20-0} then commute with $\eta_{2}$ and $\mathsf{T}_{1}$.
Since $O^{\left(\sigma\right)}$ are also unitary, we have $O^{\left(\sigma\right)}\in\mathcal{C}_{U}$
for all $\sigma\in S_{X}$. All examples described in Refs.\citep{li2016majorana,wei2016majorana,wei2017semigroup}
are of the on-site homogenous form $\mathsf{T}_{1}=I_{\left|X\right|}\otimes\mathsf{t}_{1},\eta_{2}=I_{\left|X\right|}\otimes e_{2}$.

As in our discussion of $\mathsf{T}$ in Sec.\ref{subsec:Example:-time-reversal},
the locality of $\mathsf{T}_{1}=I_{\left|X\right|}\otimes\mathsf{t}_{1}$
means that it can be applied term-wise, by symmetrizing the local
terms $h_{\phi;\mathbf{x}}\mapsto\frac{1}{2}\left(h_{\phi;\mathbf{x}}+\mathsf{T}_{1}h_{\phi;\mathbf{x}}\mathsf{T}_{1}^{-1}\right)$.
A similar procedure for $\eta_{2}$ is only possible if the inequality
in Eq.\eqref{eq:eta2h} holds as an equality (as in Ref.\citep{li2016majorana}).
The contraction semi-group defined by $\eta_{2}$ then reduces to
an orthogonal group, and one can enforce the term-wise relations $\eta_{2}h_{\phi;\mathbf{x}}+h_{\phi;\mathbf{x}}^{\dagger}\eta_{2}=0$
by $h_{\phi;\mathbf{x}}\mapsto\frac{1}{2}\left(h_{\phi;x}-\eta_{2}h_{\phi;x}^{\dagger}\eta_{2}\right)$.

Collecting the above, we see that if $\mathsf{T}_{1},\eta_{2}$ can
be brought to the form $\mathsf{T}_{1}=I_{\left|X\right|}\otimes\mathsf{t}_{1},\eta_{2}=I_{\left|X\right|}\otimes e_{2}$
by the same single-fermion local unitary $u$, then a DQMC representation
which is $\mathsf{T}_{1}$-symmetric, and respects Eq.\eqref{eq:eta2h}
terms-wise, leads to a locally-sign free DQMC simulation.

\paragraph*{Split orthogonal group}

Another recently discovered design principle is defined in terms of
the split orthogonal group $O\left(n,n\right)$ \citep{wang2015split}:
if $U_{\phi}\in O\left(n,n\right)$, then the sign of $\text{Det}\left(I+U_{\phi}\right)$
depends only on the connected component of $O\left(n,n\right)$ to
which $U_{\phi}$ belongs. If the sign of $e^{-S_{\phi}}$ is manifestly
compatible with the connected component of $U_{\phi}$ in $O\left(n,n\right)$,
one has $p\left(\phi\right)=e^{-S_{\phi}}\text{Det}\left(I+U_{\phi}\right)\geq0$.
More explicitly, the statement $U_{\phi}\in O\left(n,n\right)$ implies
that $U_{\phi}$ is a real matrix and $\eta-U_{\phi}^{T}\eta U_{\phi}=0$,
where $\eta=\text{diag}\left(I_{n},-I_{n}\right)$. Restricting to
the identity component $O_{0}\left(n,n\right)$, this amounts to the
statements that $h_{\phi}$ is in the Lie algebra $o\left(n,n\right)$:
it is real and satisfies $\eta h_{\phi}+h_{\phi}^{T}\eta=0$.

In a basis independent formulation, the data that defines the design
principle is an anti-unitary $\tilde{\mathsf{T}}$, such that $\tilde{\mathsf{T}}^{2}=I$,
and a Hermitian metric $\tilde{\eta}$ with canonical form $\eta$,
such that $\left[\tilde{\mathsf{T}},\tilde{\eta}\right]=0$. The
set $\mathcal{C}_{h}$ is then given by matrices $h_{\phi}$ satisfying
\begin{align}
\left[\tilde{\mathsf{T}},h_{\phi}\right]=0,\;\; & \tilde{\eta}h_{\phi}+h_{\phi}^{\dagger}\tilde{\eta}=0,
\end{align}
while $\mathcal{C}_{U}$ is defined by 
\begin{align}
\left[\tilde{\mathsf{T}},U_{\phi}\right]=0,\;\; & \tilde{\eta}-U_{\phi}^{\dagger}\tilde{\eta}U_{\phi}=0.
\end{align}
The analogy with \eqref{eq:eta2h}-\eqref{eq:eta2} is now manifest,
with the inequalities strengthened to equalities. Accordingly, the
$O\left(n,n\right)$ design-principle is on-site homogeneous if $\tilde{\mathsf{T}}=I_{\left|X\right|}\otimes\tilde{\mathsf{t}}$
and $\tilde{\eta}=I_{\left|X\right|}\otimes\tilde{e}$. If these forms
can be obtained by conjugation of $\tilde{\mathsf{T}},\tilde{\eta}$
with the same single-fermion local unitary $u$, then a DQMC representation
which is sign-free due to $\tilde{\mathsf{T}},\tilde{\eta}$, leads
to a locally-sign free DQMC simulation. 

The above statements hold for $U_{\phi}$ in the identity component
$O_{0}\left(n,n\right)$, which is always the case when $h_{\phi}\in o\left(n,n\right)$
and $U_{\phi}=\text{TO}e^{-\int_{0}^{\beta}h_{\phi\left(\tau\right)}\text{d}\tau}$.
Time evolutions in the additional three connected components of $O\left(n,n\right)$
can be obtained by operator insertions generalizing $U_{\phi}=U_{k}\cdots U_{2}U_{1}$
to $U_{k}\cdots O_{2}U_{2}O_{1}U_{1}$, where $O\in O\left(n,n\right)/O_{0}\left(n,n\right)$
\citep{wang2015split}. These can be incorporated into the framework
of Sec.\ref{sec:Determinantal-quantum-Monte}, if each $O_{k}$ is
supported on a disk of radius $w$ around a site $\mathbf{x}_{k}$,
i.e $\left(O_{k}\right)_{\mathbf{x},\mathbf{y}}=\delta_{\mathbf{x},\mathbf{y}}$
if $\left|\mathbf{x}-\mathbf{x}_{k}\right|>w$ or $\left|\mathbf{y}-\mathbf{x}_{k}\right|>w$.
With this generalization, all sign-free examples described in Ref.\citep{wang2015split}
amount to locally sign-free DQMC.

\paragraph*{Solvable fermionic and bosonic actions}

Reference \citep{chandrasekharan2013fermion} described a design principle
that nontrivially relates the fermionic action $S_{\psi,\phi}=\overline{\psi}D_{\phi}\psi$
and bosonic action $S_{\phi}$. A fermionic action was termed 'solvable'
if $D_{\phi}$ has the form
\begin{align}
 & D_{\phi}=\begin{pmatrix}0 & M_{\phi}\\
-M_{\phi}^{\dagger} & 0
\end{pmatrix},\label{eq:58}
\end{align}
which clearly implies $\text{Det}\left(D_{\phi}\right)=\left|\text{Det}\left(M_{\phi}\right)\right|^{2}\geq0$.
Here the imaginary time circle $\mathbb{R}/\beta\mathbb{Z}$ is discretized
to $\mathbb{Z}_{\beta}=\mathbb{Z}/\beta\mathbb{Z}$, and $D_{\phi}$
is treated as a matrix on $\mathbb{C}^{\beta}\times\mathcal{H}_{1\text{F}}=\mathbb{C}^{\beta}\times\mathbb{C}^{\left|X\right|}\times\mathbb{C}^{\mathsf{d}_{F}}$,
with indices $\left(\tau,\mathbf{x},\alpha\right),\left(\tau',\mathbf{x}',\alpha'\right)$
for time, space, and internal degrees of freedom. For example, the
Hamiltonian form $D_{\phi}=\partial_{\tau}+h_{\phi\left(\tau\right)}$
is discretized to

\begin{align}
\left[D_{\phi}\right]_{\left(\tau,\mathbf{x},\alpha\right),\left(\tau',\mathbf{x}',\alpha'\right)}= & \left(\delta_{\tau,\tau'}-\delta_{\tau-1,\tau'}\right)\delta_{\mathbf{x},\mathbf{x}'}\delta_{\alpha,\alpha'}+\delta_{\tau-1,\tau'}\left[h_{\phi\left(\tau\right)}\right]_{\left(\mathbf{x},\alpha\right),\left(\mathbf{x}',\alpha'\right)}.\label{eq:52}
\end{align}

In a basis independent language, Eq.\eqref{eq:58} corresponds to
\begin{align}
 & \left\{ \Gamma,D_{\phi}\right\} =0,\;\;D_{\phi}^{\dagger}=-D_{\phi},\label{eq:53}
\end{align}
where $\Gamma$ is a 'chiral symmetry', $\Gamma^{2}=I,\;\Gamma=\Gamma^{\dagger}$.
Eq.\eqref{eq:16} is then obtained in a basis where  $\Gamma=\text{diag}\left(I,-I\right)$.
Note however that $\Gamma$ acts on $D_{\phi}$ rather than $h_{\phi}$,
and that the form \eqref{eq:58} requires a non-canonical transformation
away from the Hamiltonian form \eqref{eq:52}. We refer to $\Gamma$
as on-site homogeneous if it is of the form $\Gamma=I_{\beta}\otimes I_{\left|X\right|}\otimes\gamma$
, and to $D_{\phi}$ as local if $D_{\phi}=\sum_{\tau,\mathbf{x}}D_{\phi;\tau,\mathbf{x}}$
where each term $D_{\phi;\tau,\mathbf{x}}$ is supported on a disk
of radius $r$ around $\left(\tau,\mathbf{x}\right)$, and depends
on the values of $\phi$ at points within this disk. The action $D_{\phi}$
is 'term-wise solvable' if each $D_{\phi;\tau,\mathbf{x}}$ satisfies
\eqref{eq:53}. Any local $D_{\phi}$ obeying \eqref{eq:53} with
$\Gamma=I_{\beta}\otimes I_{\left|X\right|}\otimes\gamma$ can be
made term-wise solvable by replacing $D_{\phi;\tau,\mathbf{x}}\mapsto\frac{1}{2}\left(D_{\phi;\tau,\mathbf{x}}-\Gamma D_{\phi;\tau,\mathbf{x}}\Gamma\right)$
and then $D_{\phi;\tau,\mathbf{x}}\mapsto\frac{1}{2}\left(D_{\phi;\tau,\mathbf{x}}-D_{\phi;\tau,\mathbf{x}}^{\dagger}\right)$.
The twisted fermionic boundary conditions in \eqref{eq:5-1-0} are implemented
by declaring that the index '$\left(\tau=0,\mathbf{x},\alpha\right)$'
that appears in Eq.\eqref{eq:52} corresponds to $\left(\tau=\beta,x+\lambda\Theta\left(y\right),y,\alpha\right)$
with $\lambda\neq0$. Equation \eqref{eq:53} then holds for all $\lambda$
if $\Gamma=I_{\beta}\otimes I_{\left|X\right|}\otimes\gamma$. Under
these conditions, solvable fermionic actions can then be incorporated
into the definition of locally sign-free DQMC given in Sec.\ref{sec:Determinantal-quantum-Monte}.

All examples given in Ref.\citep{chandrasekharan2013fermion} have
an on-site $\Gamma$ and local $D_{\phi}$, and it follows from the
above discussion that, under these conditions, solvable fermionic
actions can then be incorporated into the definition of locally sign-free
DQMC given in Sec.\ref{sec:Determinantal-quantum-Monte}. 

A bosonic action $S_{\phi}$ for a complex valued field $\phi=\left|\phi\right|e^{i\theta}$
was termed 'solvable' in Ref.\citep{chandrasekharan2013fermion} if
\begin{align}
 & S_{\phi}=S_{\left|\phi\right|}-\sum_{u,u'}\beta_{u,u'}\left|\phi_{u}\right|\left|\phi_{u'}\right|\cos\left(\varepsilon_{u}\theta_{u}+\varepsilon_{u'}\theta_{u'}\right),
\end{align}
where $u=\left(\mathbf{x},\tau\right)$, $u=\left(\mathbf{x}',\tau'\right)$,
and $\varepsilon_{u},\varepsilon_{u'}\in\left\{ \pm1\right\} $, and
$\beta_{u,u'}\geq0$. For such actions, it was shown that all correlators
$\int D\phi e^{-S_{\phi}}\phi_{u_{1}}\cdots\phi_{u_{k}}$ are non-negative,
and therefore $\phi$ can be added to the diagonal in \eqref{eq:58}
with a positive coupling constant $g>0$, $D_{\phi}=\begin{pmatrix}g\phi & M\\
-M^{\dagger} & g\phi
\end{pmatrix}$, without introducing signs, though $D_{\phi}$ is no longer solvable.
Solvable bosonic actions are easily incorporated into the framework
of Sec.\ref{sec:Determinantal-quantum-Monte}, as long as they are
local in the sense of Eq.\eqref{eq:16}, and in particular, $\beta_{u,u'}=0$
unless the points $u$ and $u'$ are close.

\bibliographystyle{unsrt}
\bibliography{Thesis.bbl}

\end{document}